\documentclass[aps,prx,nofootinbib,twocolumn,superscriptaddress,longbibliography]{revtex4-1}

\usepackage{physics}
\usepackage{dcolumn}
\usepackage{graphicx}
\usepackage{indentfirst}
\usepackage{lipsum}
\usepackage{braket}
\usepackage{float}
\usepackage{amsmath}
\usepackage{amssymb}
\usepackage{amsfonts}
\usepackage{makecell}
\usepackage{epstopdf}
\usepackage{CJK}
\usepackage{esint}
\usepackage{color}
\usepackage[T1]{fontenc}
\usepackage{amsfonts}
\usepackage{footmisc}
\usepackage{scrextend}
\usepackage{verbatim}
\usepackage{multirow}
 \usepackage{mathtools}
 \usepackage{bbm}
 \usepackage{array}
 \usepackage{booktabs}
 \usepackage{float}
\usepackage[hyperfootnotes=false]{hyperref}
\usepackage[english]{babel}
\usepackage{url}
\usepackage{bm}
\usepackage{hyperref}

\definecolor{darkblue}{rgb}{0,0,0.5}
\hypersetup{
colorlinks=true,
linkcolor=black,
filecolor=blue,
citecolor=darkblue,  
urlcolor=black,
}

\newlength{\fighskip} \fighskip=2pt
\newlength{\figvskip} \figvskip=3pt

\newcommand*{\figbox}[2]{{
  \def\figscale{#1}
  \def\arraystretch{0.8}
  \arraycolsep=0pt
  \begin{array}{c}
    \vbox{\vskip\figscale\figvskip
      \hbox{\hskip\figscale\fighskip
        \includegraphics[scale=\figscale]{#2}}}
  \end{array}}}
  
\usepackage{enumitem,kantlipsum}
\usepackage{refcount}

\newcommand{\D}{\Delta}
\newcommand{\mB}{\mathcal{B}}

\newcommand{\f}{\frac}

\newcommand{\vep}{\varepsilon}
\newcommand{\constd}{\eta_d}

\newcommand{\be}{\begin{equation}}
\newcommand{\ee}{\end{equation}}
\newcommand{\bs}{\begin{split}}
\newcommand{\es}{\end{split}}
\newcommand{\fepr}{\text{FEPR}}
\newcommand{\epr}{\text{EPR}}
\newcommand{\tfd}{\text{TFD}}

\newcommand{\SWAP}{\text{SWAP}}
\newcommand{\Size}{\mathcal{S}}
\newcommand{\String}{R}

\newcommand{\CZ}{\text{CZ}}

\def\:={\,\raisebox{0.85pt}{.}\hspace{-2.78pt}\raisebox{2.85pt}{.}\!\!=\,}
\def\=:{\,=\!\!\raisebox{0.85pt}{.}\hspace{-2.78pt}\raisebox{2.85pt}{.}\,}

\usepackage[toc,page]{appendix}

\newcommand{\TableRegimes}{
\begin{table*}\label{table:comparison}
\renewcommand{\tabcolsep}{10pt}
{\centering
\begin{tabular}{@{\hspace{0mm}}l|cccc}
\toprule
\makecell[c]{\textbf{Model}} & \makecell{\textbf{Teleportation}\\ \textbf{mechanism}} & \makecell{\textbf{Protocol}\\ \textbf{parameters}} & \makecell{\textbf{Maximum per}\\ \textbf{qubit fidelity}} & 
\makecell{\textbf{Channel}\\ \textbf{capacity}} \\
\midrule \\ [-5pt]

\makecell[l]{All scrambling systems at late times\\ (Refs.~\cite{yoshida2017efficient,maldacena2017diving}, Section~\ref{late times})} & peaked-size & $g = \pi \text{ mod } 2\pi$ & $\sim G_\beta$  & $1$ qubit \\ [15pt]

\makecell[l]{$\geq1$D RUCs \& chaotic spin systems \\ (Sections~\ref{finite temperature},\ref{geq1D},\ref{sec:rydberg})} & peaked-size & $\constd g \Size(t) / N = \pi \text{ mod } 2\pi$ & $\sim G_\beta$ & $\sim K$ qubits \\ [15pt]

\makecell[l]{0D RUCs,  with encoding \\ (Section~\ref{0D RUCs})} & peaked-size    & $\constd g \Size(t) / N = \pi \text{ mod } 2\pi$ & $\sim 1$   & $\sim K$ qubits \\ [15pt]

\makecell[l]{High-$T$ SYK,  with encoding \\ (Ref.~\cite{gao2019traversable}, Sections~\ref{sec: SYK infinite temperature})} & peaked-size       & $\constd g \Size(t) / N = \pi \text{ mod } 2\pi$ & $\sim 1$   & $\sim K$ qubits \\ [15pt]

\makecell[l]{Low-$T$ SYK / AdS$_2$ gravity  \\ (Refs.~\cite{gao2017traversable,maldacena2017diving,gao2019traversable,brown2019quantum}, Fig.~\ref{fig: 1})} & gravitational     & \makecell{$g e^t / N \sim 1$ (SYK) \\ $g G_N e^t \sim 1 $ (AdS$_2$)}& $\sim 1$        & $\sim K$ qubits  \\ [15pt]

\makecell[l]{AdS$_2$ gravity with strong stringy \\ corrections, with encoding \\ (Section~\ref{sec:stringy})} &  peaked-size & $g \Size(t) / N \sim \pi \text{ mod } 2\pi $ & $\sim G_\beta$  & --- \\ [10pt]

\bottomrule
\end{tabular}}
\caption{Summary of our expectations for teleportation in a variety of physical models. For each model, we specify the associated teleportation mechanism, the optimal value of the coupling strength $g$, the optimal teleportation fidelity, and the channel capacity. 
Here $G_\beta$ is the imaginary time two-point function (Section~\ref{finite temperature}), $\Size(t)$ is the size of a time-evolved operator, $K$ is the number of measured qubits [Fig.~\ref{fig: 1}(a)], $\constd = 1/(1-1/d^2)$ is an order one constant determined by the local qudit dimension $d$ [Sec.~\ref{Coupling size}], and $G_N$ is Newton's constant. We refer to the Summary of Results and the cited sections for further details.}
\end{table*}}

\newcommand{\FigureOne}{
\begin{figure*}
\centering
\includegraphics[width=\textwidth]{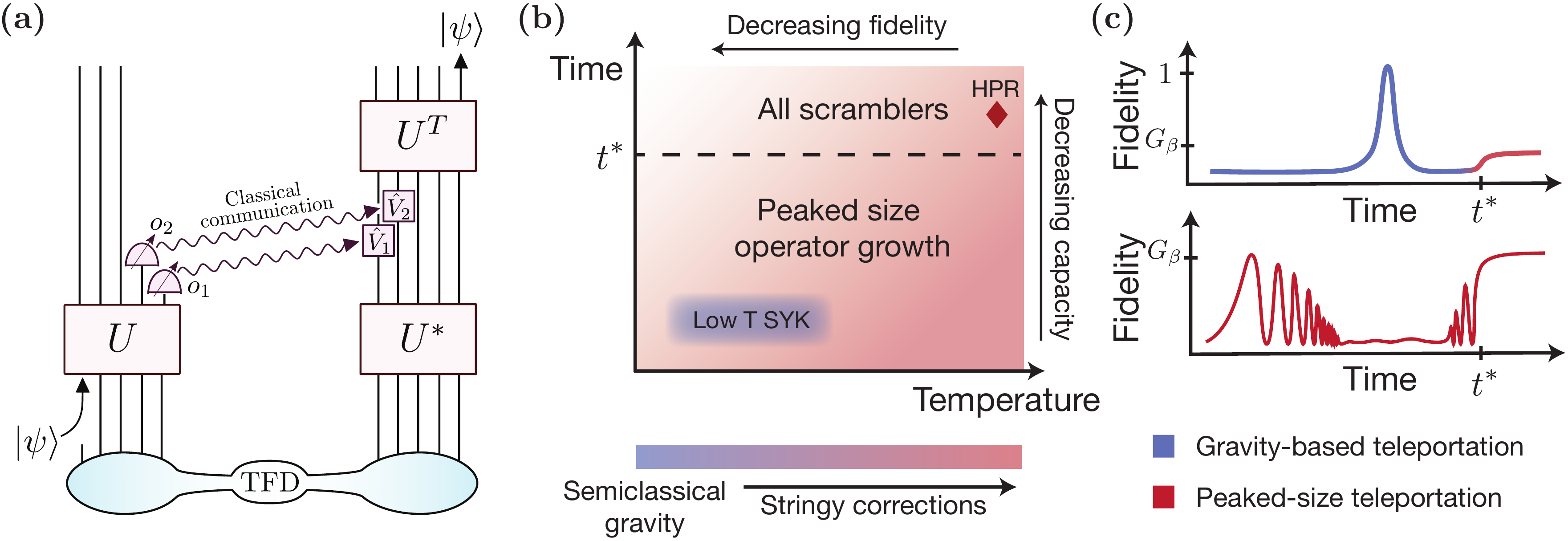}
\caption{
\textbf{(a)} Teleportation protocol, proceeding from bottom to top. 
To teleport, a subset of the left qubits are measured in the $\hat{O}_i$ basis, and operations $\hat{V}_i = e^{i g o_i \hat{O}_i / K}$ conditioned on the measurement results $o_i$ are performed on the right (purple).
\textbf{(b)} The protocol hosts two mechanisms of teleportation: peaked-size (red) and gravitational (blue).
The channel capacity of peaked-size teleportation decreases with increasing time (dark to light red), while its fidelity decreases with decreasing temperature (dark to light red, again).
At high temperature and late times, it is equivalent to teleportation in the HPR protocol (red diamond).
Gravitational teleportation occurs at low temperatures in systems dual to semiclassical gravity (e.g. the SYK model), and exhibits the same channel capacity but \emph{higher} fidelity compared to peaked-size teleportation.
Increasing the strength of stringy corrections to the gravity theory interpolates between gravitational and peaked-size teleportation.
\textbf{(c)} The two mechanisms display distinct time profiles for the teleportation fidelity at fixed coupling strength, $g$.
In systems dual to gravity (top), the fidelity features a single $\mathcal{O}(1)$ peak near the scrambling time (gravitational, blue), and a late time revival (peaked-size, red) to a fidelity suppressed by the two-point function $G_\beta$~\cite{maldacena2017diving}.
In generic thermalizing systems (bottom), the fidelity oscillates between 0 and $G_\beta$ with phase proportional to the operator size, may subsequently decay if sizes become not peaked, and revives at late times.
} 
\label{fig: 1}
\end{figure*}
}

\newcommand{\FigureRUC}{
\begin{figure*}
\centering
\includegraphics[width=1\textwidth]{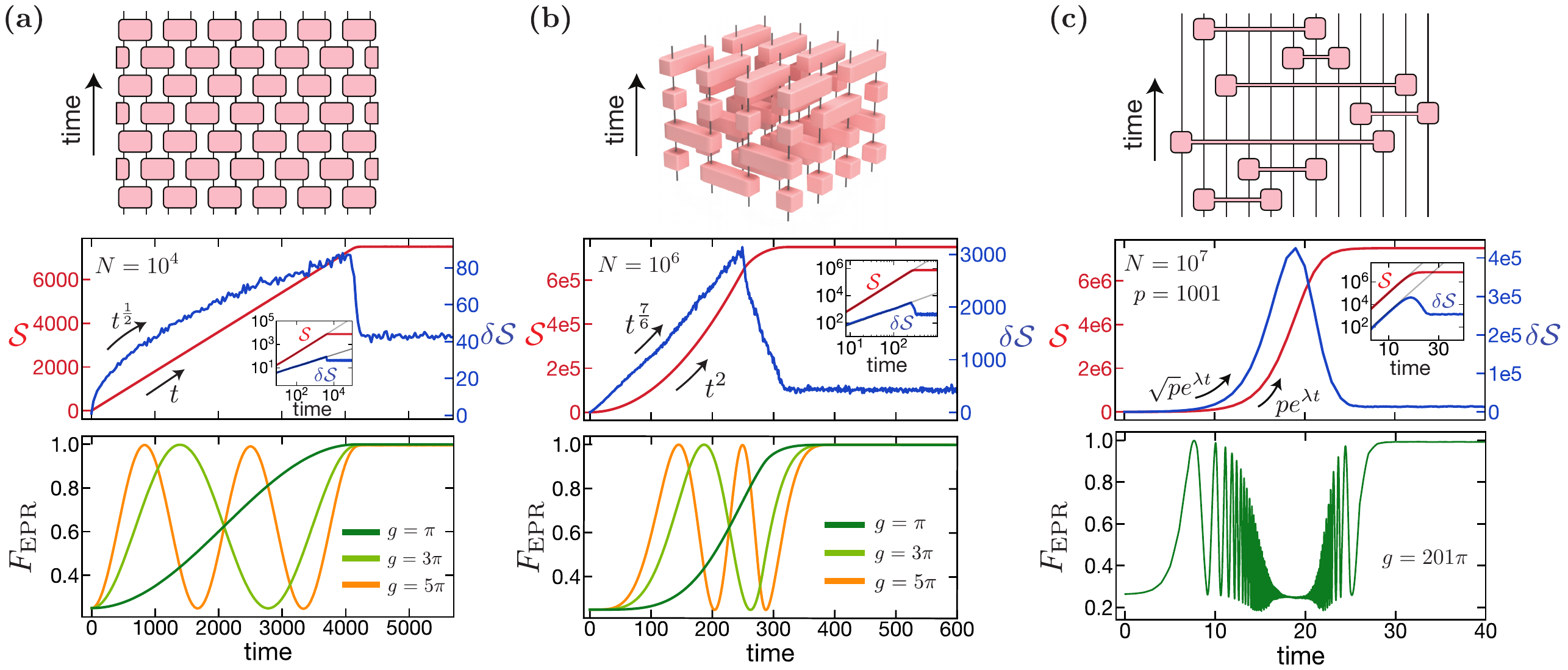}
\caption{
Numerical results for averaged operator size and teleportation fidelity of 1D, 2D, and 0D RUCs. 
\textbf{(a-b)} In 1D and 2D, sizes grow ballistically in time, while the size width grows with a slower power of $t$ and matches predictions from the KPZ universality class (Section \ref{geq1D}).
Because of the separation between the size and size width, the teleportation fidelity for a single qubit exhibits an oscillatory behavior at intermediate times, with nearly perfect maximum fidelity. 
At late times, the teleportation fidelity saturates close to $1$ for odd values of $g/\pi$, as expected for any scrambling system (Section \ref{late times}). 
\textbf{(c)} In 0D all-to-all coupled RUCs, both the size and size width grow exponentially in time and obtaining a large separation between them requires encoding the initial state into $p$-body operators. With this encoding, the teleportation fidelity displays a distinct three-regime profile for $g \gg 1$. 
In particular, as in 1D and 2D, peaked-size teleportation succeeds ($i$) at early times, with an oscillating fidelity, and ($ii$) at late times, where the fidelity saturates close to $1$ (for odd $g/\pi$). 
Between these regimes, no teleportation occurs because the size width has grown too large, $g \delta \Size / N \gtrsim 1$.
} 
\label{fig: RUC}
\end{figure*}
}

\newcommand{\FigureSizeWidth}{
\begin{figure}
\centering
\includegraphics[width=0.7\columnwidth]{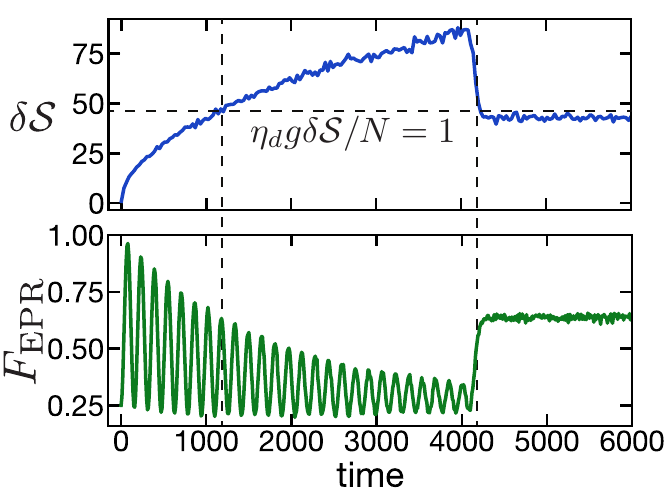}
\caption{
Probing operator size width in a 1D RUC. (top) The size width initially grows as $t^{1/2}$ and reaches a peak at the scrambling time $t^* \sim N = 10000$. (bottom) We probe this behavior by measuring the teleportation fidelity of a single qubit with a large coupling $g = 57\pi \sim \sqrt{N}$. The fidelity exhibits a distinct decay-revival profile, controlled by whether the size width has exceeded the threshold $g \delta \mathcal{S} / N \approx 1$: nearly perfect fidelity initially, power law decay towards a trivial fidelity at intermediate times, and partial revival at late times.
}
\label{fig: size width}
\end{figure}
}

\newcommand{\FigureCapacity}{
\begin{figure}
\centering
\includegraphics[width=0.35\textwidth]{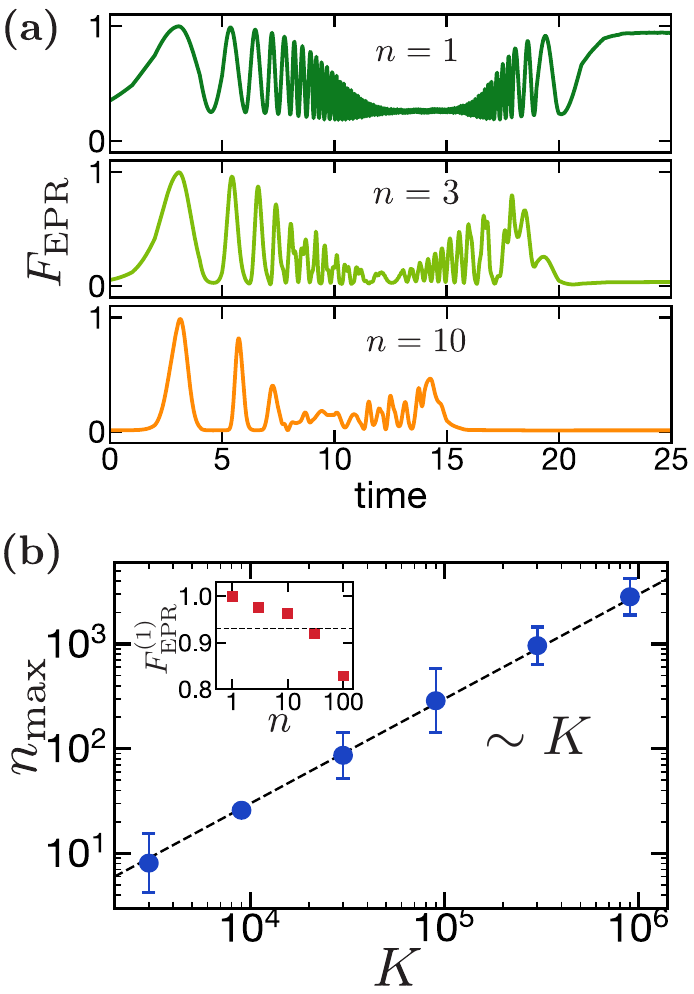}
\caption{
Teleportation of multiple qubits in 0D RUCs.
\textbf{(a)} Many-body teleportation fidelity, $F_\textrm{EPR}$, as a function of time for teleporting $n = 1, 3, 10$ qubits with fixed coupling strength ($g=177\pi$). Compared to a single qubit, the decay-revival profile for multiple qubits is shifted to earlier times, since multi-qubit operators both have a larger size width and saturate the system size earlier. Moreover, multi-qubit teleportation is not possible at late times, resulting in a trivial late-time fidelity (Sec.~\ref{sec: late-times}).
\textbf{(b)} Numerical results for the channel capacity $n_{\text{max}}$ as function of the number of coupled qubits $K$, which exhibit a clear linear scaling. 
To determine the channel capacity, we compute the maximum \emph{per qubit} fidelity $F^{(1)}_\textrm{EPR}$ for a fixed number of qubits, $n$, and couplings, $K$, while allowing the coupling strength, $g$, and evolution time to vary.
For fixed $K$, $F^{(1)}_\textrm{EPR}$ decreases  as the number of qubits $n$ is increased, as depicted in the inset for $K=9000$.
The channel capacity $n_{\text{max}}$ is defined as the maximum number of qubits for which the fidelity is above a fixed threshold (dashed line).
} 
\label{fig: capacity}
\end{figure}
}

\newcommand{\FigureOneSided}{
\begin{figure}
\centering
\includegraphics[width=\columnwidth]{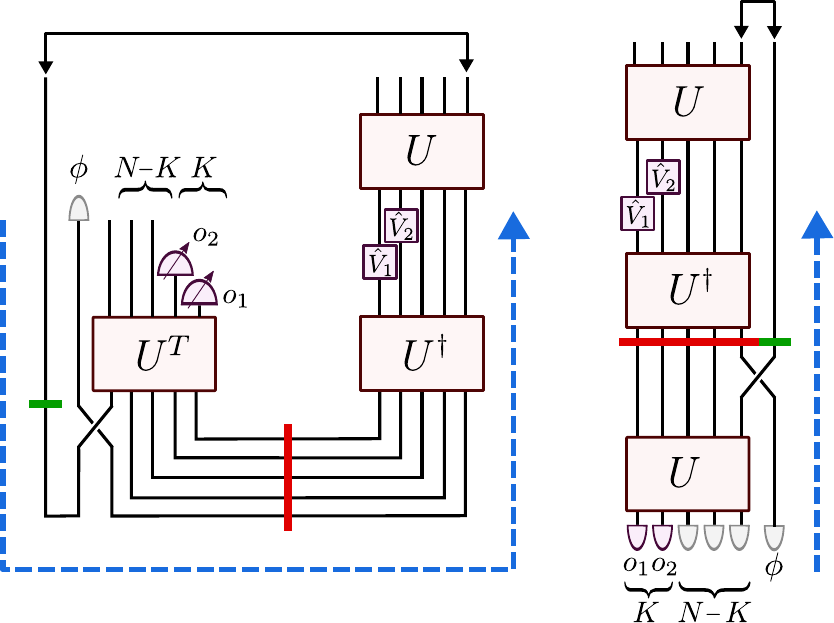}
\caption{
One-sided implementation (right) of the original two-sided teleportation protocol (left), derived using repeated applications of Eq.~(\ref{eq: O slide}) [replacing $U \rightarrow U^T$ for convenience, compared to Fig.~\ref{fig: 1}(a)]. 
Blue arrows denote the sequence of operations in the one-sided protocol, the green band marks the teleported qubit and its corresponding component in the one-sided protocol, and the red band marks the initial EPR state and its corresponding component.
}
\label{fig:one-sided}
\end{figure}
}

\newcommand{\FigureIons}{
\begin{figure*}[t]
\centering
\includegraphics[width=\textwidth]{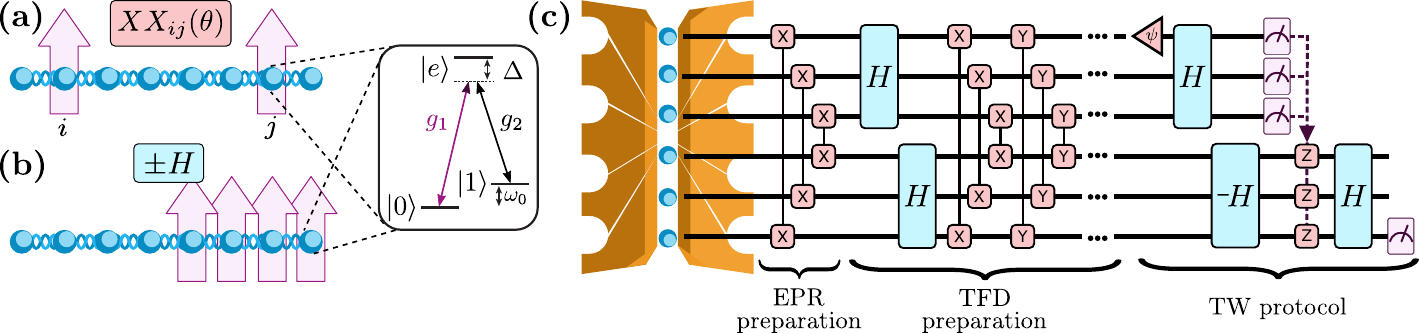}
\caption{
\textbf{(a-b)} Chain of atomic ions, with qubit states $\ket{0}$, $\ket{1}$ represented by hyperfine ground states.
The states are coupled by a pair of laser beams, one with individual addressing (with strength $g_1$, purple) and one applied globally (with strength $g_2$). 
Each beam is strongly detuned from an excited state $\ket{e}$ by an amount $\Delta$. 
The coherent beatnote between the beams, at frequency $\omega_0$, drives stimulated Raman transitions between the qubit levels with an effective Rabi frequency $g_1 g_2/2\Delta$, and also modulates the Coulomb interaction between qubits to give rise to an effective Ising interaction.  
\textbf{(a)} A two-qubit entangling gate, $XX_{ij}(\theta)$, (red) is performed by addressing only ions $i$ and $j$ with the first beam.
\textbf{(b)} Half of the qubits are addressed, which leads to analog time-evolution under the Hamiltonian Eq.~(\ref{trapped ion H}) (blue) for all addressed spins.
\textbf{(c)} Quantum circuit implementation of the teleportation protocol at finite temperature.
EPR pairs are formed using two-qubit gates.
The TFD state is then prepared via a QAOA approach by iterating multiple times between two-qubit gates coupling the sides and analog time-evolution on both sides individually~\cite{wu2019variational,zhu2019TFD}.
The state $\ket{\psi}$ is inserted either by projectively measuring the designated qubit and preparing the state, or by digitally swapping in an additional qubit (not shown).
Finally, teleportation is implemented using similar ingredients as well as feed-forward measurements (purple dotted lines).}
\label{fig:ions}
\end{figure*}
}

\newcommand{\FigureRydberg}{
\begin{figure*}[t]
\centering
\includegraphics[width=0.8\textwidth]{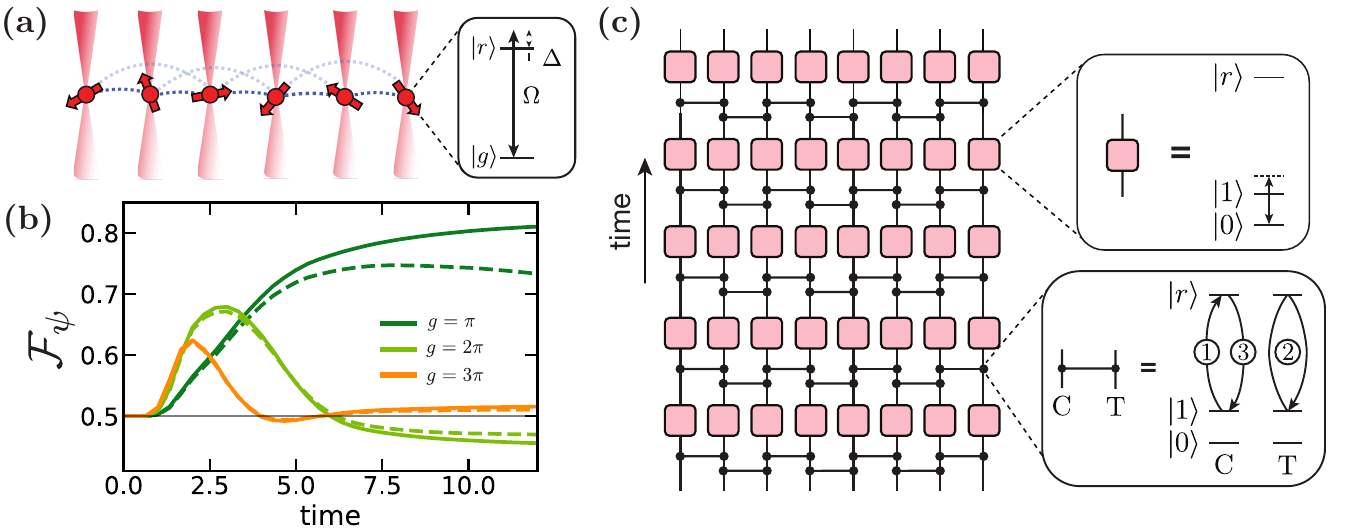}
\caption{
\textbf{(a)} In the proposed analog Rydberg teleportation protocol, qubits are encoded in a ground state $|g\rangle$ and a Rydberg state $|r\rangle$.
Nearest-neighbor interactions (dark blue) can be time-reversed, but next-nearest neighbor interactions (light blue) cannot. 
\textbf{(b)} Numerical results comparing the average state teleportation fidelity for single-qubit teleportation with perfectly reversed time-evolution (solid) with the proposed, imperfect time-reversal (dashed). In particular, we implement the one-sided protocol using $N = 20$ total spins; $K = N-1$ `measured' spins (i.e.~all except the spin encoding $\ket \psi$), whose single-qubit rotations are generated by $\hat O_i = \hat Z_i$; and time evolution under the analog Rydberg Hamiltonian [Eq.\eqref{eq:ryd-h}] with parameters $\Omega_i = .9$, $\Delta_i = -1.5$, $J_0 = 1$ (for all $i$).
\textbf{(c)} Implementation of $U$ or $U^\dagger$ in the digital protocol, consisting of alternating layers of controlled-phase gates (horizontal black lines) between nearest neighbor atoms and single-qubit rotations (red boxes).
Here, qubits are encoded in two hyperfine ground states. Insets show possible pulse sequences to implement the controlled-phase gate and the single-qubit rotations~\cite{Jaksch00}.
The full TW protocol is obtained by inserting this gate sequence (and its Hermitian conjugate) in place of $U$, $U^\dagger$ in Fig. 5.}
\label{fig:rydberg-implementation}
\end{figure*}
}

\begin{document}
\title{Many-body quantum teleportation via operator spreading in the traversable wormhole protocol}



\author{Thomas Schuster}
\thanks{These authors contributed equally to this work.}
\affiliation{Department of Physics, University of California, Berkeley, California 94720 USA}
\author{Bryce Kobrin}
\thanks{These authors contributed equally to this work.}
\affiliation{Department of Physics, University of California, Berkeley, California 94720 USA}
\affiliation{Materials Sciences Division, Lawrence Berkeley National Laboratory, Berkeley, CA 94720, USA}
\author{Ping Gao}
\affiliation{Center for Theoretical Physics, Massachusetts Institute of Technology,
Cambridge, MA 02139, USA}
\author{Iris Cong}
\affiliation{Department of Physics, Harvard University, Cambridge, Massachusetts 02138, USA}
\author{Emil T. Khabiboulline}
\affiliation{Department of Physics, Harvard University, Cambridge, Massachusetts 02138, USA}
\author{Norbert M. Linke}
\affiliation{Joint Quantum Institute and Department of Physics, University of Maryland, College Park, MD 20742 USA}
\author{Mikhail D. Lukin}
\affiliation{Department of Physics, Harvard University, Cambridge, Massachusetts 02138, USA}
\author{Christopher Monroe}
\affiliation{Joint Quantum Institute and Department of Physics, University of Maryland, College Park, MD 20742 USA}
\author{Beni Yoshida}
\affiliation{Perimeter Institute for Theoretical Physics, Waterloo, Ontario N2L 2Y5, Canada}
\author{Norman Y. Yao}
\affiliation{Department of Physics, University of California, Berkeley, California 94720 USA}
\affiliation{Materials Sciences Division, Lawrence Berkeley National Laboratory, Berkeley, CA 94720, USA}

\date{\today}

\begin{abstract}

By leveraging shared entanglement between a pair of qubits, one can teleport a quantum state from one particle to another. 
Recent advances have uncovered an intrinsically \emph{many-body} generalization of quantum teleportation, with an elegant and surprising connection to gravity. 
In particular, the teleportation of quantum information relies on many-body dynamics, which originate from strongly-interacting systems that are  holographically dual to gravity; 
from the gravitational perspective, such quantum teleportation can be understood as the transmission of information through a traversable wormhole.
Here, we propose and analyze a new mechanism for many-body quantum teleportation---dubbed peaked-size teleportation.
Intriguingly, peaked-size teleportation utilizes precisely the same type of quantum circuit as traversable wormhole teleportation, yet 
has a completely distinct microscopic origin: it relies upon the spreading of local operators under generic thermalizing dynamics and not gravitational physics.
We demonstrate the ubiquity of peaked-size teleportation, both analytically and numerically,  across a diverse landscape of physical systems, including random unitary circuits, the Sachdev-Ye-Kitaev model (at high temperatures), one-dimensional spin chains and a bulk theory of gravity with stringy corrections.
Our results pave the way towards using many-body quantum teleportation as a powerful experimental tool for: ($i$) characterizing the size distributions of operators in strongly-correlated systems and ($ii$) distinguishing between generic and intrinsically gravitational scrambling dynamics.
To this end, we provide a detailed experimental blueprint for realizing many-body quantum teleportation in both trapped ions and Rydberg atom arrays; effects of decoherence and experimental imperfections are analyzed.

\end{abstract}
\maketitle 



\FigureOne

Quantum teleportation leverages entanglement to transmit quantum information between distant locations~\cite{bennett1993teleporting,barrett2004deterministic,riebe2004deterministic,Olmschenk2009,ren2017ground}.
Typically, one thinks about teleportation in the context of a few, well-controlled degrees of freedom.
For example, two distant observers might share a pair of maximally entangled qubits (i.e.~an Einstein-Podolski-Rosen (EPR) pair~\cite{nielsen2002quantum}), enabling a measurement by one observer to teleport an unknown quantum state to the other.  

Recently, a confluence of seminal results has unveiled several novel instances of teleportation in strongly-interacting, \emph{many-body} systems~\cite{gao2017traversable,maldacena2017diving,bao2018traversable,maldacena2018eternal,yoshida2017efficient,yoshida2019disentangling,landsman2019verified,blok2020quantum,brown2019quantum,brown2020quantum,gao2019traversable}. 
Similar to conventional quantum teleportation, these protocols utilize shared entanglement  as well as measurement and classical communication.
However, they differ from conventional quantum teleportation in a few key aspects.
%
%
Most notably, prior to teleportation, the initial quantum state is \emph{scrambled} by the application of a many-body unitary.
%
At first glance, this coexistence of scrambling---broadly speaking, the increasing complexity of initially simple quantum information under many-body time dynamics~\cite{sekino2008fast,shenker2014black,roberts2015localized,maldacena2016bound,hosur2016chaos}---and teleportation might seem counterintuitive.
Indeed, one often thinks of teleportation as a directed quantum channel moving information between two specific locations; in contrast, scrambling disperses quantum information across all of the degrees of freedom in a system. 
The most natural way to reconcile these two perspectives is through the language of quantum error correction~\cite{hayden2007black}: by encoding, via scrambling, one observer's local information into non-local correlations across a many-body system, one can in fact teleport this information with access only to any few of the system's qubits.

The most notable example of many-body teleportation is the so-called traversable wormhole (TW) protocol, discovered in the context of quantum gravity~\cite{gao2017traversable,maldacena2017diving,bao2018traversable,brown2019quantum,brown2020quantum,gao2019traversable}.
From the bulk gravitational perspective, this protocol consists of a particle traveling from one side of a wormhole geometry to the other; the wormhole is  rendered traversable by the application of a coupling between the two sides. 
In the boundary theory, the wormhole geometry corresponds to a highly entangled thermofield double (TFD) state shared between two copies of a many-body system, and the coupling is implemented via measurement and feed-forward operations [Fig.~\ref{fig: 1}(a)]. 
Crucially, for this bulk-boundary correspondence to hold, the Hamiltonian describing the boundary system must exhibit ``coherent'', gravitational scrambling dynamics---this is realized, most notably, in the Sachdev-Ye-Kitaev (SYK) model at low temperatures~\cite{kitaev2015simple,maldacena2016remarks}.

Interestingly, recent work has  uncovered a number of instances of many-body teleportation \emph{without} gravitational dynamics. 
For example, teleportation in the TW protocol was recently demonstrated analytically in the SYK model at \emph{high} temperatures~\cite{gao2019traversable}, and numerically in chaotic spin chains at late times~\cite{brown2019quantum,brown2020quantum}; in both cases,  the microscopic mechanism for teleportation remains an outstanding puzzle. 
In addition to the TW protocol, an alternate many-body teleportation protocol was introduced in the context of the  Hayden-Preskill variant of the  black hole information paradox~\cite{hayden2007black,yoshida2017efficient}.
This so-called Hayden-Preskill recovery (HPR) protocol allows for many-body teleportation via \emph{generic} scrambling  dynamics.
Although the two protocols bear some structural similarity, the HPR protocol is exponentially less efficient for teleporting multiple qubits. 
%
%
To this end, understanding the precise relationship between these protocols remains an essential open question. 

In this work, we present a unified framework for many-body teleportation from the perspective of the growth of operators under scrambling time-evolution.
%
%
Most significantly, this framework leads to the identification of a new teleportation mechanism---dubbed \emph{peaked-size teleportation}---which succeeds for a wide variety of physical systems and encapsulates all known examples of many-body teleportation outside of the gravitational regime.  
We emphasize that peaked-size teleportation represents a \emph{distinct} teleportation mechanism compared to ``gravitational'' teleportation.
Although the same TW protocol can host either mechanism, the features of peaked-size teleportation differ markedly from those of gravitational teleportation [Fig.~\ref{fig: 1}(c), Table~I]. 
Crucially, this distinction implies that the TW protocol can act as a litmus test for identifying intrinsically gravitational dynamics.
More broadly, our results pave the way towards utilizing the TW protocol as a powerful experimental tool for characterizing the growth of operators in strongly interacting systems.

\section{Summary of results}\label{sec: summary}

We now provide a technical overview of our main results and the organization of our manuscript. 
This summary is intended to introduce the overarching concepts of our work, such that the remaining sections are self-contained and can be read according to individual preference. 
%
%
A more detailed, section-by-section guide to the reader is included at the end of this summary.

In \textbf{Section \ref{intro diagrams}}, we begin with a general description of the TW protocol [Fig.~1(a)].
%
In this protocol, locally encoded quantum information is inserted into one side of an entangled thermofield double (TFD) state and teleported to the other side through a combination of ($i$) unitary evolution of each side individually, and ($ii$) a simple two-sided coupling that acts on a large  subsystem of each side.
%
%
The coupling is quite flexible in form, and corresponds to unitary evolution, $e^{igV}$, under a two-sided interaction 
\begin{equation}\label{eq: V}
    V = \frac{1}{K} \sum_{i=1}^K O_{i,l} O^*_{i,r},
\end{equation}
where $\{ O_{i} \}$ are \emph{any} set of $K$ local, non-identity operators applied to the left ($l$) and right ($r$) side of the system.
This coupling can be performed as either a quantum gate, or through local measurements of $O_i$ on the left side, followed by classical communication and feed-forward operations on the right side [Fig.~1(a)]. 

In \textbf{Section \ref{sec: requirements}}, we discuss the general requirements for successful teleportation in the TW circuit.
In particular, we relate the teleportation fidelity to the following correlation functions of the two-sided system~\cite{maldacena2017diving}:
\begin{equation}\label{eq: CQ}
    C_Q(t) \equiv \bra{\tfd} Q_r(-t) e^{igV} Q_l(t) \ket{\tfd} 
\end{equation}
where $Q(\pm t)$ is a time-evolved operator initially acting on the qubit(s) to be teleported.
%
%
Our analysis leads to two conditions on these correlators that, when combined, are necessary and sufficient for  teleportation to succeed with unit fidelity: 
\begin{enumerate}
\item The magnitudes of the correlators must be maximal for every $Q$.
\item The phases of the correlators must be the same for every $Q$.
\end{enumerate}
Here, $Q$ runs over a complete basis of operators on the qubits to be teleported.
We show that Condition 1 is naturally satisfied, even without the coupling $V$, if the TFD state is at infinite temperature, in which case it reduces to an extensive set of maximally entangled EPR pairs.
On the other hand, Condition 2 requires that the coupling acts non-trivially on the operators $Q$.


In \textbf{Section \ref{size}}, we describe the relation between the coupling, $V$, and the growth of time-evolved operators, $Q(t)$. 
For the purposes of teleportation, this growth is characterized by the \emph{size distribution} of the operators~\cite{roberts2018operator,qi2019quantum,qi2019measuring}, which provides a finer-grained measure of quantum information scrambling compared to more conventional quantities such as out-of-time-ordered correlators (OTOCs)~\cite{shenker2014black,maldacena2016bound,larkin1969quasiclassical}.
%
%
%
Specifically, writing $Q(t)$ as a sum over Pauli strings, $Q(t) = \sum_{\String} c_{\String}(t) \String$, we define the size distribution as:
\begin{equation}
    P(\Size) = \sum_{\Size[\String] = \Size} |c_\String(t)|^2,
\end{equation}
where the sum is over Pauli strings, $\String$, of size, $\Size$ (equal to the string's number of non-identity components).
By probing correlations between the two sides of the doubled Hilbert space, the coupling $V$ directly measures the operator size~\cite{qi2019quantum}.

In \textbf{Section \ref{peaked-size}}, we introduce the \emph{peaked-size} mechanism for many-body teleportation.
This mechanism is possible whenever the size distributions of time-evolved operators, $Q(t)$, are tightly peaked about their average size.
In this scenario, the exponentiated coupling, $e^{igV}$, applies approximately the same phase, proportional to the size, to each coefficient, $c_\String$, and therefore to the entire operator, $Q(t)$.
We show that these applied phases are sufficient to align the correlators' phases for all operators $Q$, thereby achieving Condition 2.
We also demonstrate that the magnitudes of the correlators are unchanged by the coupling when size distributions are tightly peaked.
This implies that peaked-size teleportation achieves perfect fidelity at infinite temperature, where Condition 1 is automatically satisfied; at finite temperature, peaked-size teleportation can still occur, but  with a reduced fidelity (Table I).

%
%

In \textbf{Sections \ref{late times}-\ref{sec: intermediate time}}, we analyze examples of peaked-size teleportation across a wide variety of interacting, many-body dynamics. 
We demonstrate that the capabilities of peaked-size teleportation---most notably, the fidelity  and the number of qubits that can be sent (i.e. the channel capacity)---depend on the temperature, coupling strength,  evolution time, and the specific scrambling dynamics of the model under study (Table I). 
%
%

%
%
%
More specifically, in \textbf{Section \ref{late times}}, we provide general arguments that all scrambling systems exhibit peaked-size teleportation at \emph{late times}, after the system's scrambling time ($t \gtrsim t_s$).
In this regime, operators have become fully delocalized across the system, so that their size distributions are peaked about a typical, extensive value.
We also show that late time peaked-size teleportation is limited to transmitting only a single qubit.

\TableRegimes

In \textbf{Section \ref{sec: intermediate time}}, we show that many scrambling quantum systems also feature peaked-size teleportation at \emph{intermediate times}, i.e.~after the local thermalization time but before the scrambling time ($t_{th} \lesssim t \lesssim t_s$).
We begin with ergodic short-range interacting systems in $\ge$1D, which we show naturally possess peaked-size distributions due to thermalization within the bulk of a time-evolved operator's light cone.
In contrast, the size distributions of operators in all-to-all coupled (0D) systems are not intrinsically peaked; nevertheless, peaked sizes can be engineered by non-locally encoding the quantum information before insertion into the teleportation circuit.
Interestingly, in both of these classes of dynamics, we find that multiple ($\sim \mathcal{O}(K)$) qubits can be teleported simultaneously via the peaked-size mechanism, in contrast with the unit channel capacity of late time teleportation.
We substantiate these claims through extensive numerical and analytic studies on a variety of physical models: random unitary circuits (RUCs) in dimensions $d = 0,1,$ and $2$~\cite{nahum2018operator}, the SYK model, and (in Section \ref{sec:rydberg}) experimentally relevant spin chain Hamiltonians~\cite{Bernien17}. 

In \textbf{Section \ref{interplay}}, we discuss the interplay between peaked-size and gravitational teleportation.
Notably, we expect gravitational teleportation to occur only at low temperatures, where certain quantum mechanical models (e.g.~the SYK model) are known to possess a dual description in terms of conformal matter coupled to gravitational dynamics.
From the perspective of operator growth, the unique feature of gravitational teleportation is the presence of non-trivial phase \emph{winding} in a variant of the size distribution~\cite{brown2019quantum}.
Crucially, this effect enables gravitational teleportation to satisfy Condition 1, and thereby achieve high teleportation fidelity at low temperatures, in sharp contrast with peaked-size teleportation (Table I).

Intriguingly, while it may seem that there is a sharp distinction between peaked-size and gravitational teleportation, we find that this is not always this case. 
In particular, we show that varying the temperature of the SYK model provides a continuous interpolation between gravitational teleportation at low temperature and peaked-size teleportation at high temperature. 
In the dual picture, perturbing away from the low temperature limit corresponds to adding \emph{stringy} corrections to the gravity theory~\cite{shenker2015stringy,maldacena2016remarks,gu2019relation}.
Following this intuition, we show that teleportation in a gravity theory with strong stringy corrections~\cite{maldacena2017diving} bears a remarkable qualitative similarity to peaked-size teleportation, thus providing a first step towards a bulk understanding of this phenomenon.



Finally, in \textbf{Section \ref{experiment}}, we discuss experimental applications of the TW protocol for probing many-body dynamics.
In particular, we demonstrate that the protocol can function as a  diagnostic tool for scrambling dynamics in near-term quantum simulators, enabling one to starkly distinguish between generic thermalizing systems and gravitational dynamics. 
To this end, we provide detailed blueprints for realizing the protocol in two complementary experimental platforms---Rydberg atom arrays~\cite{Maller15,Labuhn16,Bernien17,Graham19,Madjarov20,Wilson19} and trapped ions~\cite{wineland2008entangled,monroe2013scaling,Ballance:2016, Gaebler2016high,cetina2020}.
Specifically, the observation of a high teleportation fidelity at low temperatures would be a tantalizing experimental indicator of gravitational scrambling dynamics.
In addition, gravitational dynamics exhibit unique qualitative features as a function of both evolution time and protocol parameters [Fig.~\ref{fig: 1}(c), Table~\ref{table:comparison}].
More broadly, our analysis suggests that the TW protocol can provide  insights into many-body dynamics outside the gravitational regime. 
In particular, we demonstrate that the fidelity of peaked-size teleportation probes higher moments of operator size distributions~\cite{qi2019measuring}.




\emph{Guide to the reader}---Considering the wide scope of results presented in this work, we encourage readers to skip to sections that align with their specific interests and refer to the above summary for context. 
To this end, we highlight below the nature of each section and provide recommendations for readers of different backgrounds.
%
%
Sections \ref{intro diagrams}-\ref{size} introduce the formal tools and derivations necessary for rigorously understanding our results. 
These sections will be of interest to readers with a background in quantum information who wish to understand the precise connection between teleportation and operator sizes. 
Sections \ref{peaked-size}-\ref{sec: intermediate time} introduce peaked-size teleportation and analyze its realization in several example systems. 
Since many these systems are experimentally accessible, these sections will be most relevant to members of the quantum simulation and many-body physics communities. 
Section \ref{interplay} focuses on the interplay of peaked-size teleportation and gravitational physics, both in the SYK model and from a bulk gravitational perspective.
For brevity, background material on gravitational physics is relegated to references, making this section best suited for experts at the intersection of quantum information and quantum gravity.
Finally, Section \ref{experiment} contains a summary of the experimental signatures of the TW protocol, detailed blueprints for Rydberg atom and trapped ion implementations, and a discussion of the protocol's behavior under experimental error.
This section will be of interest to AMO experimentalists and all readers interested in near-term realizations of many-body quantum teleportation~\cite{preskill2018quantum}.

\subsection{Relation to previous works}

To further elaborate on the broad context of our results, a brief summary of the relevant prior studies and their relation to our work is provided as follows. 

\emph{Gravitational teleportation in the TW protocol---}Traversable wormhole teleportation was originally introduced in Refs.~\cite{gao2017traversable,maldacena2017diving} in the context of gravitational physics, where it was realized that a coupling of the form $V$ enables a traversable channel between the boundaries of a two-sided black hole. 
The explicit quantum mechanical circuit implementing this teleportation [Fig.~\ref{fig: 1}(a)] was later introduced in Refs.~\cite{gao2019traversable,brown2019quantum}, alongside exact calculations for the teleportation fidelity in the large-$q$ SYK model~\cite{gao2019traversable}.
While the emphasis of our work is not on the bulk interpretation of gravitational teleportation---indeed, the peaked-size teleportation mechanism is intended to \emph{contrast} with the gravitational mechanism---it will be helpful to recall the main results from the gravitational perspective.

We focus on the specific case of two-dimensional anti-de Sitter space, which is the bulk dual of the SYK model at low temperatures~\cite{kitaev2018soft,maldacena2017diving}. 
In the simplest case (ignoring gravitational backreaction), the two-sided correlator, Eq.~\ref{eq: CQ}, can be explicitly calculated and is given by~\cite{maldacena2017diving}:
\begin{equation} \label{CQ gravity}
    C_Q(t) = \left( \frac{1}{2 - g \frac{\Delta_O}{2^{2\Delta_O+1}} G_N e^{2\pi t/\beta}} \right)^{2\Delta_Q}.
\end{equation}
Here, $G_N$ is Newton's constant, $\beta = 1/T$ is the inverse temperature of the black hole, $\Delta_O$ is the conformal dimension of the coupling operators $O_i$ [Eq.~(\ref{eq: V})], and $\Delta_Q$ is the conformal dimension of the operator $Q$.
In the context of the SYK model, $G_N$ is inversely proportional to the number of Majorana fermions, $N$, and the black hole temperature is equal to the temperature of the TFD state~\cite{maldacena2017diving,gao2019traversable}. 

For our purposes, the most notable feature of the correlator is that it exhibits a sharp peak at time $t \approx G_N \log(g)$ [Fig.~\ref{fig: 1}(c)], corresponding to the moment a particle inserted on one side of the black hole emerges on the other side.
While in the above formula [Eq.~(\ref{CQ gravity})], the correlator diverges at this time, in the large-$q$ SYK model, this divergence is regularized and the correlator peaks at its maximal value of unity~\cite{gao2019traversable}.
Thus, at time $t \approx G_N \log(g)$, the correlator  satisfies  Condition 1 for successful teleportation; in Ref.~\cite{gao2019traversable}, it was shown that Condition 2 is also satisfied for certain conformal dimensions of the operators $Q$.
In combination, this leads to unit teleportation fidelity.

Another notable feature of gravitational teleportation is the ability to teleport multiple qubits simultaneously, as discussed in Ref.~\cite{maldacena2017diving}.
In the gravitational picture, multi-qubit teleportation has an intuitive explanation: particles corresponding to different qubits pass through the black hole in parallel, without interacting with one another.
However, for sufficiently many qubits, the effects of gravitational backreaction become important, leading to a predicted channel capacity of $\mathcal{O}(K)$.

\emph{HPR teleportation---}An independent, but closely related, set of protocols for many-body teleportation was introduced in Ref.~\cite{yoshida2017efficient} for the recovery of information in the Hayden-Preskill thought experiment~\cite{hayden2007black}.
Unlike previous works on traversable wormholes, in Ref.~\cite{yoshida2017efficient} teleportation succeeds for any fully scrambling unitary dynamics (i.e.~at late times, $t \gtrsim t_s$), with no reliance on gravitational physics.
However, the channel capacity of HPR teleportation is fundamentally limited: multi-qubit teleportation requires a protocol whose circuit depth grows exponentially in the number of qubits to be teleported~\cite{yoshida2017efficient}. 

In Appendix~\ref{app: YK}, we show that a deterministic variant of the HPR protocol (for single-qubit teleportation) is in fact \emph{equal} to the TW protocol in  Fig.~\ref{fig: 1}(a), restricted to infinite temperature and with a particular choice of the coupling operators, $O_i$.
Furthermore, in Section~\ref{sec: YK} we show that teleportation at \emph{late times} via the peaked-size mechanism is equivalent to this variant of HPR teleportation. 
However, peaked-size teleportation is more powerful than HPR teleportation in the sense that: ($i$) it succeeds for a much larger class of couplings, $V$, ($ii$) it can succeed at intermediate times, and ($iii$) at such times, it is capable of sending multiple qubits with no change in the protocol's complexity, an exponential improvement over the HPR protocol.


%


\emph{Previous many-body teleportation experiments---}Many-body quantum teleportation has recently been demonstrated in both trapped ion~\cite{landsman2019verified} and superconducting qutrit~\cite{blok2020quantum} experiments.
Both Refs.~\cite{landsman2019verified,blok2020quantum} implement a probabilistic variant of the HPR protocol, which differs slightly from the TW protocol, while Ref.~\cite{landsman2019verified} also implements the deterministic variant discussed above.
In all cases, the scrambling dynamics, $U$, are generated by digital quantum gates acting on a small number of qubits.
Teleportation is performed for a single qubit and a fully scrambling unitary, placing the experiments in the same physical regime as late-time, peaked-size teleportation.

Our work demonstrates that experiments in the \emph{TW protocol} at \emph{intermediate times} can access new regimes of many-body quantum teleportation, with the potential to provide more information about the scrambling dynamics under study.
Most notably, such experiments can distinguish between teleportation in generic many-body systems (via the peaked-size mechanism) versus systems with a gravity dual (via the gravitational mechanism), which is not possible in the HPR protocol.



\emph{SYK teleportation in the TW protocol---}In Ref.~\cite{gao2019traversable}, the two-sided correlator of the TW protocol [Eq.~(\ref{eq: CQ})] was calculated exactly for the large-$q$ SYK model (defined in Section~\ref{sec: SYK infinite temperature}).
As anticipated in Ref.~\cite{maldacena2017diving}, the correlator at low temperatures---where the model is dual to gravity---agrees with the gravitational result [Eq.~(\ref{CQ gravity})] up to the previously mentioned regularization.
%
%
%
More surprisingly, it was shown that teleportation with unit fidelity is also possible at high temperatures---where the model is not dual to gravity.
As we will see in Section~\ref{sec: SYK infinite temperature}, all features of high temperature teleportation in the SYK model are in precise agreement with the peaked-size mechanism; our work thus provides a microscopic understanding for this previously unexplained result. 

\emph{Gravity in the lab---}Ref.~\cite{brown2019quantum} discusses various instances of teleportation in the TW protocol.
The authors distinguish two teleportation mechanisms: ($i$) an ``operator transfer'' mechanism, which occurs at intermediate times in gravitational systems and is capable of teleporting multiple qubits, and ($ii$) a ``state transfer'' mechanism, which occurs at late times in all scrambling systems, and is capable of sending only a single qubit.
Moreover, they introduce a microscopic interpretation for the teleportation mechanism in gravitational systems, termed ``size winding'', which we connect to in Section~\ref{sec: gravity size winding}.

In our terminology, the first teleportation mechanism corresponds to gravitational teleportation, while the second mechanism corresponds to peaked-size teleportation at late times\footnote{The terminology of Ref.~\cite{brown2019quantum} can be understood using our two Conditions for teleportation.
Specifically, operator transfer corresponds to situations that satisfy  Condition 1, but not necessarily Condition 2, as occurs in gravitational teleportation [see Eq.~(\ref{CQ gravity})].
State transfer corresponds to situations that satisfy both Conditions.}.
In our work, we provide a microscopic interpretation for late time teleportation (i.e.~the peaked-size mechanism) and demonstrate that it is equivalent to teleportation in the HPR protocol.
In addition, we demonstrate that peaked-size teleportation is a more general phenomenon that also occurs at intermediate times in many systems, where we show that it is capable of teleporting multiple qubits.

In a follow-up work, Ref.~\cite{brown2020quantum}, whose pre-print was posted concurrently with that of this work, the same authors elaborate on their previous results and provide more detailed examples and calculations.
These agree with our own results in areas of overlap.

\section{Introduction to diagrammatic notation}\label{intro diagrams}

We begin by introducing a diagrammatic ``tensor network'' notation for depicting the teleportation circuit.
Adapted from Ref.~\cite{yoshida2017efficient}, this notation provides a precise visual framework for analyzing teleportation in Section~\ref{sec: requirements} and will be convenient for deriving rigorous results on the teleportation fidelity in Section~\ref{sec: rigorous}.

To begin, we represent a quantum ket $\ket{\psi}$ and bra $\bra{\psi}$ as:
\begin{align}
\figbox{.3}{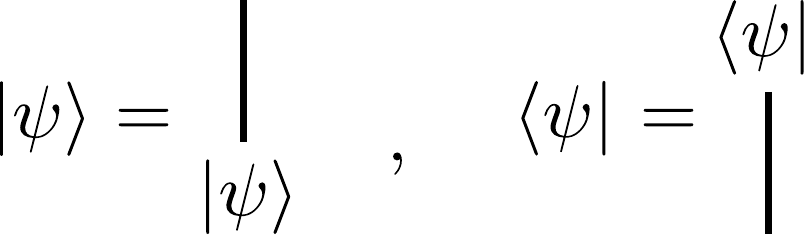} \centering
\end{align}
Note that time proceeds upwards---an initial state $\ket{\psi}$ terminates the bottom of a leg, while a final projection $\bra{\psi}$ terminates the top.
Similarly, much as in Fig.~\ref{fig: 1}(a), we represent an operator, for instance the many-body unitary $U$, as a box with input (bottom) and output (top) legs:
\begin{align}
\figbox{.3}{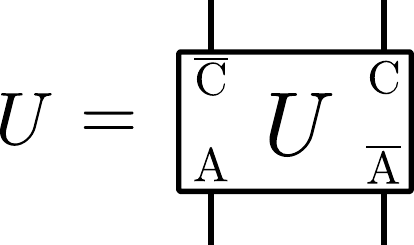} \centering
\end{align}
Here we have decomposed the input and output into two subsystems, A and its complement for the input, C and its complement for the output, in reference to the teleportation protocol.
Specifically, comparing to Fig.~\ref{fig: 1}(a), subsystem A consists of the qubits supporting the input state $\ket{\psi}$, while subsystem C consists of the coupled qubits.

The diagrammatic notation is particularly useful when working with EPR states.
The EPR state on two qubits is defined as $\ket{\epr} = (\ket{00}+\ket{11})/\sqrt{2}$; for a system of $N$ $d$-dimensional qudits, this is generalized to $\frac{1}{\sqrt{d^N}} \sum_{i=1}^{d^N} \ket{i}_l \ket{i}_r^*$.
Here $\{ i \}$ is an arbitrary $d^N$-dimensional basis, $*$ denotes time-reversal (i.e. complex conjugation), and $l$ and $r$ denote the left and right system, respectively.
In the diagrammatic notation, we represent this as:
\begin{align}
\figbox{.3}{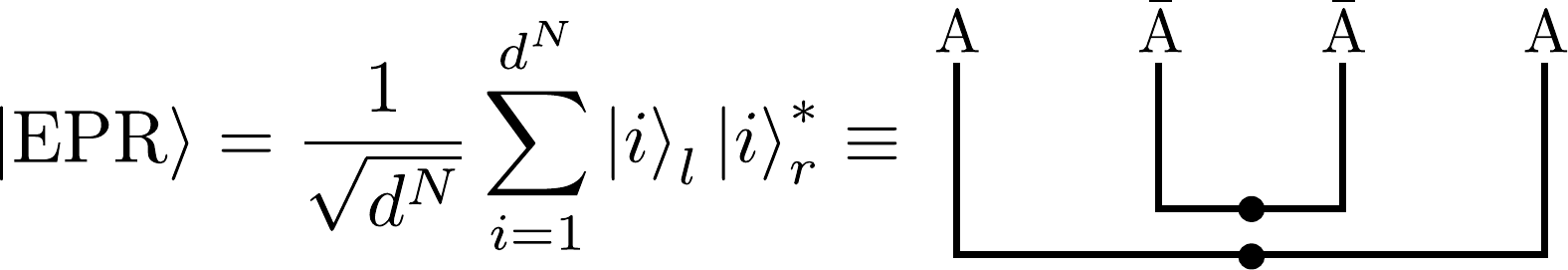} \centering \label{eq: epr}
\end{align}
We have again decomposed each system into two subsystems, A and its complement, $\bar{\text{A}}$, for convenience (subsystem A is chosen to be identical between the left and right sides).
Each dot represents a normalization factor given by the inverse square root of the subsystem's dimension.

To see the utility of the diagrammatic notation, recall that a fundamental property of the EPR state is that an operator acting on the left side is equivalent to its transpose acting on the right:
\begin{equation}
\begin{split}
    O_l \ket{\epr} & = \frac{1}{\sqrt{d^N}} \sum_{i,j} O_{ij} \ket{i}_l \ket{j^*}_r \\
    & = \frac{1}{\sqrt{d^N}} \sum_{i,j} O^T_{ij} \ket{j}_l \ket{i^*}_r =  O_r^T \ket{\epr}
\end{split}
\end{equation}
where the middle equality swaps the $i$, $j$ indices of the sum.
In diagrammatic notation, this becomes simply
\begin{align} \label{eq: O slide}
\figbox{.3}{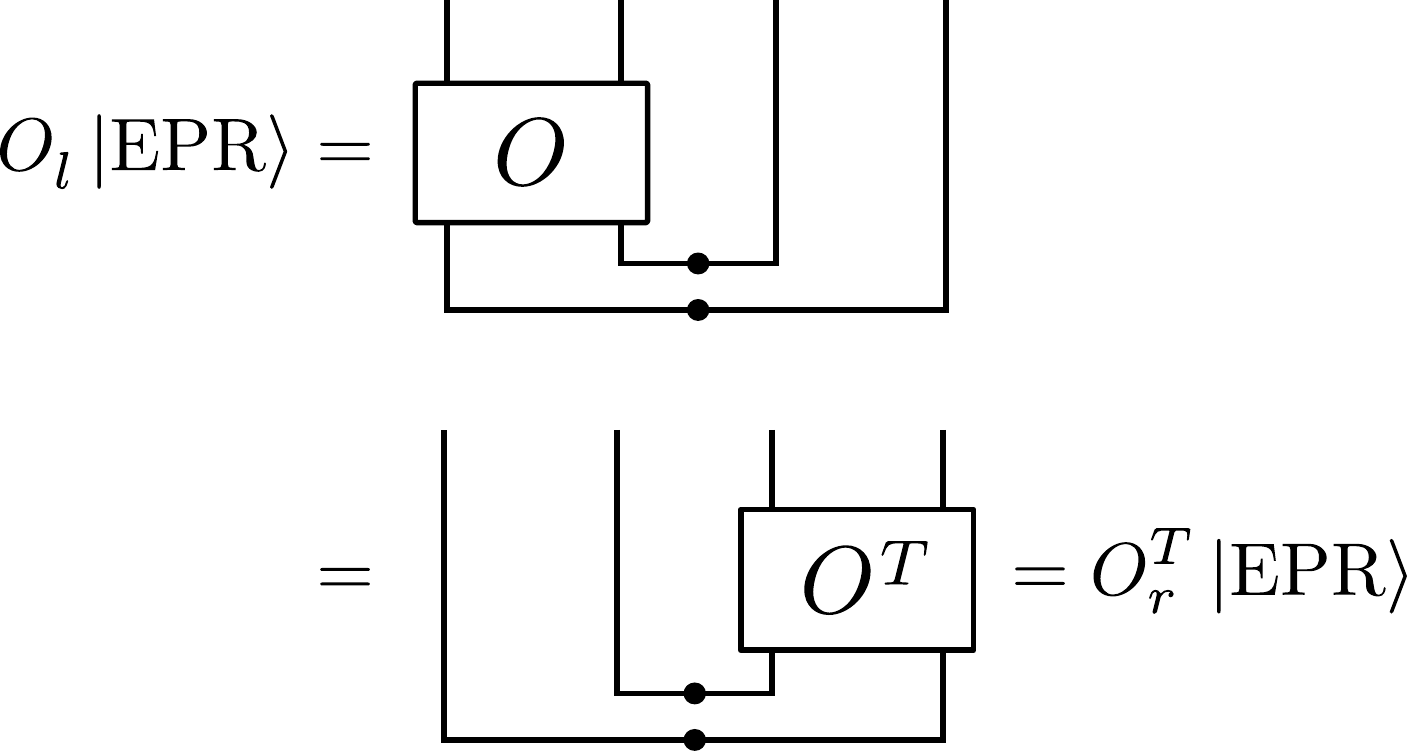} \centering 
\end{align}
i.e. the operator $O$ ``slides'' from the left to right side of the EPR pairs, with its input and output indices correspondingly transposed.
Similarly, expectation values in the EPR state can be easily computed in terms of the trace of (one-sided) operators, e.g.
\begin{align} \label{eq: AB trace}
\figbox{.28}{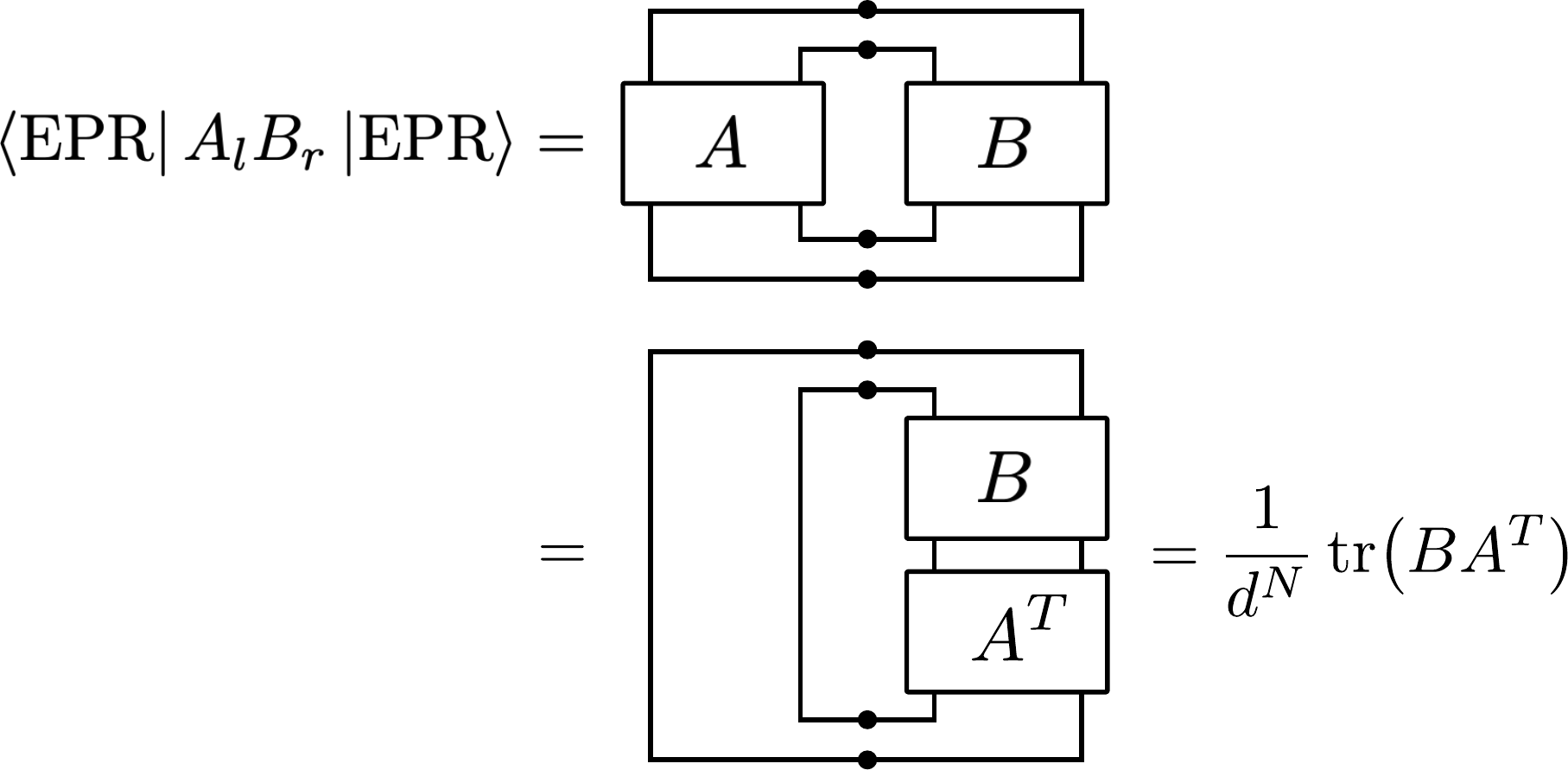} \centering 
\end{align}
where the final equality follows from $\bra{\epr} B_l A^T_l \ket{\epr} = (1/d^N)\sum_{ij} \braket{ i^* | j^* } \bra{i} B A^T \ket{j} = (1/d^N)\sum_{i} \bra{i} B A^T \ket{i}$.

The EPR state is closely related to the thermofield double (TFD) state, $\tfd \equiv  \sum_{i} e^{-\beta E_i / 2} \ket{E_i}_l \ket{E_i^*}_r / \tr(e^{-\beta H})^{1/2}$.
Here $H$ is a time-reversal symmetric Hamiltonian, $H = H^*$, with eigenstates $\ket{E_i}$, and eigenvalues $E_i$.
The TFD state is parameterized by an effective ``temperature'' $1/\beta$.
At infinite effective temperature ($\beta = 0$), the TFD and EPR states are equal.
At finite temperature, the TFD state is obtained by applying the square root of the density matrix, $\rho^{1/2} \equiv e^{-\beta H/2} /\tr(e^{-\beta H})^{1/2}$,  to either side of the EPR state, which we represent as:
\begin{align}
\figbox{.3}{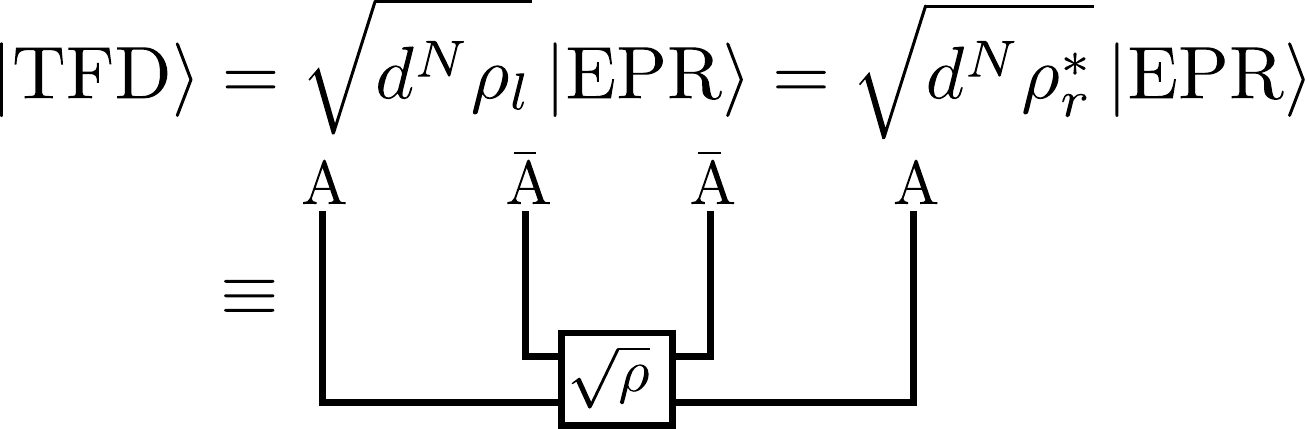} \centering \label{eq: tfd}
\end{align}

For the finite temperature TFD state, the analog of Eq.~(\ref{eq: O slide}) holds \emph{only} for operators that commute with the Hamiltonian.
Most notably, such operators include the time-evolution operator, $U = e^{-iHt}$, which thus obeys:
\begin{align} \label{eq: U slide}
\figbox{.3}{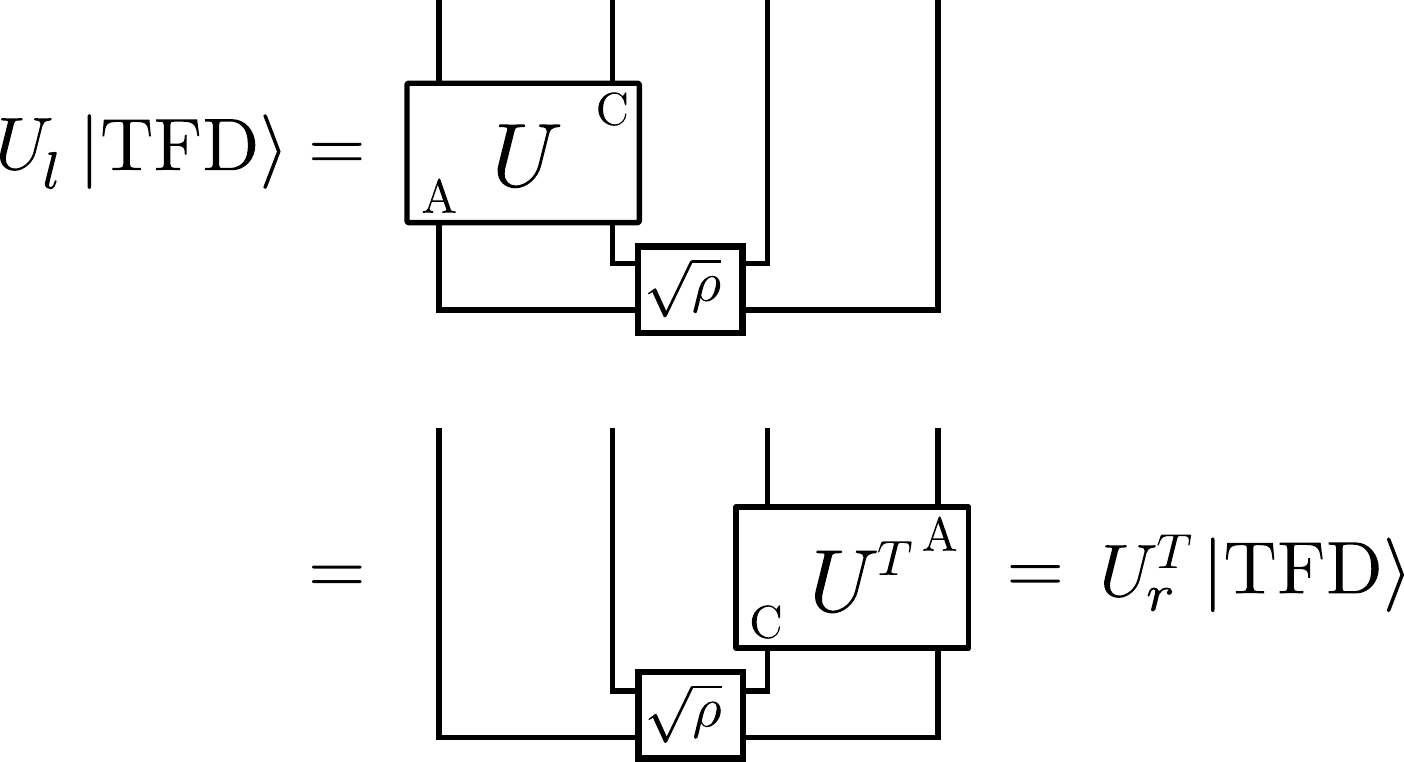} \centering 
\end{align}
Eq.~(\ref{eq: U slide}) also holds for backwards time-evolution, replacing $U \rightarrow U^\dagger, U^T \rightarrow U^*$.
We note that for time-reversal symmetric Hamiltonians, $U = U^T$.
In this case, combining Eqs.~(\ref{eq: U slide},~\ref{eq: O slide}), we have the useful identity:
\begin{equation}
    O_l(t) \ket{\tfd} = O^T_r(-t) \ket{\tfd}.
\end{equation}
Applying Eq.~(\ref{eq: AB trace}), we can again express `two-sided' expectation values in the TFD state in terms of `one-sided' correlation functions, e.g.
\begin{equation} \label{eq: AB trace TFD}
    \bra{\tfd} A_l(t) B_r(t') \ket{\tfd} = \tr ( \rho^{1/2} A^T(-t) \rho^{1/2} B(t') ).
\end{equation}

Let us now re-draw the full teleportation protocol in Fig.~\ref{fig: 1}(a) using the diagrammatic notation:
\begin{align}
\figbox{.8}{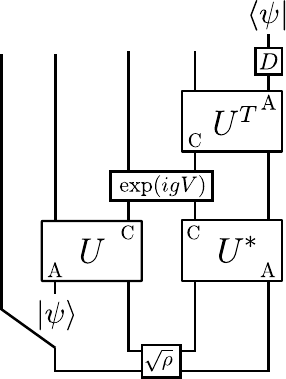} \centering \label{TWH protocol Fig1}
\end{align}
This circuit proceeds as follows: ($i$) prepare the TFD state, ($ii$) insert the state $\ket{\psi}$ on subsystem A of the left side, ($iii$) time-evolve the two sides by $U$, $U^*$, ($iv$) couple the two sides via the unitary operator $e^{igV}$, with $V$ as in Eq.~(\ref{eq: V}), ($v$) evolve the right side by $U^T$, ($vi$) apply a `decoding' operator $D$, and ($vii$) measure the output state of subsystem A on the right side.
Compared to Fig.~\ref{fig: 1}(a), we have made two modifications.
First, we have replaced the measurement and classical communication with a quantum coupling $e^{igV}$, as described in Section~\ref{sec: summary}.
Second, we now include a simple decoding operator, $D$, applied at the end of the circuit before state recovery.
We will find that $D = Y \otimes \ldots \otimes Y$ for peaked-size teleportation of a multi-qubit subsystem, where $Y$ is the single-qubit Pauli $Y$ operator (Section \ref{peaked-size}).

Finally, we note that a straightforward application of Eq.~(\ref{eq: U slide}) allows us to re-express the circuit as
\begin{align}
\figbox{.8}{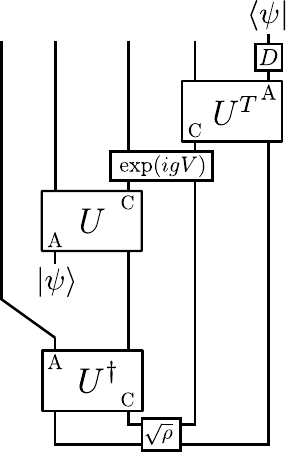} \centering \label{TWH protocol state swap}
\end{align}
This equivalent version of the protocol was introduced in Refs.~\cite{gao2019traversable,brown2019quantum} and will be more convenient for analysis from here on.

\section{General requirements for successful teleportation}\label{sec: requirements}

We now introduce heuristic arguments for when teleportation succeeds in this protocol. 
This will culminate in the two requirements for teleportation listed in Section~\ref{sec: summary}.
In Section~\ref{peaked-size}, we derive these conditions more formally by providing exact relations between the two-sided correlators in Eq.~(\ref{eq: CQ}) and the teleportation fidelity.


We begin with the protocol in Eq.~(\ref{TWH protocol state swap}).
To proceed, we insert a resolution of the identity $\mathbbm{1} = \sum_\phi \dyad{\phi}$ on the ``swapped out'' subsystem A (the output of $U^\dagger_l$)\footnote{At infinite temperature, using Eq.~(\ref{eq: O slide}), $\ket \phi$ can be understood as the counterpart of $\ket \psi$, to be teleported from right to left instead of left to right. To see this, use Eqs.~(\ref{eq: O slide}, \ref{eq: U slide}) to re-expess $\bra{\phi}_l U^\dagger_l \rightarrow U^\dagger_l \ket{\phi}_r$, and apply $D_l U^\dagger_l$ after coupling to recover $\ket{\phi}$.}:
\begin{align}
\figbox{.8}{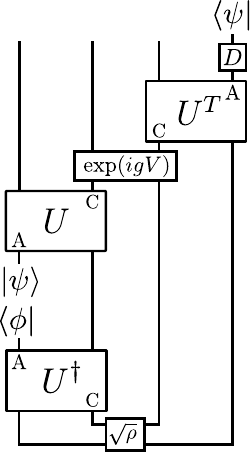} \centering \label{TWH protocol state}
\end{align}
This reformulation makes it clear that teleportation depends on the action of the coupling on states of the form $Q_{A,l}(t) \ket{\tfd}$, where $Q_A = \dyad{\psi}{\phi}$ and\footnote{Traditionally, this would be considered reverse time-evolution, and denoted $Q_A(-t)$. For brevity, we have flipped the sign of $t$ throughout the text.} $Q_A(t) \equiv U Q_A U^\dagger$.

Teleportation succeeds when the coupling ``transfers'' $\dyad{\psi(t)}{\phi(t)}$ from the left to right side of the TFD state.
More precisely, the following identity, if true for \emph{all} operators $Q_A$ on A, would guarantee successful teleportation for all states:
\begin{align}
\figbox{.4}{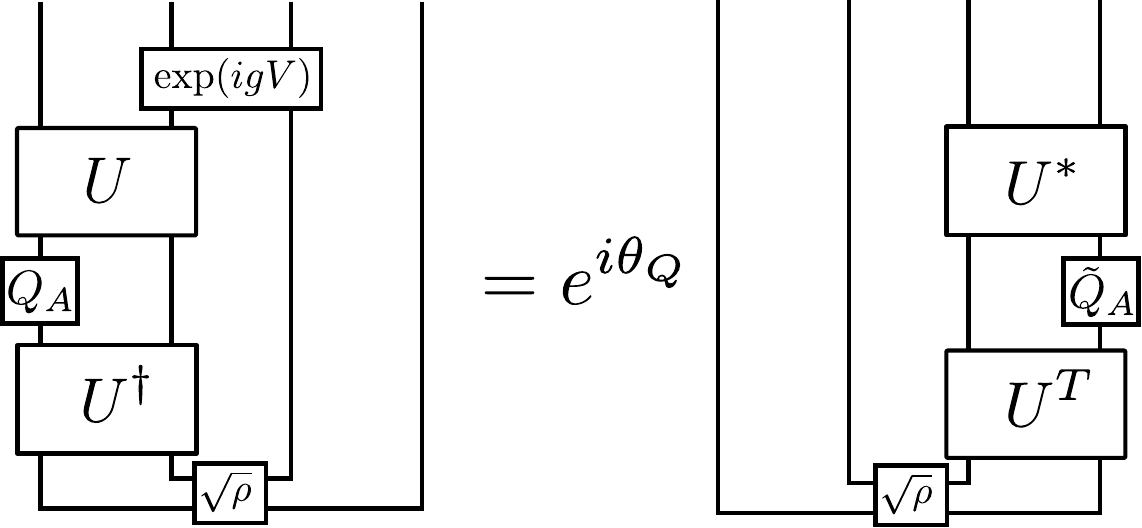} \centering \label{eq-correlator-inner-product}
\end{align}
Here $\theta_Q$ is an overall phase and we represent conjugation by the decoding operator as $\tilde{Q}_{A} \equiv D^{\dagger} Q_A D$.
One can verify this explicitly by plugging the RHS of the above equality into Eq.~(\ref{TWH protocol state}): the topmost applications of $D U^T$ and $U^* D^\dagger$ cancel, leaving $Q_A \rightarrow \dyad{\psi}{\phi}$ as the topmost operator on the right side, i.e. subsystem A is in the state $\ket{\psi}$.

To quantify whether this equality holds, we measure the inner product between the two states\footnote{For simplicity of notation and consistency with previous works~\cite{gao2017traversable,maldacena2017diving,gao2019traversable}, from here on we have assumed that the unitary is symmetric, $U^T = U, \, U^\dagger = U^*$.}:
\begin{align} \label{C_Q}
\figbox{.3}{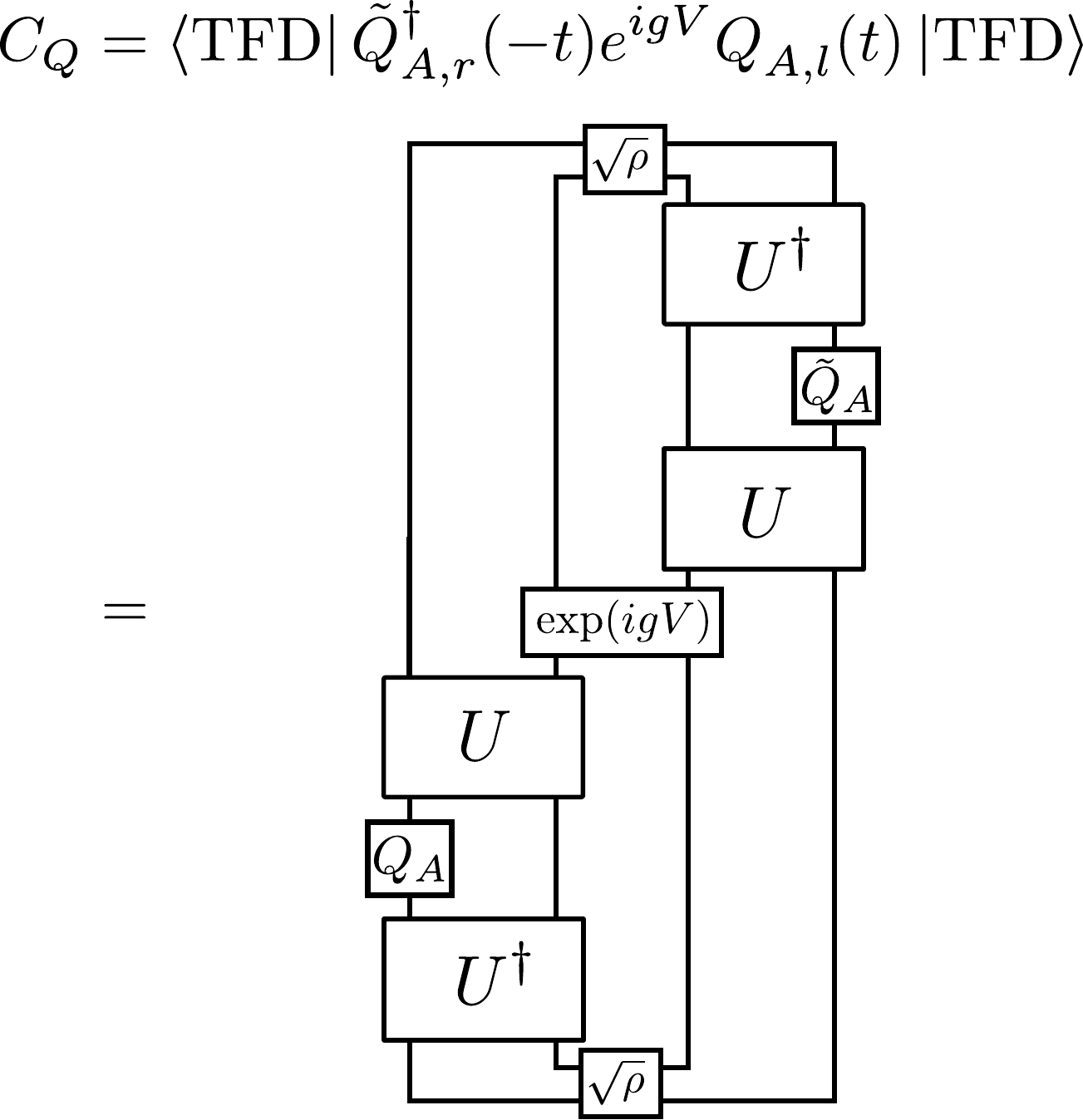} \centering 
\end{align}
This is precisely the two-sided correlation function introduced in Eq.~(\ref{eq: CQ}), now modified to include the decoding operator.
In particular, if the correlation function is maximal for all operators $Q_A$, then Eq.~(\ref{eq-correlator-inner-product}) holds and teleportation succeeds with perfect fidelity for all initial states.
%

%
%

In practice, it is sufficient to evaluate the correlators for a complete basis of operators on subsystem A (e.g. the Pauli operators).
In this case, we now have \emph{two} requirements on the operator correlators, as listed in Section~\ref{sec: summary}: ($i$) all correlators must have maximal magnitude, i.e. equal to 1, and ($ii$) all correlators must have the \emph{same phase}---if two operators both individually obey Eq.~(\ref{eq-correlator-inner-product}) but with different phases, their sum will not.



At infinite temperature, owing to Eq.~(\ref{eq: O slide}), we will see that the first requirement is satisfied even in the absence of the coupling, for any symmetric or antisymmetric operator.
To satisfy the second requirement, the role of the coupling $e^{igV}$ must be to apply a $Q_A$-dependent overall phase.
In the following section, we analyze the action of the coupling and show precisely when such an overall phase occurs.

\section{Connection to operator size} \label{size}

In this section, we outline the connection between the coupling $V$ and the operator size when $V$ is acted on states of the form:
\be \label{Q tfd}
Q_{A,l}(t) \ket{\tfd} = Q_{A,l}(t) \rho^{1/2}_l \ket{\epr}.
\ee
This connection was discovered in a number of previous works, focusing primarily on a specific bilinear coupling in fermionic systems~\cite{roberts2018operator,qi2019quantum,lin2019symmetries,susskind2019complexity,lin2019complexity,brown2019quantum,brown2020quantum}.
In the following, we introduce this connection in the context of bosonic systems and argue that it applies to a good approximation for any generic, local couplings.
From this, we then show that the action of the exponentiated coupling, $e^{igV}$, is particularly simple---it applies an overall phase---whenever operator size distributions are tightly peaked.

\subsection{Coupling measures size}\label{Coupling size}



In bosonic qudit systems, we define the size of a Pauli string as its number of non-identity elements~\cite{roberts2018operator}.
For instance, the Pauli string 
\be
\mathbbm{1} \otimes X \otimes \mathbbm{1} \otimes \mathbbm{1} \otimes Z \otimes X \otimes \mathbbm{1}
\ee
has size 3.
A more general operator can be written as a sum of Pauli strings, $R$:
\be \label{expand pauli strings}
Q_A(t) \rho^{1/2} = \sum_{\String} c_{\String}(t) \String,
\ee
and possesses a corresponding \emph{size distribution}~\cite{roberts2018operator,qi2019quantum}\footnote{We note that, at finite temperature, the coefficients $c_{\String}(t)$ will generally be complex.
Their phases thus carry information beyond that captured by the size distribution, which we discuss in Section~\ref{interplay}.}:
\begin{equation}\label{size distribution}
    P(\Size) = \sum_{\Size[\String] = \Size} |c_{\String}(t)|^2.
\end{equation}
The distribution is normalized to 1 if $Q_A$ is unitary,
\begin{equation}
    \sum_\Size P(\Size) = \sum_{\String} |c_{\String}(t)|^2 = \tr(Q_A^\dagger Q_A \rho) = 1.
\end{equation}
One can naturally characterize the size distribution via its moments---for instance, the average size, $\Size[Q_A(t) \rho^{1/2}] \equiv \sum_\Size P(\Size) \Size$ (when context is clear, we denote this simply as $\Size$), and the size width, $\delta \Size$.

We will now show that the coupling $V$ approximately measures the operator size, in the sense that it acts on states of the form Eq.~(\ref{Q tfd}) as:
\be \label{V size}
V Q_{A,l}(t) \ket{\tfd} \approx d^{N/2} \sum_{\String} \bigg( 1 - \constd \frac{\Size[\String]}{N} \bigg) c_{\String}(t) \String_l \ket{\epr},
\ee
where $\constd \equiv 1/(1-1/d^2)$ is an order one constant determined by the local qudit dimension, $d$. 
%
Expectation values of $V$ thus measure the average size, while higher powers of $V$ measure higher moments of the size distribution~\cite{roberts2018operator,qi2019quantum}.
In particular, the exponentiated coupling in the teleportation protocol applies a \emph{size-dependent phase} to each Pauli string of $Q_A(t) \rho^{1/2}$:
\be \label{coupling size phase}
\begin{split}
e^{igV} Q_{A,l}(t) \ket{\tfd} & \approx d^{N/2} \\
& e^{ig} \sum_{\String} e^{ - i \constd g \Size[\String]/N}  c_{\String}(t) \String_l \ket{\epr},
\end{split}
\ee
We derive this connection by first introducing an \emph{exact} measure of operator size in bosonic qudit systems, generalizing previous measures for Majorana fermionic systems~\cite{roberts2018operator,qi2019quantum}.
We then argue that successively more generic couplings display approximately the same behavior, when acted on time-evolved operators in generic many-body scrambling dynamics.

%
In bosonic qudit systems, we find that the operator size is precisely measured by a sum of individual EPR projectors on each qudit $i$:
\be \label{size qudit}
V_s = \frac{1}{N} \sum_{i=1}^N P_{\epr,i} = \frac{1}{N d^2} \sum_{i=1}^N \sum_{P_i} P_{i,l} P_{i,r}^*,
\ee
where $d$ is the local qudit dimension, $N$ is the number of qudits, and $P_i$ form a complete basis of single-qudit operators (e.g.~for qubits $P_i \in \{\mathbbm{1},X,Y,Z\}$).
To see this, let us first analyze the action of a single EPR projector, $P_{\epr,i}$.
Writing a given Pauli string as a tensor product of single-qudit Paulis, $\String = \bigotimes_{j=1}^N \String_j$, we find
\be
P_{\epr,i} \String_{l} \ket{\epr}  = \delta_{\String_i,\mathbbm{1}} \String_l  \ket{\epr},
\ee
using Eq.~(\ref{eq: AB trace}) and $\tr_i (\String_i)/d_i = \delta_{\String_i,\mathbbm{1}}$. 
A single EPR projector thus acts as a binary variable, giving eigenvalue $1$ or $0$ if a given Pauli string is, or is not, the identity on the designated qudit.
The full coupling is a sum of these binary variables over all qudits and therefore counts the total number of non-identity elements in the Pauli string, i.e. the operator size.
Its eigenvectors are the states $\String_l \ket{\epr}$ with eigenvalues $1 -\Size[\String]/N$, as in Eq.~(\ref{V size}).

We now turn to more general local couplings. 
First, as a trivial but useful modification, we can remove the identity operators from $V_s$, since these are not included our original definition of the coupling, $V$ [Eq.~(\ref{eq: V})].
These constitute a fraction $1/d^2$ of the complete basis, $P_i$, summed in Eq.~(\ref{size qudit}). 
Removing these terms renormalizes the eigenvalues of the coupling: 
\be \label{size qudit Pauli}
\begin{split}
\bigg( \frac{1}{N(d^2-1)} & \sum_{i=1}^N \sum_{P_i \neq \mathbbm{1}} P_{i,l} P_{i,r}^* \bigg) \String_l \ket{\epr}  \\
& = \bigg[ 1 - \constd \frac{\Size[\String]}{N} \bigg] \String_l \ket{\epr},
\end{split}
\ee
which now match those quoted in Eq.~(\ref{V size}).
Note that the left side sum is now over $N(d^2-1)$ non-identity operators and normalized accordingly.

Second, we consider omitting some of the \emph{non-identity} $P_i$ at each site.
Intuitively, under thermalizing dynamics, if an operator has spread to some qudit $i$ it should not matter which Pauli operator we use to probe the operator's presence.
For example, for qubits, we could omit the $O_j = X_i, Y_i$ couplings and keep only $O_j = Z_i$.
A random Pauli string has equal probability to commute with $Z_i$ as it would with $X_i$ and $Y_i$; thus, coupling using only $Z_i$ operators is sufficient for measuring a thermalized operator's size.

%

Third, we expect even more general couplings, composed of $O_i$ that are local but not necessarily Pauli operators, to behave similarly.
%
Specifically, each individual coupling, $O_{i,l} O_{i,r}$, will asymptote to two different expectation values before and after the time-evolved operator has spread to the support of $O_i$.
Before, the coupling will maintain its expectation value in the unperturbed TFD state, $\text{tr} ( O_{i} \rho^{1/2} O^{\dagger}_{i} \rho^{1/2}  )$.
After, the spread of $Q_A(t)$ will disrupt the two-sided correlations in the TFD state that give rise to this initial expectation value, and the coupling will instead asymptote to its value in two thermal states, $\tr ( O_{i} \rho ) \cdot \tr (O_{i} \rho)$.
As before, the sum of many terms, each behaving as above, leads to an approximate measure of operator size.

Lastly, we consider the case where the coupling is restricted to act only on some subsystem C, consisting of $K$ qudits\footnote{For simplicity, this assumes that there is a single coupling per qudit in C.}.
The coupling now measures the number of non-identity elements of a Pauli string within C---we denote this as the $K$-\emph{size}, $\Size_K$, of the Pauli string.
The eigenvalues of the coupling are the same as those in Eq.~(\ref{size qudit Pauli}), with the replacement $\Size/N \rightarrow \Size_K/K$. 
For a typical Pauli operator, we expect the $K$-size distribution of an operator to be similar to its full size distribution when $K$ is large and the coupled qubits are distributed randomly.
In particular, in this scenario we expect the average $K$-size, $\Size_K$, to be proportional to the average size, $\Size$,
\be \label{K size}
\frac{\Size_K}{K} \approx \frac{\Size}{N}.
\ee
For simplicity, we will make this substitution in the remainder of the work.
However, if $C$ is a spatially local subsystem (instead of a random subsystem), then this replacement will be modified depending on the spatial extent of the operator.

As a final remark, we note that the operator size distribution is directly related to out-of-time-order correlators (OTOCs), a more familiar quantity for probing operator growth \cite{shenker2014black,maldacena2016bound,larkin1969quasiclassical}.
In particular, the average size is equal to a sum of OTOCs between $Q_A$ and $O_i$~\cite{roberts2018operator,qi2019quantum},
\begin{align}
\figbox{.28}{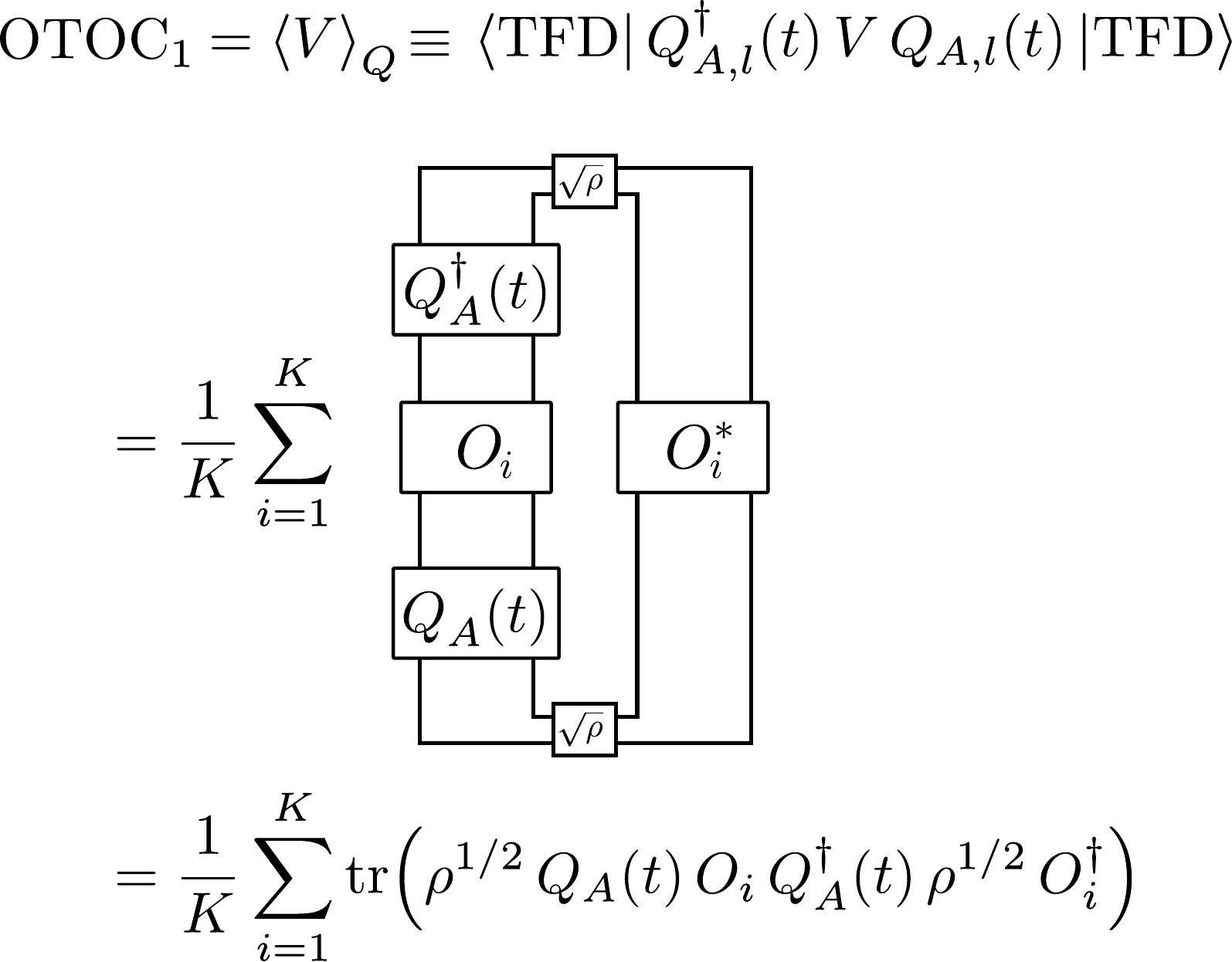} \centering \label{V OTOC}
\end{align}
using Eqs.~(\ref{eq: O slide}-\ref{eq: AB trace TFD}).
%
Higher moments of the size distribution can also be probed by OTOCs, now between $Q_A$ and various products of the $O_i$, e.g. $O_i O_j$ for the size width.
We discuss these relations further, paying particular attention to subtleties that arise at finite temperature, in Section~\ref{interplay}.
%




\subsection{Peaked-size distributions} \label{peaked sizes}


The exponentiated coupling [Eq.~(\ref{coupling size phase})] has a particularly action when the size distribution of $Q_A(t)\rho^{1/2}$ is \emph{tightly peaked} about its average size.
In this regime, each Pauli string gains approximately the same phase, and so the action of the coupling reduces to applying a $Q_A$-dependent overall phase,
\be \label{phase assumption}
e^{igV} Q_{A,l}(t) \ket{\tfd} \approx e^{i g \expval{V}_Q} Q_{A,l}(t) \ket{\tfd},
\ee
where the applied phase is proportional to the average $K$-size [see Eq.~(\ref{size qudit Pauli},~\ref{K size})],
\be \label{V Q}
\begin{split}
g \expval{V}_Q &  = g \bra{\tfd} Q^{\dagger}_{A,l}(t)   V  Q_{A,l}(t)\ket{\tfd} \\
& \approx  g   - \constd g \frac{\Size_K[Q_{A}(t)\rho^{1/2}] }{K}, \\
\end{split}
\ee
defining $\constd \equiv 1/(1-1/d^2)$ for convenience.
%

Corrections to this behavior are controlled by higher moments of the size distribution. 
Focusing on the overlap of the coupled and uncoupled states, the leading order correction is equal to the $K$-size variance, $\delta S_K^2/K^2 = \expval{V^2}_Q - \expval{V}_Q^2$, multiplied by $g^2$:
\be \label{eigV expansion}
\begin{split}
\expval{e^{igV}}_Q =  & \expval{1 + i g V - \frac{1}{2} g^2 V^2 + \ldots}_Q \\
 = &  \bigg( 1 + i g \expval{V}_Q - \frac{1}{2} g^2 \expval{V}_Q^2 + \ldots \bigg) \\
 & - \frac{1}{2} g^2 \bigg(\expval{V^2}_Q - \expval{V}_Q^2 \bigg) + \ldots \\
 =  & \exp \bigg( i g \expval{V}_Q \bigg) - \frac{1}{2} (\constd g)^2 \delta \Size_K^2 / K^2  + \ldots
\end{split}
\ee
The $K$-size variance receives contributions from two sources: the variance of the full size distribution, $\delta \Size^2$, and a statistical error from sampling only $K$ of $N$ qubits for the $K$-size.
If the $K$ qubits are distributed randomly, these errors scale as $\delta \Size_K \sim \delta \Size \cdot (K/N)$
and $\delta \Size_K \sim \sqrt{\Size_K} \approx \sqrt{\Size K / N }$, respectively (see Appendix~\ref{app: RUC} for a detailed derivation of the latter).
These are small compared to the average $K$-size whenever $\delta \Size \ll \Size$ and $1 \ll \Size_K$.

In Appendix~\ref{app: bounds}, we go beyond these leading order corrections and provide quantitative bounds on when the peaked-size approximation in Eq.~\eqref{phase assumption} is valid.
In general, we can strictly prove that this approximation holds whenever there is a parametric separation between an asymptotic size width, defined in the appendix, and the average size.



\section{Peaked-size teleportation} \label{peaked-size}

%
%

Having established general conditions for successful teleportation (Section \ref{sec: requirements}) as well as the connection between the coupling in the TW protocol and operator size distributions (Section \ref{size}), we are now ready to introduce the peaked-size mechanism for teleportation.
%
In this section, we first demonstrate peaked-size teleportation in its simplest context: teleportation of a single qubit at infinite temperature.
We then show that the fidelity of peaked-size teleportation is necessarily suppressed at finite temperature.
For ease of reading, we relegate rigorous results supporting each of the above arguments to the end of the section.
We turn to specific physical systems realizing peaked-size teleportation in the following sections: in Section \ref{late times} we show that peaked-size teleportation of a single qubit occurs in all scrambling systems at late times, while in Section \ref{sec: intermediate time} we show that peaked-size teleportation of multiple qubits occurs in certain systems at intermediate times.

\subsection{Single-qubit teleportation} \label{phase based}


To analyze teleportation of a single qubit, we turn to the two-sided correlators in Eq.~(\ref{C_Q}), 
with $Q_A \in \{\mathbbm{1},X,Y,Z\}$ running over the single-qubit Pauli operators. 
%
We recall that the requirements for teleportation are for all $C_Q$ to have ($i$) maximal magnitude and ($ii$) the same phase.

The first requirement is naturally satisfied at infinite temperature even \emph{before coupling and decoding} but the second requirement is not.
In particular, the four correlators with $D = \mathbbm{1}, \, g = 0$ are:
\begin{center}
\begin{tabular}{ |c|c| } 
\hline
$Q_A$ & $C_Q$ \\
\hline
$\mathbbm{1}$ & $+1$  \\ 
$X$ & $+1$  \\ 
$Y$ & $-1$  \\ 
$Z$ & $+1$  \\ 
 \hline
 \multicolumn{2}{c}{$(D=\mathbbm{1})$} \\
\multicolumn{2}{c}{$(g=0)$}
\end{tabular}
\end{center}
where the left entries are qubit operators, $Q_A$, and the right entries are the correlators, $C_Q$.
The correlators have maximal magnitude because each operator can be transferred perfectly from left to right using Eq.~(\ref{eq: O slide}).
However, the $Y$ operator picks up an overall minus sign during this process, since $Y^T = - Y$, and so the correlator phases are not aligned.
One can verify the resulting teleportation fidelity is indeed trivial.
Our goal will be to show that the action of the coupling in Eq.~(\ref{phase assumption}), as well as a simple decoding operation, are sufficient to align the four phases.

To begin, we assume that all time-evolved Pauli operators have a tightly peaked size distribution and that the average size $\Size$ is the same for all non-identity operators.
From Eqs.~(\ref{phase assumption}-\ref{V Q}), we have that the coupling applies a total phase difference $\constd g \Size/N$ between the thermofield double state (the identity operator; size zero) and all perturbed states (time-evolved Pauli operators; size $\Size$).
Our table of correlator phases is thus modified to:
\begin{center}
\begin{tabular}{ |c|c| } 
\hline
$Q_A$ & $C_Q$ \\
\hline
$\mathbbm{1}$ & $+1$  \\ 
$X$ & $+1$  \\ 
$Y$ & $-1$  \\ 
$Z$ & $+1$  \\ 
 \hline
\multicolumn{2}{c}{$(D=\mathbbm{1})$} \\
\multicolumn{2}{c}{$(g=0)$} \\
\end{tabular}
$\longrightarrow$
\begin{tabular}{ |c|c| } 
\hline
$Q_A$ & $C_Q$ \\
\hline
$\mathbbm{1}$ & $e^{-i\constd g \Size / N}$  \\ 
$X$ & $+1$  \\ 
$Y$ & $-1$  \\ 
$Z$ & $+1$  \\ 
 \hline
 \multicolumn{2}{c}{$(D=\mathbbm{1})$} \\
\multicolumn{2}{c}{$(g\neq0)$}
\end{tabular}\\
\end{center}
We again do not achieve perfect phase alignment.
However, we can now correct the misaligned phases using the decoding operator, $D = Y$.
This applies an additional minus sign to the $X$ and $Z$ correlators:
\begin{center}
\begin{tabular}{ |c|c| } 
\hline
$Q_A$ & $C_Q$ \\
\hline
$\mathbbm{1}$ & $+1$  \\ 
$X$ & $+1$  \\ 
$Y$ & $-1$  \\ 
$Z$ & $+1$  \\ 
 \hline
  \multicolumn{2}{c}{$(D=\mathbbm{1})$} \\
\multicolumn{2}{c}{$(g=0)$}
\end{tabular}
$\longrightarrow$
\begin{tabular}{ |c|c| } 
\hline
$Q_A$ & $C_Q$ \\
\hline
$\mathbbm{1}$ & $e^{-i\constd g \Size / N}$  \\ 
$X$ & $+1$  \\ 
$Y$ & $-1$  \\ 
$Z$ & $+1$  \\ 
 \hline
  \multicolumn{2}{c}{$(D=\mathbbm{1})$} \\
\multicolumn{2}{c}{$(g\neq0)$}
\end{tabular}
$\longrightarrow$
\begin{tabular}{ |c|c| } 
\hline
$Q_A$ & $C_Q$ \\
\hline
$\mathbbm{1}$ & $e^{-i\constd g \Size / N}$  \\ 
$X$ & $-1$  \\ 
$Y$ & $-1$  \\ 
$Z$ & $-1$  \\ 
 \hline
  \multicolumn{2}{c}{$(D=Y)$} \\
\multicolumn{2}{c}{$(g\neq0)$}
\end{tabular}
\end{center}
The correlator phases are now aligned whenever
\be \label{eq:phase-matching}
\constd g  \, \frac{\Size}{N} = \pi \text{ mod } 2\pi,
\ee
leading to perfect teleportation at these values.

%

\subsection{Peaked-size teleportation at finite temperature} \label{finite temperature}

%
There are two important modifications to peaked-size teleportation at finite temperature.
First, the relevant notion of operator size is modified~\cite{qi2019quantum}.
In particular, in the peaked-size regime,  the difference in phase applied between the identity and non-identity Pauli operators is modified to
\be \label{finite temperature size}
\Size[Q_A(t)] \rightarrow \Size[Q_A(t) \rho^{1/2}] - \Size[\rho^{1/2}].
\ee

Second, the maximal fidelity of peaked-size teleportation is reduced at finite temperature.
%
%
In particular, when sizes are tightly peaked, the two-sided correlators factorize into a constant magnitude multipled by an overall phase:
\be \label{C2}
\begin{split}
C_Q & = \bra{\tfd} \tilde Q^{\dagger}_{A,r} Q_{A,l}\ket{\tfd} e^{i(g- \constd g \Size_K[Q_A(t) \rho^{1/2}]/K)}\\
 & =G_\beta(Q_A) \cdot e^{i \theta_Q} \\
\end{split}
\ee
where $\theta_Q$ combines the effects of transposition, coupling, and decoding, and the correlator magnitude corresponds to an imaginary-time Green's function, 
\begin{equation}\label{Gbeta}
    G_\beta(Q_A) \equiv \text{tr}(  Q_A^{\dagger} \, \rho^{1/2} Q_A \, \rho^{1/2} ) \leq 1.
\end{equation}
\vspace{2mm}
This Green's function is unity at infinite temperature and generically decreases at finite temperatures, due to the reduced entanglement of the TFD state.
This violates the maximal magnitude requirement for teleportation, and therefore leads to a corresponding decrease in the teleportation fidelity.

The astute reader will recall that finite temperature teleportation is known to succeed with $\mathcal{O}(1)$ fidelities (i.e. higher than $G_\beta$) in theories with a gravity dual~\cite{gao2017traversable,maldacena2017diving,gao2019traversable}; this is a signature of physics outside the peaked-size regime, which we connect to in Section~\ref{interplay}.

\subsection{Rigorous expressions for teleportation fidelity} \label{sec: rigorous}

We now derive formal expressions of the teleportation fidelity for $n$ teleported qubits as a function of the correlator phases.
To do so, we consider a variant of the protocol where instead of teleporting a quantum state we attempt to distill an EPR pair:
\begin{equation}
\figbox{.7}{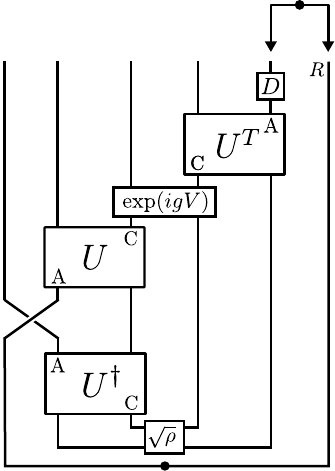} \centering \label{TWH protocol epr}
\end{equation}
Here state insertion is replaced by swapping in one ``half'' of an EPR pair with a reference subsystem R (far right) into subsystem A of the left side.
When subsystem A is teleported from left to right, the circuit results in an EPR pair between the reference subsystem R and subsystem A of the right (top arrows).
The fidelity of EPR distillation is precisely related to the average fidelity of state teleportation~\cite{yoshida2019disentangling}, $F_{\text{EPR}} = [(d_A+1) \langle F_\psi \rangle - 1]/d_A$, where $d_A = 2^{n}$ is the dimension of subsystem A when teleporting $n$ qubits.

We calculate the teleportation fidelity by Pauli decomposing the SWAP operator as $\SWAP = \sum_{Q_A} Q_A \otimes Q_A^{\dagger} / d_A$.
This gives:
\begin{widetext}
\begin{align}
\figbox{.49}{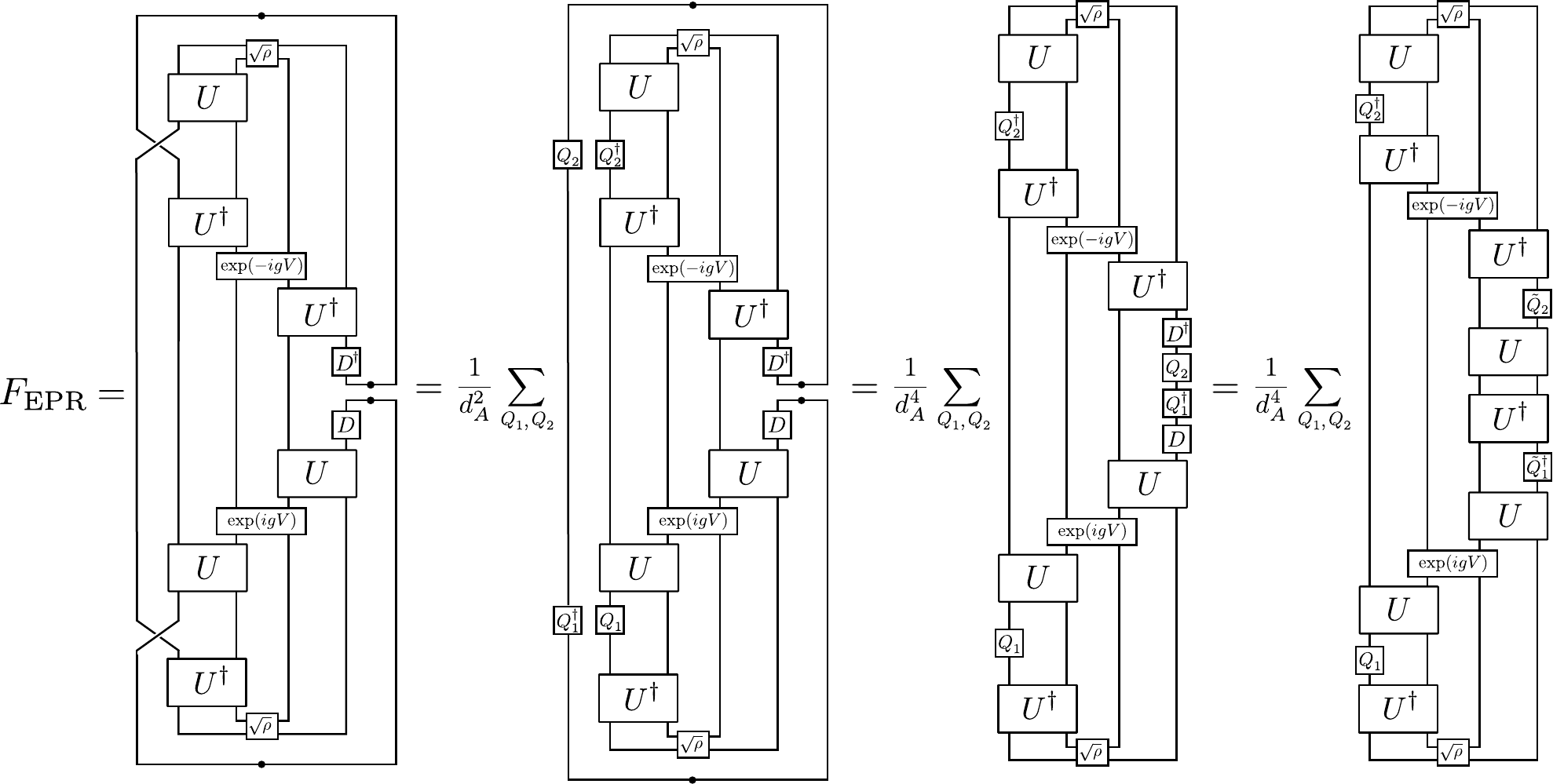} \centering \label{eq: EPR-fidelity}.
\end{align}
\end{widetext}
where the third equality utilizes the diagrammatic identities Eqs.~(\ref{eq: O slide},~\ref{eq: AB trace}), and the fourth equality inserts the identity, $\mathbbm{1} = D_r U_r U_r^\dagger D_r^\dagger$, in the center of the right side (recall our notation $\tilde{Q}_{1/2} = D^\dagger Q_{1/2} D$).
Writing the rightmost diagram as an equation, we have:
\be \label{F_EPR eq}
\begin{split}
F_{\epr}  = & \frac{1}{d_A^4} \sum_{Q_1,Q_2} \bra{\tfd} Q_{2,l}^{\dagger}(t)\,e^{-igV} \,  \tilde{Q}_{2,r}(-t) \\
& \times \tilde{Q}_{1,r}^{\dagger}(-t) \, e^{igV}\,Q_{1,l}(t) \ket{\tfd}.
\end{split}
\ee
Similar expressions for teleportation of quantum states are contained in Appendix~\ref{app: state fidelity}.

In general, the teleportation fidelity and two-sided correlators are related only by a lower bound,\footnote{Under special circumstances, namely large-$N$ models, one may be able to factorize the above expression in terms of  correlators of the form Eq.~(\ref{C_Q})~\cite{gao2019traversable}.}
\be \label{F bound}
F_{\epr} \geq \bigg| \frac{1}{d_A^2} \sum_{Q_A} C_{Q} \bigg|^2.
\ee
This is obtained diagrammatically by inserting the projector, $\dyad{\tfd}$, into the center of Eq.~(\ref{eq: EPR-fidelity}):
\begin{equation}
\figbox{.53}{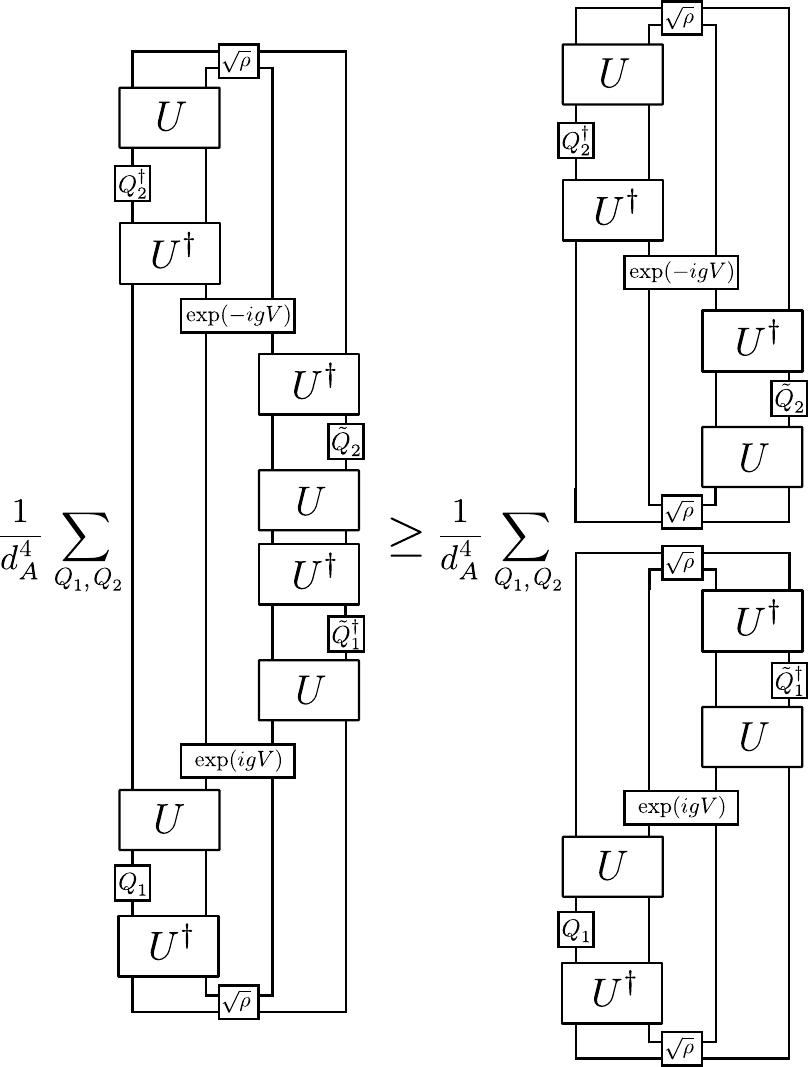} \centering \label{fidelity correlator}
\end{equation}
A similar bound was obtained in Ref.~\cite{brown2019quantum,brown2020quantum}, conditional on certain assumptions about operators' size distributions.

At infinite temperature \emph{in the peaked-size regime}, we have $C_Q = e^{ i \theta_Q}$ and the fidelity is equal to the lower bound:
\be \label{F phases}
\begin{split}
F_{\epr}  
 = \frac{1}{d_A^4} \sum_{Q_1,Q_2} e^{i(\theta_{Q_1}-\theta_{Q_2})} 
 = \bigg| \frac{1}{d_A^2} \sum_{Q_A} e^{i\theta_{Q}} \bigg|^2.  \\
\end{split}
\ee
The sum is over $d_A^2$ terms, and is unity only when all the operators' phases are the same.
In the case of a single-qubit teleportation at infinite temperature in the peaked-size regime, plugging the final table of Section \ref{phase based} into the above equation gives a fidelity:
\begin{equation} \label{eq: F peaked-size}
    F_{\epr} = \frac{5}{8} - \frac{3}{8} \cos( \constd g \Size / N),
\end{equation}
which oscillates between trivial fidelity ($F_{\text{EPR}} = 1/4$) and unity as a function of the operators' size.
At finite temperature in the peaked-size regime, we instead find
\be
\begin{split}
F_{\epr}  & = \frac{1}{d_A^4} \sum_{Q_1,Q_2}  e^{i(\theta_{Q_1}-\theta_{Q_2})}  \tr (  Q_2^{\dagger} Q_1 \, \rho^{1/2}  Q_1^{\dagger} Q_2 \, \rho^{1/2} ) \\
& \leq \frac{1}{d_A^4} \sum_{Q_1,Q_2}  \tr (  Q_2^{\dagger} Q_1 \, \rho^{1/2}  Q_1^{\dagger} Q_2 \, \rho^{1/2} ) \\
& =  \frac{1}{d_A^2} \sum_{Q_A}  \tr (  Q_A \, \rho^{1/2}  Q_A^{\dagger} \, \rho^{1/2} ) \\
& =  \frac{1}{d_A^2} \sum_{Q_A}  G_\beta(Q_A).  
\end{split}
\ee
where the maximum fidelity is again achieved when the correlator phases align. 
However, its value is now less than unity, and instead is equal to a sum of various imaginary time Green's functions, i.e. the correlator magnitudes [Section \ref{finite temperature}, Eq.~(\ref{Gbeta})].

%

\section{Peaked-size teleportation at late times}\label{late times}

We now introduce the simplest physical example of peaked-size teleportation: teleportation in any scrambling system at late times (after the scrambling time).
There are two distinguishing features of this regime: ($i$) the circuit can only teleport a single qubit, i.e. the channel capacity is one, and ($ii$) as for all peaked-size teleportation, the teleportation fidelity is suppressed at low temperatures.
We also demonstrate that this regime of peaked-size teleportation, as well as the full quantum circuit implementing the TW protocol, are equivalent to HPR teleportation of a single qubit.
In Section~\ref{sec: intermediate time}, we will demonstrate that the single-qubit late time channel capacity can be overcome at intermediate times in many scrambling systems.

\subsection{Teleportation at late times}\label{sec: late-times}

At late times, the dynamics of a scrambling system can be approximated by a Haar random unitary\footnote{This approximation is modified in systems with a conserved quantity. Size distributions in such systems have been considered in Refs.~\cite{rakovszky2018diffusive,khemani2018operator,rakovszky2020dissipation}; at late times (after conserved quantities have diffused across the entire system), they are expected to be similar to size distributions without a conserved quantity, up to corrections $\sim 1/N$.}~\cite{hayden2007black,roberts2017chaos}.
In this case, each time-evolved operator, $Q_A(t)$, becomes a sum of random Pauli strings, each with probability $1/d^2$ to be the identity at any individual site. 
As a result, time-evolved operators have an average size,
\be
\Size \approx (1 - 1/d^2) N,
\ee
and a size width,
\be \label{delta size late time}
\delta \Size \sim \sqrt{N},
\ee
where the scaling is based on the central limit theorem.
The $K$-size distribution takes the same form, replacing $N$ with $K$, and is tightly peaked as long as $K$ is large (specifically, $ g \delta \Size_K / K \approx g / \sqrt{K} \ll 1$).

For simplicity, we will focus on late time teleportation at infinite temperature; finite temperature modifications follow according to Section~\ref{finite temperature}.
Using Eqs.~(\ref{phase assumption}-\ref{V Q}), we find that the coupling applies a relative phase $e^{i g}$ between the identity operator (size zero) and all non-identity Pauli operators (size above)~\cite{maldacena2017diving}:
\be \label{interference regime}
\begin{split}
e^{i g V} \ket{\epr} & = e^{i g} \ket{\epr} \\
e^{i g V} Q_{A,l}(t) \ket{\epr} & = Q_{A,l}(t) \ket{\epr}.
\end{split}
\ee
The lack of an applied phase for non-identity Pauli operators corresponds to the vanishing of $\expval{V}_Q$ at late times, when OTOCs have decayed to zero [see Eq.~(\ref{V Q})].
From Section~\ref{phase based}, we see that whenever
\be
 g = \pi \text{ mod } 2\pi,
\ee
single-qubit teleportation succeeds.

%
A brief argument shows that late time teleportion of higher dimensional quantum states is not possible.
Consider teleportation of a $d$-dimensional qudit, with a basis of states $\ket{i}$, $i = 0, \ldots , d - 1$.
The qudit Pauli operators are generated by the `clock' and `shift' operators: $Z \ket{ i } = e^{i \omega} \ket{i}$, with $\omega = 2\pi/d$, and $X \ket{i} = \ket{i + 1}$.
The two generators obey the commutation relation, $X Z  = e^{-i \omega} Z X$.
After transposition, each Pauli operator, $X^p Z^q$, becomes
\be
(X^p Z^q)^T = Z^{T,q} X^{T,p} = Z^{q} X^{-p} = e^{ - i p q \omega} X^{-p} Z^q.
\ee
Meanwhile, late time dynamics ensure that the coupling applies an overall phase only to the identity operator.
For teleportation to be successful, we would therefore require a decoding operation, $D$, that acts as $D X^{-p} Z^q D^{\dagger} \sim X^p Z^q$.
%
%
Suppose there was such a unitary operator\footnote{The astute reader may note that this operation is in fact implemented by the \emph{anti-unitary} operator, $D \ket{i} = \ket{-i \text{ mod } d}^*$. However, if one decomposes state insertion in terms of Pauli operators as $\dyad{\psi}{\phi} = \sum_{Q_A} c_Q Q_A$ (see Section~\ref{sec: requirements}), one desires that the entire operator $\dyad{\psi}{\phi}$ be transferred from left to right for all possible $\bra{\phi}$. The preceding anti-unitary operator will complex conjugate the coefficients $c_Q$, thus spoiling teleportation for any $\bra{\phi}$ where these are complex.}, and consider its action on the generators: $D X D^{\dagger} = X^{-1}$ and $D Z D^{\dagger} = Z$.
The above action implies that commuting the two generators gives a different phase before and after decoding: $D X Z D^{\dagger} = e^{- i \omega} D Z X D^{\dagger} = e^{- i \omega} Z X^{-1}$ and $D X Z D^{\dagger} = X^{-1} Z = e^{+ i \omega} Z X^{-1}$.
This is a contradiction whenever $e^{+ i \omega} \neq e^{-i \omega}$, i.e. whenever $d > 2$.


\subsection{Equivalence to HPR protocol} \label{sec: YK}

We now turn to the equivalence between peaked-size teleportation and teleportation in the HPR protocol.
The latter was originally introduced to recover information in the Hayden-Preskill thought experiment~\cite{hayden2007black,yoshida2017efficient}, and is reviewed in detail in Appendix~\ref{app: YK}.

Here, we restrict our attention to teleportation in the deterministic variant of the protocol, of a single qubit at infinite temperature~\cite{yoshida2017efficient,landsman2019verified}.
The protocol takes the form:
\begin{align}
\figbox{.8}{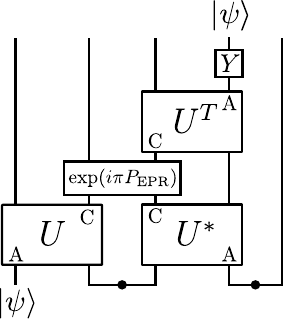} \centering \label{eq: Grover one step simplified}
\end{align}
where $P_{\epr}$ projects onto an EPR pair between subsystems C on the left and right sides.
%

%
%
%
%

%

The equivalence between this protocol and the TW protocol [Eq.~(\ref{TWH protocol Fig1})] is manifest, with the only difference being the locality of the coupling.
Specifically, the HPR coupling is of the same general form as the TW coupling [Eq.~(\ref{eq: V})]:
\be \label{V YK}
g V \equiv \pi P_{\epr} = \frac{\pi}{d_C^2} \sum_{P_C} P_{C,l} \, P^{*}_{C,r},
\ee
where the sum is over of a complete basis of $d_C^2$ Pauli operators on C.
However, the operators $P_C$ are typically non-local across C, whereas the coupling considered in the TW protocol was restricted to local operators. 
%
%
%
As a consequence, the HPR coupling functions as a binary variable measuring whether or not an operator has support on subsystem C (see Section~\ref{size}).
%
In contrast, the TW coupling measures the operator size within C, which takes an approximately continuous range of values when C is large.
Crucially, at late times under scrambling dynamics, the effect of both couplings will be the same: to apply an overall phase to non-identity operators.

A few additional remarks are in order.
First, while the leading order effect of the HPR and TW couplings is the same, they lead to different finite-size corrections.
In particular, in a fully scrambled system, the variance in the phases applied by the HPR coupling is equal to the probability of a random Pauli string not having support on C, which is suppressed exponentially in the size of C, i.e.~$1/d_C^2$.
On the other hand, the variance in phases applied by the TW coupling is suppressed only polynomially, by $\sim  g^2 \delta \Size_K^2 / K^2 \sim g^2 \Size_K / K^2 \sim g^2 / K$ [see Eq.~(\ref{delta size late time}) and the discussion below Eq.~(\ref{eigV expansion})].
These enhanced phase fluctuations are relevant for finite-size implementations of the TW protocol, as discussed further in Section~\ref{experiment}.

Second, it has previously been shown that an extended version of the HPR protocol allows for teleportation of \emph{multiple} qubits at late times~\cite{yoshida2017efficient}.
Because of the equivalence between the protocols, this extension would also allow for multi-qubit teleportation via the peaked-size mechanism.
However, the enhanced channel capacity comes with a trade-off: the circuit complexity (measured by the number of applications of the unitary $U$) grows exponentially in the number of qubits to be teleported.
As we will see in the following section, this limitation can be overcome by peaked-size teleportation in the TW protocol at intermediate times, owing to the locality of the TW coupling. 



\section{Peaked-size teleportation at intermediate times} \label{sec: intermediate time}

\FigureRUC

We now turn to analyzing the behavior of peaked-size teleportation at intermediate times, i.e.~before the scrambling time.
In this regime, multiple qubits can be teleported given a certain condition on the growth of time-evolved operators, namely when the overlap of the operators' support is sufficiently small.

We explicitly demonstrate that this condition is satisfied, and multi-qubit teleporation is possible, in a wide variety of physical systems at infinite temperature.
These include random unitary circuits (RUCs) in $\ge$1D, for which peaked sizes naturally occur due to local thermalization within each operator's light cone, and time-evolved operators are non-overlapping due to spatial locality.
More surprisingly, we show that multi-qubit peaked-size teleportation can also be achieved in `fast scrambling', all-to-all coupled systems, including 0D random unitary circuits and the SYK model (at infinite temperature)~\cite{hayden2007black,sekino2008fast}.
In this case, operators are \emph{not} spatially separated at any nonzero time; nonetheless, the overlap of their size distributions remains \emph{probabilistically} small at sufficiently early times.
Furthermore, we demonstrate that while size distributions of local operators are generically \emph{not} tightly peaked in all-to-all systems, peaked size distributions can be engineered in the TW protocol by \emph{encoding} one's initial state into large $p$-body operators.

Finally, we consider the channel capacity---i.e.~the maximum number of qubits that can be teleported (allowing both $g$ and $t$ to vary)---of peaked-size teleportation in all-to-all coupled systems.
This is an essential question for comparing the capabilities of peaked-size teleportation with those of gravitational teleportation in traversable wormholes~\cite{maldacena2017diving}.
Remarkably, we provide analytic and numerical evidence that the channel capacity of peaked-size teleportation in 0D RUCs, a quite simple microscopic system, is asymptotically equivalent to that of the gravitational mechanism! Namely, the number of qubits $n$ that can be teleported scales with the number of couplings in the protocol, $n \sim K$.

\subsection{Multi-qubit teleportation: additive operator sizes} \label{decomposition}

We begin with a few simple examples of multi-qubit teleportation to build intuition.
First, consider a unitary $U$ that factorizes as $U=U_1 \otimes \cdots \otimes U_n$, where each $U_i$ acts on a disjoint subsystem. 
If we insert $n$ qubits individually into the $n$ different subsystems, then the entire protocol decouples into $n$ independent channels and there is no restriction on sending multiple qubits.
This trivial example relies on the fact that $U$ does not scramble information across the entire system but only within each disjoint subsystem.
We see that full scrambling of information by $U$ in fact \emph{inhibits} the teleportation protocol's channel capacity (considered for a fixed set of qubits and dynamics).

A similar situation occurs even when the dynamics are not factorizable, as long as the teleported qubits are in causally separated regions. 
For example, consider a $(\mathcal{D} \geq 1)$-dimensional system with short-range interactions, where the inserted qubits are spatially separated.
At intermediate times, the time-evolved qubit operators will have support within a local `light cone' about their initial location, but will continue to act on disjoint subsystems. 
This scenario is therefore no different from the previous example and multi-qubit teleportation remains possible, as long as
($i$) the size distributions of each operator is tightly peaked,
($ii$) the coupling $V$ has support within each qubit's light cone,
and ($iii$) the light cones of each qubit are non-overlapping.
This final requirement constrains the number of qubits that can be sent at a given time $t$.
In particular, the light cone of each operator will have a radius $v_B t$ where $v_B$ is the butterfly velocity.
The maximum number of non-overlapping light cones---equal to the total number of qubits $n$ that can be teleported---is therefore $n \lesssim N/(v_B t)^\mathcal{D}$, where $N$ is the total system volume.

More formally, we can analyze the success of $n$-qubit teleportation using the two-sided correlators, $C_Q$.
We are concerned with $n$-qubit operators $Q(t) = Q_1(t) \ldots Q_n(t)$, where each $Q_i \in \{I,X,Y,Z\}$ is a single-qubit Pauli on the $i^{\text{th}}$ teleported qubit.
We work at infinite temperature and assume that sizes are tightly peaked.
Teleportation therefore succeeds whenever all correlators have the same phase.

Inspired by the example of $n$ decoupled protocols, we will take the decoding operator to be the tensor product, $D = Y \otimes \ldots \otimes Y$. 
The combination of transposition and conjugation by $D$ thus applies a minus sign to every single-qubit non-identity Pauli operator. 
An additional phase is applied by coupling proportional to the size of each operator. 
For example, for $n=2$ qubits, we have:
\begin{center}
\begin{tabular}{ |c|c| } 
\hline
$Q_A$ & $C_Q$ \\
\hline
$\mathbbm{1} \otimes \mathbbm{1}$ & $1$  \\ 
$Q_1  \otimes \mathbbm{1} $ & $-1 \times e^{-i \constd g \Size_1 / N}$  \\ 
$\mathbbm{1} \otimes Q_2$ & $-1 \times e^{-i \constd g \Size_2 / N}$  \\ 
$Q_1 \otimes Q_2$ & $(-1)^2 \times e^{-i \constd g \Size_{12} / N}$  \\ 
 \hline
\end{tabular}
\end{center}
where $\Size_i$ and $\Size_{ij}$ are shorthand for $\Size[Q_i(t)]$ and $\Size[Q_{i}(t)Q_{j}(t)]$.
In order for all correlators to have the same phase, we require that $\constd g \Size_1 / N= \constd g \Size_2 / N = \pi \text{ mod } 2\pi$, and that the operator sizes \emph{add},
such that $e^{-i \constd g \Size_{12} / N} \approx e^{-i\constd g(\Size_{1}+\Size_{2})/N} = e^{i(\pi+\pi)} = 1$.
%

This requirements generalize straightforwardly to $n$ qubits.
Specifically, teleportation succeeds whenever the single-qubit operator sizes obey $\constd g \Size_i / N = \pi \text{ mod } 2\pi$ and the multi-qubit operator sizes \emph{add} under operator multiplication:
\be \label{eq: size decompose}
\begin{split}
\Size[Q_1(t) & Q_2(t) \ldots Q_n(t)] \\
& \approx \Size[Q_1(t)] + \Size[Q_2(t)] + \ldots + \Size[Q_n(t)].
\end{split}
\ee
This latter requirement implies that the phases applied by the coupling, $e^{igV}$, factorize, and allows the $n$ qubits to be teleported `in parallel' as in the previous simple examples. 

The size addition requirement naturally bounds the channel capacity in terms of the number of couplings, $K$.
%
Specifically, the $K$-size takes integer values between $1$ and $K$.
However, the requirement that all three single-qubit Pauli operators have the same $K$-size increases the minimum $K$-size to $2$.
From Eq.~(\ref{eq: size decompose}), this implies that an $n$-qubit operator has a $K$-size of at least $2n$, which is only possible if
\be \label{strict cc}
2n \leq K.
\ee
Indeed, this strict upper bound can also be understood from an information theoretic perspective: teleporting $n$ qubits requires an increase of $2n$ in the mutual information between the left and right sides of the system.
Each of the $K$ classical bits sent from left to right in Fig.~\ref{fig: 1}(a) increases the mutual information by at most $1$, so at least $2n$ bits are required.

\subsection{$\geq$1D random unitary circuits}\label{geq1D}


As a first concrete example of intermediate time peaked-size teleportation, we consider a random unitary circuit (RUC) applied to a lattice of $N$ qubits in one or higher dimensions.
At each time step, pairs of neighboring qubits are evolved via independent Haar random unitaries arranged in a `brick-layer' fashion, with periodic boundary conditions [Fig.~\ref{fig: RUC}(a,b)].
Operator growth in such systems has been studied at great length, and is believed to be a good model for many aspects of information scrambling under Hamiltonian dynamics~\cite{nahum2018operator, von2018operator,khemani2018operator,rakovszky2018diffusive,li2018quantum,skinner2019measurement}.
We extend these previous studies by demonstrating new results on the behavior of the operator size \emph{width}---i.e.~power-law scaling at intermediate times and suppression at late times---which we show can be detected by the teleportation fidelity (Fig.~\ref{fig: size width}).
%
%

A key property of Haar random unitary circuits is that the expectation values of many circuit quantities can be computed by replacing the Haar random unitaries with randomly chosen \emph{Clifford} unitaries, thereby enabling efficient classical simulation~\cite{dankert2009exact,nahum2018operator}. 
Generally, this equivalence holds for any quantity that contains no more than \emph{two} copies each of $U$ and $U^\dagger$ (e.g. the Renyi-2 entropy, or the OTOC); however, for systems of qubits, this property holds for up to three copies~\cite{webb2015clifford,kueng2015qubit,zhu2015multiqubit}.
From Eq.~\eqref{F_EPR eq}, we see that the teleportation fidelity contains three copies of $U$ and $U^{\dagger}$, so the average fidelity is efficiently simulable\footnote{For higher-dimensional qudits, while we cannot efficiently simulate the teleportation fidelity, we can still calculate the correlators Eq.~(\ref{C_Q}), which lower bound the fidelity via Eq.~(\ref{F bound}).}. 
Moreover, by definition, the size distributions of operators under Clifford dynamics are perfectly tightly-peaked, since a Pauli operator $Q_A$ evolved under a Clifford unitary remains a single Pauli string.
Hence, the teleportation fidelity can be computed using the simplified expression given in Eq.~(\ref{F phases}). 
%

In more detail, we calculate the average EPR fidelity for teleporting $n$ qubits through the following procedure.
First, we choose a particular realization of $U$ by sampling each 2-qubit unitary from a uniform distribution of 2-qubit Clifford unitaries.
Second, we determine the $K$-size of $U Q_A U^\dagger$ for each $n$-qubit Pauli operator, $Q_A$, or, if $n$ is large, for a random subset of these operators; such simulations can be performed efficiently with a time cost that scales linearly with the circuit depth.
Third, we compute the fidelity for a given coupling $g$ using Eq.~\eqref{F phases}, with the phases $\theta_Q = \constd g \mathcal{S}_K / K + \pi \Size[Q_A(0)]$, where the latter term captures the fact that decoding and transposition apply a minus sign for each non-identity element of the initial $Q_A$.
Finally, we average the EPR fidelity over multiple realizations of $U$.

The results of these simulations for $n=1$ qubit in 1D and 2D are shown in Fig.~\ref{fig: RUC}(a,b).
As expected, the average operator size grows ballistically, $\Size \propto t^\mathcal{D}$, until the operator's light cone reaches the edge of the system, at which point the size saturates to $3/4N$. 
While the behavior of the size width is more complex, in both dimensionalities it grows more slowly than the average size. 
This implies that the size distribution is tightly-peaked and the teleportation fidelity can be approximated by $F= \frac 5 8 - \frac 3 8 \cos(\constd g\Size/N)$ [Eq.~\eqref{eq: F peaked-size}].
We verify that the time profile of the fidelity follows this prediction, and nearly perfect fidelity is achieved when $\constd g \Size/N = \pi \mod 2\pi$.
In Appendix~\ref{app: RUC-numerics}, we also demonstrate that teleportation of $n > 1$ qubits is also possible at intermediate times, as long as their light cones do not overlap.
        
\FigureSizeWidth

%
\emph{Probing the size width---}Let us now turn to the time profile of the size width, which exhibits a \emph{peak} near the scrambling time in both 1D and 2D.
Qualitatively, this behavior arises from fact that the size width receives contributions from two sources: the interior of the light cone, and the boundary of the light cone.
Within the light cone, we expect a $\geq$1D system with a small local Hilbert space to `locally thermalize' as the operator spreads.
This implies that the bulk's contribution to the size width scales as $\delta \mathcal{S}_\textrm{bulk} \propto \sqrt{\mathcal{S}} \propto t^{\mathcal{D}/2}$ and saturates at the scrambling time.
Second, the size width also receives contributions from the light cone's boundary, which has not yet thermalized.
%
At late times, the boundary of the light cone reaches the edge of the system and these additional contributions subside, leading to the peak in the size width at the scrambling time. 

To quantify these effects, we note that the growth of operators in $\geq$1D RUCs is predicted to fall in the 
Kardar–Parisi–Zhang (KPZ) universality class~\cite{kardar1986dynamic,nahum2018operator}. 
In 1D, fluctuations in the light cone boundary have been verified numerically to have a growing width $\sim \! t^{\alpha}$ with the KPZ growth exponent $\alpha = 1/2$~\cite{nahum2018operator}.
This implies that the contribution of the boundary to the size width is $\delta S_\textrm{boundary} \propto t^{1/2}$, and the full width is 
\begin{align} \label{eq:kpz_1d}
\delta S = \left\{ \begin{array}{cc} 
                (\alpha_\textrm{bulk} + \alpha_\textrm{boundary})t^{1/2}, & \hspace{5mm} t\lesssim t_\textrm{scr} \\
                \alpha_\textrm{bulk}t_\textrm{scr}^{1/2}, & \hspace{5mm} t \gtrsim t_\textrm{scr}
                \end{array} \right.
\end{align}
We note that the maximum size width relative to the late-time size width is a constant set by $(\alpha_\textrm{bulk} + \alpha_\textrm{boundary})/\alpha_\textrm{bulk}$.
Comparing the size width of multiple system sizes, we observe excellent agreement with predicted scalings over a wide range of system sizes (Appendix~\ref{app: RUC-numerics}). 
%

The time profile of the size width is directly observable in the peaked-size teleportation fidelity if we scale $g \sim t_{\text{scr}}^{1/2} \sim N^{1/2}$.
In particular, by setting $N/g$ to lie between the maximum size width and the late time size width, we observe a distinct decay-revival profile for the teleportation fidelity (Fig.~\ref{fig: size width}).
At early times, we observe successful teleportation with an oscillating fidelity.
%
The fidelity decays slowly, as a power law in time, as it receives corrections proportional to the growing size variance $\sim g^2 \delta \Size^2 /N^2$.
After the scrambling time, we see a revival in the teleportation fidelity as the size width narrows.
The lack of a parametric separation between the maximum and late time size widths means that late time teleportation will also have some finite error for this value of $g$.

In 2D, we find that the scaling of the size width also matches predictions from the KPZ universality class.
In this case, the width of the boundary scales as $\sim t^\alpha$, with $\alpha = 1/3$~\cite{nahum2018operator}. However, to calculate the boundary's contribution to the size width, one must take into account two additional considerations.
First, the boundary is 1-dimensional, so its length trivially grows in time as $\sim t$.
Second, fluctuations of the boundary are expected to have a finite correlation length, $\xi \sim t^{1/z}$, where $z = 3/2$ is the KPZ dynamic exponent~\cite{corwin2012kardar}.
Thus, the boundary can be modeled as $n_\xi \sim t/\xi=t^{1/3}$ uncorrelated regions, each of length $\xi$.
Each region contributes $\sim \xi t^\alpha$ to the size width; adding the uncorrelated contributions from all regions yields a total size width $\delta \Size \sim \sqrt {n_\xi} \, \xi \, t^\alpha = t^{1/6+2/3+1/3} = t^{7/6}$.

The time profile of the size width in 2D is thus given by
\begin{align} \label{eq:kpz_2d}
\delta S = \left\{ \begin{array}{cc} 
                \beta_\textrm{bulk}t + \beta_\textrm{boundary}t^{7/6}, & \hspace{5mm} t\lesssim t_\textrm{scr} \\
                \beta_\textrm{bulk}t_\textrm{scr}, & \hspace{5mm} t \gtrsim t_\textrm{scr}
                \end{array} \right.
\end{align}
We confirm these scalings in our numerics (Fig.~\ref{fig: RUC}(b) and Appendix~\ref{app: RUC-numerics}).
Notably, the size width is now dominated by the boundary contribution at intermediate times, such that the ratio of the maximum size width to the late time size width scales as $t^{1/6}_\textrm{scr} \sim N^{1/12}$.
%
%
As in 1D, one can probe this behavior using the peaked-size teleportation fidelity, now with $g \sim N / t_{\text{scr}}^{7/6} \sim N^{5/12}$.
We emphasize that in 2D, the scaling of the size width is determined by \emph{correlations} between different points on the light-cone boundary.
This goes beyond the behavior studied in previous works on RUCs, which focus on quantities probed by local OTOCs.

\subsection{0D random unitary circuits}\label{0D RUCs}


We now turn to random unitary circuits in zero dimensions, a prototypical model for `fast scramblers'~\cite{hayden2007black,sekino2008fast}.
These circuits are constructued as follows: at each time-step, we partition the $N$ qubits into randomly chosen pairs, and apply independent Haar random 2-qubit unitaries to each pair.

Below we analyze such circuits using theoretical arguments, in combination with numerical simulations via Clifford circuits. 
As the later parts of our analysis are rather technical, we briefly summarize the main results: ($i$) peaked size teleportation remains possible but only if the input state is initially encoded in non-local, $p$-body operators; ($ii$) even though there is no complete separation of operator light cones, size addition still occurs at intermediate times in a probabilistic sense and enables mutli-qubit teleportation; and ($iii$) the maximum channel capacity is linear in the number of coupled qubits, $K$.
These results are depicted numerically in Fig.~\ref{fig: RUC}(c) and \ref{fig: capacity}.



\emph{Peaked sizes}---In all-to-all coupled systems, operators are generally expected to grow exponentially in time, $\mathcal{S} \sim e^{\lambda t}$, where $\lambda$ is the Lyapunov exponent~\cite{sekino2008fast}. 
The reason is simple: at each time step, every term in an operator---rather than just those on a `light-cone' boundary---has a fixed probability of spreading under  random pairwise unitaries. 
A somewhat less intuitive expectation is that the size width also generally grows exponentially~\cite{qi2019quantum}.
One way of understanding this is by imagining two realizations of the dynamics: in one realization the initial operator doubles at the first time and in the other it does not. 
In effect, the latter system now lags behind the former by one time step, $\Delta t$, and the difference in their sizes at later times will be exponentially magnified, to $e^{\lambda t}( 1 - e^{-\lambda \Delta t})$. 
%

The lack of separation between the size and size width seems to preclude the possibility of peaked-size teleportation at intermediate times.
Nevertheless, we can engineer such a separation by \emph{encoding} the information of each input qubit into $p$-body operators, with $p \gg 1$~\cite{gao2019traversable}.
As an example, consider encoding a single qubit into $p = 5$ qubit operators via
\be
\begin{split}
E ( X \otimes \mathbbm{1} \otimes \mathbbm{1} \otimes \mathbbm{1} \otimes \mathbbm{1}) E^{\dagger} & = Z \otimes X \otimes X \otimes Y \otimes Z  \\ 
E ( Y \otimes \mathbbm{1} \otimes \mathbbm{1} \otimes \mathbbm{1} \otimes \mathbbm{1}) E^{\dagger} & = Y \otimes Z \otimes Z \otimes X \otimes Y  \\ 
E ( Z \otimes \mathbbm{1} \otimes \mathbbm{1} \otimes \mathbbm{1} \otimes \mathbbm{1}) E^{\dagger} & = X \otimes Y \otimes Y \otimes Z \otimes X,  \\ 
\end{split}
\ee
Here, $E$ is a Clifford unitary encoding operation that conjugates state insertion and decoding [explicitly, replacing $U \rightarrow U E, U^* \rightarrow U^* E^*$, and $U^T \rightarrow E^T U^T$ in Fig.~\ref{fig: 1}(a)].
The success of teleportation is now dependent on the size distributions of time-evolved $p$-body operators, $Q_A(t) = U E P E^\dagger U^\dagger$, where $P$ runs over the initial unencoded single-qubit Pauli operators.
As we will soon verify explicitly, before the scrambling time the support of each of the $p$ operators composing $Q_A$ will be approximately non-overlapping, so that their size distributions will convolve.
Thus, the total operator size is multiplied by a factor of $p$ but, through the central limit theorem, the size width is multiplied only by $\sqrt{p}$.
 
In more detail, consider the size growth of an operator, $Q_A$, with initial size  $\Size_0 = p$.
During a single time step, each qubit $i$ in the support of $Q_A(t)$ is paired with another random qubit; for simplicity, we assume the second qubit is outside the support of $Q_A(t)$, which should be valid at times well before the scrambling time.
%
%
Under random two-qubit Clifford time-evolution, $Q_{A}(t)$ grows to have support on both qubits with probability $\nu = 1 - 2(d^2-1)/(d^4-1)$ ($9/15$ for qubits).
The operator size, $\Size_t$, therefore grows stochastically in time, according to
\be
\begin{split}
\Size_{t+1} & = \Size_t + \sum_{i=0}^{\Size_t} s_i \\
& = \Size_t + \text{Bi}_t(\Size_t, \nu) \\
& \approx (1+ \nu) \Size_t + \sqrt{\Size_t \nu (1-\nu)} \, \mathcal{N}_t(0,1)
\end{split}
\ee
where each $s_i$ is a binary random variable that increases the size by $1$ with probability $\nu$ and $0$ with probability $1-\nu$, and $\text{Bi}_t(\Size_t, \nu)$ denotes the binomial distribution with $\Size_t$ trials and probability $\nu$, which we can approximate as a normal distribution, $\mathcal{N}_t(\nu\Size_t,\sqrt{\Size_t \nu (1-\nu)})$.
%
The size at time $t$ can thus be written as a sum of random variables drawn at each time step:
\be
\begin{split}
\Size_{t} \approx & (1+\nu)^t p \\
& + \sqrt{\nu (1-\nu)} \sum_{t'=0}^{t-1}  (1+\nu)^{t-t'-1}  \sqrt{\Size_{t'}} \, \mathcal{N}_{t'}(0,1) \\
\end{split}
\ee
from which we see that the average size grows exponentially in time with Lyapunov exponent $e^{\lambda} = 1 + \nu$.
Deviations arise at each time step $t'$, with typical magnitude $(1+\nu)^{t-t'-1}  \sqrt{\Size_{t'}} \approx (1+\nu)^{t-1-t'/2} \sqrt{p}$.
Since this decays exponentially in $t'$, we can approximate the total variation, $\delta \Size_t$, as the largest term in the sum ($t'=0$), which has magnitude
\be
\delta \Size_t \sim (1+\nu)^{t-1} \sqrt{p} \approx \frac{\Size_t}{\sqrt{p}}.
\ee
As anticipated, the size width is dominated by early time errors that have exponentially grown in time, so that the ratio of the size width to the size remains constant at $\sim 1/\sqrt{p}$ (after some period of growth from its initial value, $0$). 
%
%

To support these claims, we numerically simulate the time-evolved size distribution of operators with an initial size $p \approx 1000$ [Fig.~\ref{fig: RUC}(c)].
As expected, we observe that the average size grows exponentially as $\sim p e^{\lambda t}$ and saturates at a timescale $t^* \sim \log(N/p)$. 
Moreover, the size width grows at the same exponential rate but its magnitude is suppressed by a factor of $\sqrt p$ compared to the average size.  

To verify that this allows for teleportation, we next compute the fidelity for teleporting a single qubit, in the regime $g \gg 1$.
As shown in Fig.~\ref{fig: RUC}(c), teleportation occurs with near perfect fidelity beginning at $t \approx t^* - \log({gp})$, corresponding $g \Size/N \approx 1$. 
Thereafter, the teleportation fidelity decreases exponentially in time, consistent with the increase of the size width.
At time $t \approx t^* - \log({g\sqrt p})$, teleportation stops succeeding entirely, since the size width has reached the limit $\delta \Size/N \sim 1$.
Finally, at late times $t \approx t^* - \log (p)$, the fidelity revives as the system becomes fully scrambled and the operator size width narrows to $\delta \Size \sim \sqrt{\Size}$.

\textit{Size addition---}We now turn to the possibility of teleporting multiple qubits in 0D RUCs. 
Within the peaked-size regime, this reduces to the question of whether operator sizes add according to Eq.~(\ref{eq: size decompose}). 
Satisfying this requirement in all-to-all coupled systems is not as trivial as in $\geq1$D, since time-evolved operators typically act on overlapping subsystems at any finite time.
%
%
Nevertheless, we now provide a simple argument for why size addition holds despite this.

To do so, we model each time-evolved Pauli operator $Q_i(t)$ as an independent random Pauli string of size $\Size[Q_i]$. 
%
Consider two such strings, $P_1$ and $P_2$, with support on regions $A_1$ and $A_2$ and sizes $\Size[P_1] = |A_1|$ and $\Size[P_2] = |A_2|$. The size of the product, $P_1 P_2$,  is the size of the union $A_1 \cup A_2$, minus the number of sites where the two strings overlap and have the same single-qubit Pauli operator. This occurs with probability $1/(d^2-1) = 1/3$ at each site in the region $A_1 \cap A_2$, giving
\be \label{size addition approx}
\begin{split}
\Size[P_1 P_2] & \approx | A_1 \cup A_2 | - \frac{1}{3}  | A_1 \cap A_2 | \\
 & = \Size[P_1] + \Size[P_2] - \frac{4}{3}  | A_1 \cap A_2 |. \\
\end{split}
\ee
The deviation from the simple additive rule $\Size[P_1 P_2] = \Size[P_1] + \Size[P_2]$ is thus controlled by $| A_1 \cap A_2 |$. 
If the Pauli strings $P_1, P_2$ have independently random areas of support, the size of this intersection scales as:
\begin{equation} \label{intersection}
    | A_1 \cap A_2 | \sim \Size[P_1] \Size[P_2] / N,
\end{equation}
which is subleading to $\Size[P_i]$ at intermediate times (when $\Size/N \ll 1$). To derive this, note that the probability for \emph{both} strings to have support on a given qubit is $\sim (\Size[P_1]/N)(\Size[P_2] / N)$; summing over $N$ qubits gives the above result.

For $n$-qubit teleportation, one must consider the combined size, $\Size[P_1\ldots P_m]$, of $m$ independent Pauli strings, where $m$ takes a typical value $m \approx 3n/4$ (a typical $n$-qubit operator has non-identity support on $3n/4$ qubits).
In general, this quantity will receive corrections from $\binom{m}{k}$ different $k$-way intersections of the strings, for all $2 \leq k \leq m$. 
For random Pauli strings, the expected size of these intersections scales as $N \overline{|A_1 \cap \ldots \cap A_k |} = \prod_{i=1}^k \frac{|A_i|}{N} \sim \Size^k/N^{k-1}$, where $\Size \sim |A_i|$ is the typical size of a single Pauli string [see Eq.~(\ref{intersection}) above].
For a given $k$, the correction to size addition will be the sum of $\binom{m}{k} \sim m^k$ different intersections and therefore scales as $m \Size (m \Size/N)^{k-1}$.
%
These corrections can be neglected if they are small compared to the total size; this occurs when $m \Size  \ll N$, which corresponds to a timescale much less than the scrambling time. 

To demonstrate this claim, we numerically simulate the teleportation protocol with $n>1$ qubits in the regime $1 \ll p, np \ll K$ [Fig.~\ref{fig: capacity}].
Analogous to single-qubit teleportation, the teleportation fidelity exhibits oscillations beginning at $t \approx t^* - \log ({gp})$, and vanishes at $t \approx t^* - \log({g\sqrt{pn}})$ due to the growth of the combined size width.
However, in contrast to the single-qubit case, teleportation of multiple qubits is not possible at late times, $t \gtrsim t^* - \log ({gpn})$, as predicted in Section~\ref{late times}.
Interestingly, between these two regimes, we observe a partial revival of the fidelity: this indicates that the operator size widths begin to narrow before the additive condition is completely invalidated.

\emph{Error analysis---}While we have confirmed that multi-qubit teleportation can be achieved in certain ideal limits, a key question remains: how does the maximum number of qubits that can be teleported scale as a function of $K$, i.e. what is the protocol's channel capacity?
%
%
%
To answer this question, we now estimate how deviations from these ideal limits lead to errors in peaked-size teleportation and ultimately constrain the channel capacity.
Throughout this discussion, we assume that the size, $\Size$, is extensive, but $K$ is not; this is the natural regime for probing the channel capacity of the protocol at intermediate times, and is the physical scenario in the context of traversable wormholes~\cite{maldacena2017diving}.
The details of this and the following subsection are quite technical in nature, and may be skipped by most readers.

In summary, we identify four distinct sources of error in the multi-qubit teleportation fidelity, $F = 1 - \epsilon$:
\begin{enumerate}
    \item Errors due to finite $p$: $\epsilon \sim n g^2\Size_K^2/K^2 p$
    \item Errors due to finite $K$: $\epsilon \sim n g^2\Size_K/K^2$
    \item Errors due to imperfect size addition: $\epsilon \sim \big[ n^2 g^2 \Size_K^4/K^4 + \ldots \big]$, where ellipses indicate higher orders in $(n\Size_K/K)^2$
    \item Errors due to fluctuations in size addition: $\epsilon \sim \big[ n^2 g^2 \Size_K^2/K^3 + \ldots \big]$, where ellipses indicate higher orders in $n\Size_K/K$
\end{enumerate}
We discuss each of these errors in detail below.

The first and second sources of error are due to imperfectly peaked $K$-size distributions.
The $K$-size width receives contributions from finite-$p$ corrections, $\sim \Size_K / \sqrt{p}$, and finite-$K$ corrections, $\sim \sqrt{\Size_K}$ [see the discussion below Eq.~(\ref{eigV expansion})].
To translate these into errors in the teleportation fidelity, we multiply the size width by $g/K$ and take the square.
This gives fidelity errors $\sim g^2 \Size_K^2 / p K^2$ and $\sim g^2 \Size_K / K^2$ per teleported qubit.

The third and fourth sources of error arise from imperfect size addition.
This leads both to `systematic' errors, due to the average overlap of operators, as well as `sampling' errors, due to random fluctuations in this overlap.
We begin with the systematic errors: as we recall, the size addition of $m$ time-evolved operators receives corrections from $k$-way overlaps of the operators, each scaling as $\sim m \Size_K (m \Size_K/K)^{k-1}$, for $2 \leq k \leq m$ (rescaling our previous results to the $K$-size instead of the size).
The nonlinear dependence on $m$ indicates that sizes do not add perfectly.
Nevertheless, when teleporting an $n$-qubit initial state for large $n$, we can correct for the above effect at leading order by using a linear approximation for $m^k$ about its typical value, $(3n/4)^k$.
This leads to an effectively smaller operator size, which can be observed in the reduced frequency of the fidelity oscillations for 10-qubit teleportation compared to 1-,3-qubit teleportation in Fig.~\ref{fig: capacity}(a).
The leading errors after this shift are quadratic in $\delta m \equiv m - 3n/4$, which has a typical magnitude $\delta m \sim \sqrt{n}$.
Multiplying by $g/K$ and taking the square, we therefore find multi-qubit fidelity errors $\sim (g \Size_K /K)^2 (n \Size_K / K)^{2k-2}$; at leading order $k = 2$, this gives $\sim n^2 g^2 \Size_K^4 / K^4$.

Finally, each intersection above is subject to additional random fluctuations about its average value.
When operator sizes are much smaller than the system size, we can treat each intersection as arising from a binomial process, in which case fluctuations are proportional to the square root of the intersection's average size (see Appendix~\ref{app: RUC} for a detailed accounting).
These add in quadrature for $\sim n^k$ overlaps, producing a total fidelity error $\sim (g^2/K) (n \Size_K / K )^{k}$.

\emph{Channel capacity---} 
To define the channel capacity of the teleportation protocol, we fix a per qubit error threshold $\epsilon_{\text{th}}$, and determine the maximum number of qubits that can be sent while maintaining a multi-qubit fidelity above this threshold\footnote{We note that this definition of channel capacity differs from more conventional definitions~\cite{nielsen2002quantum}; we do not expect this difference to qualitatively affect the scaling of the channel capacity with $K$, as the fidelity drops off steeply above the capacity [Fig.~\ref{fig: capacity}(b)].}, i.e.~$F \geq 1 - n \, \epsilon_{\text{th}}$.
We are interested in how the channel capacity scales with the number of couplings, $K$, while allowing both $g$ and $\Size_K$ (determined by the evolution time) to vary.

In 0D RUCs, all errors increase with $g$, so it is optimal to set $g$ to its minimal value, $\constd g \Size / N = \pi$.
This gives a per qubit error
\begin{equation}
\begin{split}
    \frac{\epsilon}{n} \sim & \frac{1}{p} + \frac{1}{\Size_K}  + \bigg[ \frac{n \Size_K^2}{K^2} + \ldots \bigg] + \bigg[ \frac{n^2}{K} + \ldots \bigg].
\end{split}
\end{equation}
The first term is negligible in the large $p$ limit and so we will neglect it from here on.

We minimize the remaining terms with respect to $\Size_K$. 
There are two relevant regimes.
For $n \lesssim \sqrt K$, the minimum is determined entirely by the leading order contributions in $n \Size_K / K$ to the error (i.e.~neglecting the ellipses).
Taking the derivative and setting to zero, we have the minimum at $\Size^{(1)}_K \sim K^{2/3}/n^{1/3}$.
As we increase $n$, the optimal size approaches the value $\Size_K^{(2)} \sim K/n$.
At this point, size addition errors of all orders (i.e.~the ellipses) become large, and so the true minimum becomes fixed just below $\Size^{(2)}_K$.
This crossover between these two minima occurs at $n \sim \sqrt{K}$, at which $\Size^{(1)}_K \sim \Size^{(2)}_K$.

The above minima give two distinct scalings for the per qubit error and thus the channel capacity.
The first minimum has a per qubit error $\epsilon^{(1)}/n \sim (n/K^2)^{1/3}$, which gives rise to a superlinear channel capacity, $n \lesssim \epsilon_{\text{th}}^3 K^2$.
However, as we increase $K$, this capacity eventually surpasses the value $\sqrt{K}$.
Above this, the optimal size is given by the second minimum, which has an error $\epsilon^{(2)}/n \sim n/K$, and thus the channel features an asymptotically linear capacity,
\be
n \lesssim \epsilon_{\text{th}} K.
\ee
This is a stronger instance of the strict general bound Eq.~(\ref{strict cc}).
Intuitively, this channel capacity arises because the individual $K$-sizes must be large, $\Size_K \gg 1$, for the $K$-size to be tightly peaked, while at same time the combined $K$-size must be much smaller than $K$, $n \Size_K \ll K$, for the $K$-sizes to add; hence $n \ll K$. 

We test this scaling numerically by simulating the teleportation protocol and measuring the per qubit fidelity, $F^{(1)}_{\textrm{EPR}}$, as a function of $n$ and $K$.
Specifically, for each value of $K$, we sweep the number of qubits $n$ and determine the maximum qubits that can be sent before the infidelity exceeds a threshold, $1 - F^{(1)}_\textrm{EPR} = \epsilon_\textrm{th}$. 
These results are shown in Fig.~\ref{fig: capacity}(b) and exhibit a clear linear trend across two orders of magnitude, confirming our prediction of a linear channel capacity.

A few final remarks are in order.
First, while in principle the per qubit fidelity can be calculated by taking the $n^{\text{th}}$ root of the full $n$-body fidelity, this approach is numerically unstable for large $n$.
Thus, we instead compute the fidelity of a \emph{single} qubit, while trying to send multiple qubits, using an approach derived in Appendix~\ref{app: RUC-numerics}. 
This amounts to performing a sum analogous to Eq.~\eqref{F phases}, but only including pairs of $Q_1$ and $Q_2$ that are equal on all sites except for one.

Second, the range of system parameters that lie within the linear scaling regime is ultimately constrained by the finite total system size, $N=10^8$.
In particular, to maximize the linear scaling regime, we choose $p = 101$ and $\epsilon_\textrm{th} = 0.07$.
The former ensures that finite-$p$ errors are negligible, while the latter allows the number of qubits at the threshold to be large enough to access the $n \gtrsim \sqrt K$ regime but small enough that the operators are initially dilute, i.e.~$n \ll N/p$.

\FigureCapacity

\subsection{Large-$q$ SYK model: infinite temperature} \label{sec: SYK infinite temperature}

We now demonstrate peaked-size teleportation in a 0D Hamiltonian system, the large-$q$ SYK model, at infinite temperature.
While teleportation at low temperatures in the SYK model is known to succeed via the gravitational mechanism, teleportation at infinite temperature was discovered only recently~\cite{gao2019traversable}.
In addition to showing that this mechanism is in fact peaked-size teleportation, we also find that, remarkably, \emph{all} qualitative aspects of this teleportation match those of 0D RUCs.

The large-$q$ SYK model is defined by the Hamiltonian \cite{maldacena2016remarks,qi2019quantum}:
\be
H = i^{q/2} \sum_{1 \leq j_1 \leq \ldots \leq j_q} J_{j_1, \ldots, j_q} \psi_{j_1} \ldots \psi_{j_q}, 
\ee
where $\psi_i$ are Majorana fermions, $\{ \psi_i, \psi_j \} = 2 \delta_{ij}$, and the couplings are drawn independently from a Gaussian distribution with zero mean and a variance $\langle J_{j_1, \ldots, j_q}^2 \rangle = J^2/2q \binom{N-1}{q-1}$.
This model is exactly solvable at all temperatures in the large-$q$, large-$N$ limit~\cite{maldacena2016remarks,qi2019quantum}. 

To construct the teleportation protocol for the SYK model, we first define the $N$-fermion EPR state,
\be
 \psi_{j,l} \ket{\fepr} \equiv  -i\psi_{j,r} \ket{\fepr}, \,\,\, \forall \, j = 1, \ldots, N
\ee
From this, the TFD state is obtained as before,
\be
 \ket{\tfd} \equiv e^{-\beta H_l / 2} \ket{\fepr}.
\ee
For the two-sided coupling, we consider the simple bilinear interaction, 
\be \label{V fermions}
V = \frac{1}{2qN} \sum_{j=0}^N i \psi_{j,l} \psi_{j,r},
\ee
which measures the size of operators in the Majorana string basis, divided by $q N$~\cite{roberts2018operator,qi2019quantum}. 

As in 0D RUCs, the size and size width of time-evolved operators in the SYK model increase exponentially in time, and exhibit a large separation only when initially encoded in $p$-body operators.
To see this, we can generalize previous computations of size distributions in the large-$q$ SYK model~\cite{qi2019quantum} to initial $p$-body operators, $\psi = \psi_1 \psi_2 \ldots \psi_p$; this relies on the factorization of SYK correlation functions in the large-$N$ limit~\cite{gao2019traversable}. 
After the relaxation time ($t \gtrsim 1/J$), but before the scrambling time ($t \lesssim \log(N/p)/J$), the size and size width are:
\be \label{eq: SYK size}
\Size \approx \frac{p}{2} e^{2 J t}, \,\,\,\,\,\,\,\, \delta \Size \approx \frac{\sqrt{2qp}}{4}  e^{2 J t}.
\ee
The scaling $\delta \Size \sim \Size / \sqrt{p}$ matches that found for 0D RUCs; in particular, ensuring a large separation between the size and size width requires $p \gg q$.
Note that our condition for peaked size distributions depends on the (large) parameter $q$, through the size width.

This large separation suggests that peaked-size teleportation is possible at early times in the large-$p$ limit.
To verify this, we analyze the two-sided correlator, which is given by \cite{gao2017traversable}
\be \label{infinite temperature SYK correlator}
\begin{split}
C_\psi(t) & = \langle e^{-igV} \psi_r(-t) e^{igV} \psi_l(t) \rangle \\
& =  \bigg( \frac{1}{ 1 + i \frac{g}{N} \frac{1}{4} e^{2 J t} } \bigg)^{2p/q} \\
\end{split}
\ee
at infinite temperature before the scrambling time\footnote{The inclusion of $e^{-igV}$ in the correlator applies a phase $e^{-ig}$ to the bra on the left side, which conveniently subtracts off the constant term in $V$'s relation to operator size [Eq.~(\ref{V size})].}.
For large $p$ and early times, we can approximate the correlator as
\be \label{infinite temperature SYK phase}
\begin{split}
C_\psi(t) & \approx \exp( - i \frac{g}{qN} \frac{p}{2} e^{2 J t} ), \\
\end{split}
\ee
using $(1 + i x)^m \approx e^{imx}$, valid when $mx^2 \equiv \frac{2p}{q} \big( \frac{g}{N} \frac{1}{4} e^{2 J t} \big)^2 \ll 1$.
We refer to this regime as the ``early time regime'', and analyze its analog in large-$N$ systems at finite temperature in Section \ref{sec: early time}.

Crucially, as expected for peaked-size teleportation, the early time correlator consists of an overall phase equal the average operator size, Eq.~(\ref{eq: SYK size}), multiplied by $g/qN$.
This indicates that teleportation succeeds with nearly maximal fidelity beginning when $g\Size/qN \approx 1$.
Based on its similarity with 0D RUCs, we expect that teleportation in this regime is capable of teleporting $\mathcal{O}(K)$ qubits (Table~\ref{table:comparison}); however, we do not calculate this explicitly.
Teleportation continues to succeed until the above approximation breaks down, which occurs when the size width, $\delta \Size$, becomes of order $(g/qN)^{-1}$.
As for all scrambling systems, the two-sided correlator is expected to revive at late times, $t \gtrsim \log(N/p)/J$, at which point the sizes saturate the entire system~\cite{gao2017traversable,maldacena2017diving} (see Section~\ref{late times}); this is not reflected in Eq.~(\ref{infinite temperature SYK correlator}), which is valid only before the scrambling time. 

\section{Interplay between peaked-size and gravitational teleportation} \label{interplay}

In this section, we seek to understand the interplay between peaked-size and gravitational teleportation.
A central theme in this understanding is a comparison between the size distribution introduced in Section~\ref{size}, and the \emph{winding size distribution} introduced in Ref.~\cite{brown2019quantum,brown2020quantum}.

To illustrate the distinction between these distributions, consider a time-evolved Majorana fermion operator, decomposed in a basis of Majorana strings, $\chi$~\cite{roberts2018operator,qi2019quantum}:
\begin{equation} \label{eq: Majorana decomposition}
    \psi(t) \rho^{1/2} = \sum_\chi c_\chi \chi.
\end{equation} 
From this decomposition, one defines the size distribution~\cite{roberts2018operator,qi2019quantum}, 
\begin{equation} \label{Majorana size distribution}
    P(\Size) = \sum_{\chi \,: \, \Size[\chi] = \Size} |c_\chi|^2,
\end{equation}
and the winding size distribution~\cite{brown2019quantum,brown2020quantum},
\begin{equation} \label{winding size distribution}
    f(\Size) = \sum_{\chi \,: \, \Size[\chi] = \Size} c_\chi^2,
\end{equation}
where $\Size[\chi]$ is the size of the string $\chi$.
Note that the size distribution is real-valued, while the winding size distribution may be complex.

The teleportation correlators [under coupling Eq.~(\ref{V fermions})] are, in fact, directly related to the winding size distribution~\cite{brown2019quantum,brown2020quantum}: 
\be  \label{eq: C winding size}
C_\psi(t) =  -i \sum_{\Size=0}^{\infty} e^{-i g \Size / q N} f(\Size),
\ee 
which can be derived by explicitly plugging Eq.~(\ref{eq: Majorana decomposition}) into the teleportation correlator.
The size distribution, by contrast, is related to ``one-sided'' correlation functions, e.g. Eq.~(\ref{V OTOC}), where both instances of the time-evolved operator appear on the same side of the TFD state~\cite{qi2019quantum}.

Despite this distinction, we have so far been able to analyze teleportation using the size distribution, as opposed to the winding size distribution, because the two are equal in two circumstances. 
%
The first is at infinite temperature, where the coefficients $c_\chi$ are real because $\psi(t)$ is Hermitian.
The second has been precisely our focus: when size distributions are perfectly tightly peaked, in which case both distributions approach a delta function.

In what follows, we describe several scenarios in which the distinction between the two distributions becomes relevant.
First we begin in large-$N$ systems, where large-$N$ factorization provides a precise relation between the teleportation correlator and the OTOC at early times.
We find that, even in the presence of the large-$p$ encoding, the correlator deviates from the peaked-size prediction whenever the OTOC contains an imaginary part.
Large-$N$ systems encompass both peaked-size and gravitational teleportation---our results suggest that the former occurs in systems where the OTOC is real (e.g. at infinite temperature with large-$p$ encoding, see Section \ref{sec: intermediate time}), while the latter occurs where the OTOC is imaginary (e.g. at low temperature in SYK)~\cite{kitaev2018soft,gu2019relation}.
Second, we review recent results showing that this deviation eventually leads an $\mathcal{O}(1)$ correlator magnitude when the winding size distribution takes a particular form, thereby enabling teleportation with unit fidelity (see Section \ref{sec: requirements}).
This is conjectured to be the microscopic origin of gravitational teleportation~\cite{brown2019quantum,brown2020quantum}, and so we expect it to occur only in low temperature models with a gravity dual.
Third, we return to teleportation in the large-$q$ SYK model and show that this model interpolates between gravitational teleportation at low temperatures and peaked-size teleportation at high temperatures.
Surprisingly, this interpolation occurs despite the fact that the large-$p$ encoding ensures a large separation between the size and size width, i.e.~the size distribution naively appears tightly peaked, even at low temperatures.
Finally, motivated by this smooth interpolation, we conclude this section by searching for a `dual' description of peaked-size teleportation in a bulk gravitational theory.
In particular, we argue that strong stringy effects lead to the same qualitative features as peaked-size teleportation.

\subsection{Early time teleportation in large-$N$ systems}\label{sec: early time}

In Section \ref{size}, we saw that for peaked-size operators the teleportation correlator depends only on the first moment of the size distribution, i.e.~the average size [Eq.~(\ref{V OTOC})].
We will now show that a more general relationship holds for large-$N$ systems at early times, where we substitute the average size with the first moment of the \emph{winding} size distribution.
Specifically, using Eqs.~(\ref{eq: O slide}-\ref{eq: AB trace TFD}), the first moment of the winding size is given by a two-sided OTOC: 
\begin{align}
\figbox{.3}{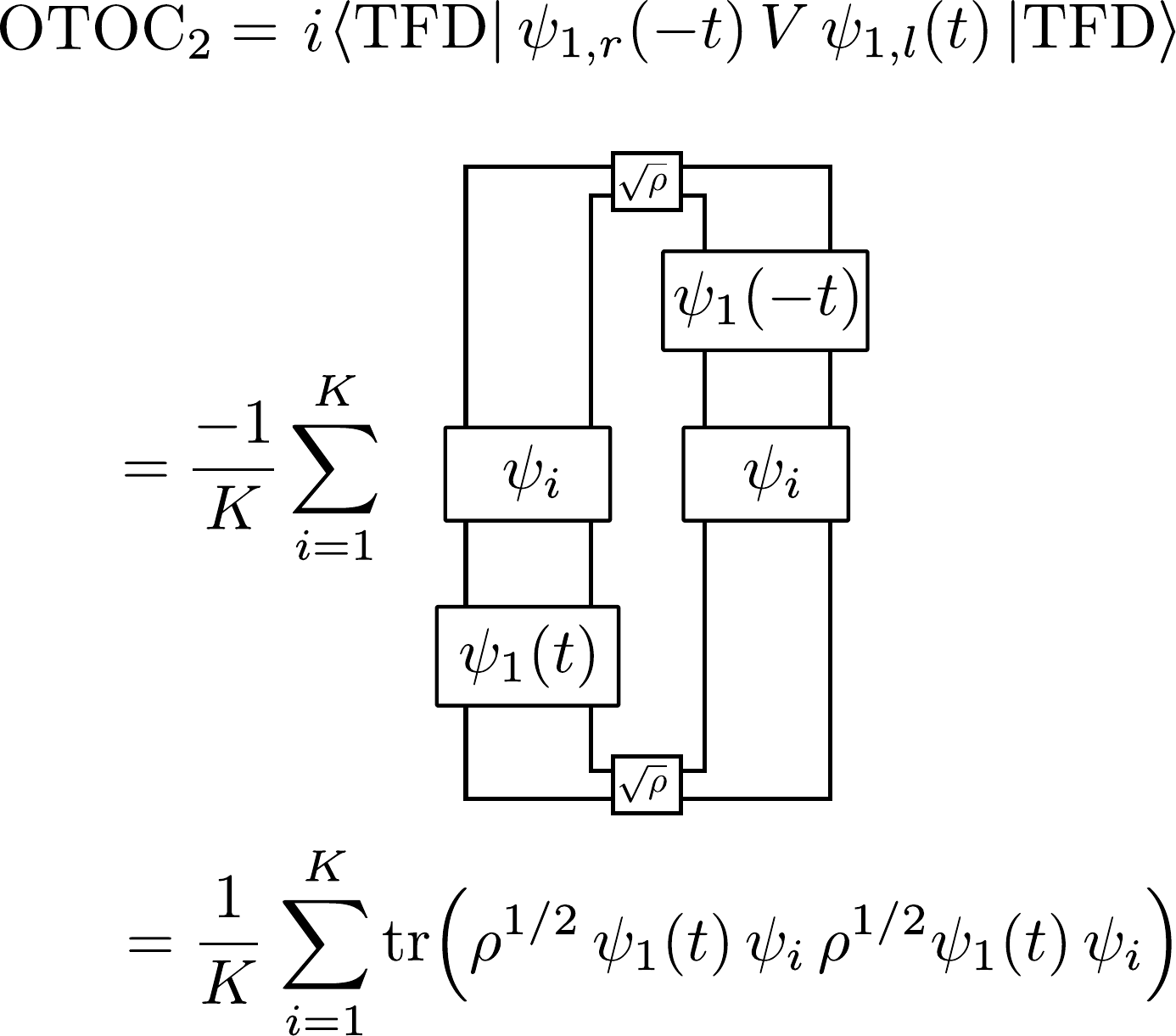} \centering \label{OTOC reg}
\end{align}
using Eqs.~(\ref{eq: O slide}-\ref{eq: AB trace TFD}).
This differs from the one-sided OTOC, for probing the average size [Eq.~(\ref{V OTOC})], in terms of the placement of the thermal density matrix.

To relate the OTOC and the teleportation fidelity, we consider two simplifying assumptions.
%
%
%
First, we focus on 0D large-$N$ systems, e.g.~the SYK model, with a $p$-body initial encoding.
In such systems, the teleportation correlator in fact factorizes into a product of single-body correlators (up to $1/N$ corrections)~\cite{gao2019traversable}:
\be
\begin{split} \label{correlator OTOC}
C_\psi(t)  & = \left < e^{-igV} \psi_r(-t) e^{igV} \psi_l(t) \right > \\
& \approx  \bigg[\langle e^{-igV} \psi_{1,r}(-t)e^{i g V} \psi_{1,l}(t) \rangle \bigg]^p, \\
\end{split}
\ee
where $\psi_1$ is a single-body operator.

Second, generalizing Eqs.~(\ref{eigV expansion},~\ref{infinite temperature SYK phase}), we consider sufficiently early times to work at leading order in $g$\footnote{In the first line, we use the fact that the thermofield double state has peaked size~\cite{qi2019quantum} to pull $e^{-igV}$ outside the correlator. In the second line we use the expansion $(1+ix)^m \approx e^{imx}$. In the third line we use $\langle V \rangle = i \langle \psi_{1,l} \psi_{1,r} \rangle / 2q = -G_\beta/2q$.}:
\be
\begin{split} \label{eq:59}
C_\psi(t)  \approx &\, e^{-igp \expval{V}} \bigg[ \langle \psi_{1,r}  \psi_{1,l} \rangle + i g \langle \psi_{1,r} \, V \, \psi_{1,l} \rangle + \ldots \bigg]^p \\
\approx &\,  e^{-igp \expval{V}} \langle \psi_{1,r}  \psi_{1,l} \rangle^p  \bigg[ \exp( i g p  \frac{ \langle \psi_{1,r} \, V \, \psi_{1,l} \rangle }{\langle \psi_{1,r}  \psi_{1,l} \rangle} )  + \ldots \bigg]\\
= &\,  (-i G_\beta)^p \exp( -i \frac{g p}{2q}  \bigg[\frac{ \text{OTOC}_2 }{G_\beta} - G_\beta \bigg] )  +\ldots \\
= &\,  (-i G_\beta)^p \exp( -i \frac{g p}{2qN}  G_\beta  \mathcal{F}_2(t)  )  + \ldots
\end{split}
\ee
where $G_\beta =  i \langle \psi_{1,r}  \psi_{1,l} \rangle = \tr( \rho^{1/2} \psi_1 \rho^{1/2} \psi_1)$ is the imaginary time Green's function, and $\mathcal{F}_2(t)$ is the first-order, connected component of the two-sided OTOC [Eq.~\eqref{OTOC reg}], 
\be
\begin{split} \label{OTOC}
\textrm{OTOC}_2&\approx G_\beta^2 \left (1+\frac 1 N \mathcal{F}_2(t) + \cdots \right).
\end{split}
\ee
Similar to Eq. \eqref{eigV expansion}, the leading correction to Eq.~\eqref{eq:59} is $ \sim p g^2 [ \expval{V^2}_\psi  - \expval{V}_\psi^2/ G_\beta] $, and the approximation holds when this is small.


Let us now consider the behavior of the teleportation correlator, Eq.~\eqref{eq:59}, under different physical scenarios.
We focus on chaotic systems during the so-called Lyapunov regime, which occurs between the thermalization time, $t \sim \mathcal{O}(1)$, and the scrambling time, $t \sim \mathcal{O}(\log N)$.
In this regime, the connected OTOC is characterized by a simple exponential $\mathcal{F}_2(t) \sim e^{\lambda t}$ with a prefactor that is generally complex.
As a result, we expect the teleportation correlator to exhibit two distinct effects: ($i$) the real part of $\mathcal{F}_2(t)$ causes rapid phase oscillations in the teleportation correlator, while ($ii$) the imaginary part increases/decreases the teleportation correlator magnitude, depending on the sign of the coupling $g$. 
%
%
 
At infinite temperature, $\mathcal{F}_2(t)$ is strictly real and thus only effect ($i$) can occur.
Indeed, in this case, the two-sided OTOC directly measures the operator size and Eq.~\eqref{eq:59} is equivalent to Eq.~\eqref{eigV expansion}.
%
It follows that peaked-size teleportation can be achieved with perfect fidelity: the teleportation correlator magnitudes are equal to one due to the infinite temperature, and their phases can be aligned by tuning $g$ or $t$.
More generally, at finite temperature, $\mathcal{F}_2(t)$ contains both a real and imaginary part, and the real part---which leads to effect ($i$)---is formally distinct from the first moment of the size distribution.
%
Rather, recent work has shown that $\text{Re} \{\mathcal{F}_2(t) \}$ is computable via a ladder diagram identity and is physically interpreted as a `branching time'~\cite{gu2019relation,zhang2020obstacle}.
Here teleportation is similarly possible by tuning $g$ or $t$ to align the correlator phases, however the teleportation fidelity is bounded from above if the correlators do not have magnitude one (Section \ref{sec: requirements}).

At the opposite extreme, effect ($ii$) is dominant in systems with a gravity dual~\cite{kitaev2018soft,gu2019relation} (as well as other maximally chaotic systems, e.g.~maximally chaotic 2D CFTs with a large central charge~\cite{gao2018regenesis}).
In such cases, $\mathcal{F}_2(t)$ is mostly imaginary and leads to the growth (or decay) in the magnitude of the correlator. 
This opens the door to magnitudes \emph{greater} than the two-point function, $|C_\psi(t)| > G_\beta$, which is not possible in peaked-size teleportation (Section \ref{finite temperature}).
%
%
Interpolating between the two above limits, it has been conjectured that the prefactor of $\mathcal{F}_2(t)$ is proportional to $e^{i\lambda \beta /4 \pi}$~\cite{kitaev2018soft,gu2019relation}.
This would imply that the imaginary part is dominant if and only if $\lambda \approx 2\pi\beta$, i.e.~the system approaches the bound on chaos \cite{maldacena2016bound}.

\subsection{Gravitational teleportation and the size-winding mechanism} \label{sec: gravity size winding}

We now move beyond early times and provide a brief review of how the correlator can achieve its maximal magnitude, $1$, at finite temperatures.
This occurs via the `size winding' phenomenon introduced in Ref.~\cite{brown2019quantum,brown2020quantum} as the microscopic mechanism for gravitational teleportation~\cite{gao2017traversable,maldacena2017diving}.
We refer the reader to Ref.~\cite{brown2020quantum} for a complete discussion of this mechanism, including its connection to physical quantities in the bulk gravity theory.
As we emphasize in Section~\ref{sec: requirements}, maximizing the magnitude of the correlators is necessary for high fidelity teleportation, but it is not sufficient: we must also align the correlator phases, for every operator on the subspace to be teleported.
%
%

To begin, note that the winding size distribution is normalized to the two-point function, $G_\beta \leq 1$, in contrast to the size distribution, which is normalized to $1$.
From Eq.~(\ref{winding size distribution}), we see that this norm being less than one implies that the phases of the coefficients $c_\chi$ are not perfectly aligned for different strings $\chi$.
It is convenient to separate this misalignment into two classes: first, when  coefficients of strings of the same size $\Size$ are misaligned, which manifests in the magnitude of $f(\Size)$ being less than maximal for a given $\Size$, and second, when the phases of $f(\Size)$ for different sizes $\Size$ do not align with each other.

We focus on the latter case and, more specifically, consider an ansatz in which the coefficients' phases \emph{wind} with the size~\cite{brown2019quantum,brown2020quantum}:
\be
\begin{split}
c_\chi = e^{-i \alpha \Size[\chi] / q} |c_\chi|,
\end{split}
\ee
In this case, the coupling of the teleportation protocol, by applying a phase that is also proportional to the size, can serve to unwind the phases of $f(\Size)$ at the value $g/N = -2\alpha$ [see Eq.~(\ref{eq: C winding size})].
This increases the teleportation correlator magnitude from its initial value, $G_\beta$, to unity.
%
%
Although seemingly artificial, in the following subsection we show that this ansatz holds exactly for the SYK model at low temperatures.

\subsection{Large-$q$ SYK model: finite temperature}\label{SYK finite}

We now turn to explore the interplay between peaked-size and gravitational teleportation in an explicit example: the large-$q$ SYK model at finite temperature and large-$p$ encoding~\cite{qi2019quantum}.
Despite the fact that this model features a large separation between the size and size width, we show that teleportation is \emph{not} governed by the peaked-size mechanism at low temperatures, due to the presence of strong size winding. 
%

To begin, let us consider the finite-temperature teleportation correlator, given by~\cite{gao2019traversable}:
\be
\begin{split} \label{ping correlator}
C_\psi(t)
& = (-i G_\beta)^p \bigg( \frac{1}{ 1 - \frac{g}{N} \frac{J}{2\lambda} e^{\lambda t} \sin( \lambda \beta / 4) + i \frac{g}{N} \frac{1}{4} e^{\lambda t} } \bigg)^{2p/q}, \\
\end{split}
\ee
where $(G_\beta)^p = i^p \expval{\psi_r \psi_l } = (\lambda/2J)^{2p/q}$ is the $p$-body two-point function, and the Lyapunov exponent $\lambda$ corresponds to the solution of
\begin{equation} \label{lyapunov SYK}
\beta \lambda=2\beta J \cos (\lambda\beta/4)
\end{equation}
and interpolates between $2\pi/\beta$ at low temperatures and $2J$ at high temperatures.
At infinite temperature, the correlator reduces to Eq.~(\ref{infinite temperature SYK correlator}), and follows our expectations for peaked-size teleportation (see Section~\ref{sec: SYK infinite temperature}).
At low temperatures, where the model is known to possess a gravitational dual~\cite{kitaev2015simple,maldacena2016remarks,kitaev2018soft}, the correlator behaves substantially differently; most notably, its magnitude increases from $G_\beta^p$ at time zero to unity when $gJe^{\lambda t}/2 \lambda N = 1$ [illustrated in Fig.~\ref{fig: 1}(c)].

From this correlator, we can verify the two predictions made in Sections~\ref{sec: early time} and~\ref{sec: gravity size winding}: ($i$) the early time behavior is governed by the two-sided OTOC, and ($ii$) the size winding mechanism is responsible for the $\mathcal{O}(1)$ peak in the correlator magnitude at low temperatures.
To see the former, we expand the correlator in the early time regime:
\be
\begin{split} \label{ping correlator early}
C_\psi(t)
& \approx (-i G_\beta)^p \exp( -\frac{i g p}{2 q N} \bigg[ i \frac{2J}{\lambda} e^{\lambda t} \sin( \lambda \beta / 4) + e^{\lambda t} \bigg] ).
\end{split}
\ee
Indeed, the term in the exponent is directly proportional to the connected piece of the two-sided OTOC~\cite{gu2019relation},
\be
\mathcal{F}_2(t) = i \frac{2J}{\lambda} e^{\lambda t} \sin( \lambda \beta / 4) +   e^{\lambda t}, \\
\ee
matching Eq.~\eqref{eq:59}\footnote{More precisely, the correlator in Eq.~(\ref{ping correlator early}) is missing a factor of $G_\beta^p$ compared to Eq.~(\ref{eq:59}). This same mismatch is noted in Ref.~\cite{qi2019quantum}, and is attributed to the large-$q$ limit utilized for the calculation, since in this limit $G_\beta$ approaches 1.}.
At high temperatures this OTOC is equal to two times the operator size [Eq.~(\ref{eq: SYK size})], resulting in phase oscillations, whereas at low temperatures the OTOC rotates to become predominantly imaginary, leading to an exponential growth in the correlator magnitude.

Next, to understand the role of size winding, we must analyze the full winding size distribution. 
We can derive this distribution by expanding the teleportation correlator in powers of $e^{-ig/qN}$ to match Eq.~(\ref{eq: C winding size})~\cite{qi2019quantum,brown2019quantum,brown2020quantum}.
To do so, it is convenient to consider the exact correlator (before a $g/N \ll 1$ approximation)~\cite{qi2019quantum,gao2019traversable}:
\be
\begin{split} \label{ping correlator full}
C_\psi&(t) = \\
& \!\!\!\!\! (-i G_\beta)^p \bigg( \frac{e^{-ig/2N}}{ 1 + i (1-e^{-ig/N}) [ \frac{J}{2\lambda} \sin( \lambda \beta / 4) -  \frac{i}{4} ] e^{\lambda t} } \bigg)^{2p/q} \\
\end{split}
\ee
Rewriting this correlator using Eq.~(\ref{lyapunov SYK}) and the Taylor expansion,
\be
\begin{split}
& \bigg( \frac{1}{1 + (1-e^{-\mu}) x} \bigg)^{2p/q} \\
&\,\,\,\,\,\,\,\,\, = \frac{1}{(1+x)^{2p/q}} \sum_{n=0}^{\infty} e^{-n \mu} \binom{n + \frac{2p}{q} - 1}{n}  \frac{1}{(1 + 1/x)^n},
\end{split}
\ee
and identifying the $n^{\text{th}}$ coefficient with the winding size distribution, we have:
\be
\begin{split}
f(qn+p) = & -\frac{ (-i G_\beta)^p }{(1+ \frac{J}{2\lambda}e^{\lambda t} e^{i\lambda \beta / 4})^{2p/q}} \\
& \times \binom{n + \frac{2p}{q} - 1}{n}
\frac{1}{(1 + \frac{2\lambda}{J}e^{-\lambda t} e^{-i\lambda \beta / 4})^n}.
\end{split}
\ee
At intermediate times and large $p$, the distribution takes a particularly simple form,
\be
\begin{split}
f(qn+p) \approx (-i G_\beta)^p \frac{(\gamma + i2\alpha)^{2p/q}}{\Gamma(\frac{2p}{q})} n^{\frac{2p}{q}-1} e^{- \gamma n} e^{ -i 2 \alpha n}
\end{split}
\ee
where we define the size decay rate, $\gamma$, as
\be
\gamma =  \frac{2\lambda}{J} e^{-\lambda t} \cos(\lambda \beta / 4) = \left( \frac{\lambda}{J} \right)^2 e^{-\lambda t},
\ee
using Eq.~(\ref{lyapunov SYK}), and the size winding coefficient, $\alpha$, as
\be
2 \alpha = -\frac{2\lambda}{J} e^{-\lambda t} \sin(\lambda \beta / 4).
\ee
The above expression holds when $(2p/q)^2 \ll n \ll 1/\gamma^2, 1/\alpha^2$.
Crucially, the distribution follows the size winding ansatz, $f(n) = |f(n)|e^{-i2\alpha n}$.
Thus, we recognize that the maximum in the correlator magnitude occurs when the coupling has unwound the phases of $f(n)$, at $g/N = -2\alpha$, as expected from Section~\ref{sec: gravity size winding}~\cite{brown2019quantum,brown2020quantum}.



The fact that the correlator magnitude increases in time, and moreover reaches an $\mathcal{O}(1)$ value at low temperatures, is a hallmark of gravitational teleportation and signals physics outside the peaked-size regime.
Naively, this result is surprising, as we expect the $p$-body encoding to ensure a peaked size distribution.
Indeed, the average size and size width remain separated by $\sqrt{p}$ at all temperatures~\cite{qi2019quantum}:
\be \label{finite temperature SYK size}
\Size[ \psi(t) \rho^{1/2} ] - \Size[\rho^{1/2}] \approx \frac{p}{2} \left( \frac{2J}{\lambda} \right)^2 e^{\lambda t} = \frac{2p}{\gamma},
\ee
\be
 \delta \Size[ \psi(t) \rho^{1/2} ] \approx \frac{\sqrt{2qp}}{4} \left( \frac{2J}{\lambda} \right)^2 e^{\lambda t} = \frac{\sqrt{2qp}}{\gamma}.
\ee
This demonstrates that our simple intuition, of judging a size distribution to be tightly peaked if the ratio between the size width and average size is small, is not always correct.
Rather, in Appendix~\ref{app: bounds}, we provide a more precise condition for when peaked-size teleportation holds, and explicitly show that this condition breaks down for the SYK model at finite temperature (but remains satisfied at infinite temperature).
%

Let us now provide intuition for \emph{how} peaked-size teleportation is modified by size winding at low temperatures. 
To this end, we express the SYK correlator in terms of the winding size distribution parameters:
\be \label{corr size winding}
\begin{split}
C_\psi(t) 
\approx & \, (-i G_\beta)^p \frac{(\gamma + i2\alpha)^{2p/q}}{\Gamma(\frac{2p}{q})} \\
& \times \int_{0}^{\infty} dn  \, n^{\frac{2p}{q}-1} \exp( - \gamma n  ) \exp(-i [g/N + 2 \alpha] n).  \\
= &  (-i G_\beta)^p \bigg[\frac{ \gamma + i2 \alpha }{  \gamma + i 2 \alpha + i g/N  }\bigg]^{2p/q}  \\
\end{split}
\ee
At early times, this integral can be solved using a saddle-point approximation. 
At infinite temperature, the saddle point, $n_s$, occurs precisely at the average size, $n_s = (2p/q)/\gamma = \Size/q$, giving the peaked-size correlator, $C_\psi = (-i G_\beta)^p \cdot \exp( - i g \Size / qN )$.
In contrast, at finite temperature, the size winding $\alpha$ shifts the saddle point in the imaginary direction of the complex plane, giving $n_s = (2p/q)/(\gamma+2i\alpha)$ and a correlator $C_\psi = (-i G_\beta)^p \cdot \exp( - i g n_s / qN )$.
From this, we recognize the saddle point as precisely the two-sided OTOC, $n_s = \frac{p}{2q} \mathcal{F}_2(t)$.

The inclusion of the size winding in the low temperature saddle point thus has two effects.
First, it contributes an imaginary part to the OTOC and thereby increases the magnitude of the teleportation correlator.
More subtly, it also alters the \emph{real} part of the OTOC.
At low temperatures, $\alpha/\gamma \approx \beta J \gg 1$, and we can approximate the saddle as $n_s \approx (2p/q)/(2i\alpha) + (2p/q \gamma) (\gamma/2\alpha)^2$.
Recognizing $\mathcal{S} = 2p/\gamma$, we see that the real part of the OTOC now corresponds to the average size suppressed by two factors of the ratio $(\alpha/\gamma)^2$.

\begin{figure*}[t]
\centering
\includegraphics[width=0.8\textwidth]{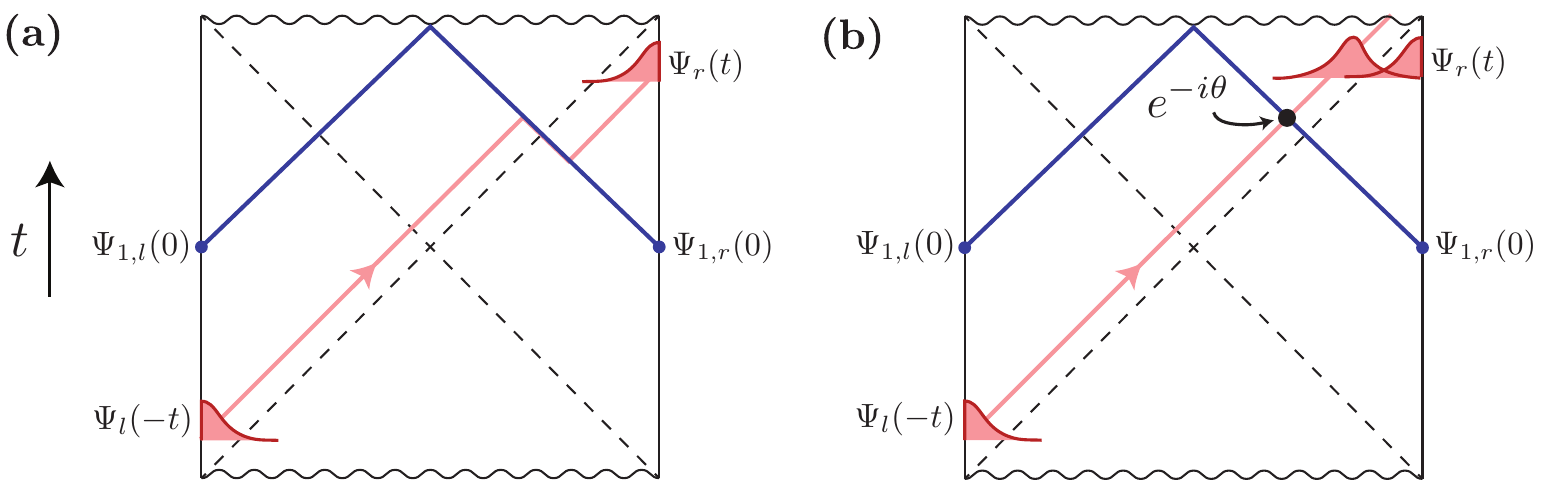}
\caption{Schematic of the teleportation protocol from the bulk gravitational perspective in AdS$_2$, under both \textbf{(a)} semiclassical gravity, and \textbf{(b)} strong stringy corrections.
The TFD state corresponds to a two-sided black hole.
Local quantum mechanical operators, $\psi_{l/r}$, create or annihilate particles near the two boundaries, with wavefunctions $\Psi_{l/r}$ (red). 
The protocol begins by inserting a particle on the left side, with wavefunction $\Psi_l$ (red, bottom left), at time $-t$, which then falls towards the interior of the geometry during time-evolution (red line). 
The two-sided coupling, $\frac{g}{N} \sum_i \psi_{i,l} \psi_{i,r}$, is then applied, producing a shock wave (blue) that interacts with the in-falling particle~\cite{gao2017traversable,maldacena2017diving}. 
\textbf{(a)} In the semiclassical limit, the shock wave shifts the position the in-falling particle outside of the right horizon (dashed), which enables the particle to reemerge near the right boundary (red, top right)~\cite{gao2017traversable,maldacena2017diving}. 
\textbf{(b)} When stringy effects are present, the scattering amplitude between the in-falling particle and the shock wave is modified according to Eq.~\eqref{eq:82} \cite{maldacena2017diving, shenker2015stringy}.
In the highly stringy limit and at early times, the interaction results in an overall phase shift, $\theta = g G_N A_{\varepsilon} (\D/2)^\varepsilon e^{\varepsilon t}$ [Eq.~\eqref{eq:c_stringy_saddle}].
The overlap between the in-falling particle and a particle at the right boundary is nevertheless non-zero (red, top right), and is given by the unperturbed two-point function, $G_\beta = i\left <\psi_l \psi_r\right >$. 
[Note that stringy effects may also modify the initial wavefunctions of $\Psi_{l/r}$, as we discuss in the context of Eq.~\eqref{eq: stringy wf}.]
}
\label{fig:stringy_schematic}
\end{figure*}

\subsection{Gravity with stringy effects} \label{sec:stringy}

While the bulk of this paper approaches teleportation firmly through the lens of quantum mechanics, we would be remiss not to explore the analog of peaked-size teleportation in gravitational physics.
Specifically, we would like to ask: is there a teleportation mechanism in gravitational systems that shares the same features as peaked-size teleportation? 
Such a connection might seem surprising, given the prevalence of peaked-size teleportation in quantum mechanical models with no apparent connection to gravity.
Nonetheless, the smooth blending between gravitational teleportation and peaked-size teleportation in the SYK model suggests a positive answer.

Here, we demonstrate---in a particular gravitational geometry, AdS$_2$---that an analog of peaked-size teleportation indeed occurs when strong stringy corrections~\cite{shenker2015stringy,maldacena2017diving} are included in the gravitational theory\footnote{We are grateful to Zhenbin Yang and Douglas Stanford for discussions leading to this connection.}.
Intuitively, our results are consistent with our previous analysis of the SYK model, where, in the dual gravitational theory, increasing the temperature is known to add stringy effects~\cite{kitaev2018soft}. 

Our derivation closely follows that of Ref.~\cite{maldacena2017diving} and assumes a background familiarity with the gravitational description of teleportation in AdS$_2$ (a thorough summary of which can be found in the seminal works of Refs.~\cite{gao2017traversable,maldacena2017diving}).
%
%
In this setting, the teleportation correlator can be calculated explicitly by considering gravitational scattering in a wormhole geometry [Fig.~\ref{fig:stringy_schematic}].
We will maintain our SYK notation, so that $V$ consists of $K$ single-body fermion operators, $\psi_i$, and our input operator is a $p$-body fermion, $\psi$. 
The correlator can be solved for by decomposing the fermion operators in a momentum basis and applying the scattering matrix: 
\be \label{eq:77}
\begin{split}
C^{\text{sc}}_\psi(t) & = e^{-ig \langle V\rangle}\int dk \Psi_{r}(k,t)\Psi_{l}^*(k,-t) \\
& \times \exp \left( ig \int ds e^{i\delta(k,s)} i\Psi_{1,r}(s,0)\Psi_{1,l}^*(s,0) \right)
\end{split}
\ee
where $\Psi_{l/r}(k,t)$ is the wavefunction for the $p$-body operator inserted on the left/right boundary with in-falling momentum $k$ (and similarly $\Psi_{1,l/r}(s,0)$ for any single-body operator in $V$), and $e^{i\delta(k,s)}$ is the scattering matrix element between $\psi(t)$ and $\psi_1(0)$. 
In pure gravity, i.e.~in the absence of stringy effects, these quantities take the form~\cite{maldacena2017diving}:
\begin{align}
\Psi_{r}(k,t)\Psi_{l}^*(k,-t)&=\frac{(2ike^{-t})^{2\D}e^{-4ike^{-t}}}{i\Gamma(2\D)(-k)}\Theta(-k)\\
\delta(k,s)&=G_N k s
\end{align}
where we have set $\beta = 2\pi$ for convenience, $\Theta(x)$ is the Heavyside function, and $\D = p/q$ is the conformal weight of $\psi$. The single-body wavefunction, $\Psi_{1}(s,0)$, is obtained by setting $t=0$ and replacing $\D\rightarrow \D_1=1/q$ (i.e. the conformal weight of a single fermion). 

In the semiclassical limit, we can evaluate the correlator by expanding $e^{i\delta}$ to linear order in $G_N$~\cite{maldacena2017diving}. We find:
\be
\begin{split}
C^{\text{sc}}_\psi(t) & = \langle \psi_l \psi_r \rangle \ \frac{(-i)4^{2\Delta}}{\Gamma(2\Delta)} \\
\times & \int^{\infty}_0 d{k} \, (-ik)^{2 \Delta - 1} \exp \left( - i (\tilde{g} G_N e^{t}-4) k \right),
\end{split}
\ee
where $\tilde{g} \equiv g 4^{-\D_1}\D_1/2$. This expression is almost identical to the large-$q$ SYK correlator of Eq.~(\ref{corr size winding}), setting the size decay rate to zero, $\gamma = 0$, and identifying the momentum $k$ in the gravitational calculation with the size $n$ in the SYK model~\cite{susskind2018things}. 
Notably, the correlator diverges at the teleportation time, $4 = \tilde{g} G_N e^t$. 
In bulk gravity, this divergence is exactly the light-cone pole between the left and right sides of the traversable wormhole, and is regulated by including higher order terms in $G_N$ or stringy corrections~\cite{maldacena2017diving}.

While the full effects of stringy scattering in an AdS background are not known, we will take a phenomenological treatment as in Ref.~\cite{maldacena2017diving, shenker2015stringy}.
Here, the total effect of stringy corrections is to change the scattering amplitude to
\be \label{eq:82}
\delta(k,s)=iG_N (-iks)^{\vep},\; 0\leq \vep\leq 1,
\ee
where $\vep$ controls the strength of stringy effects, and varies from $1$ in pure gravity to $0$ in the highly stringy limit. 

Again expanding $e^{i\delta}$ to leading order in $G_N$, and Wick rotating $k\rightarrow -ik$, we can write the correlator as
\be
\begin{split} \label{eq:c_stringy}
C^{\text{stringy}}_\psi(t) & = \langle \psi_l \psi_r \rangle \ \frac{4^{2\Delta}}{\Gamma(2\Delta)} \\
\times & \int d{k} \, k^{2 \Delta - 1} e^{- 4 k} \exp( - i^{1+\varepsilon} g G_N A_{\varepsilon} k^\varepsilon e^{\varepsilon t})
\end{split}
\ee
where $A_{\varepsilon}$ is a constant of order 1. 
Note that the $k$-dependence in front of exponential is a Poisson distribution with a saddle point at $k_{s}\approx \D/2$ in the heavy particle limit, $\D = p/q \gg 1$. 
At early times, $e^{\vep t}G_N\ll 1$, and for strong stringy effects, $\varepsilon \rightarrow 0$, the change in this saddle point from the scattering, $g$, is negligible.
In these limits, the saddle point approximation thus gives the correlator:
\be \label{eq:c_stringy_saddle}
C^{\text{stringy}}_\psi(t) \approx \langle \psi_l \psi_r \rangle \exp( - i g G_N A_{\varepsilon} (\D/2)^\varepsilon e^{\varepsilon t}),
\ee
which has exactly the same form as in peaked-size teleportation [Eq.~\eqref{C2}]\footnote{Note that the phase in Eq.~\eqref{eq:c_stringy_saddle} becomes order-one within the Lyapunov regime, i.e.~$t \lesssim 1/\epsilon \log(1/G_N)$, but at sufficiently early times to satisfy $G_N e^{\varepsilon t} \ll 1$. These conditions are consistent as long as $\Delta = p/q$ is sufficiently large to ensure $A_{\varepsilon} (\D/2)^\varepsilon \gg 1$.}!
Specifically, the correlator is equal to the two-point function, $G_\beta = i \langle \psi_l \psi_r \rangle,$ multiplied by a pure phase.
Tentatively, this suggests interpreting the phase as the operator size in a dual boundary theory.
This size,
\be
\Size/N \sim G_N A_{\varepsilon} (\D/2)^\varepsilon e^{\varepsilon t},
\ee
grows exponentially in time with a non-maximal Lyapunov exponent, $2\pi\vep/\beta$.

A few remarks are in order. First, while in the above treatment the strength of stringy effects depends on a `free' parameter $\vep$, we expect that in a UV complete theory $\vep$ would in turn depend on the temperature (and other physical parameters). 
In particular, we expect $\vep \rightarrow 1$ at low temperature in theories that are dual to pure gravity, and $\vep \rightarrow 0$ at high temperature, where stringy, UV effects should play an important role. 
This statement also follows from the point of view of the boundary field theory, since the scattering matrix is proportional to an OTOC of the boundary theory, which is real at infinite temperature.

Second, if we would like to recover the infinite temperature SYK correlator, Eq. \eqref{infinite temperature SYK correlator}, from the scattering computation, choosing a proper $\vep$ as a function of $\beta$ is not enough. 
One also needs to modify the \emph{wavefunction} of $\psi$, to:
\be \label{eq: stringy wf}
\Psi_{r}(k,t)\Psi_{l}^*(k,-t)=\frac{\vep(2ik^\vep e^{-\vep t})^{2\D}e^{-4ik^\vep e^{-\vep t}}}{i\Gamma(2\D)(-k)}\Theta(-k)
\ee
Such a wavefunction modification due to UV data should be model dependent, and it would be interesting to understand how to derive this `stringy-corrected' wavefunction from the bulk point of view.
Nevertheless, one particular feature of the modified wavefunction has a clear motivation from the boundary perspective.
Specifically, Wick rotating Eq.~\eqref{eq: stringy wf}, $k \rightarrow -ik$, leads to a distribution whose width, $\delta k \sim \Delta^{1/\vep}$, \emph{broadens} as $\vep \rightarrow 0$.
This broadening increases the phase variations in the exponential of Eq.~\eqref{eq:c_stringy} and results in the decay of the correlator at the timescale $e^{\vep t}G_N/\sqrt{\Delta} \approx 1$ for small $\vep$.
From the boundary point of view, this decay corresponds to the requirement that the size width must be small, $g\delta \Size/N \lesssim 1$, for peaked-size teleportation, as we saw for 0D RUCs and infinite temperature SYK (Section \ref{sec: intermediate time}).
We expect this decay to be common to many 0D quantum systems at high temperatures, which suggests that the broadening of the bulk stringy wavefunction as $\vep \rightarrow 0$ might also be a general feature.

Finally, the most obvious effect of a non-unity $\vep$ is to change the scattering phase, $\delta(k,s)$, from being real-valued to complex. 
Indeed, in the strong stringy limit, $\delta(k,s)$ becomes purely imaginary.
In general scattering theory, a complex $\delta$ means that the scattering matrix, $e^{i\delta}$, is no longer normalized, and implies the existence of inelastic scattering~\cite{shenker2015stringy}. 
Since peaked-size teleportation is replicated in the limit $\vep \rightarrow 0$, this suggests a more general relationship between peaked sizes and inelastic scattering. 
In Appendix~\ref{app: inelastic}, we demonstrate that these two phenomena also coincide at infinite temperature, for arbitrary wavefunctions and scattering amplitudes.

\section{Experimental proposals}\label{experiment}

Having illustrated the wide breadth of physics that enters into the TW protocol, in this section we outline explicitly how one can probe this physics in the laboratory.
We begin with a summary of the key signatures of teleportation, and how they can be applied towards ($i$) characterizing operator size distributions in generic scrambling dynamics, and ($ii$) distinguishing generic vs. gravitational scrambling dynamics.
For ($i$), we show that the TW protocol can be simplified dramatically at infinite temperature, where an equivalent `one-sided' protocol eliminates the need to experimentally prepare the thermofield double state.
%
%
We next present two near-term experimental realizations of the protocol: first with neutral atoms and second with trapped ions.
The fundamental requirement is the ability to time-evolve forwards and backwards under many-body scrambling dynamics;  recent experimental progress has demonstrated this in a number of quantum simulation platforms~\cite{garttner2017measuring,li2017measuring,arute2019quantum,wei2019emergent,meier2019exploring}. 
%
%
%
We conclude with a discussion of the effect of experimental error, and a comparison of the TW protocol with other diagnostics of scrambling physics.

\subsection{Signatures of the TW protocol} \label{sec: signatures}


We begin by reviewing the key signatures of the TW protocol, as discussed in the previous sections and summarized in Table~\ref{table:comparison}.
We first recall that the simplest experimental signal---that is, any non-trivial teleportation fidelity of a single qubit---has already been demonstrated experimentally in the closely-related HPR protocol~\cite{landsman2019verified,blok2020quantum}.
As discussed in Section~\ref{late times}, this signifies that the implemented unitary is scrambling but does not distinguish between peaked-size or gravitational teleportation.
In what follows, we discuss two more refined applications of the TW protocol.

\emph{Characterizing size distributions in generic scrambling dynamics}---The dynamics of the teleportation fidelity within the TW protocol can be used  to  probe the size distributions of time-evolved operators.
This approach relies on the peaked-size teleportation mechanism and thus applies to generic scrambling systems, including the examples analyzed in Section \ref{sec: intermediate time} (e.g.~RUCs, spin chains, high T SYK).
%

Specifically, the teleportation fidelity as a function of time exhibits three relevant features. 
First, since peaked-size teleportation relies on the \emph{width} of the size distribution being small, $ g \delta \Size / N \lesssim 1$, its success or failure indicates whether the width has surpassed the tunable value, $N /  g$.
Depending on the model and the value of $g$, this leads to a temporal profile that exhibits  three regimes: initial teleportation when the size width is small, no teleportation when $\delta \Size \gtrsim N / g$, and late time teleportation once the size width converges to its small final value in a finite-size system [as depicted schematically in Fig.~\ref{fig: 1}(c) and observed numerically in 0D RUCs in Fig.~\ref{fig: RUC}(c)].

Second, within the peaked-size regime, oscillations in the teleportation fidelity as a function of time, $F = \frac 5 8 - \frac 3 8 \cos( \constd g \Size(t) / N )$ [Eq.~\eqref{eq: F peaked-size}], provide a direct measurement of the growth in operator size. In particular, setting $g = 2\pi n + \pi$, one expects to see $n$ oscillations in the teleportation fidelity before it reaches its late time plateau.
The peaks in these oscillations give the operator size as a function of time: $\Size = (m / n)(1-1/d^2) N$ at the $m^{\text{th}}$ peak.

Third, the teleportation of multiple qubits demonstrates the equivalent channel capacities of peaked-size and gravitational teleportation (Section~\ref{sec: intermediate time}).
Formally, multi-qubit teleportation probes whether the sizes of time-evolved operators \emph{add} under operator composition.
While this is trivial when the operators are causally separated, determining the requirements for size addition under more general dynamics---e.g.~all-to-all or power-law interactions---remains an open question\footnote{Indeed, recent work has indicated that, in theories with a gravitational dual, the lack of size addition is related to a scattering event among infalling particles \cite{haehl2021six}.}.

\emph{Distinguishing gravitational scrambling dynamics---}The TW protocol can also be used as an experimental litmus test for gravitational dynamics.
%
%
To this end, we propose to use two experimental signatures that distinguish between gravitational and peaked-size teleportation: ($i$) the teleportation fidelity at low temperature, and ($ii$) the behavior of the teleportation fidelity as a function of time, $t$, and the coupling strength, $g$.
For ($i$), the observation of a high teleportation fidelity, $\sim \mathcal{O}(1)$, at low temperatures strongly suggests the occurrence of gravitational teleportation, since the fidelity of peaked-size teleportation is limited at such temperatures by the (small) two-point function, $G_\beta$.
For ($ii$), one observes that the qualitative profile of the teleportation fidelity as a function of time differs between the two mechanisms (see Fig.~\ref{fig: 1}(c) for a comparison between the two, and Figs.~\ref{fig: RUC},~\ref{fig: size width} for additional examples of peaked-size teleportation).
Namely, keeping $g$ fixed, the fidelity of gravitational teleportation is expected to display a single peak as a function of time, whereas the fidelity of peaked-size teleportation is highly oscillatory in time.
Furthermore, gravitational teleportation works only for a specific \emph{sign} of the coupling, $g > 0$, while the peaked-size teleportation fidelity is an even function of $g$~\cite{gao2017traversable,maldacena2017diving,brown2019quantum,brown2020quantum}.

\emph{Contrasting with finite-size effects}---Finally, we would like to distinguish many-body teleportation from spurious effects that may be seen in the TW protocol at small-size systems.
The most effective way to avoid such signals is by utilizing a coupling $gV$ [Eq.~(\ref{eq: V})] whose individual terms have a small magnitude, i.e.~$g/K \ll 1$; this is most naturally achieved by including many couplings---which requires a sufficiently large system---and setting $g \sim \mathcal{O}(1)$.
In this limit, the action of the coupling is negligible unless local operators have grown significantly under many-body dynamics, i.e.~$\mathcal{S} \sim K/g \gg 1$ (see Section \ref{size}); any teleportation signal is thus necessarily a result of scrambling dynamics.
Furthermore, we expect large-size operators to generically exhibit smooth size distributions, justifying our approximation (Section \ref{peaked sizes}) that the teleportation fidelity is governed by the distributions' first few moments.

Away from this limit, our general framework relating the teleportation fidelity to operator size distributions remains valid [e.g.~Eq.\eqref{coupling size phase}].
However, for $g/K \lesssim 1$, we expect the fidelity to be sensitive to the discrete nature of the size distributions, and our predictions based on the first few moments may no longer apply.
%
%
Fortunately, as we show in the following subsections, none of these complications are evident for experimentally relevant system sizes (e.g.~$K \sim N \sim 20$) and $g \sim \mathcal{O}(1)$ coupling strengths; indeed, our finite-size numerical results agree very well with predictions from the peaked-size teleportation framework [Fig.~\ref{fig:rydberg-implementation}(b) and~\ref{fig:rydberg_scaling}]. 


%
Lastly, in the case where $g/K \sim 1$, operator growth is no longer necessary for the coupling to have a strong effect, leading to the possibility of a teleportation signal unrelated to scrambling. 
Indeed, for $g/K = \pi$, the coupling effectively `swaps' the left and right qubits.
This is made precise for the coupling $V_s$ [Eq.~(\ref{size qudit})], where $\exp( i \pi N V_s ) = (\text{SWAP}) Y_l Y_r$.
In this case, one would observe perfect teleportation fidelity even without many-body time evolution, i.e.~$U = \mathbbm{1}$; in fact, if $U$ is perturbed away from the identity via  scrambling dynamics, the teleportation fidelity would actually become suppressed. The simplest way to see this is via Fig.~\ref{fig: 1}(a)---in particular, any subsequent time-evolution on the right side of the system is in the wrong direction to refocus the time-evolved state (one would want to apply $U^\dagger$ after the coupling, not $U^T$).
To achieve a large teleportation fidelity, the combined time-evolution, $U^T U$, would therefore need to preserve the ``teleported'' state, $\bra{\psi} U^T U \ket{\psi} \sim 1$, a situation that is only likely to occur if the dynamics are non-scrambling ($U=\mathbbm{1}$ is a special case of this) or undergo a late-time, fine-tuned, Poincare-type recurrence.  

\subsection{One-sided implementation of teleportation circuit}\label{sec:one-sided}

%



\FigureOneSided

Before proceeding to the experimental blueprints, we first introduce a simpler implementation of the teleportation protocol that works at infinite temperature (Fig.~\ref{fig:one-sided}).
%
%
The outcome of this protocol is equivalent to that of the two-sided protocol (up to experimental errors), yet it eliminates the need to prepare EPR pairs and requires half as many degrees of freedom. 
%
The cost of this simplification is two-fold: ($i$) it is restricted to simulating an infinite temperature TFD state, and ($ii$) it requires a higher depth quantum circuit.
%
 
We derive the one-sided implementation from the `two-sided' implementation [copied in Fig.~\ref{fig:one-sided} from Fig.~\ref{fig: 1}(a)] by sliding all operations from the left side of the many-body EPR pairs to the right side, using Eq.~(\ref{eq: O slide}).
%
%
The initial state of the one-sided circuit thus corresponds to the top left of the two-sided implementation.
Namely, we initialize the $K$ `measured' qubits of subsystem C in a definite outcome state, $\ket {o_1 \cdots o_K}$ (purple).
These states should be drawn from the distribution of measurement outcomes, but when teleporting an EPR pair at infinite temperature they will be uniformly distributed.
For the $N-K$ `unmeasured' qubits, we use the resolution of the identity $\mathbbm{1} \propto \sum_{s} \ket s \bra s$ to replace the unterminated legs with an initial product state in the computational basis, $\ket {o_{K+1} \cdots o_N}$ (gray).
This state should be sampled from shot-to-shot over all $2^{N-K}$ basis states, in effect preparing a maximally mixed state on these qubits.
Finally, we include one ancillary qubit for each qubit to be teleported, whose initial state is sampled over a complete basis $\ket{\phi}$ for the teleported subsystem (i.e. subsystem A in Section~\ref{intro diagrams}).
Similar to the unmeasured qubits, this corresponds to the unterminated leg of the thermofield double state when we insert the teleported qubit $\ket{\psi}$ in the two-sided implementation.

Having defined an initial pure state, we now implement the circuit starting from the top left of the two-sided implementation and proceeding counter-clockwise (Fig.~\ref{fig:one-sided}).
%
%
The circuit consists of three successive applications of $U$ or $U^\dagger$, interspersed with a swap gate exchanging subsystem A with the ancillary qubit(s), and operations $\hat{V}_i = e^{i g o_i \hat{O}_i / K}$ determined by the initial state of the `measured' qubits.
The outcome of the circuit is an EPR measurement between the ancilla qubit and subsystem A (black arrows).

As one can see in Fig.~\ref{fig:one-sided}, the one-sided implementation no longer performs teleportation, but rather prepares an EPR pair from an otherwise scrambled, many-body system.
Specifically, we know that upon swapping out, subsystem A is maximally entangled with the remaining qubits whenever the unitary, $U$, is scrambling; the one-sided circuit distills this entanglement into an output EPR pair.
This connection has been noted in gravity, where similar one-sided protocols can be interpreted as distilling the partner operators of emitted Hawking radiation~\cite{yoshida2019observer,yoshida2019firewalls} or observing behind the horizon in the SYK model~\cite{kourkoulou2017pure}.

\subsection{Preparing the thermofield double state} \label{sec: preparing TFD}

In the previous subsection, we introduced a one-sided protocol that obviates the need to prepare the highly entangled TFD state.
However, this approach was restricted to infinite temperature; at \emph{finite temperature}, one must implement the original two-sided protocol, which necessitates preparing a finite temperature TFD state.
A number of recent works have explored the preparation of TFD states variationally using quantum approximate optimization algorithms (QAOA)~\cite{wu2019variational,zhu2019TFD,su2020variational};  we note that these preparation strategies require no additional experimental capabilities beyond those already necessary for the TW protocol.
%
The optimization step within a QAOA-based TFD preparation  relies on a cost function that requires one to  measure the  entanglement entropy between the two sides~\cite{wu2019variational,zhu2019TFD}. While challenging, this can in principle be experimentally realized by either using several copies of the system~\cite{daley2012measuring,abanin2012measuring,Johri2017} or via randomized measurements~\cite{Elben2018}, both of which have been demonstrated in small-scale trapped ion experiments~\cite{Linke2018, Brydges2020}.

\subsection{Implementation with neutral Rydberg atoms} \label{sec:rydberg}

\FigureRydberg

One particularly promising platform for implementing the traversable wormhole protocol is a quantum simulator based on neutral alkali or alkaline-earth atoms held in a reconfigurable and controllable array of optical dipole traps. Recent experiments have already achieved near-deterministic trapping and loading of atoms into arbitrary geometries in one, two, and three dimensions~\cite{Xia15,Maller15,barredo2018synthetic}. 
By leveraging the  strong dipole coupling between atomic Rydberg states, high-fidelity analog quantum simulations and digital gates have also recently been demonstrated ~\cite{Maller15,Labuhn16,Bernien17,Graham19,Madjarov20,Wilson19}. 
These demonstrations have primarily used two natural schemes of encoding qubits into neutral atoms:
\begin{enumerate}
    \item A qubit can be encoded by choosing an atomic ground state $|g\rangle$ to be the $|0\rangle$ state, and a highly excited Rydberg state $|r\rangle$ with principal quantum number $n\gg 1$ as the $|1\rangle$ state [see Fig.~\ref{fig:rydberg-implementation}(a)].
    \item Alternatively, the qubit states can also be chosen as two long-lived hyperfine ground states (for alkali atoms or fermionic alkaline earth atoms) or a ground state and a metastable clock state (for bosonic alkaline earth atoms), such that the $|1\rangle$ state can be coupled to a Rydberg state to perform entangling gates [see Fig.~\ref{fig:rydberg-implementation}(c)].
\end{enumerate}

We will show how both encodings can be used to realize the teleportation protocol in feasible near-term experiments. We find that the first encoding is naturally suited to `analog' time-evolution under the native (Ising-type) Hamiltonian for a Rydberg setup, but is limited to system sizes of $\lesssim 30-35$ qubits (in one spatial dimension) due to the inability to perfectly time-reverse long-range interactions. On the other hand, the second encoding is more flexible and  allows for digital time-evolution including RUCs and Floquet dynamics. This time-evolution can be reversed exactly and is limited only by qubit and gate fidelities. 
While we will primarily consider realizations of our protocol in experimental setups where the neutral atoms are individually trapped in optical tweezers and undergo (near-)resonant excitation to Rydberg states, we also conclude by discussing how similar physics can be seen in an optical lattice setup where the atoms are primarily in ground states $|0\rangle$ and $|1\rangle$, but one of these states is `dressed' by an off-resonant laser field which couples it to a Rydberg state~\cite{Glaetzle15,Potirniche17,Zeiher17}.

\emph{Analog implementation}---We first consider the encoding where the qubit states $|0\rangle$ and $|1\rangle$ correspond to a ground state $|g\rangle$ and a highly excited Rydberg state $|r\rangle$. 
While neutral atoms are effectively non-interacting in their ground states, nearby atoms interact strongly via van der Waals interactions $\propto n^{11}/R^6$ if they are both in the Rydberg state, where $R$ is the distance between the atoms. 
If one drives the transition $|g_i\rangle\leftrightarrow |r_i\rangle$ at each site $i$ with tunable Rabi frequency $\Omega_i$ and detuning $\Delta_i$ [see Fig.~\ref{fig:rydberg-implementation}(b)], the system will undergo analog time evolution under the Hamiltonian
\begin{equation}
\label{eq:ryd-h}
H = \sum_i \frac{\Omega_i}{2}  X_i  +   \sum_i \frac{\Delta_i}{2} (1-Z_i) + \sum_{i\neq j} \frac{J_{ij}}{4} (1-Z_i)(1-Z_j)
\end{equation}
where $X_i = |g_i\rangle \langle r_i|+|r_i \rangle \langle g_i|$, $Z_i = |g_i \rangle \langle g_i | - |r_i \rangle \langle r_i |$, and $J_{ij} = J_0/|i-j|^6$ is the van der Waals interaction strength between two atoms at positions $i$ and $j$. 

The Hamiltonian in Eq.~(\ref{eq:ryd-h}) is scrambling and exhibits a scrambling time
limited by the smaller of $J_0$ and $\Omega_i$,  $t^* \sim N / \text{min}(J_0,\Omega_i)$. To minimize the total evolution time, we set $|\Omega_i| \sim J_0$, so that evolution under $H$ for a time $\sim \! N/J_0$ implements a fully scrambling unitary $U$ in the teleportation protocol. 
To implement $U^\dagger$, we reverse the nearest-neighbor interactions by conjugating time-evolution via Pauli operators $X_i$ (i.e. applying $\pi$-pulses) on every other site. 
The tunable single-site parameters $\Omega_i$ and $\Delta_i$ are then adjusted to ensure that each single-site term is also reversed.
We note that this simple scheme does \emph{not} reverse the (much weaker) next-nearest-neighbor interactions.

In a one-dimensional array, the errors in our implementation will arise from two main sources: ($i$) the finite lifetime of the Rydberg state, which gives rise to a nonzero decoherence rate at each of the $N$ sites, and ($ii$) the weak next-nearest neighbor interactions $\sim \! J_0/2^6 = J_0/64$, which cannot be time-reversed simultaneously with nearest neighbor interactions. 
%
To estimate the effect of the former, let us consider the specific case of $^{87}$Rb atoms excited to the $70S$ Rydberg state~\cite{Bernien17,Labuhn16}, which has a lifetime $\tau \approx 150$~$\mu\textrm{s}$.
Realistically achievable Rabi frequencies and interaction strengths are of order $\sim 2\pi \times 10-100$~MHz. 
The total time to implement the three scrambling unitaries of the teleportation protocol is thus $\sim 3N/|\Omega_i|$; when summed  over $N$ qubits and compared to the Rydberg lifetime, this  gives an estimated many-body error $\sim 3 N^2 / |\Omega_i| \tau$.

\begin{figure}
\centering
\includegraphics[width=.35\textwidth]{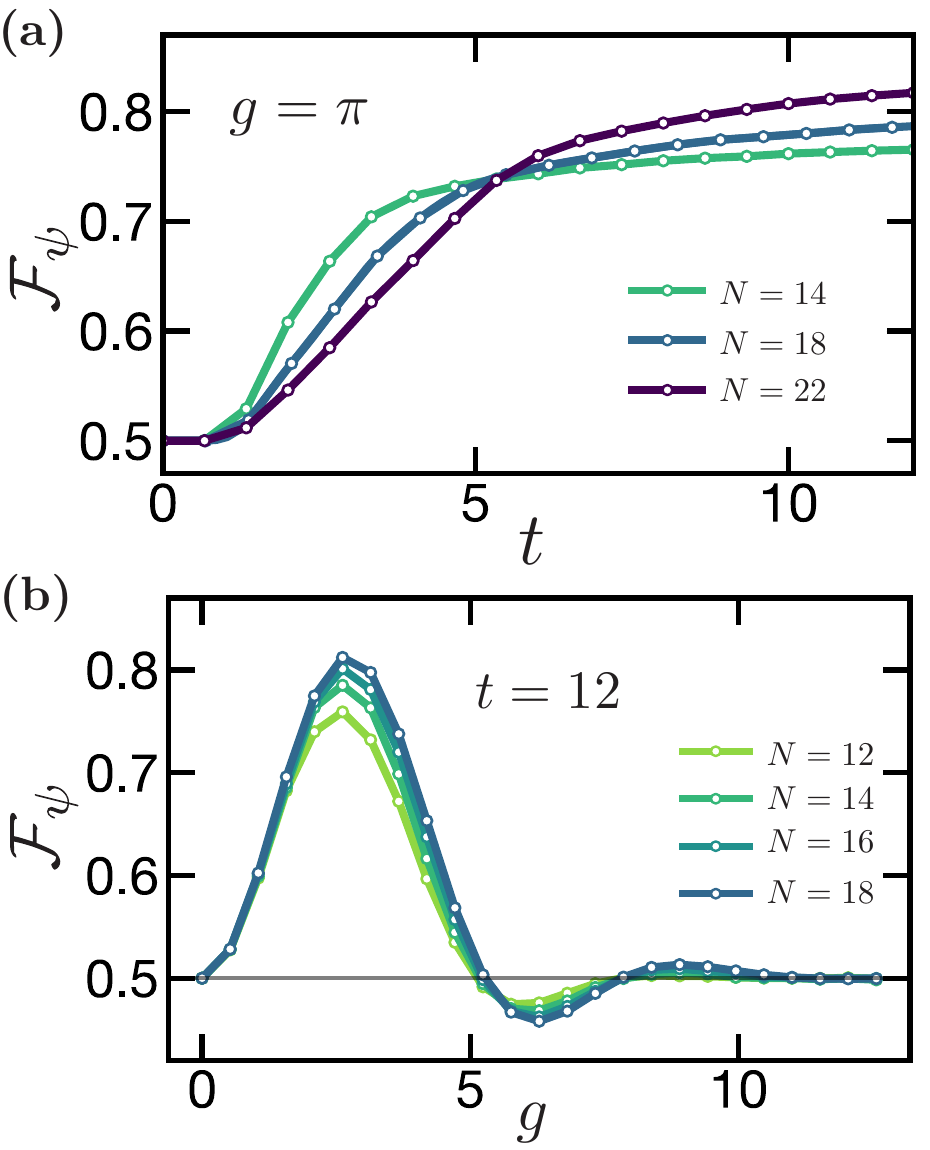}
\caption{
Finite-size scaling of the Rydberg simulations \textbf{(a)} as a function of time with $g=\pi$, and \textbf{(b)} as a function of coupling strength $g$ with $t = 12$. The system was evolved under the Rydberg Hamiltonian, Eq.~\eqref{eq:ryd-h}, with the same system parameters as in Fig.~\ref{fig:rydberg-implementation}. 
At late times, the fidelity increases for larger systems but decreases for larger values of $g$.
This is consistent with our error analysis in Section \ref{late times}; in particular, we expect the error to scale as  $ g^2 \delta S^2 / N^2$ and the size distribution to approach a binomial distribution for which $\delta S \sim \mathcal{S}/\sqrt{N}$.
In contrast, at early times, smaller systems exhibit a larger fidelity not because of the size width but because the acquired phase is $\constd g \mathcal{S}(t)/N$, where $\constd g$ is fixed and $\mathcal{S}(t)$ is initially independent of size. 
The curves in (a) intersect near the scrambling time due to the transition between the early and late time regimes. 
} 
\label{fig:rydberg_scaling}
\end{figure}

%
In order to precisely characterize the effects of imperfect backwards time-evolution, we perform large-scale numerical simulations of the teleportation protocol with the Rydberg Hamiltonian, Eq.~\eqref{eq:ryd-h} \cite{dynamite}. 
Our results are depicted in Fig.~\ref{fig:rydberg-implementation}(b) for a one-dimensional chain of $N=20$ atoms and three values of the coupling $g$. 
Analogous to our 1D RUC numerics [Fig.~\ref{fig: RUC}(a)], the fidelity increases monotonically in time for $g=\pi$; while, for $g=2\pi$ and $g=3\pi$, the fidelity oscillates in time, reaching a local maximum whenever the average size satisfies the phase-matching condition [Eq.\eqref{eq:phase-matching}].
Notably, even with \emph{perfect} time reversal, the overall fidelity is reduced from unity due to the finite width of the size distribution. 
This is a general feature of peaked-size teleportation in finite-size systems, since the relative size width scales as $\delta \mathcal{S}/\mathcal{S} \sim 1/\sqrt{N}$ (Section \ref{late times}). 
Indeed, in Fig.~\ref{fig:rydberg_scaling}, we confirm that the fidelity improves with increasing system size and is consistent with our peaked-size error analysis [e.g.~see Eq.~\eqref{eigV expansion}]. 


%
With \emph{imperfect} time reversal, we observe an additional $\sim 10\%$ reduction in the fidelity compared to the ideal case at the scrambling time [Fig.~\ref{fig:rydberg-implementation}(b)]. 
%
We can estimate the magnitude of this effect by assuming errors due to the next-nearest-neighbor interactions add coherently over time-intervals $\delta t \sim 1/J_0$ (the local thermalization time), and incoherently at larger time-scales.
Within each $\delta t$, each atom accumulates an error $\sim (\delta t \, J_0/64)^2$; summed over $N$ atoms and total time $3 t^* \approx 3 N \delta t$, this gives a total many-body error $\sim 3 N^2 / 64^2$.
%
Thus, the error due to imperfect time reversal is magnified at larger system sizes and will eventually outweigh the improvement in fidelity from the narrowing of the size distribution. 

Combined with the Rydberg lifetime error, this suggests that near-term experiments should be able to implement peaked-size teleportation in systems of $N \sim 35$ qubits.
We note that in higher dimensions, the smaller relative distance of next-nearest neighbor atoms gives rise to a larger error contribution from imperfect time-reversal. 


\emph{Digital implementation}---To implement the protocol in larger systems, higher dimensions, and at finite temperature, we propose a digital scheme, using the second type of qubit encoding (i.e.~hyperfine ground states) [Fig.~\ref{fig:rydberg-implementation}(c)]. 
In this approach, we envision time-evolution to be formed from alternating layers of nearest-neighbor controlled-phase gates  and single-qubit rotations. 
Here, the controlled-phase gates can be implemented by applying a simple pulse sequence to excite and de-excite qubits from the $|1\rangle$ state to the $|r\rangle$ state, so that the wavefunction acquires a phase of $-1$ if either of the two qubits are in the $|1\rangle$ state, but not if both qubits are in the $|0\rangle$ state [see Fig.~\ref{fig:rydberg-implementation}(c) insets]~\cite{Jaksch00}.
As demonstrated in recent experiments~\cite{Levine19}, these Rydberg-mediated controlled-phase gates can be performed in parallel for sufficiently well-separated pairs of qubits, and non-nearest neighbor interactions can be avoided by slightly reducing the parallelism within each layer of controlled-phase gates. 
Single-qubit rotations can be performed with sufficiently high fidelity such that the overall circuit fidelity is primarily limited by the entangling gates~\cite{Xia15,Levine18}. 

For a generic choice of gates, the circuit will be fully scrambling when $U$ is composed of $\sim N$ layers of controlled-phase gates.
The fidelity of the overall implementation is limited by the finite lifetime of the Rydberg state, which is populated for time $\sim 1/J_0$ during each controlled-phase gate. 
Assuming the same experimental parameters as in the analog case, one expects to be able to perform approximately $\Omega \tau \sim 10^3-10^4$ controlled-phase gates within the decoherence time-scale. 
Thus, in the digital approach, one expects that the teleportation protocol can naturally be implemented for $N \sim 200$ qubits.

The digital approach can also be adapted to experiments using Rydberg-dressed neutral atoms in an optical lattice~\cite{Glaetzle15,Potirniche17,Zeiher17}.
In such a setup, qubits are again encoded in hyperfine ground states and strong Ising-like interactions are generated by coupling the qubit state $|1\rangle$ to a Rydberg state with a far-detuned laser field. 
In this way, the Rydberg interaction gives rise to an energy shift for two neighboring atoms both in the $|1\rangle$ state. 
Analogous to our previous discussion, a simple scrambling unitary could consist of alternating layers of Rydberg-dressed interactions and single-qubit rotations.
While the total accumulated error  in the Rydberg-dressing approach is comparable to the gate-based protocol, one potential advantage is an increased tunability of the interactions~\cite{van2014quantum,de2019observation}.

In addition to scrambling time evolution, there are three ingredients to implement the one-sided teleportation circuit (Fig.~\ref{fig:one-sided}): ($i$) the ability to `swap' in the qubit $\ket{\phi}$, ($ii$) single-qubit rotations, $V_i = e^{\pm i g Z_i / K}$, and ($iii$) the final measurement in the EPR basis.
In both digital setups, these are easily accomplished by combining controlled-phase gates, arbitrary single-qubit rotations, and local measurements.
In the analog setup, we propose to temporarily `turn off' the Hamiltonian by transferring each Rydberg state $\ket{r}$ to a hyperfine ground state (e.g. the state used as $\ket{1}$ in the \emph{digital} protocol) using a resonant laser pulse.
Once this is done, all of the above operations can be performed identically as in the digital setup.
Afterwards, an additional resonant laser pulse returns the system to the analog encoding.
The ancillary qubit can be decoupled from the system qubits during Hamiltonian time-evolution in two ways: ($i$) by physically positioning the ancillary qubit far from the system, or ($ii$) by encoding the ancillary qubit in the hyperfine subspace throughout time-evolution.

The two-sided, finite temperature TW protocol can be achieved by combining the above techniques with TFD preparation as in Section~\ref{sec: preparing TFD}.
A particularly natural geometry for such a realization would be two parallel chains of Rydberg atoms, with each chain forming one side of the TFD state.
The coupling between the two sides is naturally realized by the atoms' Ising interactions.
This coupling can be applied independently from the one-sided Hamiltonian using either full digital control or by manipulating the inter- vs. intra-chain atomic distance.
%

\FigureIons

\subsection{Implementation with trapped ions} \label{sec: trapped}

A second experimental platform that naturally enables the implementation of the TW protocol is arrays of individual trapped atomic ions~\cite{bohnet2016quantum,vermersch2019probing,zhang2017observation}.
Trapped ion qubits feature near-perfect replicability, negligible idle errors, and the ability to implement both  a universal set of reconfigurable quantum gates~\cite{cetina2020} as well as analog long-range spin Hamiltonians~\cite{wineland2008entangled,monroe2013scaling}. 
%
Entangling quantum gates have been demonstrated between isolated pairs of trapped ions with fidelities exceeding $99.9\%$ \cite{Ballance:2016, Gaebler2016high}.
%
Teleportation protocols---including the HPR protocol~\cite{landsman2019verified}---involving gate operations, partial measurement and feedforward operations, have been experimentally realized in a number of contexts~\cite{riebe2004deterministic, barrett2003sypathetic, Olmschenk2009, landsman2019verified}.

Compared to Rydberg atom arrays, trapped ions offer two new regimes for exploring many-body teleportation.
First, trapped ions naturally interact via a long-range analog Hamiltonian, whose time-evolution can be fully reversed within certain experimental regimes~\cite{korenblit2012quantum,teoh2020machine}.
Implementing the TW protocol in this setting would  provide a window into operator spreading and size distributions under such long-range dynamics~\cite{else2020improved,zhou2020operator}.
Second, when operated digitally, the same long-range interaction has already been demonstrated to enable the preparation of thermofield double states~\cite{martyn2019product,zhu2019TFD,su2020variational,wu2019variational}, a crucial step towards realizing the two-sided TW protocol at finite temperature (see Section~\ref{sec: preparing TFD}).

We begin by outlining the analog and digital forms of time-evolution that are possible in trapped ion systems. 
Interactions between qubits typically stem from state-dependent optical dipole forces that off-resonantly drive motional sidebands of the qubit~\cite{cirac1995quantum,molmer_multiparticle_1999}. 
These sideband operations mediate entanglement and give rise to an effective Ising coupling. 
When the optical forces are symmetrically detuned far from the upper and lower sidebands, the motion is only virtually excited, resulting in a long-range Ising Hamiltonian [Fig. \ref{fig:ions}(b)]:
\begin{equation}
    H = \sum_{i<j} J_{ij} X_i X_j + B_z \sum_i Z_i,
\label{trapped ion H}
\end{equation} 
where $J_{ij} \approx J_0/|i-j|^\alpha$, with $0<\alpha<3$ and $J_0 \lesssim 1$ kHz,
and the effective magnetic field $B_z$ can be realized by slightly asymmetrically detuning the driving field~\cite{monroe2019programmable}.
The sign of the couplings can be reversed by changing the detuning of the optical forces from the motional sidebands~\cite{korenblit2012quantum,teoh2020machine}.


On the other hand, when the optical dipole forces are closer to resonances of the motional  modes, one can mediate interactions significantly faster, allowing for the execution of rapid, entangling quantum gates between pairs of illuminated ion qubits [Fig.~\ref{fig:ions}(a)]~\cite{zhu2006trapped, DebnathQC:2016}.
The native entangling gates are based upon Ising interactions between any selected pair of ions with a tunable interaction angle; in particular, both $XX_{ij}(\theta) = e^{-i\theta X_i X_j/2}$ and $YY_{ij}(\theta) = e^{-i\theta Y_i Y_j /2}$ gates are available and $\theta=\pi/2$ naturally creates an EPR pair~\cite{Landsman2019arb, Wright:2019}.  
%
%
Typical entangling operations have duration $1/J_{\text{ent}} \sim 100$~$\mu$s, while decoherence time-scales are on the order of  $\tau \sim 400$~ms~\cite{debnath2016programmable}.
Following the estimates of Section~\ref{sec:rydberg} and requiring $3 N^2 / J_{\text{ent}} \tau \lesssim 1$, we estimate that near-term state-of-the-art experiments can support high-fidelity many-body teleportation for up to $N \sim 35$ qubits.

Let us now describe an implementation of the one-sided TW protocol (Fig.~\ref{fig:one-sided}).
We first focus on the ability to implement both $U$ and its inverse $U^\dag$. 
For analog time-evolution [Eq.~\eqref{trapped ion H}], $U^\dag$ can be implemented by changing the sign of the detuning~\cite{garttner2017measuring}, while for digital time-evolution, one can directly invert and reverse the ordering of the quantum gates.

The one-sided protocol also requires the ability to locally address a sub-extensive number of individual qubits. 
%
In particular, a subset $K$ of the qubits must be initially prepared in a product state, $|o_1, \ldots, o_K \rangle$ and later rotated by  $\hat{V}_i = e^{i g o_i \hat{O}_i / K}$. 
These rotations can be achieved by taking $\hat{O}_i = \hat{Z}_i$ and individually addressing the target ions using an auxiliary ``poke'' laser beam~\cite{smith2016many,zhang2017observation}.

Following  the first application of $U$, one must swap out  the qubit(s) corresponding to the teleported subsystem.
This swap can be implemented either digitally by applying a SWAP-gate, or physically, by exchanging the two ions via a modulation of the ion trap's axial fields \cite{hensinger2006t,monroe2013scaling,kaufmann2017fast}.

Extending this implementation to the two-sided protocol [Fig.~\ref{fig: 1}(a)] is  straightforward. 
Initialization into EPR pairs (for infinite temperature) can be accomplished via simple Ising gates at the input of the circuit [Fig.~\ref{fig:ions}(a,c)], while the TFD state (for finite temperature) can be prepared via variational methods (Section~\ref{sec: preparing TFD}). 
Time-evolution can again take the form of either digital quantum gates [Fig.~\ref{fig:ions}(a)] or analog Hamiltonian dynamics. 
To separately implement analog dynamics on the two sides of the system, one would illuminate only \emph{half} of the ion chain at any given time  [Fig.~\ref{fig:ions}(b)]; this has the added benefit of avoiding unwanted coupling between the left and right sides, but implies that the time-evolution must be performed serially [Fig.~\ref{fig:ions}(c)]. 

Finally, in the two-sided protocol, one must perform projective measurements on $K$ qubits that feed-forward to the conditional rotations, $\hat{V}_i$. 
These partial measurements can be accomplished by using multiple ion species (i.e.~different elements or isotopes) \cite{barrett2003sypathetic}, or  alternatively, this entire procedure can be replaced with a specific interaction, $e^{igV}$, between the two sides; this interaction is naturally realized via an $XX_{ij}(\theta)$ gate with $\theta = 2g/K$.

\subsection{Effects of experimental error and relation to quantum error correction}

We now turn to the effect of experimental error on the TW protocol.
We find that teleportation is robust to nearly all errors that occur on the left side of the TFD state after time-evolution by $U$, but is strongly sensitive to errors at nearly all other locations in the protocol. 
These two extremes are emblematic of two different relations between scrambling and error: the former corresponds to interpretations of scrambling as an error-correcting code~\cite{hayden2007black}, while the latter reflects recent results showing that the effect of errors on scrambling measurements is enhanced proportional to the size, $\Size$, of time-evolved operators~\cite{schuster2021operator}.
In the following discussion, we demonstrate each of these points through simple but representative examples of experimental error.

We begin with the first case: consider errors occurring on the left side of the TFD state after application of $U$ but before measurement/coupling.
Recall that, in the absence of error, one can perform teleportation by using any $K \sim \mathcal{O}(1)$ qubits of the left side.
This implies that teleportation is robust to \emph{any} errors that affect only $N-K$ qubits: as long as one has knowledge of at least $K$ qubits that are unaffected, measuring these qubits performs teleportation identically to the error-free case.

This robustness reflects previously noted connections between scrambling and quantum error correction~\cite{hayden2007black}.
In particular, we note that many-body teleportation can be understood as an especially generic example of entanglement-assisted quantum error correction (EAQEC)~\cite{brun2006correcting}. 
Indeed, the setup for EAQEC is identical to that of the teleportation protocol: two parties, Alice and Bob, share entanglement (the TFD state), Alice applies an encoding circuit to her share of qubits (the left unitary, $U$), and decoding is achieved by teleporting Alice's quantum state to Bob's share of qubits (via the coupling, $V$, and unitaries on the right).
%
%
Previous schemes for EAQEC have focused primarily on encodings via Clifford unitaries.
In contrast, many-body teleportation, and more specifically peaked-size teleportation, succeeds for a \emph{vastly} broader class of encoding procedures---i.e.~scrambling many-body time dynamics---indicating that naturally occurring, strongly interacting systems offer novel methods of EAQEC.
%

On the other hand, errors that occur \emph{during} encoding or decoding---i.e. during the application of $U$ on the left side or at any point on the right side---strongly inhibit teleportation.
As a first example, consider a single local error, $W_1$, occurring with probability $\varepsilon$ on the right side after coupling but before $U^T$ (i.e. just before decoding).
If the error, $W_1$, grows to have a size, $\Size$, after $U^T$ is applied, one estimates that it will decrease the teleportation fidelity by an amount, $1-F \sim \varepsilon \Size / N$, proportional to the probability that $W_1$ has support on the teleported qubit after time-evolution.
If we sum over such errors on all $N$ qubits, we have $1-F \sim \varepsilon \Size$.

As a second example, consider a local error, $W_2$, occurring with probability $\varepsilon$ on the left side simultaneously with state insertion (e.g. a damaged TFD state in Fig.~\ref{fig: 1}).
In effect, this error shifts the correlator operators [Eq.~(\ref{eq: CQ})], $Q \rightarrow Q \otimes W_2$; following the arguments of Section~\ref{sec: intermediate time}, one then requires that the sizes add for teleportation to succeed, $\Size[QW_2] = \Size[Q] + \Size[W_2]$.
In a 1D short-range system (Section~\ref{geq1D}), this condition holds if and only if the light cones of $W_2$ and $Q$ do not overlap.
For $\mathcal{O}(\varepsilon N)$ randomly distributed errors, we expect this to hold as long as the spacing between errors, $1/\varepsilon$, is much larger than the size of the light cone, $\varepsilon \Size \ll 1$.
A similar scaling holds in 0D (Section~\ref{0D RUCs}).
Here, we expect size addition to hold as long as the size of the total error (corresponding to a time-evolved product of $\sim \varepsilon N$ initially local operators), is much smaller than the system size, $N$.
Once again, this requires $\varepsilon N \Size \ll N$, or $\varepsilon \Size \ll 1$.

The two previous examples are straightforwardly generalized to errors that accumulate continuously throughout time-evolution.
To do so, we replace the error probability with an error rate, $\varepsilon$ (now with units of inverse time).
The total effect of the error is then given by the integral of the error rate multiplied by the size over time, $\varepsilon \int_0^t dt' \, \Size(t')$~\cite{schuster2021operator}.
In one-dimensional systems  evolved up to the scrambling time, i.e. $\Size \sim J t$ and $t_s \sim N / J$ for a local interaction strength $J$, we thus estimate a total error, $\varepsilon \int_0^{t_s} dt' J t' \sim \varepsilon \Size t_s \sim \varepsilon N^2 / J$, in agreement with our rough estimates in Sections~\ref{sec:rydberg} and \ref{sec: trapped}.

Finally, we consider a particular form of error that may be relevant for analog time-evolution: mismatches between the evolution times of $U$, $U^*$, and $U^T$.
We denote these three evolution times as $t_1, t_2, t_3$, respectively, and their mismatches as $\Delta t_{12} = t_2 - t_1$ and $\Delta t_{13} = t_3 - t_1$.
We can characterize the mismatches' effect on the teleportation fidelity using the correlators, $C_Q(t_1,t_2,t_3) = \langle  U_r^*(t_3) Q_r U_r^T(t_3) e^{igV} U_l(t_1) Q_l U^\dagger_l(t_2) \rangle$ (Section \ref{sec: requirements}).
From this, we anticipate that the protocol is relatively insensitive to mismatches between $t_3$ and $t_1, t_2$: teleportation succeeds as long as the mismatch is small compared to the local interaction strength, $J$, i.e. $J \Delta t_{13} \lesssim 1$.
To estimate this, we set $g=0$ and $t_1 = t_2$ in which case the correlator magnitude is given by an autocorrelation function, $C_Q = \langle Q(t_1) Q(t_3) \rangle = G(\Delta t_{13})$.
The teleportation fidelity is bounded above by this expression, which decays on a time-scale $\sim 1/J$.
On the other hand, teleportation is more strongly sensitive to the mismatch between $t_1$ and $t_2$.
To estimate this, we treat the difference in time-evolution between $U$ and $U^*$ as a product of $\sim (J \Delta t_{12})^2 N$ local errors occurring simultaneous with state insertion (to motivate this scaling, note that one can approximate $U(\Delta t_{12})$ as a product of $\sim \! N$ local unitaries for small $\Delta t_{12}$, and we expect the error to be an even function of $\Delta t_{12}$).
Following our previous analysis, teleportation is successful as long as $\Size (J \Delta t_{12})^2 N \ll N$, or $\Size (J \Delta t_{12})^2 \ll 1$. 

%
%
%

\subsection{Directly measuring the size distribution}

In Section~\ref{sec: signatures}, we discussed that the time profile of the teleportation fidelity reveals important features of the operators' size distributions, including the average operator size and the size width.
We now demonstrate that a more precise characterization of the operator size distribution can be obtained by sweeping the coupling strength, $g$, at a fixed time, $t$.

For simplicity, we restrict to infinite temperature\footnote{At finite temperature, a similar procedure to what follows determines the winding size distribution discussed in Section~\ref{sec: gravity size winding}~\cite{brown2020quantum}. The size distribution can be determined by moving the final measurement of the TW protocol to the left side.} and the coupling $V_s$ in Eq.~(\ref{size qudit}), which precisely measures the operator size.
In this case, the two-sided correlator [Eq.~(\ref{eq: CQ})] is equal to the characteristic function, $\Phi_\Size(g)$, of the size:
\begin{equation}
    C_Q(t) =
    e^{ig} \sum_\Size P(\Size) e^{-i g \Size / N} \equiv e^{ig} \Phi_\Size(g) 
\end{equation}
from which the size distribution can be obtained by a Fourier transform in $g$.

More precisely, to measure the \emph{real part} of the characteristic function (i.e. the teleportation correlator), we perform the teleportation protocol with two small modifications: ($i$) we replace state insertion with the specific projection operator, $(1+Q)/2$, and ($ii$) we measure the expectation value of $Q$ applied to the right side, instead of the teleportation fidelity.
This yields the quantity:
\begin{equation}
\begin{split}
    & \bra{\epr} \frac{1+Q_{l}(t)}{2} e^{-igV_s}  Q_{r}(-t) e^{igV_s} \frac{1+ Q_{l}(t)}{2} \ket{\epr} \\
    & = \text{Re} \big[ \bra{\epr} e^{-igV_s} Q_{r}(-t) e^{igV_s} Q_{l}(t) \ket{\epr} \big] \\
    & = \text{Re} \big[ \varphi_\Size(g) \big], \\
\end{split}
\end{equation}
where in the second line we use that the ``diagonal'' terms between the two copies of $(1+Q)/2$ vanish at infinite temperature.
The imaginary part of the characteristic function can be obtained similarly, by replacing state insertion, $(1+Q)/2$, with an application of the unitary operator, $(1+iQ)/\sqrt{2}$.
Analogous to Fig.~\ref{fig:one-sided}, both of these measurement schemes can be adapted into one-sided protocols using Eq.~(\ref{eq: O slide}) whenever the coupling $V$ is classical (i.e. composed of terms $O_{i,l} O_{i,r}^*$, where $\{ O_i \}$ mutually commute).
While such couplings do not measure the exact size distribution, we expect their behavior to be similar in most cases (Section \ref{Coupling size}).

For completeness, we also note an alternate method to measure the size distribution: one prepares the state $Q_l(t) \ket{\epr}$ and directly measures the two-sided coupling $V_s$. The probability distribution of the measurement results gives the size distribution [see the discussion below Eq.~(\ref{size qudit})].


Let us now compare these two protocols to other schemes for characterizing the size distribution of operators. 
First, we recall that a sum of local OTOCs yields the average operator size [Eq.~\eqref{V OTOC}]. 
Hence, many existing protocols for measuring local OTOCs~\cite{swingle2016measuring,yao2016interferometric} can be straightforwardly adapted to measuring the average size. 
Higher order moments of the size distribution can similarly be obtained from local OTOCs, using Eq.~(\ref{size qudit}):
\begin{equation}
\begin{split}
    \langle (1 & -\Size/N)^n \rangle = \left<V_s^n\right>_Q \\
    & = \frac{1}{N^n} \sum_{P_{i_1},\ldots,P_{i_n}} \tr (Q(t)  \, \prod_{k=1}^n P_{i_k} \, Q^{\dagger}(t) \, \prod_{k=n}^1 P_{i_k}^\dagger ).
\end{split}
\end{equation}
where the sum is over every possible combination of $n$ single-qubit Pauli operators $P_{i_1},\ldots,P_{i_n}$. 
%
%
Based on this approach, however, the number of measurements required to determine the $n^{\text{th}}$ moment scales as $\mathcal{O}(N^n)$.
In certain situations, this scaling may be reduced through sampling, though this depends on the nature of the size distribution and the desired degree of precision. 
Furthermore, reconstructing the full profile of the size distribution from a finite number of moments is generally a difficult numerical task~\cite{john2007techniques}. 
In contrast to these limitations, our proposal directly yields the full size distribution, and can recover its moments with a number of measurements independent of the system size\footnote{This is simplest to see in the protocol which measures the two-sided coupling $V_s$.  Here, the error in one's measurement of the $n^{\text{th}}$ moment is equal to the expectation value of the moment's variance divided by the number of measurements, $(\delta \overline{\langle \Size^n \rangle})^2 = (\langle \Size^{2n} \rangle - \langle \Size^n \rangle^2)/M$. If one wishes to resolve the moment to within a relative error $\varepsilon$, i.e. $\delta \overline{\langle \Size^n \rangle} < \varepsilon \overline{\langle \Size^n \rangle}$, one requires $M \sim \frac{\langle \Size^{2n} \rangle - \langle \Size^n \rangle^2}{\varepsilon^2 \langle \Size^n \rangle^2}$ measurements. This number does not scale with $N$ since it contains the same powers of $\Size$ in the numerator and denominator.}.

We can also compare our proposal to an independent protocol for measuring the size distribution introduced in Ref.~\cite{qi2019measuring}.
The protocol of Ref.~\cite{qi2019measuring} is experimentally simpler than our own, and in particular involves only a single application of time-evolution by $U$ (and no backwards time-evolution).
However, this simplicity comes at a cost: resolving high-size components of the distribution requires a number of measurements that scales \emph{exponentially} with size.

\section{Outlook} \label{discussion}

In this work, we developed a unified framework for understanding many-body teleportation from the perspective of operator growth under scrambling dynamics. 
The unifying concept within this framework is the size distribution of time-evolved operators~\cite{roberts2018operator,qi2019quantum,qi2019measuring,brown2019quantum,brown2020quantum}: these form the backbone of peaked-size teleportation, and provide a more fine-grained measure of operator growth compared to the average operator size (as given by the expectation value of OTOCs).

Our work suggests several future directions for applying and building upon this framework. 
First, while we have studied the size distributions in 0D and $\geq1$D RUCs, it would be interesting to extend this analysis to a multitude of other physical systems, where one expects to find qualitatively distinct behavior.
These include long-range interacting systems~\cite{else2018improved,tran2020hierarchy}, interacting and non-interacting integrable systems~\cite{qi2019measuring}, $\geq1$D systems with a large on-site Hilbert space~\cite{gu2017local}, 0D systems with sparse couplings~\cite{bentsen2019fast}, and systems with conserved quantities~\cite{khemani2018operator}.

Another set of open questions concerns the notion of operator size at finite temperature.
In systems with peaked size distributions, we found that the phase of the two-sided teleportation correlator was directly proportional to the conventional definition of operator size~\cite{qi2019quantum}.
Surprisingly, we observed that this relationship did not hold in the finite temperature SYK model; rather, the phase was given by the real part of the two-sided OTOC. 
Unlike the conventional size, this OTOC is not UV divergent, and is thus expected to be inherently independent of the microscopic Hilbert space.
Recent work has shown that its real part isolates an incoherent component of operator spreading in large-$N$ models~\cite{gu2019relation}; further work is needed to establish and expand this framework.
Related to these considerations, one may hope to better understand the bulk analogue of operator size in theories dual to gravity with strong stringy effects.
While we have seen that stringy effects can mimic peaked-size teleportation, developing a physical interpretation of this correspondence would be extremely exciting.

Third, we have shown that a promising application of the teleportation protocol is to distinguish between different classes of scrambling dynamics.
%
%
In particular, we have focused on two classes of scramblers---generic thermalizing systems and those with gravitational duals---and demonstrated that the key distinction between them is their teleportation fidelity at low temperatures. 
It is intriguing to ask whether the fidelity increase associated with gravitational teleportation may also occur in other systems, without a gravitational dual. 
For instance, recently the teleportation correlator magnitude was observed to increase slightly above $G_\beta$ in non-local random Hamiltonian systems~\cite{brown2019quantum,brown2020quantum}; generalizing this to other physical models would be of tremendous interest.

One may also wonder what role an extensive low temperature entropy---a key feature of the SYK model~\cite{maldacena2016remarks}---plays in the teleportation process.
In particular, how well can systems with extensive low temperature entropy but no known gravitational dual teleport~\cite{salberger2017deformed,alexander2018exact}?
We conjecture that an extensive entropy would allow one to \emph{locally} encode each qubit into low-energy degrees of freedom (i.e. operators with an $\mathcal{O}(1)$ two-point function), since one would only require $\mathcal{O}(1)$ qubits on the left side of the TFD in order to have one qubit of mutual information with the right side.
Such an encoding would allow low temperature teleportation with perfect fidelity if operator sizes were peaked, naturally motivating the study of operator size distributions in such models.

Finally, we would like to discuss the relation between our results on the TW protocol and the eternal traversable wormhole (ETW) introduced in Ref.~\cite{maldacena2018eternal}.
In the latter, the coupling, $V$, has an $\mathcal{O}(1)$ coefficient and, moreover, is applied \emph{simultaneously} with single-sided Hamiltonian evolution (i.e. the full system evolves under a Hamiltonian, $H_l + H_r + g \sum_j O_{j,l} O_{j,r}^*$).
Under these conditions, Refs.~\cite{maldacena2018eternal,plugge2020revival} find that the ETW teleportation fidelity oscillates in time under gravitational dynamics, indicating that information is transmitted back and forth between the two boundaries.
%
%
Intriguingly, unlike the TW protocol, the ETW oscillations occur at a time-scale given by the single-sided thermalization time ($\sim \! \beta$, the inverse effective temperature), and \emph{not} the scrambling time.
Developing a microscopic understanding of the ETW in terms of operator spreading, as well as exploring analogous physics in more generic many-body systems, remains an exciting open direction.

\emph{Note added}---After this work had been completed, we learned of an independent investigation of gravitational many-body teleportation by Nezami, Lin, Brown, Gharibyan, Leichenauer, Salton, Susskind, Swingle, and Walker, which will appear in the same arXiv posting.

\emph{Acknowledgments}---We gratefully acknowledge discussions with Zhenbin Yang, Douglas Stanford, Sepehr Nezami, Yingfei Gu, Xiangyu Cao, Jaewon Kim, Yimu Bao, Hannes Pichler, Alexander Keesling, Harry Levine, Geoffrey Pennington, Maxwell Block, Sagar Vijay, and Daniel Jafferis.
This work was supported by the U.S. Department of Energy through the Quantum Information Science Enabled Discovery (QuantISED) for High Energy
Physics (KA2401032) and through the GeoFlow Grant No. de-sc0019380. 
This research used resources of the National Energy Research Scientific Computing Center, a U.S. Department of Energy Office of Science User Facility operated under Contract No. DE-AC02-05CH11231.
The code used for our numerical simulations is available on the Zenodo public database \cite{zenodo}.
For the exact dynamical simulations, we utilize the \texttt{dynamite} Python frontend \cite{dynamite}, which supports a matrix-free implementation of Krylov subspace methods based on the \texttt{PETSc} and \texttt{SLEPc} packages.
T.S. acknowledges support from the National Science Foundation Graduate Research Fellowship Program under Grant No. DGE 1752814.
P.G. acknowledges support by the US Department of Energy grants DE-SC0018944 and
DE-SC0019127, and also the Simons foundation as a member of the \emph{It from Qubit} collaboration.
I.C. acknowledges support from the Alfred Spector and Rhonda Kost Fellowship of the Hertz Foundation and the Department of Defense through the National Defense Science and Engineering Graduate Fellowship Program. 
E.T.K. acknowledges support from the National Science Foundation Graduate Research Fellowship Program under Grant Nos. DGE1144152 and DGE1745303.
N.M.L. acknowledges support from the Maryland---Army-Research-Lab Quantum Partnership under Grant No. W911NF1920181.

\bibliography{refs_traversable_wormhole}

\begin{thebibliography}{130}%
\makeatletter
\providecommand \@ifxundefined [1]{%
 \@ifx{#1\undefined}
}%
\providecommand \@ifnum [1]{%
 \ifnum #1\expandafter \@firstoftwo
 \else \expandafter \@secondoftwo
 \fi
}%
\providecommand \@ifx [1]{%
 \ifx #1\expandafter \@firstoftwo
 \else \expandafter \@secondoftwo
 \fi
}%
\providecommand \natexlab [1]{#1}%
\providecommand \enquote  [1]{``#1''}%
\providecommand \bibnamefont  [1]{#1}%
\providecommand \bibfnamefont [1]{#1}%
\providecommand \citenamefont [1]{#1}%
\providecommand \href@noop [0]{\@secondoftwo}%
\providecommand \href [0]{\begingroup \@sanitize@url \@href}%
\providecommand \@href[1]{\@@startlink{#1}\@@href}%
\providecommand \@@href[1]{\endgroup#1\@@endlink}%
\providecommand \@sanitize@url [0]{\catcode `\\12\catcode `\$12\catcode
  `\&12\catcode `\#12\catcode `\^12\catcode `\_12\catcode `\%12\relax}%
\providecommand \@@startlink[1]{}%
\providecommand \@@endlink[0]{}%
\providecommand \url  [0]{\begingroup\@sanitize@url \@url }%
\providecommand \@url [1]{\endgroup\@href {#1}{\urlprefix }}%
\providecommand \urlprefix  [0]{URL }%
\providecommand \Eprint [0]{\href }%
\providecommand \doibase [0]{http://dx.doi.org/}%
\providecommand \selectlanguage [0]{\@gobble}%
\providecommand \bibinfo  [0]{\@secondoftwo}%
\providecommand \bibfield  [0]{\@secondoftwo}%
\providecommand \translation [1]{[#1]}%
\providecommand \BibitemOpen [0]{}%
\providecommand \bibitemStop [0]{}%
\providecommand \bibitemNoStop [0]{.\EOS\space}%
\providecommand \EOS [0]{\spacefactor3000\relax}%
\providecommand \BibitemShut  [1]{\csname bibitem#1\endcsname}%
\let\auto@bib@innerbib\@empty
\bibitem [{\citenamefont {Maldacena}\ \emph {et~al.}(2017)\citenamefont
  {Maldacena}, \citenamefont {Stanford},\ and\ \citenamefont
  {Yang}}]{maldacena2017diving}%
  \BibitemOpen
  \bibfield  {author} {\bibinfo {author} {\bibfnamefont {Juan}\ \bibnamefont
  {Maldacena}}, \bibinfo {author} {\bibfnamefont {Douglas}\ \bibnamefont
  {Stanford}}, \ and\ \bibinfo {author} {\bibfnamefont {Zhenbin}\ \bibnamefont
  {Yang}},\ }\bibfield  {title} {\enquote {\bibinfo {title} {Diving into
  traversable wormholes},}\ }\href@noop {} {\bibfield  {journal} {\bibinfo
  {journal} {Fortschritte der Physik}\ }\textbf {\bibinfo {volume} {65}},\
  \bibinfo {pages} {1700034} (\bibinfo {year} {2017})}\BibitemShut {NoStop}%
\bibitem [{\citenamefont {Bennett}\ \emph {et~al.}(1993)\citenamefont
  {Bennett}, \citenamefont {Brassard}, \citenamefont {Cr{\'e}peau},
  \citenamefont {Jozsa}, \citenamefont {Peres},\ and\ \citenamefont
  {Wootters}}]{bennett1993teleporting}%
  \BibitemOpen
  \bibfield  {author} {\bibinfo {author} {\bibfnamefont {Charles~H}\
  \bibnamefont {Bennett}}, \bibinfo {author} {\bibfnamefont {Gilles}\
  \bibnamefont {Brassard}}, \bibinfo {author} {\bibfnamefont {Claude}\
  \bibnamefont {Cr{\'e}peau}}, \bibinfo {author} {\bibfnamefont {Richard}\
  \bibnamefont {Jozsa}}, \bibinfo {author} {\bibfnamefont {Asher}\ \bibnamefont
  {Peres}}, \ and\ \bibinfo {author} {\bibfnamefont {William~K}\ \bibnamefont
  {Wootters}},\ }\bibfield  {title} {\enquote {\bibinfo {title} {Teleporting an
  unknown quantum state via dual classical and einstein-podolsky-rosen
  channels},}\ }\href@noop {} {\bibfield  {journal} {\bibinfo  {journal}
  {Physical review letters}\ }\textbf {\bibinfo {volume} {70}},\ \bibinfo
  {pages} {1895} (\bibinfo {year} {1993})}\BibitemShut {NoStop}%
\bibitem [{\citenamefont {Barrett}\ \emph {et~al.}(2004)\citenamefont
  {Barrett}, \citenamefont {Chiaverini}, \citenamefont {Schaetz}, \citenamefont
  {Britton}, \citenamefont {Itano}, \citenamefont {Jost}, \citenamefont
  {Knill}, \citenamefont {Langer}, \citenamefont {Leibfried}, \citenamefont
  {Ozeri} \emph {et~al.}}]{barrett2004deterministic}%
  \BibitemOpen
  \bibfield  {author} {\bibinfo {author} {\bibfnamefont {MD}~\bibnamefont
  {Barrett}}, \bibinfo {author} {\bibfnamefont {J}~\bibnamefont {Chiaverini}},
  \bibinfo {author} {\bibfnamefont {T}~\bibnamefont {Schaetz}}, \bibinfo
  {author} {\bibfnamefont {J}~\bibnamefont {Britton}}, \bibinfo {author}
  {\bibfnamefont {WM}~\bibnamefont {Itano}}, \bibinfo {author} {\bibfnamefont
  {JD}~\bibnamefont {Jost}}, \bibinfo {author} {\bibfnamefont {E}~\bibnamefont
  {Knill}}, \bibinfo {author} {\bibfnamefont {C}~\bibnamefont {Langer}},
  \bibinfo {author} {\bibfnamefont {D}~\bibnamefont {Leibfried}}, \bibinfo
  {author} {\bibfnamefont {R}~\bibnamefont {Ozeri}},  \emph {et~al.},\
  }\bibfield  {title} {\enquote {\bibinfo {title} {Deterministic quantum
  teleportation of atomic qubits},}\ }\href@noop {} {\bibfield  {journal}
  {\bibinfo  {journal} {Nature}\ }\textbf {\bibinfo {volume} {429}},\ \bibinfo
  {pages} {737--739} (\bibinfo {year} {2004})}\BibitemShut {NoStop}%
\bibitem [{\citenamefont {Riebe}\ \emph {et~al.}(2004)\citenamefont {Riebe},
  \citenamefont {H{\"a}ffner}, \citenamefont {Roos}, \citenamefont
  {H{\"a}nsel}, \citenamefont {Benhelm}, \citenamefont {Lancaster},
  \citenamefont {K{\"o}rber}, \citenamefont {Becher}, \citenamefont
  {Schmidt-Kaler}, \citenamefont {James} \emph
  {et~al.}}]{riebe2004deterministic}%
  \BibitemOpen
  \bibfield  {author} {\bibinfo {author} {\bibfnamefont {Mark}\ \bibnamefont
  {Riebe}}, \bibinfo {author} {\bibfnamefont {H}~\bibnamefont {H{\"a}ffner}},
  \bibinfo {author} {\bibfnamefont {CF}~\bibnamefont {Roos}}, \bibinfo {author}
  {\bibfnamefont {W}~\bibnamefont {H{\"a}nsel}}, \bibinfo {author}
  {\bibfnamefont {J}~\bibnamefont {Benhelm}}, \bibinfo {author} {\bibfnamefont
  {GPT}\ \bibnamefont {Lancaster}}, \bibinfo {author} {\bibfnamefont
  {TW}~\bibnamefont {K{\"o}rber}}, \bibinfo {author} {\bibfnamefont
  {C}~\bibnamefont {Becher}}, \bibinfo {author} {\bibfnamefont {Ferdinand}\
  \bibnamefont {Schmidt-Kaler}}, \bibinfo {author} {\bibfnamefont {DFV}\
  \bibnamefont {James}},  \emph {et~al.},\ }\bibfield  {title} {\enquote
  {\bibinfo {title} {Deterministic quantum teleportation with atoms},}\
  }\href@noop {} {\bibfield  {journal} {\bibinfo  {journal} {Nature}\ }\textbf
  {\bibinfo {volume} {429}},\ \bibinfo {pages} {734--737} (\bibinfo {year}
  {2004})}\BibitemShut {NoStop}%
\bibitem [{\citenamefont {Olmschenk}\ \emph {et~al.}(2009)\citenamefont
  {Olmschenk}, \citenamefont {Matsukevich}, \citenamefont {Maunz},
  \citenamefont {Hayes}, \citenamefont {Duan},\ and\ \citenamefont
  {Monroe}}]{Olmschenk2009}%
  \BibitemOpen
  \bibfield  {author} {\bibinfo {author} {\bibfnamefont {S.}~\bibnamefont
  {Olmschenk}}, \bibinfo {author} {\bibfnamefont {D.~N.}\ \bibnamefont
  {Matsukevich}}, \bibinfo {author} {\bibfnamefont {P.}~\bibnamefont {Maunz}},
  \bibinfo {author} {\bibfnamefont {D.}~\bibnamefont {Hayes}}, \bibinfo
  {author} {\bibfnamefont {L.-M.}\ \bibnamefont {Duan}}, \ and\ \bibinfo
  {author} {\bibfnamefont {C.}~\bibnamefont {Monroe}},\ }\bibfield  {title}
  {\enquote {\bibinfo {title} {Quantum teleportation between distant matter
  qubits},}\ }\href {\doibase 10.1126/science.1167209} {\bibfield  {journal}
  {\bibinfo  {journal} {Science}\ }\textbf {\bibinfo {volume} {323}},\ \bibinfo
  {pages} {486--489} (\bibinfo {year} {2009})}\BibitemShut {NoStop}%
\bibitem [{\citenamefont {Ren}\ \emph {et~al.}(2017)\citenamefont {Ren},
  \citenamefont {Xu}, \citenamefont {Yong}, \citenamefont {Zhang},
  \citenamefont {Liao}, \citenamefont {Yin}, \citenamefont {Liu}, \citenamefont
  {Cai}, \citenamefont {Yang}, \citenamefont {Li} \emph
  {et~al.}}]{ren2017ground}%
  \BibitemOpen
  \bibfield  {author} {\bibinfo {author} {\bibfnamefont {Ji-Gang}\ \bibnamefont
  {Ren}}, \bibinfo {author} {\bibfnamefont {Ping}\ \bibnamefont {Xu}}, \bibinfo
  {author} {\bibfnamefont {Hai-Lin}\ \bibnamefont {Yong}}, \bibinfo {author}
  {\bibfnamefont {Liang}\ \bibnamefont {Zhang}}, \bibinfo {author}
  {\bibfnamefont {Sheng-Kai}\ \bibnamefont {Liao}}, \bibinfo {author}
  {\bibfnamefont {Juan}\ \bibnamefont {Yin}}, \bibinfo {author} {\bibfnamefont
  {Wei-Yue}\ \bibnamefont {Liu}}, \bibinfo {author} {\bibfnamefont {Wen-Qi}\
  \bibnamefont {Cai}}, \bibinfo {author} {\bibfnamefont {Meng}\ \bibnamefont
  {Yang}}, \bibinfo {author} {\bibfnamefont {Li}~\bibnamefont {Li}},  \emph
  {et~al.},\ }\bibfield  {title} {\enquote {\bibinfo {title}
  {Ground-to-satellite quantum teleportation},}\ }\href@noop {} {\bibfield
  {journal} {\bibinfo  {journal} {Nature}\ }\textbf {\bibinfo {volume} {549}},\
  \bibinfo {pages} {70} (\bibinfo {year} {2017})}\BibitemShut {NoStop}%
\bibitem [{\citenamefont {Nielsen}\ and\ \citenamefont
  {Chuang}(2002)}]{nielsen2002quantum}%
  \BibitemOpen
  \bibfield  {author} {\bibinfo {author} {\bibfnamefont {Michael~A}\
  \bibnamefont {Nielsen}}\ and\ \bibinfo {author} {\bibfnamefont {Isaac}\
  \bibnamefont {Chuang}},\ }\href@noop {} {\enquote {\bibinfo {title} {Quantum
  computation and quantum information},}\ } (\bibinfo {year}
  {2002})\BibitemShut {NoStop}%
\bibitem [{\citenamefont {Gao}\ \emph {et~al.}(2017)\citenamefont {Gao},
  \citenamefont {Jafferis},\ and\ \citenamefont {Wall}}]{gao2017traversable}%
  \BibitemOpen
  \bibfield  {author} {\bibinfo {author} {\bibfnamefont {Ping}\ \bibnamefont
  {Gao}}, \bibinfo {author} {\bibfnamefont {Daniel~Louis}\ \bibnamefont
  {Jafferis}}, \ and\ \bibinfo {author} {\bibfnamefont {Aron~C}\ \bibnamefont
  {Wall}},\ }\bibfield  {title} {\enquote {\bibinfo {title} {Traversable
  wormholes via a double trace deformation},}\ }\href@noop {} {\ \textbf
  {\bibinfo {volume} {2017}},\ \bibinfo {pages} {151} (\bibinfo {year}
  {2017})}\BibitemShut {NoStop}%
\bibitem [{\citenamefont {Bao}\ \emph {et~al.}(2018)\citenamefont {Bao},
  \citenamefont {Chatwin-Davies}, \citenamefont {Pollack},\ and\ \citenamefont
  {Remmen}}]{bao2018traversable}%
  \BibitemOpen
  \bibfield  {author} {\bibinfo {author} {\bibfnamefont {Ning}\ \bibnamefont
  {Bao}}, \bibinfo {author} {\bibfnamefont {Aidan}\ \bibnamefont
  {Chatwin-Davies}}, \bibinfo {author} {\bibfnamefont {Jason}\ \bibnamefont
  {Pollack}}, \ and\ \bibinfo {author} {\bibfnamefont {Grant~N}\ \bibnamefont
  {Remmen}},\ }\bibfield  {title} {\enquote {\bibinfo {title} {Traversable
  wormholes as quantum channels: exploring cft entanglement structure and
  channel capacity in holography},}\ }\href@noop {} {\ \textbf {\bibinfo
  {volume} {2018}},\ \bibinfo {pages} {71} (\bibinfo {year}
  {2018})}\BibitemShut {NoStop}%
\bibitem [{\citenamefont {Maldacena}\ and\ \citenamefont
  {Qi}(2018)}]{maldacena2018eternal}%
  \BibitemOpen
  \bibfield  {author} {\bibinfo {author} {\bibfnamefont {Juan}\ \bibnamefont
  {Maldacena}}\ and\ \bibinfo {author} {\bibfnamefont {Xiao-Liang}\
  \bibnamefont {Qi}},\ }\bibfield  {title} {\enquote {\bibinfo {title} {Eternal
  traversable wormhole},}\ }\href@noop {} {\bibfield  {journal} {\bibinfo
  {journal} {arXiv preprint arXiv:1804.00491}\ } (\bibinfo {year}
  {2018})}\BibitemShut {NoStop}%
\bibitem [{\citenamefont {Yoshida}\ and\ \citenamefont
  {Kitaev}(2017)}]{yoshida2017efficient}%
  \BibitemOpen
  \bibfield  {author} {\bibinfo {author} {\bibfnamefont {Beni}\ \bibnamefont
  {Yoshida}}\ and\ \bibinfo {author} {\bibfnamefont {Alexei}\ \bibnamefont
  {Kitaev}},\ }\bibfield  {title} {\enquote {\bibinfo {title} {Efficient
  decoding for the hayden-preskill protocol},}\ }\href@noop {} {\bibfield
  {journal} {\bibinfo  {journal} {arXiv preprint arXiv:1710.03363}\ } (\bibinfo
  {year} {2017})}\BibitemShut {NoStop}%
\bibitem [{\citenamefont {Yoshida}\ and\ \citenamefont
  {Yao}(2019)}]{yoshida2019disentangling}%
  \BibitemOpen
  \bibfield  {author} {\bibinfo {author} {\bibfnamefont {Beni}\ \bibnamefont
  {Yoshida}}\ and\ \bibinfo {author} {\bibfnamefont {Norman~Y}\ \bibnamefont
  {Yao}},\ }\bibfield  {title} {\enquote {\bibinfo {title} {Disentangling
  scrambling and decoherence via quantum teleportation},}\ }\href@noop {}
  {\bibfield  {journal} {\bibinfo  {journal} {Physical Review X}\ }\textbf
  {\bibinfo {volume} {9}},\ \bibinfo {pages} {011006} (\bibinfo {year}
  {2019})}\BibitemShut {NoStop}%
\bibitem [{\citenamefont {Landsman}\ \emph
  {et~al.}(2019{\natexlab{a}})\citenamefont {Landsman}, \citenamefont
  {Figgatt}, \citenamefont {Schuster}, \citenamefont {Linke}, \citenamefont
  {Yoshida}, \citenamefont {Yao},\ and\ \citenamefont
  {Monroe}}]{landsman2019verified}%
  \BibitemOpen
  \bibfield  {author} {\bibinfo {author} {\bibfnamefont {Kevin~A}\ \bibnamefont
  {Landsman}}, \bibinfo {author} {\bibfnamefont {Caroline}\ \bibnamefont
  {Figgatt}}, \bibinfo {author} {\bibfnamefont {Thomas}\ \bibnamefont
  {Schuster}}, \bibinfo {author} {\bibfnamefont {Norbert~M}\ \bibnamefont
  {Linke}}, \bibinfo {author} {\bibfnamefont {Beni}\ \bibnamefont {Yoshida}},
  \bibinfo {author} {\bibfnamefont {Norman~Y}\ \bibnamefont {Yao}}, \ and\
  \bibinfo {author} {\bibfnamefont {Christopher}\ \bibnamefont {Monroe}},\
  }\bibfield  {title} {\enquote {\bibinfo {title} {Verified quantum information
  scrambling},}\ }\href@noop {} {\bibfield  {journal} {\bibinfo  {journal}
  {Nature}\ }\textbf {\bibinfo {volume} {567}},\ \bibinfo {pages} {61}
  (\bibinfo {year} {2019}{\natexlab{a}})}\BibitemShut {NoStop}%
\bibitem [{\citenamefont {Blok}\ \emph {et~al.}(2020)\citenamefont {Blok},
  \citenamefont {Ramasesh}, \citenamefont {Schuster}, \citenamefont {O'Brien},
  \citenamefont {Kreikebaum}, \citenamefont {Dahlen}, \citenamefont {Morvan},
  \citenamefont {Yoshida}, \citenamefont {Yao},\ and\ \citenamefont
  {Siddiqi}}]{blok2020quantum}%
  \BibitemOpen
  \bibfield  {author} {\bibinfo {author} {\bibfnamefont {M.~S.}\ \bibnamefont
  {Blok}}, \bibinfo {author} {\bibfnamefont {V.~V.}\ \bibnamefont {Ramasesh}},
  \bibinfo {author} {\bibfnamefont {T.}~\bibnamefont {Schuster}}, \bibinfo
  {author} {\bibfnamefont {K.}~\bibnamefont {O'Brien}}, \bibinfo {author}
  {\bibfnamefont {J.~M.}\ \bibnamefont {Kreikebaum}}, \bibinfo {author}
  {\bibfnamefont {D.}~\bibnamefont {Dahlen}}, \bibinfo {author} {\bibfnamefont
  {A.}~\bibnamefont {Morvan}}, \bibinfo {author} {\bibfnamefont
  {B.}~\bibnamefont {Yoshida}}, \bibinfo {author} {\bibfnamefont {N.~Y.}\
  \bibnamefont {Yao}}, \ and\ \bibinfo {author} {\bibfnamefont
  {I.}~\bibnamefont {Siddiqi}},\ }\bibfield  {title} {\enquote {\bibinfo
  {title} {Quantum information scrambling in a superconducting qutrit
  processor},}\ }\href@noop {} {\bibfield  {journal} {\bibinfo  {journal} {to
  appear}\ } (\bibinfo {year} {2020})}\BibitemShut {NoStop}%
\bibitem [{\citenamefont {Brown}\ \emph {et~al.}(2019)\citenamefont {Brown},
  \citenamefont {Gharibyan}, \citenamefont {Leichenauer}, \citenamefont {Lin},
  \citenamefont {Nezami}, \citenamefont {Salton}, \citenamefont {Susskind},
  \citenamefont {Swingle},\ and\ \citenamefont {Walter}}]{brown2019quantum}%
  \BibitemOpen
  \bibfield  {author} {\bibinfo {author} {\bibfnamefont {Adam~R}\ \bibnamefont
  {Brown}}, \bibinfo {author} {\bibfnamefont {Hrant}\ \bibnamefont
  {Gharibyan}}, \bibinfo {author} {\bibfnamefont {Stefan}\ \bibnamefont
  {Leichenauer}}, \bibinfo {author} {\bibfnamefont {Henry~W}\ \bibnamefont
  {Lin}}, \bibinfo {author} {\bibfnamefont {Sepehr}\ \bibnamefont {Nezami}},
  \bibinfo {author} {\bibfnamefont {Grant}\ \bibnamefont {Salton}}, \bibinfo
  {author} {\bibfnamefont {Leonard}\ \bibnamefont {Susskind}}, \bibinfo
  {author} {\bibfnamefont {Brian}\ \bibnamefont {Swingle}}, \ and\ \bibinfo
  {author} {\bibfnamefont {Michael}\ \bibnamefont {Walter}},\ }\bibfield
  {title} {\enquote {\bibinfo {title} {Quantum gravity in the lab:
  Teleportation by size and traversable wormholes},}\ }\href@noop {} {\bibfield
   {journal} {\bibinfo  {journal} {arXiv preprint arXiv:1911.06314}\ }
  (\bibinfo {year} {2019})}\BibitemShut {NoStop}%
\bibitem [{\citenamefont {Nezami}\ \emph {et~al.}(2021)\citenamefont {Nezami},
  \citenamefont {Lin}, \citenamefont {Brown}, \citenamefont {Gharibyan},
  \citenamefont {Leichenauer}, \citenamefont {Salton}, \citenamefont
  {Susskind}, \citenamefont {Swingle},\ and\ \citenamefont
  {Walter}}]{brown2020quantum}%
  \BibitemOpen
  \bibfield  {author} {\bibinfo {author} {\bibfnamefont {Sepehr}\ \bibnamefont
  {Nezami}}, \bibinfo {author} {\bibfnamefont {Henry~W}\ \bibnamefont {Lin}},
  \bibinfo {author} {\bibfnamefont {Adam~R}\ \bibnamefont {Brown}}, \bibinfo
  {author} {\bibfnamefont {Hrant}\ \bibnamefont {Gharibyan}}, \bibinfo {author}
  {\bibfnamefont {Stefan}\ \bibnamefont {Leichenauer}}, \bibinfo {author}
  {\bibfnamefont {Grant}\ \bibnamefont {Salton}}, \bibinfo {author}
  {\bibfnamefont {Leonard}\ \bibnamefont {Susskind}}, \bibinfo {author}
  {\bibfnamefont {Brian}\ \bibnamefont {Swingle}}, \ and\ \bibinfo {author}
  {\bibfnamefont {Michael}\ \bibnamefont {Walter}},\ }\bibfield  {title}
  {\enquote {\bibinfo {title} {Quantum gravity in the lab: teleportation by
  size and traversable wormholes, part ii},}\ }\href@noop {} {\bibfield
  {journal} {\bibinfo  {journal} {arXiv preprint arXiv:2102.01064}\ } (\bibinfo
  {year} {2021})}\BibitemShut {NoStop}%
\bibitem [{\citenamefont {Gao}\ and\ \citenamefont
  {Jafferis}(2019)}]{gao2019traversable}%
  \BibitemOpen
  \bibfield  {author} {\bibinfo {author} {\bibfnamefont {Ping}\ \bibnamefont
  {Gao}}\ and\ \bibinfo {author} {\bibfnamefont {Daniel~Louis}\ \bibnamefont
  {Jafferis}},\ }\bibfield  {title} {\enquote {\bibinfo {title} {A traversable
  wormhole teleportation protocol in the syk model},}\ }\href@noop {}
  {\bibfield  {journal} {\bibinfo  {journal} {arXiv preprint arXiv:1911.07416}\
  } (\bibinfo {year} {2019})}\BibitemShut {NoStop}%
\bibitem [{\citenamefont {Sekino}\ and\ \citenamefont
  {Susskind}(2008)}]{sekino2008fast}%
  \BibitemOpen
  \bibfield  {author} {\bibinfo {author} {\bibfnamefont {Yasuhiro}\
  \bibnamefont {Sekino}}\ and\ \bibinfo {author} {\bibfnamefont {Leonard}\
  \bibnamefont {Susskind}},\ }\bibfield  {title} {\enquote {\bibinfo {title}
  {Fast scramblers},}\ }\href@noop {} {\ \textbf {\bibinfo {volume} {2008}},\
  \bibinfo {pages} {065} (\bibinfo {year} {2008})}\BibitemShut {NoStop}%
\bibitem [{\citenamefont {Shenker}\ and\ \citenamefont
  {Stanford}(2014)}]{shenker2014black}%
  \BibitemOpen
  \bibfield  {author} {\bibinfo {author} {\bibfnamefont {Stephen~H}\
  \bibnamefont {Shenker}}\ and\ \bibinfo {author} {\bibfnamefont {Douglas}\
  \bibnamefont {Stanford}},\ }\bibfield  {title} {\enquote {\bibinfo {title}
  {Black holes and the butterfly effect},}\ }\href@noop {} {\ \textbf {\bibinfo
  {volume} {2014}},\ \bibinfo {pages} {67} (\bibinfo {year}
  {2014})}\BibitemShut {NoStop}%
\bibitem [{\citenamefont {Roberts}\ \emph {et~al.}(2015)\citenamefont
  {Roberts}, \citenamefont {Stanford},\ and\ \citenamefont
  {Susskind}}]{roberts2015localized}%
  \BibitemOpen
  \bibfield  {author} {\bibinfo {author} {\bibfnamefont {Daniel~A}\
  \bibnamefont {Roberts}}, \bibinfo {author} {\bibfnamefont {Douglas}\
  \bibnamefont {Stanford}}, \ and\ \bibinfo {author} {\bibfnamefont {Leonard}\
  \bibnamefont {Susskind}},\ }\bibfield  {title} {\enquote {\bibinfo {title}
  {Localized shocks},}\ }\href@noop {} {\ \textbf {\bibinfo {volume} {2015}},\
  \bibinfo {pages} {51} (\bibinfo {year} {2015})}\BibitemShut {NoStop}%
\bibitem [{\citenamefont {Maldacena}\ \emph {et~al.}(2016)\citenamefont
  {Maldacena}, \citenamefont {Shenker},\ and\ \citenamefont
  {Stanford}}]{maldacena2016bound}%
  \BibitemOpen
  \bibfield  {author} {\bibinfo {author} {\bibfnamefont {Juan}\ \bibnamefont
  {Maldacena}}, \bibinfo {author} {\bibfnamefont {Stephen~H}\ \bibnamefont
  {Shenker}}, \ and\ \bibinfo {author} {\bibfnamefont {Douglas}\ \bibnamefont
  {Stanford}},\ }\bibfield  {title} {\enquote {\bibinfo {title} {A bound on
  chaos},}\ }\href@noop {} {\ \textbf {\bibinfo {volume} {2016}},\ \bibinfo
  {pages} {106} (\bibinfo {year} {2016})}\BibitemShut {NoStop}%
\bibitem [{\citenamefont {Hosur}\ \emph {et~al.}(2016)\citenamefont {Hosur},
  \citenamefont {Qi}, \citenamefont {Roberts},\ and\ \citenamefont
  {Yoshida}}]{hosur2016chaos}%
  \BibitemOpen
  \bibfield  {author} {\bibinfo {author} {\bibfnamefont {Pavan}\ \bibnamefont
  {Hosur}}, \bibinfo {author} {\bibfnamefont {Xiao-Liang}\ \bibnamefont {Qi}},
  \bibinfo {author} {\bibfnamefont {Daniel~A}\ \bibnamefont {Roberts}}, \ and\
  \bibinfo {author} {\bibfnamefont {Beni}\ \bibnamefont {Yoshida}},\ }\bibfield
   {title} {\enquote {\bibinfo {title} {Chaos in quantum channels},}\
  }\href@noop {} {\ \textbf {\bibinfo {volume} {2016}},\ \bibinfo {pages} {4}
  (\bibinfo {year} {2016})}\BibitemShut {NoStop}%
\bibitem [{\citenamefont {Hayden}\ and\ \citenamefont
  {Preskill}(2007)}]{hayden2007black}%
  \BibitemOpen
  \bibfield  {author} {\bibinfo {author} {\bibfnamefont {Patrick}\ \bibnamefont
  {Hayden}}\ and\ \bibinfo {author} {\bibfnamefont {John}\ \bibnamefont
  {Preskill}},\ }\bibfield  {title} {\enquote {\bibinfo {title} {Black holes as
  mirrors: quantum information in random subsystems},}\ }\href@noop {} {\
  \textbf {\bibinfo {volume} {2007}},\ \bibinfo {pages} {120} (\bibinfo {year}
  {2007})}\BibitemShut {NoStop}%
\bibitem [{\citenamefont {Kitaev}(2015)}]{kitaev2015simple}%
  \BibitemOpen
  \bibfield  {author} {\bibinfo {author} {\bibfnamefont {Alexei}\ \bibnamefont
  {Kitaev}},\ }\href@noop {} {\enquote {\bibinfo {title} {A simple model of
  quantum holography},}\ } (\bibinfo {year} {2015})\BibitemShut {NoStop}%
\bibitem [{\citenamefont {Maldacena}\ and\ \citenamefont
  {Stanford}(2016)}]{maldacena2016remarks}%
  \BibitemOpen
  \bibfield  {author} {\bibinfo {author} {\bibfnamefont {Juan}\ \bibnamefont
  {Maldacena}}\ and\ \bibinfo {author} {\bibfnamefont {Douglas}\ \bibnamefont
  {Stanford}},\ }\bibfield  {title} {\enquote {\bibinfo {title} {Remarks on the
  sachdev-ye-kitaev model},}\ }\href@noop {} {\bibfield  {journal} {\bibinfo
  {journal} {Physical Review D}\ }\textbf {\bibinfo {volume} {94}},\ \bibinfo
  {pages} {106002} (\bibinfo {year} {2016})}\BibitemShut {NoStop}%
\bibitem [{\citenamefont {Roberts}\ \emph {et~al.}(2018)\citenamefont
  {Roberts}, \citenamefont {Stanford},\ and\ \citenamefont
  {Streicher}}]{roberts2018operator}%
  \BibitemOpen
  \bibfield  {author} {\bibinfo {author} {\bibfnamefont {Daniel~A}\
  \bibnamefont {Roberts}}, \bibinfo {author} {\bibfnamefont {Douglas}\
  \bibnamefont {Stanford}}, \ and\ \bibinfo {author} {\bibfnamefont
  {Alexandre}\ \bibnamefont {Streicher}},\ }\bibfield  {title} {\enquote
  {\bibinfo {title} {Operator growth in the syk model},}\ }\href@noop {} {\
  \textbf {\bibinfo {volume} {2018}},\ \bibinfo {pages} {122} (\bibinfo {year}
  {2018})}\BibitemShut {NoStop}%
\bibitem [{\citenamefont {Qi}\ and\ \citenamefont
  {Streicher}(2019)}]{qi2019quantum}%
  \BibitemOpen
  \bibfield  {author} {\bibinfo {author} {\bibfnamefont {Xiao-Liang}\
  \bibnamefont {Qi}}\ and\ \bibinfo {author} {\bibfnamefont {Alexandre}\
  \bibnamefont {Streicher}},\ }\bibfield  {title} {\enquote {\bibinfo {title}
  {Quantum epidemiology: operator growth, thermal effects, and syk},}\
  }\href@noop {} {\ \textbf {\bibinfo {volume} {2019}},\ \bibinfo {pages} {12}
  (\bibinfo {year} {2019})}\BibitemShut {NoStop}%
\bibitem [{\citenamefont {Qi}\ \emph {et~al.}(2019)\citenamefont {Qi},
  \citenamefont {Davis}, \citenamefont {Periwal},\ and\ \citenamefont
  {Schleier-Smith}}]{qi2019measuring}%
  \BibitemOpen
  \bibfield  {author} {\bibinfo {author} {\bibfnamefont {Xiao-Liang}\
  \bibnamefont {Qi}}, \bibinfo {author} {\bibfnamefont {Emily~J}\ \bibnamefont
  {Davis}}, \bibinfo {author} {\bibfnamefont {Avikar}\ \bibnamefont {Periwal}},
  \ and\ \bibinfo {author} {\bibfnamefont {Monika}\ \bibnamefont
  {Schleier-Smith}},\ }\bibfield  {title} {\enquote {\bibinfo {title}
  {Measuring operator size growth in quantum quench experiments},}\ }\href@noop
  {} {\bibfield  {journal} {\bibinfo  {journal} {arXiv preprint
  arXiv:1906.00524}\ } (\bibinfo {year} {2019})}\BibitemShut {NoStop}%
\bibitem [{\citenamefont {Larkin}\ and\ \citenamefont
  {Ovchinnikov}(1969)}]{larkin1969quasiclassical}%
  \BibitemOpen
  \bibfield  {author} {\bibinfo {author} {\bibfnamefont {AI}~\bibnamefont
  {Larkin}}\ and\ \bibinfo {author} {\bibfnamefont {Yu~N}\ \bibnamefont
  {Ovchinnikov}},\ }\bibfield  {title} {\enquote {\bibinfo {title}
  {Quasiclassical method in the theory of superconductivity},}\ }\href@noop {}
  {\bibfield  {journal} {\bibinfo  {journal} {Sov Phys JETP}\ }\textbf
  {\bibinfo {volume} {28}},\ \bibinfo {pages} {1200--1205} (\bibinfo {year}
  {1969})}\BibitemShut {NoStop}%
\bibitem [{\citenamefont {Nahum}\ \emph {et~al.}(2018)\citenamefont {Nahum},
  \citenamefont {Vijay},\ and\ \citenamefont {Haah}}]{nahum2018operator}%
  \BibitemOpen
  \bibfield  {author} {\bibinfo {author} {\bibfnamefont {Adam}\ \bibnamefont
  {Nahum}}, \bibinfo {author} {\bibfnamefont {Sagar}\ \bibnamefont {Vijay}}, \
  and\ \bibinfo {author} {\bibfnamefont {Jeongwan}\ \bibnamefont {Haah}},\
  }\bibfield  {title} {\enquote {\bibinfo {title} {Operator spreading in random
  unitary circuits},}\ }\href@noop {} {\bibfield  {journal} {\bibinfo
  {journal} {Physical Review X}\ }\textbf {\bibinfo {volume} {8}},\ \bibinfo
  {pages} {021014} (\bibinfo {year} {2018})}\BibitemShut {NoStop}%
\bibitem [{\citenamefont {Bernien}\ \emph {et~al.}(2017)\citenamefont
  {Bernien}, \citenamefont {Schwartz}, \citenamefont {Keesling}, \citenamefont
  {Levine}, \citenamefont {Omran}, \citenamefont {Pichler}, \citenamefont
  {Choi}, \citenamefont {Zibrov}, \citenamefont {Endres}, \citenamefont
  {Greiner}, \citenamefont {Vuletic},\ and\ \citenamefont {Lukin}}]{Bernien17}%
  \BibitemOpen
  \bibfield  {author} {\bibinfo {author} {\bibfnamefont {H.}~\bibnamefont
  {Bernien}}, \bibinfo {author} {\bibfnamefont {S.}~\bibnamefont {Schwartz}},
  \bibinfo {author} {\bibfnamefont {A.}~\bibnamefont {Keesling}}, \bibinfo
  {author} {\bibfnamefont {H.}~\bibnamefont {Levine}}, \bibinfo {author}
  {\bibfnamefont {A.}~\bibnamefont {Omran}}, \bibinfo {author} {\bibfnamefont
  {H.}~\bibnamefont {Pichler}}, \bibinfo {author} {\bibfnamefont
  {S.}~\bibnamefont {Choi}}, \bibinfo {author} {\bibfnamefont {A.~S.}\
  \bibnamefont {Zibrov}}, \bibinfo {author} {\bibfnamefont {M.}~\bibnamefont
  {Endres}}, \bibinfo {author} {\bibfnamefont {M.}~\bibnamefont {Greiner}},
  \bibinfo {author} {\bibfnamefont {V.}~\bibnamefont {Vuletic}}, \ and\
  \bibinfo {author} {\bibfnamefont {M.~D.}\ \bibnamefont {Lukin}},\ }\bibfield
  {title} {\enquote {\bibinfo {title} {Probing many-body dynamics on a 51-atom
  quantum simulator},}\ }\href@noop {} {\bibfield  {journal} {\bibinfo
  {journal} {Nature}\ }\textbf {\bibinfo {volume} {551}},\ \bibinfo {pages}
  {579--584} (\bibinfo {year} {2017})}\BibitemShut {NoStop}%
\bibitem [{\citenamefont {Shenker}\ and\ \citenamefont
  {Stanford}(2015)}]{shenker2015stringy}%
  \BibitemOpen
  \bibfield  {author} {\bibinfo {author} {\bibfnamefont {Stephen~H}\
  \bibnamefont {Shenker}}\ and\ \bibinfo {author} {\bibfnamefont {Douglas}\
  \bibnamefont {Stanford}},\ }\bibfield  {title} {\enquote {\bibinfo {title}
  {Stringy effects in scrambling},}\ }\href@noop {} {\ \textbf {\bibinfo
  {volume} {2015}},\ \bibinfo {pages} {132} (\bibinfo {year}
  {2015})}\BibitemShut {NoStop}%
\bibitem [{\citenamefont {Gu}\ and\ \citenamefont
  {Kitaev}(2019)}]{gu2019relation}%
  \BibitemOpen
  \bibfield  {author} {\bibinfo {author} {\bibfnamefont {Yingfei}\ \bibnamefont
  {Gu}}\ and\ \bibinfo {author} {\bibfnamefont {Alexei}\ \bibnamefont
  {Kitaev}},\ }\bibfield  {title} {\enquote {\bibinfo {title} {On the relation
  between the magnitude and exponent of otocs},}\ }\href@noop {} {\ \textbf
  {\bibinfo {volume} {2019}},\ \bibinfo {pages} {75} (\bibinfo {year}
  {2019})}\BibitemShut {NoStop}%
\bibitem [{\citenamefont {Maller}\ \emph {et~al.}(2015)\citenamefont {Maller},
  \citenamefont {Lichtman}, \citenamefont {Xi}, \citenamefont {Sun},
  \citenamefont {Piotrowicz}, \citenamefont {Carr}, \citenamefont {Isenhower},\
  and\ \citenamefont {Saffman}}]{Maller15}%
  \BibitemOpen
  \bibfield  {author} {\bibinfo {author} {\bibfnamefont {K.~M.}\ \bibnamefont
  {Maller}}, \bibinfo {author} {\bibfnamefont {M.~T.}\ \bibnamefont
  {Lichtman}}, \bibinfo {author} {\bibfnamefont {T.}~\bibnamefont {Xi}},
  \bibinfo {author} {\bibfnamefont {Y.}~\bibnamefont {Sun}}, \bibinfo {author}
  {\bibfnamefont {M.~J.}\ \bibnamefont {Piotrowicz}}, \bibinfo {author}
  {\bibfnamefont {A.~W.}\ \bibnamefont {Carr}}, \bibinfo {author}
  {\bibfnamefont {L.}~\bibnamefont {Isenhower}}, \ and\ \bibinfo {author}
  {\bibfnamefont {M.}~\bibnamefont {Saffman}},\ }\bibfield  {title} {\enquote
  {\bibinfo {title} {Rydberg-blockade controlled-not gate and entanglement in a
  two-dimensional array of neutral-atom qubits},}\ }\href@noop {} {\bibfield
  {journal} {\bibinfo  {journal} {Phys. Rev. A}\ }\textbf {\bibinfo {volume}
  {92}},\ \bibinfo {pages} {022336} (\bibinfo {year} {2015})}\BibitemShut
  {NoStop}%
\bibitem [{\citenamefont {Labuhn}\ \emph {et~al.}(2016)\citenamefont {Labuhn},
  \citenamefont {Barredo}, \citenamefont {Ravets}, \citenamefont
  {de~D.~L{\'e}s{\'e}leuc}, \citenamefont {Macri}, \citenamefont {Lahaye},\
  and\ \citenamefont {Browaeys}}]{Labuhn16}%
  \BibitemOpen
  \bibfield  {author} {\bibinfo {author} {\bibfnamefont {H.}~\bibnamefont
  {Labuhn}}, \bibinfo {author} {\bibfnamefont {D.}~\bibnamefont {Barredo}},
  \bibinfo {author} {\bibfnamefont {S.}~\bibnamefont {Ravets}}, \bibinfo
  {author} {\bibfnamefont {S.}~\bibnamefont {de~D.~L{\'e}s{\'e}leuc}}, \bibinfo
  {author} {\bibfnamefont {M.}~\bibnamefont {Macri}}, \bibinfo {author}
  {\bibfnamefont {T.}~\bibnamefont {Lahaye}}, \ and\ \bibinfo {author}
  {\bibfnamefont {A.}~\bibnamefont {Browaeys}},\ }\bibfield  {title} {\enquote
  {\bibinfo {title} {Tunable two-dimensional arrays of single {R}ydberg atoms
  for realizing quantum ising models},}\ }\href@noop {} {\bibfield  {journal}
  {\bibinfo  {journal} {Nature}\ }\textbf {\bibinfo {volume} {534}},\ \bibinfo
  {pages} {667--70} (\bibinfo {year} {2016})}\BibitemShut {NoStop}%
\bibitem [{\citenamefont {Graham}\ \emph {et~al.}(2019)\citenamefont {Graham},
  \citenamefont {Kwon}, \citenamefont {Grinkemeyer}, \citenamefont {Marra},
  \citenamefont {Jiang}, \citenamefont {Lichtman}, \citenamefont {Sun},
  \citenamefont {Ebert},\ and\ \citenamefont {Saffman}}]{Graham19}%
  \BibitemOpen
  \bibfield  {author} {\bibinfo {author} {\bibfnamefont {T.~M.}\ \bibnamefont
  {Graham}}, \bibinfo {author} {\bibfnamefont {M.}~\bibnamefont {Kwon}},
  \bibinfo {author} {\bibfnamefont {B.}~\bibnamefont {Grinkemeyer}}, \bibinfo
  {author} {\bibfnamefont {Z.}~\bibnamefont {Marra}}, \bibinfo {author}
  {\bibfnamefont {X.}~\bibnamefont {Jiang}}, \bibinfo {author} {\bibfnamefont
  {M.~T.}\ \bibnamefont {Lichtman}}, \bibinfo {author} {\bibfnamefont
  {Y.}~\bibnamefont {Sun}}, \bibinfo {author} {\bibfnamefont {M.}~\bibnamefont
  {Ebert}}, \ and\ \bibinfo {author} {\bibfnamefont {M.}~\bibnamefont
  {Saffman}},\ }\bibfield  {title} {\enquote {\bibinfo {title} {Rydberg
  mediated entanglement in a two-dimensional neutral atom qubit array},}\
  }\href@noop {} {\bibfield  {journal} {\bibinfo  {journal} {Phys. Rev. Lett.}\
  }\textbf {\bibinfo {volume} {123}},\ \bibinfo {pages} {230501} (\bibinfo
  {year} {2019})}\BibitemShut {NoStop}%
\bibitem [{\citenamefont {Madjarov}\ \emph {et~al.}(2020)\citenamefont
  {Madjarov}, \citenamefont {Covey}, \citenamefont {Shaw}, \citenamefont
  {Choi}, \citenamefont {Kale}, \citenamefont {Cooper}, \citenamefont {Picher},
  \citenamefont {Schkolnik}, \citenamefont {Williams},\ and\ \citenamefont
  {Endres}}]{Madjarov20}%
  \BibitemOpen
  \bibfield  {author} {\bibinfo {author} {\bibfnamefont {I.~S.}\ \bibnamefont
  {Madjarov}}, \bibinfo {author} {\bibfnamefont {J.~P.}\ \bibnamefont {Covey}},
  \bibinfo {author} {\bibfnamefont {A.~L.}\ \bibnamefont {Shaw}}, \bibinfo
  {author} {\bibfnamefont {J.}~\bibnamefont {Choi}}, \bibinfo {author}
  {\bibfnamefont {A.}~\bibnamefont {Kale}}, \bibinfo {author} {\bibfnamefont
  {A.}~\bibnamefont {Cooper}}, \bibinfo {author} {\bibfnamefont
  {H.}~\bibnamefont {Picher}}, \bibinfo {author} {\bibfnamefont
  {V.}~\bibnamefont {Schkolnik}}, \bibinfo {author} {\bibfnamefont {J.~R.}\
  \bibnamefont {Williams}}, \ and\ \bibinfo {author} {\bibfnamefont
  {M.}~\bibnamefont {Endres}},\ }\bibfield  {title} {\enquote {\bibinfo {title}
  {High-fidelity entanglement and detection of alkaline-earth {R}ydberg
  atoms},}\ }\href@noop {} {\bibfield  {journal} {\bibinfo  {journal} {Nat.
  Phys.}\ } (\bibinfo {year} {2020})}\BibitemShut {NoStop}%
\bibitem [{\citenamefont {Wilson}\ \emph {et~al.}(2018)\citenamefont {Wilson},
  \citenamefont {Saskin}, \citenamefont {Meng}, \citenamefont {Ma},
  \citenamefont {Dilip}, \citenamefont {Burgers},\ and\ \citenamefont
  {Thompson}}]{Wilson19}%
  \BibitemOpen
  \bibfield  {author} {\bibinfo {author} {\bibfnamefont {J.}~\bibnamefont
  {Wilson}}, \bibinfo {author} {\bibfnamefont {S.}~\bibnamefont {Saskin}},
  \bibinfo {author} {\bibfnamefont {Y.}~\bibnamefont {Meng}}, \bibinfo {author}
  {\bibfnamefont {S.}~\bibnamefont {Ma}}, \bibinfo {author} {\bibfnamefont
  {R.}~\bibnamefont {Dilip}}, \bibinfo {author} {\bibfnamefont
  {A.}~\bibnamefont {Burgers}}, \ and\ \bibinfo {author} {\bibfnamefont
  {J.}~\bibnamefont {Thompson}},\ }\bibfield  {title} {\enquote {\bibinfo
  {title} {Trapped arrays of alkaline earth rydberg atoms in optical
  tweezers},}\ }\href@noop {} {\bibfield  {journal} {\bibinfo  {journal} {arXiv
  preprint arXiv:1912.08754}\ } (\bibinfo {year} {2018})}\BibitemShut {NoStop}%
\bibitem [{\citenamefont {Blatt}\ and\ \citenamefont
  {Wineland}(2008)}]{wineland2008entangled}%
  \BibitemOpen
  \bibfield  {author} {\bibinfo {author} {\bibfnamefont {R.}~\bibnamefont
  {Blatt}}\ and\ \bibinfo {author} {\bibfnamefont {D.}~\bibnamefont
  {Wineland}},\ }\bibfield  {title} {\enquote {\bibinfo {title} {Entangled
  states of trapped atomic ions},}\ }\href
  {https://www.nature.com/articles/nature07125} {\bibfield  {journal} {\bibinfo
   {journal} {Nature}\ }\textbf {\bibinfo {volume} {453}},\ \bibinfo {pages}
  {1008--1014} (\bibinfo {year} {2008})}\BibitemShut {NoStop}%
\bibitem [{\citenamefont {Monroe}\ and\ \citenamefont
  {Kim}(2013)}]{monroe2013scaling}%
  \BibitemOpen
  \bibfield  {author} {\bibinfo {author} {\bibfnamefont {Christopher}\
  \bibnamefont {Monroe}}\ and\ \bibinfo {author} {\bibfnamefont {Jungsang}\
  \bibnamefont {Kim}},\ }\bibfield  {title} {\enquote {\bibinfo {title}
  {Scaling the ion trap quantum processor},}\ }\href@noop {} {\bibfield
  {journal} {\bibinfo  {journal} {Science}\ }\textbf {\bibinfo {volume}
  {339}},\ \bibinfo {pages} {1164--1169} (\bibinfo {year} {2013})}\BibitemShut
  {NoStop}%
\bibitem [{\citenamefont {Ballance}\ \emph {et~al.}(2016)\citenamefont
  {Ballance}, \citenamefont {Harty}, \citenamefont {Linke}, \citenamefont
  {Sepiol},\ and\ \citenamefont {Lucas}}]{Ballance:2016}%
  \BibitemOpen
  \bibfield  {author} {\bibinfo {author} {\bibfnamefont {C.~J.}\ \bibnamefont
  {Ballance}}, \bibinfo {author} {\bibfnamefont {T.~P.}\ \bibnamefont {Harty}},
  \bibinfo {author} {\bibfnamefont {N.~M.}\ \bibnamefont {Linke}}, \bibinfo
  {author} {\bibfnamefont {M.~A.}\ \bibnamefont {Sepiol}}, \ and\ \bibinfo
  {author} {\bibfnamefont {D.~M.}\ \bibnamefont {Lucas}},\ }\bibfield  {title}
  {\enquote {\bibinfo {title} {High-fidelity quantum logic gates using
  trapped-ion hyperfine qubits},}\ }\href {\doibase
  10.1103/PhysRevLett.117.060504} {\bibfield  {journal} {\bibinfo  {journal}
  {Phys. Rev. Lett.}\ }\textbf {\bibinfo {volume} {117}},\ \bibinfo {pages}
  {060504} (\bibinfo {year} {2016})}\BibitemShut {NoStop}%
\bibitem [{\citenamefont {Gaebler}\ \emph {et~al.}(2016)\citenamefont
  {Gaebler}, \citenamefont {Tan}, \citenamefont {Lin}, \citenamefont {Wan},
  \citenamefont {Bowler}, \citenamefont {Keith}, \citenamefont {Glancy},
  \citenamefont {Coakley}, \citenamefont {Knill}, \citenamefont {Leibfried},\
  and\ \citenamefont {Wineland}}]{Gaebler2016high}%
  \BibitemOpen
  \bibfield  {author} {\bibinfo {author} {\bibfnamefont {J.~P.}\ \bibnamefont
  {Gaebler}}, \bibinfo {author} {\bibfnamefont {T.~R.}\ \bibnamefont {Tan}},
  \bibinfo {author} {\bibfnamefont {Y.}~\bibnamefont {Lin}}, \bibinfo {author}
  {\bibfnamefont {Y.}~\bibnamefont {Wan}}, \bibinfo {author} {\bibfnamefont
  {R.}~\bibnamefont {Bowler}}, \bibinfo {author} {\bibfnamefont {A.~C.}\
  \bibnamefont {Keith}}, \bibinfo {author} {\bibfnamefont {S.}~\bibnamefont
  {Glancy}}, \bibinfo {author} {\bibfnamefont {K.}~\bibnamefont {Coakley}},
  \bibinfo {author} {\bibfnamefont {E.}~\bibnamefont {Knill}}, \bibinfo
  {author} {\bibfnamefont {D.}~\bibnamefont {Leibfried}}, \ and\ \bibinfo
  {author} {\bibfnamefont {D.~J.}\ \bibnamefont {Wineland}},\ }\bibfield
  {title} {\enquote {\bibinfo {title} {High-fidelity universal gate set for
  ${^{9}\mathrm{Be}}^{+}$ ion qubits},}\ }\href {\doibase
  10.1103/PhysRevLett.117.060505} {\bibfield  {journal} {\bibinfo  {journal}
  {Phys. Rev. Lett.}\ }\textbf {\bibinfo {volume} {117}},\ \bibinfo {pages}
  {060505} (\bibinfo {year} {2016})}\BibitemShut {NoStop}%
\bibitem [{\citenamefont {Cetina}\ \emph {et~al.}(2020)\citenamefont {Cetina},
  \citenamefont {Egan}, \citenamefont {Noel}, \citenamefont {Goldman},
  \citenamefont {Risinger}, \citenamefont {Zhu}, \citenamefont {Biswas},\ and\
  \citenamefont {Monroe}}]{cetina2020}%
  \BibitemOpen
  \bibfield  {author} {\bibinfo {author} {\bibfnamefont {M.}~\bibnamefont
  {Cetina}}, \bibinfo {author} {\bibfnamefont {L.~N.}\ \bibnamefont {Egan}},
  \bibinfo {author} {\bibfnamefont {C.~A.}\ \bibnamefont {Noel}}, \bibinfo
  {author} {\bibfnamefont {M.~L.}\ \bibnamefont {Goldman}}, \bibinfo {author}
  {\bibfnamefont {A.~R.}\ \bibnamefont {Risinger}}, \bibinfo {author}
  {\bibfnamefont {D.}~\bibnamefont {Zhu}}, \bibinfo {author} {\bibfnamefont
  {D.}~\bibnamefont {Biswas}}, \ and\ \bibinfo {author} {\bibfnamefont
  {C.}~\bibnamefont {Monroe}},\ }\bibfield  {title} {\enquote {\bibinfo {title}
  {Quantum gates on individually-addressed atomic qubits subject to noisy
  transverse motion},}\ }\href@noop {} {\bibfield  {journal} {\bibinfo
  {journal} {arXiv preprint arXiv:2007.06768}\ } (\bibinfo {year}
  {2020})}\BibitemShut {NoStop}%
\bibitem [{\citenamefont {Preskill}(2018)}]{preskill2018quantum}%
  \BibitemOpen
  \bibfield  {author} {\bibinfo {author} {\bibfnamefont {John}\ \bibnamefont
  {Preskill}},\ }\bibfield  {title} {\enquote {\bibinfo {title} {Quantum
  computing in the nisq era and beyond},}\ }\href@noop {} {\bibfield  {journal}
  {\bibinfo  {journal} {Quantum}\ }\textbf {\bibinfo {volume} {2}},\ \bibinfo
  {pages} {79} (\bibinfo {year} {2018})}\BibitemShut {NoStop}%
\bibitem [{\citenamefont {Kitaev}\ and\ \citenamefont
  {Suh}(2018)}]{kitaev2018soft}%
  \BibitemOpen
  \bibfield  {author} {\bibinfo {author} {\bibfnamefont {Alexei}\ \bibnamefont
  {Kitaev}}\ and\ \bibinfo {author} {\bibfnamefont {S~Josephine}\ \bibnamefont
  {Suh}},\ }\bibfield  {title} {\enquote {\bibinfo {title} {The soft mode in
  the sachdev-ye-kitaev model and its gravity dual},}\ }\href@noop {} {\
  \textbf {\bibinfo {volume} {2018}},\ \bibinfo {pages} {183} (\bibinfo {year}
  {2018})}\BibitemShut {NoStop}%
\bibitem [{\citenamefont {Lin}\ \emph {et~al.}(2019)\citenamefont {Lin},
  \citenamefont {Maldacena},\ and\ \citenamefont {Zhao}}]{lin2019symmetries}%
  \BibitemOpen
  \bibfield  {author} {\bibinfo {author} {\bibfnamefont {Henry~W}\ \bibnamefont
  {Lin}}, \bibinfo {author} {\bibfnamefont {Juan}\ \bibnamefont {Maldacena}}, \
  and\ \bibinfo {author} {\bibfnamefont {Ying}\ \bibnamefont {Zhao}},\
  }\bibfield  {title} {\enquote {\bibinfo {title} {Symmetries near the
  horizon},}\ }\href@noop {} {\bibfield  {journal} {\bibinfo  {journal} {arXiv
  preprint arXiv:1904.12820}\ } (\bibinfo {year} {2019})}\BibitemShut {NoStop}%
\bibitem [{\citenamefont {Susskind}(2019)}]{susskind2019complexity}%
  \BibitemOpen
  \bibfield  {author} {\bibinfo {author} {\bibfnamefont {Leonard}\ \bibnamefont
  {Susskind}},\ }\bibfield  {title} {\enquote {\bibinfo {title} {Complexity and
  newton's laws},}\ }\href@noop {} {\bibfield  {journal} {\bibinfo  {journal}
  {arXiv preprint arXiv:1904.12819}\ } (\bibinfo {year} {2019})}\BibitemShut
  {NoStop}%
\bibitem [{\citenamefont {Lin}\ and\ \citenamefont
  {Susskind}(2019)}]{lin2019complexity}%
  \BibitemOpen
  \bibfield  {author} {\bibinfo {author} {\bibfnamefont {Henry~W}\ \bibnamefont
  {Lin}}\ and\ \bibinfo {author} {\bibfnamefont {Leonard}\ \bibnamefont
  {Susskind}},\ }\bibfield  {title} {\enquote {\bibinfo {title} {Complexity
  geometry and schwarzian dynamics},}\ }\href@noop {} {\bibfield  {journal}
  {\bibinfo  {journal} {arXiv preprint arXiv:1911.02603}\ } (\bibinfo {year}
  {2019})}\BibitemShut {NoStop}%
\bibitem [{\citenamefont {Rakovszky}\ \emph {et~al.}(2018)\citenamefont
  {Rakovszky}, \citenamefont {Pollmann},\ and\ \citenamefont {von
  Keyserlingk}}]{rakovszky2018diffusive}%
  \BibitemOpen
  \bibfield  {author} {\bibinfo {author} {\bibfnamefont {Tibor}\ \bibnamefont
  {Rakovszky}}, \bibinfo {author} {\bibfnamefont {Frank}\ \bibnamefont
  {Pollmann}}, \ and\ \bibinfo {author} {\bibfnamefont {CW}~\bibnamefont {von
  Keyserlingk}},\ }\bibfield  {title} {\enquote {\bibinfo {title} {Diffusive
  hydrodynamics of out-of-time-ordered correlators with charge conservation},}\
  }\href@noop {} {\bibfield  {journal} {\bibinfo  {journal} {Physical Review
  X}\ }\textbf {\bibinfo {volume} {8}},\ \bibinfo {pages} {031058} (\bibinfo
  {year} {2018})}\BibitemShut {NoStop}%
\bibitem [{\citenamefont {Khemani}\ \emph {et~al.}(2018)\citenamefont
  {Khemani}, \citenamefont {Vishwanath},\ and\ \citenamefont
  {Huse}}]{khemani2018operator}%
  \BibitemOpen
  \bibfield  {author} {\bibinfo {author} {\bibfnamefont {Vedika}\ \bibnamefont
  {Khemani}}, \bibinfo {author} {\bibfnamefont {Ashvin}\ \bibnamefont
  {Vishwanath}}, \ and\ \bibinfo {author} {\bibfnamefont {David~A}\
  \bibnamefont {Huse}},\ }\bibfield  {title} {\enquote {\bibinfo {title}
  {Operator spreading and the emergence of dissipative hydrodynamics under
  unitary evolution with conservation laws},}\ }\href@noop {} {\bibfield
  {journal} {\bibinfo  {journal} {Physical Review X}\ }\textbf {\bibinfo
  {volume} {8}},\ \bibinfo {pages} {031057} (\bibinfo {year}
  {2018})}\BibitemShut {NoStop}%
\bibitem [{\citenamefont {Rakovszky}\ \emph {et~al.}(2020)\citenamefont
  {Rakovszky}, \citenamefont {von Keyserlingk},\ and\ \citenamefont
  {Pollmann}}]{rakovszky2020dissipation}%
  \BibitemOpen
  \bibfield  {author} {\bibinfo {author} {\bibfnamefont {Tibor}\ \bibnamefont
  {Rakovszky}}, \bibinfo {author} {\bibfnamefont {CW}~\bibnamefont {von
  Keyserlingk}}, \ and\ \bibinfo {author} {\bibfnamefont {Frank}\ \bibnamefont
  {Pollmann}},\ }\bibfield  {title} {\enquote {\bibinfo {title}
  {Dissipation-assisted operator evolution method for capturing hydrodynamic
  transport},}\ }\href@noop {} {\bibfield  {journal} {\bibinfo  {journal}
  {arXiv preprint arXiv:2004.05177}\ } (\bibinfo {year} {2020})}\BibitemShut
  {NoStop}%
\bibitem [{\citenamefont {Roberts}\ and\ \citenamefont
  {Yoshida}(2017)}]{roberts2017chaos}%
  \BibitemOpen
  \bibfield  {author} {\bibinfo {author} {\bibfnamefont {Daniel~A}\
  \bibnamefont {Roberts}}\ and\ \bibinfo {author} {\bibfnamefont {Beni}\
  \bibnamefont {Yoshida}},\ }\bibfield  {title} {\enquote {\bibinfo {title}
  {Chaos and complexity by design},}\ }\href@noop {} {\ \textbf {\bibinfo
  {volume} {2017}},\ \bibinfo {pages} {121} (\bibinfo {year}
  {2017})}\BibitemShut {NoStop}%
\bibitem [{\citenamefont {Von~Keyserlingk}\ \emph {et~al.}(2018)\citenamefont
  {Von~Keyserlingk}, \citenamefont {Rakovszky}, \citenamefont {Pollmann},\ and\
  \citenamefont {Sondhi}}]{von2018operator}%
  \BibitemOpen
  \bibfield  {author} {\bibinfo {author} {\bibfnamefont {CW}~\bibnamefont
  {Von~Keyserlingk}}, \bibinfo {author} {\bibfnamefont {Tibor}\ \bibnamefont
  {Rakovszky}}, \bibinfo {author} {\bibfnamefont {Frank}\ \bibnamefont
  {Pollmann}}, \ and\ \bibinfo {author} {\bibfnamefont {Shivaji~Lal}\
  \bibnamefont {Sondhi}},\ }\bibfield  {title} {\enquote {\bibinfo {title}
  {Operator hydrodynamics, otocs, and entanglement growth in systems without
  conservation laws},}\ }\href@noop {} {\bibfield  {journal} {\bibinfo
  {journal} {Physical Review X}\ }\textbf {\bibinfo {volume} {8}},\ \bibinfo
  {pages} {021013} (\bibinfo {year} {2018})}\BibitemShut {NoStop}%
\bibitem [{\citenamefont {Li}\ \emph {et~al.}(2018)\citenamefont {Li},
  \citenamefont {Chen},\ and\ \citenamefont {Fisher}}]{li2018quantum}%
  \BibitemOpen
  \bibfield  {author} {\bibinfo {author} {\bibfnamefont {Yaodong}\ \bibnamefont
  {Li}}, \bibinfo {author} {\bibfnamefont {Xiao}\ \bibnamefont {Chen}}, \ and\
  \bibinfo {author} {\bibfnamefont {Matthew~PA}\ \bibnamefont {Fisher}},\
  }\bibfield  {title} {\enquote {\bibinfo {title} {Quantum zeno effect and the
  many-body entanglement transition},}\ }\href@noop {} {\bibfield  {journal}
  {\bibinfo  {journal} {Physical Review B}\ }\textbf {\bibinfo {volume} {98}},\
  \bibinfo {pages} {205136} (\bibinfo {year} {2018})}\BibitemShut {NoStop}%
\bibitem [{\citenamefont {Skinner}\ \emph {et~al.}(2019)\citenamefont
  {Skinner}, \citenamefont {Ruhman},\ and\ \citenamefont
  {Nahum}}]{skinner2019measurement}%
  \BibitemOpen
  \bibfield  {author} {\bibinfo {author} {\bibfnamefont {Brian}\ \bibnamefont
  {Skinner}}, \bibinfo {author} {\bibfnamefont {Jonathan}\ \bibnamefont
  {Ruhman}}, \ and\ \bibinfo {author} {\bibfnamefont {Adam}\ \bibnamefont
  {Nahum}},\ }\bibfield  {title} {\enquote {\bibinfo {title}
  {Measurement-induced phase transitions in the dynamics of entanglement},}\
  }\href@noop {} {\bibfield  {journal} {\bibinfo  {journal} {Physical Review
  X}\ }\textbf {\bibinfo {volume} {9}},\ \bibinfo {pages} {031009} (\bibinfo
  {year} {2019})}\BibitemShut {NoStop}%
\bibitem [{\citenamefont {Dankert}\ \emph {et~al.}(2009)\citenamefont
  {Dankert}, \citenamefont {Cleve}, \citenamefont {Emerson},\ and\
  \citenamefont {Livine}}]{dankert2009exact}%
  \BibitemOpen
  \bibfield  {author} {\bibinfo {author} {\bibfnamefont {Christoph}\
  \bibnamefont {Dankert}}, \bibinfo {author} {\bibfnamefont {Richard}\
  \bibnamefont {Cleve}}, \bibinfo {author} {\bibfnamefont {Joseph}\
  \bibnamefont {Emerson}}, \ and\ \bibinfo {author} {\bibfnamefont {Etera}\
  \bibnamefont {Livine}},\ }\bibfield  {title} {\enquote {\bibinfo {title}
  {Exact and approximate unitary 2-designs and their application to fidelity
  estimation},}\ }\href@noop {} {\bibfield  {journal} {\bibinfo  {journal}
  {Physical Review A}\ }\textbf {\bibinfo {volume} {80}},\ \bibinfo {pages}
  {012304} (\bibinfo {year} {2009})}\BibitemShut {NoStop}%
\bibitem [{\citenamefont {Webb}(2015)}]{webb2015clifford}%
  \BibitemOpen
  \bibfield  {author} {\bibinfo {author} {\bibfnamefont {Zak}\ \bibnamefont
  {Webb}},\ }\bibfield  {title} {\enquote {\bibinfo {title} {The clifford group
  forms a unitary 3-design},}\ }\href@noop {} {\bibfield  {journal} {\bibinfo
  {journal} {arXiv preprint arXiv:1510.02769}\ } (\bibinfo {year}
  {2015})}\BibitemShut {NoStop}%
\bibitem [{\citenamefont {Kueng}\ and\ \citenamefont
  {Gross}(2015)}]{kueng2015qubit}%
  \BibitemOpen
  \bibfield  {author} {\bibinfo {author} {\bibfnamefont {R.}~\bibnamefont
  {Kueng}}\ and\ \bibinfo {author} {\bibfnamefont {D.}~\bibnamefont {Gross}},\
  }\bibfield  {title} {\enquote {\bibinfo {title} {Qubit stabilizer states are
  complex projective 3-designs},}\ }\href@noop {} {\bibfield  {journal}
  {\bibinfo  {journal} {ArXiv}\ }\textbf {\bibinfo {volume} {abs/1510.02767}}
  (\bibinfo {year} {2015})}\BibitemShut {NoStop}%
\bibitem [{\citenamefont {Zhu}(2017)}]{zhu2015multiqubit}%
  \BibitemOpen
  \bibfield  {author} {\bibinfo {author} {\bibfnamefont {Huangjun}\
  \bibnamefont {Zhu}},\ }\bibfield  {title} {\enquote {\bibinfo {title}
  {Multiqubit clifford groups are unitary 3-designs},}\ }\href {\doibase
  10.1103/PhysRevA.96.062336} {\bibfield  {journal} {\bibinfo  {journal} {Phys.
  Rev. A}\ }\textbf {\bibinfo {volume} {96}},\ \bibinfo {pages} {062336}
  (\bibinfo {year} {2017})}\BibitemShut {NoStop}%
\bibitem [{\citenamefont {Kardar}\ \emph {et~al.}(1986)\citenamefont {Kardar},
  \citenamefont {Parisi},\ and\ \citenamefont {Zhang}}]{kardar1986dynamic}%
  \BibitemOpen
  \bibfield  {author} {\bibinfo {author} {\bibfnamefont {Mehran}\ \bibnamefont
  {Kardar}}, \bibinfo {author} {\bibfnamefont {Giorgio}\ \bibnamefont
  {Parisi}}, \ and\ \bibinfo {author} {\bibfnamefont {Yi-Cheng}\ \bibnamefont
  {Zhang}},\ }\bibfield  {title} {\enquote {\bibinfo {title} {Dynamic scaling
  of growing interfaces},}\ }\href@noop {} {\bibfield  {journal} {\bibinfo
  {journal} {Physical Review Letters}\ }\textbf {\bibinfo {volume} {56}},\
  \bibinfo {pages} {889} (\bibinfo {year} {1986})}\BibitemShut {NoStop}%
\bibitem [{\citenamefont {Corwin}(2012)}]{corwin2012kardar}%
  \BibitemOpen
  \bibfield  {author} {\bibinfo {author} {\bibfnamefont {Ivan}\ \bibnamefont
  {Corwin}},\ }\bibfield  {title} {\enquote {\bibinfo {title} {The
  kardar--parisi--zhang equation and universality class},}\ }\href@noop {}
  {\bibfield  {journal} {\bibinfo  {journal} {Random matrices: Theory and
  applications}\ }\textbf {\bibinfo {volume} {1}},\ \bibinfo {pages} {1130001}
  (\bibinfo {year} {2012})}\BibitemShut {NoStop}%
\bibitem [{\citenamefont {Zhang}\ \emph {et~al.}(2020)\citenamefont {Zhang},
  \citenamefont {Gu},\ and\ \citenamefont {Kitaev}}]{zhang2020obstacle}%
  \BibitemOpen
  \bibfield  {author} {\bibinfo {author} {\bibfnamefont {Pengfei}\ \bibnamefont
  {Zhang}}, \bibinfo {author} {\bibfnamefont {Yingfei}\ \bibnamefont {Gu}}, \
  and\ \bibinfo {author} {\bibfnamefont {Alexei}\ \bibnamefont {Kitaev}},\
  }\bibfield  {title} {\enquote {\bibinfo {title} {An obstacle to sub-ads
  holography for syk-like models},}\ }\href@noop {} {\bibfield  {journal}
  {\bibinfo  {journal} {arXiv preprint arXiv:2012.01620}\ } (\bibinfo {year}
  {2020})}\BibitemShut {NoStop}%
\bibitem [{\citenamefont {Gao}\ and\ \citenamefont
  {Liu}(2018)}]{gao2018regenesis}%
  \BibitemOpen
  \bibfield  {author} {\bibinfo {author} {\bibfnamefont {Ping}\ \bibnamefont
  {Gao}}\ and\ \bibinfo {author} {\bibfnamefont {Hong}\ \bibnamefont {Liu}},\
  }\bibfield  {title} {\enquote {\bibinfo {title} {Regenesis and quantum
  traversable wormholes},}\ }\href@noop {} {\bibfield  {journal} {\bibinfo
  {journal} {arXiv preprint arXiv:1810.01444}\ } (\bibinfo {year}
  {2018})}\BibitemShut {NoStop}%
\bibitem [{\citenamefont {Susskind}(2018)}]{susskind2018things}%
  \BibitemOpen
  \bibfield  {author} {\bibinfo {author} {\bibfnamefont {Leonard}\ \bibnamefont
  {Susskind}},\ }\bibfield  {title} {\enquote {\bibinfo {title} {Why do things
  fall?}}\ }\href@noop {} {\bibfield  {journal} {\bibinfo  {journal} {arXiv
  preprint arXiv:1802.01198}\ } (\bibinfo {year} {2018})}\BibitemShut {NoStop}%
\bibitem [{\citenamefont {G{\"a}rttner}\ \emph {et~al.}(2017)\citenamefont
  {G{\"a}rttner}, \citenamefont {Bohnet}, \citenamefont {Safavi-Naini},
  \citenamefont {Wall}, \citenamefont {Bollinger},\ and\ \citenamefont
  {Rey}}]{garttner2017measuring}%
  \BibitemOpen
  \bibfield  {author} {\bibinfo {author} {\bibfnamefont {Martin}\ \bibnamefont
  {G{\"a}rttner}}, \bibinfo {author} {\bibfnamefont {Justin~G}\ \bibnamefont
  {Bohnet}}, \bibinfo {author} {\bibfnamefont {Arghavan}\ \bibnamefont
  {Safavi-Naini}}, \bibinfo {author} {\bibfnamefont {Michael~L}\ \bibnamefont
  {Wall}}, \bibinfo {author} {\bibfnamefont {John~J}\ \bibnamefont
  {Bollinger}}, \ and\ \bibinfo {author} {\bibfnamefont {Ana~Maria}\
  \bibnamefont {Rey}},\ }\bibfield  {title} {\enquote {\bibinfo {title}
  {Measuring out-of-time-order correlations and multiple quantum spectra in a
  trapped-ion quantum magnet},}\ }\href@noop {} {\bibfield  {journal} {\bibinfo
   {journal} {Nature Physics}\ }\textbf {\bibinfo {volume} {13}},\ \bibinfo
  {pages} {781--786} (\bibinfo {year} {2017})}\BibitemShut {NoStop}%
\bibitem [{\citenamefont {Li}\ \emph {et~al.}(2017)\citenamefont {Li},
  \citenamefont {Fan}, \citenamefont {Wang}, \citenamefont {Ye}, \citenamefont
  {Zeng}, \citenamefont {Zhai}, \citenamefont {Peng},\ and\ \citenamefont
  {Du}}]{li2017measuring}%
  \BibitemOpen
  \bibfield  {author} {\bibinfo {author} {\bibfnamefont {Jun}\ \bibnamefont
  {Li}}, \bibinfo {author} {\bibfnamefont {Ruihua}\ \bibnamefont {Fan}},
  \bibinfo {author} {\bibfnamefont {Hengyan}\ \bibnamefont {Wang}}, \bibinfo
  {author} {\bibfnamefont {Bingtian}\ \bibnamefont {Ye}}, \bibinfo {author}
  {\bibfnamefont {Bei}\ \bibnamefont {Zeng}}, \bibinfo {author} {\bibfnamefont
  {Hui}\ \bibnamefont {Zhai}}, \bibinfo {author} {\bibfnamefont {Xinhua}\
  \bibnamefont {Peng}}, \ and\ \bibinfo {author} {\bibfnamefont {Jiangfeng}\
  \bibnamefont {Du}},\ }\bibfield  {title} {\enquote {\bibinfo {title}
  {Measuring out-of-time-order correlators on a nuclear magnetic resonance
  quantum simulator},}\ }\href@noop {} {\bibfield  {journal} {\bibinfo
  {journal} {Physical Review X}\ }\textbf {\bibinfo {volume} {7}},\ \bibinfo
  {pages} {031011} (\bibinfo {year} {2017})}\BibitemShut {NoStop}%
\bibitem [{\citenamefont {Arute}\ \emph {et~al.}(2019)\citenamefont {Arute},
  \citenamefont {Arya}, \citenamefont {Babbush}, \citenamefont {Bacon},
  \citenamefont {Bardin}, \citenamefont {Barends}, \citenamefont {Biswas},
  \citenamefont {Boixo}, \citenamefont {Brandao}, \citenamefont {Buell} \emph
  {et~al.}}]{arute2019quantum}%
  \BibitemOpen
  \bibfield  {author} {\bibinfo {author} {\bibfnamefont {Frank}\ \bibnamefont
  {Arute}}, \bibinfo {author} {\bibfnamefont {Kunal}\ \bibnamefont {Arya}},
  \bibinfo {author} {\bibfnamefont {Ryan}\ \bibnamefont {Babbush}}, \bibinfo
  {author} {\bibfnamefont {Dave}\ \bibnamefont {Bacon}}, \bibinfo {author}
  {\bibfnamefont {Joseph~C}\ \bibnamefont {Bardin}}, \bibinfo {author}
  {\bibfnamefont {Rami}\ \bibnamefont {Barends}}, \bibinfo {author}
  {\bibfnamefont {Rupak}\ \bibnamefont {Biswas}}, \bibinfo {author}
  {\bibfnamefont {Sergio}\ \bibnamefont {Boixo}}, \bibinfo {author}
  {\bibfnamefont {Fernando~GSL}\ \bibnamefont {Brandao}}, \bibinfo {author}
  {\bibfnamefont {David~A}\ \bibnamefont {Buell}},  \emph {et~al.},\ }\bibfield
   {title} {\enquote {\bibinfo {title} {Quantum supremacy using a programmable
  superconducting processor},}\ }\href@noop {} {\bibfield  {journal} {\bibinfo
  {journal} {Nature}\ }\textbf {\bibinfo {volume} {574}},\ \bibinfo {pages}
  {505--510} (\bibinfo {year} {2019})}\BibitemShut {NoStop}%
\bibitem [{\citenamefont {Wei}\ \emph {et~al.}(2019)\citenamefont {Wei},
  \citenamefont {Peng}, \citenamefont {Shtanko}, \citenamefont {Marvian},
  \citenamefont {Lloyd}, \citenamefont {Ramanathan}, \citenamefont {Cappellaro}
  \emph {et~al.}}]{wei2019emergent}%
  \BibitemOpen
  \bibfield  {author} {\bibinfo {author} {\bibfnamefont {Ken~Xuan}\
  \bibnamefont {Wei}}, \bibinfo {author} {\bibfnamefont {Pai}\ \bibnamefont
  {Peng}}, \bibinfo {author} {\bibfnamefont {Oles}\ \bibnamefont {Shtanko}},
  \bibinfo {author} {\bibfnamefont {Iman}\ \bibnamefont {Marvian}}, \bibinfo
  {author} {\bibfnamefont {Seth}\ \bibnamefont {Lloyd}}, \bibinfo {author}
  {\bibfnamefont {Chandrasekhar}\ \bibnamefont {Ramanathan}}, \bibinfo {author}
  {\bibfnamefont {Paola}\ \bibnamefont {Cappellaro}},  \emph {et~al.},\
  }\bibfield  {title} {\enquote {\bibinfo {title} {Emergent prethermalization
  signatures in out-of-time ordered correlations},}\ }\href@noop {} {\bibfield
  {journal} {\bibinfo  {journal} {Physical review letters}\ }\textbf {\bibinfo
  {volume} {123}},\ \bibinfo {pages} {090605} (\bibinfo {year}
  {2019})}\BibitemShut {NoStop}%
\bibitem [{\citenamefont {Meier}\ \emph {et~al.}(2019)\citenamefont {Meier},
  \citenamefont {An}, \citenamefont {Gadway} \emph
  {et~al.}}]{meier2019exploring}%
  \BibitemOpen
  \bibfield  {author} {\bibinfo {author} {\bibfnamefont {Eric~J}\ \bibnamefont
  {Meier}}, \bibinfo {author} {\bibfnamefont {Fangzhao~Alex}\ \bibnamefont
  {An}}, \bibinfo {author} {\bibfnamefont {Bryce}\ \bibnamefont {Gadway}},
  \emph {et~al.},\ }\bibfield  {title} {\enquote {\bibinfo {title} {Exploring
  quantum signatures of chaos on a floquet synthetic lattice},}\ }\href@noop {}
  {\bibfield  {journal} {\bibinfo  {journal} {Physical Review A}\ }\textbf
  {\bibinfo {volume} {100}},\ \bibinfo {pages} {013623} (\bibinfo {year}
  {2019})}\BibitemShut {NoStop}%
\bibitem [{\citenamefont {Haehl}\ \emph {et~al.}(2021)\citenamefont {Haehl},
  \citenamefont {Streicher},\ and\ \citenamefont {Zhao}}]{haehl2021six}%
  \BibitemOpen
  \bibfield  {author} {\bibinfo {author} {\bibfnamefont {Felix~M}\ \bibnamefont
  {Haehl}}, \bibinfo {author} {\bibfnamefont {Alexandre}\ \bibnamefont
  {Streicher}}, \ and\ \bibinfo {author} {\bibfnamefont {Ying}\ \bibnamefont
  {Zhao}},\ }\bibfield  {title} {\enquote {\bibinfo {title} {Six-point
  functions and collisions in the black hole interior},}\ }\href@noop {}
  {\bibfield  {journal} {\bibinfo  {journal} {arXiv preprint arXiv:2105.12755}\
  } (\bibinfo {year} {2021})}\BibitemShut {NoStop}%
\bibitem [{\citenamefont {Yoshida}(2019{\natexlab{a}})}]{yoshida2019observer}%
  \BibitemOpen
  \bibfield  {author} {\bibinfo {author} {\bibfnamefont {Beni}\ \bibnamefont
  {Yoshida}},\ }\bibfield  {title} {\enquote {\bibinfo {title}
  {Observer-dependent black hole interior from operator collision},}\
  }\href@noop {} {\bibfield  {journal} {\bibinfo  {journal} {arXiv preprint
  arXiv:1910.11346}\ } (\bibinfo {year} {2019}{\natexlab{a}})}\BibitemShut
  {NoStop}%
\bibitem [{\citenamefont {Yoshida}(2019{\natexlab{b}})}]{yoshida2019firewalls}%
  \BibitemOpen
  \bibfield  {author} {\bibinfo {author} {\bibfnamefont {Beni}\ \bibnamefont
  {Yoshida}},\ }\bibfield  {title} {\enquote {\bibinfo {title} {Firewalls vs.
  scrambling},}\ }\href@noop {} {\ \textbf {\bibinfo {volume} {2019}},\
  \bibinfo {pages} {132} (\bibinfo {year} {2019}{\natexlab{b}})}\BibitemShut
  {NoStop}%
\bibitem [{\citenamefont {Kourkoulou}\ and\ \citenamefont
  {Maldacena}(2017)}]{kourkoulou2017pure}%
  \BibitemOpen
  \bibfield  {author} {\bibinfo {author} {\bibfnamefont {Ioanna}\ \bibnamefont
  {Kourkoulou}}\ and\ \bibinfo {author} {\bibfnamefont {Juan}\ \bibnamefont
  {Maldacena}},\ }\bibfield  {title} {\enquote {\bibinfo {title} {Pure states
  in the syk model and nearly-$ ads\_2 $ gravity},}\ }\href@noop {} {\bibfield
  {journal} {\bibinfo  {journal} {arXiv preprint arXiv:1707.02325}\ } (\bibinfo
  {year} {2017})}\BibitemShut {NoStop}%
\bibitem [{\citenamefont {Wu}\ and\ \citenamefont
  {Hsieh}(2019)}]{wu2019variational}%
  \BibitemOpen
  \bibfield  {author} {\bibinfo {author} {\bibfnamefont {Jingxiang}\
  \bibnamefont {Wu}}\ and\ \bibinfo {author} {\bibfnamefont {Timothy~H.}\
  \bibnamefont {Hsieh}},\ }\bibfield  {title} {\enquote {\bibinfo {title}
  {Variational thermal quantum simulation via thermofield double states},}\
  }\href {\doibase 10.1103/PhysRevLett.123.220502} {\bibfield  {journal}
  {\bibinfo  {journal} {Phys. Rev. Lett.}\ }\textbf {\bibinfo {volume} {123}},\
  \bibinfo {pages} {220502} (\bibinfo {year} {2019})}\BibitemShut {NoStop}%
\bibitem [{\citenamefont {Zhu}\ \emph {et~al.}(2019)\citenamefont {Zhu},
  \citenamefont {Johri}, \citenamefont {Linke}, \citenamefont {Landsman},
  \citenamefont {Nguyen}, \citenamefont {Alderete}, \citenamefont {Matsuura},
  \citenamefont {Hsieh},\ and\ \citenamefont {Monroe}}]{zhu2019TFD}%
  \BibitemOpen
  \bibfield  {author} {\bibinfo {author} {\bibfnamefont {D.}~\bibnamefont
  {Zhu}}, \bibinfo {author} {\bibfnamefont {S.}~\bibnamefont {Johri}}, \bibinfo
  {author} {\bibfnamefont {N.~M.}\ \bibnamefont {Linke}}, \bibinfo {author}
  {\bibfnamefont {K.~A.}\ \bibnamefont {Landsman}}, \bibinfo {author}
  {\bibfnamefont {N.~H.}\ \bibnamefont {Nguyen}}, \bibinfo {author}
  {\bibfnamefont {C.~H.}\ \bibnamefont {Alderete}}, \bibinfo {author}
  {\bibfnamefont {A.~Y.}\ \bibnamefont {Matsuura}}, \bibinfo {author}
  {\bibfnamefont {T.~H.}\ \bibnamefont {Hsieh}}, \ and\ \bibinfo {author}
  {\bibfnamefont {C.}~\bibnamefont {Monroe}},\ }\bibfield  {title} {\enquote
  {\bibinfo {title} {Generation of thermofield double states and critical
  ground states with a quantum computer},}\ }\href@noop {} {\bibfield
  {journal} {\bibinfo  {journal} {arXiv preprint arXiv:1906.02699}\ } (\bibinfo
  {year} {2019})}\BibitemShut {NoStop}%
\bibitem [{\citenamefont {Su}(2020)}]{su2020variational}%
  \BibitemOpen
  \bibfield  {author} {\bibinfo {author} {\bibfnamefont {Vincent~Paul}\
  \bibnamefont {Su}},\ }\bibfield  {title} {\enquote {\bibinfo {title}
  {Variational preparation of the sachdev-ye-kitaev thermofield double},}\
  }\href@noop {} {\bibfield  {journal} {\bibinfo  {journal} {arXiv preprint
  arXiv:2009.04488}\ } (\bibinfo {year} {2020})}\BibitemShut {NoStop}%
\bibitem [{\citenamefont {Daley}\ \emph {et~al.}(2012)\citenamefont {Daley},
  \citenamefont {Pichler}, \citenamefont {Schachenmayer},\ and\ \citenamefont
  {Zoller}}]{daley2012measuring}%
  \BibitemOpen
  \bibfield  {author} {\bibinfo {author} {\bibfnamefont {AJ}~\bibnamefont
  {Daley}}, \bibinfo {author} {\bibfnamefont {H}~\bibnamefont {Pichler}},
  \bibinfo {author} {\bibfnamefont {J}~\bibnamefont {Schachenmayer}}, \ and\
  \bibinfo {author} {\bibfnamefont {P}~\bibnamefont {Zoller}},\ }\bibfield
  {title} {\enquote {\bibinfo {title} {Measuring entanglement growth in quench
  dynamics of bosons in an optical lattice},}\ }\href@noop {} {\bibfield
  {journal} {\bibinfo  {journal} {Physical review letters}\ }\textbf {\bibinfo
  {volume} {109}},\ \bibinfo {pages} {020505} (\bibinfo {year}
  {2012})}\BibitemShut {NoStop}%
\bibitem [{\citenamefont {Abanin}\ and\ \citenamefont
  {Demler}(2012)}]{abanin2012measuring}%
  \BibitemOpen
  \bibfield  {author} {\bibinfo {author} {\bibfnamefont {Dmitry~A}\
  \bibnamefont {Abanin}}\ and\ \bibinfo {author} {\bibfnamefont {Eugene}\
  \bibnamefont {Demler}},\ }\bibfield  {title} {\enquote {\bibinfo {title}
  {Measuring entanglement entropy of a generic many-body system with a quantum
  switch},}\ }\href@noop {} {\bibfield  {journal} {\bibinfo  {journal}
  {Physical review letters}\ }\textbf {\bibinfo {volume} {109}},\ \bibinfo
  {pages} {020504} (\bibinfo {year} {2012})}\BibitemShut {NoStop}%
\bibitem [{\citenamefont {Johri}\ \emph {et~al.}(2017)\citenamefont {Johri},
  \citenamefont {Steiger},\ and\ \citenamefont {Troyer}}]{Johri2017}%
  \BibitemOpen
  \bibfield  {author} {\bibinfo {author} {\bibfnamefont {Sonika}\ \bibnamefont
  {Johri}}, \bibinfo {author} {\bibfnamefont {Damian~S.}\ \bibnamefont
  {Steiger}}, \ and\ \bibinfo {author} {\bibfnamefont {Matthias}\ \bibnamefont
  {Troyer}},\ }\bibfield  {title} {\enquote {\bibinfo {title} {Entanglement
  spectroscopy on a quantum computer},}\ }\href {\doibase
  10.1103/PhysRevB.96.195136} {\bibfield  {journal} {\bibinfo  {journal} {Phys.
  Rev. B}\ }\textbf {\bibinfo {volume} {96}},\ \bibinfo {pages} {195136}
  (\bibinfo {year} {2017})}\BibitemShut {NoStop}%
\bibitem [{\citenamefont {Elben}\ \emph {et~al.}(2018)\citenamefont {Elben},
  \citenamefont {Vermersch}, \citenamefont {Dalmonte}, \citenamefont {Cirac},\
  and\ \citenamefont {Zoller}}]{Elben2018}%
  \BibitemOpen
  \bibfield  {author} {\bibinfo {author} {\bibfnamefont {A.}~\bibnamefont
  {Elben}}, \bibinfo {author} {\bibfnamefont {B.}~\bibnamefont {Vermersch}},
  \bibinfo {author} {\bibfnamefont {M.}~\bibnamefont {Dalmonte}}, \bibinfo
  {author} {\bibfnamefont {J.~I.}\ \bibnamefont {Cirac}}, \ and\ \bibinfo
  {author} {\bibfnamefont {P.}~\bibnamefont {Zoller}},\ }\bibfield  {title}
  {\enquote {\bibinfo {title} {R\'enyi entropies from random quenches in atomic
  hubbard and spin models},}\ }\href {\doibase 10.1103/PhysRevLett.120.050406}
  {\bibfield  {journal} {\bibinfo  {journal} {Phys. Rev. Lett.}\ }\textbf
  {\bibinfo {volume} {120}},\ \bibinfo {pages} {050406} (\bibinfo {year}
  {2018})}\BibitemShut {NoStop}%
\bibitem [{\citenamefont {Linke}\ \emph {et~al.}(2018)\citenamefont {Linke},
  \citenamefont {Johri}, \citenamefont {Figgatt}, \citenamefont {Landsman},
  \citenamefont {Matsuura},\ and\ \citenamefont {Monroe}}]{Linke2018}%
  \BibitemOpen
  \bibfield  {author} {\bibinfo {author} {\bibfnamefont {N.~M.}\ \bibnamefont
  {Linke}}, \bibinfo {author} {\bibfnamefont {S.}~\bibnamefont {Johri}},
  \bibinfo {author} {\bibfnamefont {C.}~\bibnamefont {Figgatt}}, \bibinfo
  {author} {\bibfnamefont {K.~A.}\ \bibnamefont {Landsman}}, \bibinfo {author}
  {\bibfnamefont {A.~Y.}\ \bibnamefont {Matsuura}}, \ and\ \bibinfo {author}
  {\bibfnamefont {C.}~\bibnamefont {Monroe}},\ }\bibfield  {title} {\enquote
  {\bibinfo {title} {Measuring the r\'enyi entropy of a two-site fermi-hubbard
  model on a trapped ion quantum computer},}\ }\href {\doibase
  10.1103/PhysRevA.98.052334} {\bibfield  {journal} {\bibinfo  {journal} {Phys.
  Rev. A}\ }\textbf {\bibinfo {volume} {98}},\ \bibinfo {pages} {052334}
  (\bibinfo {year} {2018})}\BibitemShut {NoStop}%
\bibitem [{\citenamefont {Brydges}\ \emph {et~al.}(2019)\citenamefont
  {Brydges}, \citenamefont {Elben}, \citenamefont {Jurcevic}, \citenamefont
  {Vermersch}, \citenamefont {Maier}, \citenamefont {Lanyon}, \citenamefont
  {Zoller}, \citenamefont {Blatt},\ and\ \citenamefont {Roos}}]{Brydges2020}%
  \BibitemOpen
  \bibfield  {author} {\bibinfo {author} {\bibfnamefont {Tiff}\ \bibnamefont
  {Brydges}}, \bibinfo {author} {\bibfnamefont {Andreas}\ \bibnamefont
  {Elben}}, \bibinfo {author} {\bibfnamefont {Petar}\ \bibnamefont {Jurcevic}},
  \bibinfo {author} {\bibfnamefont {Beno{\^\i}t}\ \bibnamefont {Vermersch}},
  \bibinfo {author} {\bibfnamefont {Christine}\ \bibnamefont {Maier}}, \bibinfo
  {author} {\bibfnamefont {Ben~P.}\ \bibnamefont {Lanyon}}, \bibinfo {author}
  {\bibfnamefont {Peter}\ \bibnamefont {Zoller}}, \bibinfo {author}
  {\bibfnamefont {Rainer}\ \bibnamefont {Blatt}}, \ and\ \bibinfo {author}
  {\bibfnamefont {Christian~F.}\ \bibnamefont {Roos}},\ }\bibfield  {title}
  {\enquote {\bibinfo {title} {Probing r{\'e}nyi entanglement entropy via
  randomized measurements},}\ }\href {\doibase 10.1126/science.aau4963}
  {\bibfield  {journal} {\bibinfo  {journal} {Science}\ }\textbf {\bibinfo
  {volume} {364}},\ \bibinfo {pages} {260--263} (\bibinfo {year} {2019})},\
  \Eprint
  {http://arxiv.org/abs/https://science.sciencemag.org/content/364/6437/260.full.pdf}
  {https://science.sciencemag.org/content/364/6437/260.full.pdf} \BibitemShut
  {NoStop}%
\bibitem [{\citenamefont {Jaksch}\ \emph {et~al.}(2000)\citenamefont {Jaksch},
  \citenamefont {Cirac}, \citenamefont {Zoller}, \citenamefont {Rolston},
  \citenamefont {Cote},\ and\ \citenamefont {Lukin}}]{Jaksch00}%
  \BibitemOpen
  \bibfield  {author} {\bibinfo {author} {\bibfnamefont {D.}~\bibnamefont
  {Jaksch}}, \bibinfo {author} {\bibfnamefont {J.~I.}\ \bibnamefont {Cirac}},
  \bibinfo {author} {\bibfnamefont {P.}~\bibnamefont {Zoller}}, \bibinfo
  {author} {\bibfnamefont {S.~L.}\ \bibnamefont {Rolston}}, \bibinfo {author}
  {\bibfnamefont {R.}~\bibnamefont {Cote}}, \ and\ \bibinfo {author}
  {\bibfnamefont {M.~D.}\ \bibnamefont {Lukin}},\ }\bibfield  {title} {\enquote
  {\bibinfo {title} {Fast quantum gates for neutral atoms},}\ }\href@noop {}
  {\bibfield  {journal} {\bibinfo  {journal} {Phys. Rev. Lett.}\ }\textbf
  {\bibinfo {volume} {85}},\ \bibinfo {pages} {2208} (\bibinfo {year}
  {2000})}\BibitemShut {NoStop}%
\bibitem [{\citenamefont {Xia}\ \emph {et~al.}(2015)\citenamefont {Xia},
  \citenamefont {Lichtman}, \citenamefont {Maller}, \citenamefont {Carr},
  \citenamefont {Piotrowicz}, \citenamefont {Isenhower},\ and\ \citenamefont
  {Saffman}}]{Xia15}%
  \BibitemOpen
  \bibfield  {author} {\bibinfo {author} {\bibfnamefont {T.}~\bibnamefont
  {Xia}}, \bibinfo {author} {\bibfnamefont {M.}~\bibnamefont {Lichtman}},
  \bibinfo {author} {\bibfnamefont {K.}~\bibnamefont {Maller}}, \bibinfo
  {author} {\bibfnamefont {A.~W.}\ \bibnamefont {Carr}}, \bibinfo {author}
  {\bibfnamefont {M.~J.}\ \bibnamefont {Piotrowicz}}, \bibinfo {author}
  {\bibfnamefont {L.}~\bibnamefont {Isenhower}}, \ and\ \bibinfo {author}
  {\bibfnamefont {M.}~\bibnamefont {Saffman}},\ }\bibfield  {title} {\enquote
  {\bibinfo {title} {Randomized benchmarking of single-qubit gates in a 2d
  array of neutral-atom qubits},}\ }\href@noop {} {\bibfield  {journal}
  {\bibinfo  {journal} {Phys. Rev. Lett.}\ }\textbf {\bibinfo {volume} {114}},\
  \bibinfo {pages} {100503} (\bibinfo {year} {2015})}\BibitemShut {NoStop}%
\bibitem [{\citenamefont {Barredo}\ \emph {et~al.}(2018)\citenamefont
  {Barredo}, \citenamefont {Lienhard}, \citenamefont {De~Leseleuc},
  \citenamefont {Lahaye},\ and\ \citenamefont
  {Browaeys}}]{barredo2018synthetic}%
  \BibitemOpen
  \bibfield  {author} {\bibinfo {author} {\bibfnamefont {Daniel}\ \bibnamefont
  {Barredo}}, \bibinfo {author} {\bibfnamefont {Vincent}\ \bibnamefont
  {Lienhard}}, \bibinfo {author} {\bibfnamefont {Sylvain}\ \bibnamefont
  {De~Leseleuc}}, \bibinfo {author} {\bibfnamefont {Thierry}\ \bibnamefont
  {Lahaye}}, \ and\ \bibinfo {author} {\bibfnamefont {Antoine}\ \bibnamefont
  {Browaeys}},\ }\bibfield  {title} {\enquote {\bibinfo {title} {Synthetic
  three-dimensional atomic structures assembled atom by atom},}\ }\href@noop {}
  {\bibfield  {journal} {\bibinfo  {journal} {Nature}\ }\textbf {\bibinfo
  {volume} {561}},\ \bibinfo {pages} {79--82} (\bibinfo {year}
  {2018})}\BibitemShut {NoStop}%
\bibitem [{\citenamefont {Glaetzle}\ \emph {et~al.}(2015)\citenamefont
  {Glaetzle}, \citenamefont {Dalmonte}, \citenamefont {Nath}, \citenamefont
  {Gross}, \citenamefont {Bloch},\ and\ \citenamefont {Zoller}}]{Glaetzle15}%
  \BibitemOpen
  \bibfield  {author} {\bibinfo {author} {\bibfnamefont {A.~W.}\ \bibnamefont
  {Glaetzle}}, \bibinfo {author} {\bibfnamefont {M.}~\bibnamefont {Dalmonte}},
  \bibinfo {author} {\bibfnamefont {R.}~\bibnamefont {Nath}}, \bibinfo {author}
  {\bibfnamefont {C.}~\bibnamefont {Gross}}, \bibinfo {author} {\bibfnamefont
  {I.}~\bibnamefont {Bloch}}, \ and\ \bibinfo {author} {\bibfnamefont
  {P.}~\bibnamefont {Zoller}},\ }\bibfield  {title} {\enquote {\bibinfo {title}
  {Designing frustrated quantum magnets with laser-dressed rydberg atoms},}\
  }\href@noop {} {\bibfield  {journal} {\bibinfo  {journal} {Phys. Rev. Lett.}\
  }\textbf {\bibinfo {volume} {114}},\ \bibinfo {pages} {173002} (\bibinfo
  {year} {2015})}\BibitemShut {NoStop}%
\bibitem [{\citenamefont {Potirniche}\ \emph {et~al.}(2017)\citenamefont
  {Potirniche}, \citenamefont {Potter}, \citenamefont {Schleier-Smith},
  \citenamefont {A.Vishwanath},\ and\ \citenamefont {Yao}}]{Potirniche17}%
  \BibitemOpen
  \bibfield  {author} {\bibinfo {author} {\bibfnamefont {I.-D.}\ \bibnamefont
  {Potirniche}}, \bibinfo {author} {\bibfnamefont {A.~C.}\ \bibnamefont
  {Potter}}, \bibinfo {author} {\bibfnamefont {M.}~\bibnamefont
  {Schleier-Smith}}, \bibinfo {author} {\bibnamefont {A.Vishwanath}}, \ and\
  \bibinfo {author} {\bibfnamefont {N.}~\bibnamefont {Yao}},\ }\bibfield
  {title} {\enquote {\bibinfo {title} {Floquet symmetry-protected topological
  phases in cold-atom systems},}\ }\href@noop {} {\bibfield  {journal}
  {\bibinfo  {journal} {Phys. Rev. Lett.}\ }\textbf {\bibinfo {volume} {119}},\
  \bibinfo {pages} {123601} (\bibinfo {year} {2017})}\BibitemShut {NoStop}%
\bibitem [{\citenamefont {Zeiher}\ \emph {et~al.}(2017)\citenamefont {Zeiher},
  \citenamefont {y.~Choi}, \citenamefont {Rubio-Abadal}, \citenamefont {Pohl},
  \citenamefont {van Bijnen}, \citenamefont {Bloch},\ and\ \citenamefont
  {Gross}}]{Zeiher17}%
  \BibitemOpen
  \bibfield  {author} {\bibinfo {author} {\bibfnamefont {J.}~\bibnamefont
  {Zeiher}}, \bibinfo {author} {\bibfnamefont {J.}~\bibnamefont {y.~Choi}},
  \bibinfo {author} {\bibfnamefont {A.}~\bibnamefont {Rubio-Abadal}}, \bibinfo
  {author} {\bibfnamefont {T.}~\bibnamefont {Pohl}}, \bibinfo {author}
  {\bibfnamefont {R.}~\bibnamefont {van Bijnen}}, \bibinfo {author}
  {\bibfnamefont {I.}~\bibnamefont {Bloch}}, \ and\ \bibinfo {author}
  {\bibfnamefont {C.}~\bibnamefont {Gross}},\ }\bibfield  {title} {\enquote
  {\bibinfo {title} {Coherent many-body spin dynamics in a long-range
  interacting ising chain},}\ }\href@noop {} {\bibfield  {journal} {\bibinfo
  {journal} {Phys. Rev. X}\ }\textbf {\bibinfo {volume} {7}},\ \bibinfo {pages}
  {041063} (\bibinfo {year} {2017})}\BibitemShut {NoStop}%
\bibitem [{dyn()}]{dynamite}%
  \BibitemOpen
  \href@noop {} {}\bibinfo {note} {Our parallelized dynamics code is available
  open-source as the package \texttt{dynamite}:
  \url{https://dynamite.readthedocs.io/}
  DOI:10.5281/zenodo.3606826}\BibitemShut {NoStop}%
\bibitem [{\citenamefont {Levine}\ \emph {et~al.}(2019)\citenamefont {Levine},
  \citenamefont {Keesling}, \citenamefont {Semeghini}, \citenamefont {Omran},
  \citenamefont {Wang}, \citenamefont {Ebadi}, \citenamefont {Bernien},
  \citenamefont {Greiner}, \citenamefont {Vuletic}, \citenamefont {Pichler},\
  and\ \citenamefont {Lukin}}]{Levine19}%
  \BibitemOpen
  \bibfield  {author} {\bibinfo {author} {\bibfnamefont {H.}~\bibnamefont
  {Levine}}, \bibinfo {author} {\bibfnamefont {A.}~\bibnamefont {Keesling}},
  \bibinfo {author} {\bibfnamefont {G.}~\bibnamefont {Semeghini}}, \bibinfo
  {author} {\bibfnamefont {A.}~\bibnamefont {Omran}}, \bibinfo {author}
  {\bibfnamefont {T.~T.}\ \bibnamefont {Wang}}, \bibinfo {author}
  {\bibfnamefont {S.}~\bibnamefont {Ebadi}}, \bibinfo {author} {\bibfnamefont
  {H.}~\bibnamefont {Bernien}}, \bibinfo {author} {\bibfnamefont
  {M.}~\bibnamefont {Greiner}}, \bibinfo {author} {\bibfnamefont
  {V.}~\bibnamefont {Vuletic}}, \bibinfo {author} {\bibfnamefont
  {H.}~\bibnamefont {Pichler}}, \ and\ \bibinfo {author} {\bibfnamefont
  {M.~D.}\ \bibnamefont {Lukin}},\ }\bibfield  {title} {\enquote {\bibinfo
  {title} {Parallel implementation of high-fidelity multi-qubit gates with
  neutral atoms},}\ }\href@noop {} {\bibfield  {journal} {\bibinfo  {journal}
  {arXiv preprint arXiv:1908.06101}\ } (\bibinfo {year} {2019})}\BibitemShut
  {NoStop}%
\bibitem [{\citenamefont {Levine}\ \emph {et~al.}(2018)\citenamefont {Levine},
  \citenamefont {Keesling}, \citenamefont {Omran}, \citenamefont {Bernien},
  \citenamefont {Schwartz}, \citenamefont {Zibrov}, \citenamefont {Endres},
  \citenamefont {Greiner}, \citenamefont {Vuletic},\ and\ \citenamefont
  {Lukin}}]{Levine18}%
  \BibitemOpen
  \bibfield  {author} {\bibinfo {author} {\bibfnamefont {H.}~\bibnamefont
  {Levine}}, \bibinfo {author} {\bibfnamefont {A.}~\bibnamefont {Keesling}},
  \bibinfo {author} {\bibfnamefont {A.}~\bibnamefont {Omran}}, \bibinfo
  {author} {\bibfnamefont {H.}~\bibnamefont {Bernien}}, \bibinfo {author}
  {\bibfnamefont {S.}~\bibnamefont {Schwartz}}, \bibinfo {author}
  {\bibfnamefont {A.~S.}\ \bibnamefont {Zibrov}}, \bibinfo {author}
  {\bibfnamefont {M.}~\bibnamefont {Endres}}, \bibinfo {author} {\bibfnamefont
  {M.}~\bibnamefont {Greiner}}, \bibinfo {author} {\bibfnamefont
  {V.}~\bibnamefont {Vuletic}}, \ and\ \bibinfo {author} {\bibfnamefont
  {M.~D.}\ \bibnamefont {Lukin}},\ }\bibfield  {title} {\enquote {\bibinfo
  {title} {High-fidelity control and entanglement of {R}ydberg atom qubits},}\
  }\href@noop {} {\bibfield  {journal} {\bibinfo  {journal} {Phys. Rev. Lett.}\
  }\textbf {\bibinfo {volume} {121}},\ \bibinfo {pages} {123603} (\bibinfo
  {year} {2018})}\BibitemShut {NoStop}%
\bibitem [{\citenamefont {van Bijnen}\ and\ \citenamefont
  {Pohl}(2014)}]{van2014quantum}%
  \BibitemOpen
  \bibfield  {author} {\bibinfo {author} {\bibfnamefont {RMW}\ \bibnamefont
  {van Bijnen}}\ and\ \bibinfo {author} {\bibfnamefont {T}~\bibnamefont
  {Pohl}},\ }\bibfield  {title} {\enquote {\bibinfo {title} {Quantum magnetism
  and topological ordering via enhanced rydberg-dressing near
  forster-resonances},}\ }\href@noop {} {\bibfield  {journal} {\bibinfo
  {journal} {arXiv preprint arXiv:1411.3118}\ } (\bibinfo {year}
  {2014})}\BibitemShut {NoStop}%
\bibitem [{\citenamefont {de~L{\'e}s{\'e}leuc}\ \emph
  {et~al.}(2019)\citenamefont {de~L{\'e}s{\'e}leuc}, \citenamefont {Lienhard},
  \citenamefont {Scholl}, \citenamefont {Barredo}, \citenamefont {Weber},
  \citenamefont {Lang}, \citenamefont {B{\"u}chler}, \citenamefont {Lahaye},\
  and\ \citenamefont {Browaeys}}]{de2019observation}%
  \BibitemOpen
  \bibfield  {author} {\bibinfo {author} {\bibfnamefont {Sylvain}\ \bibnamefont
  {de~L{\'e}s{\'e}leuc}}, \bibinfo {author} {\bibfnamefont {Vincent}\
  \bibnamefont {Lienhard}}, \bibinfo {author} {\bibfnamefont {Pascal}\
  \bibnamefont {Scholl}}, \bibinfo {author} {\bibfnamefont {Daniel}\
  \bibnamefont {Barredo}}, \bibinfo {author} {\bibfnamefont {Sebastian}\
  \bibnamefont {Weber}}, \bibinfo {author} {\bibfnamefont {Nicolai}\
  \bibnamefont {Lang}}, \bibinfo {author} {\bibfnamefont {Hans~Peter}\
  \bibnamefont {B{\"u}chler}}, \bibinfo {author} {\bibfnamefont {Thierry}\
  \bibnamefont {Lahaye}}, \ and\ \bibinfo {author} {\bibfnamefont {Antoine}\
  \bibnamefont {Browaeys}},\ }\bibfield  {title} {\enquote {\bibinfo {title}
  {Observation of a symmetry-protected topological phase of interacting bosons
  with rydberg atoms},}\ }\href@noop {} {\bibfield  {journal} {\bibinfo
  {journal} {Science}\ }\textbf {\bibinfo {volume} {365}},\ \bibinfo {pages}
  {775--780} (\bibinfo {year} {2019})}\BibitemShut {NoStop}%
\bibitem [{\citenamefont {Bohnet}\ \emph {et~al.}(2016)\citenamefont {Bohnet},
  \citenamefont {Sawyer}, \citenamefont {Britton}, \citenamefont {Wall},
  \citenamefont {Rey}, \citenamefont {Foss-Feig},\ and\ \citenamefont
  {Bollinger}}]{bohnet2016quantum}%
  \BibitemOpen
  \bibfield  {author} {\bibinfo {author} {\bibfnamefont {Justin~G}\
  \bibnamefont {Bohnet}}, \bibinfo {author} {\bibfnamefont {Brian~C}\
  \bibnamefont {Sawyer}}, \bibinfo {author} {\bibfnamefont {Joseph~W}\
  \bibnamefont {Britton}}, \bibinfo {author} {\bibfnamefont {Michael~L}\
  \bibnamefont {Wall}}, \bibinfo {author} {\bibfnamefont {Ana~Maria}\
  \bibnamefont {Rey}}, \bibinfo {author} {\bibfnamefont {Michael}\ \bibnamefont
  {Foss-Feig}}, \ and\ \bibinfo {author} {\bibfnamefont {John~J}\ \bibnamefont
  {Bollinger}},\ }\bibfield  {title} {\enquote {\bibinfo {title} {Quantum spin
  dynamics and entanglement generation with hundreds of trapped ions},}\
  }\href@noop {} {\bibfield  {journal} {\bibinfo  {journal} {Science}\ }\textbf
  {\bibinfo {volume} {352}},\ \bibinfo {pages} {1297--1301} (\bibinfo {year}
  {2016})}\BibitemShut {NoStop}%
\bibitem [{\citenamefont {Vermersch}\ \emph {et~al.}(2019)\citenamefont
  {Vermersch}, \citenamefont {Elben}, \citenamefont {Sieberer}, \citenamefont
  {Yao},\ and\ \citenamefont {Zoller}}]{vermersch2019probing}%
  \BibitemOpen
  \bibfield  {author} {\bibinfo {author} {\bibfnamefont {Beno{\^\i}t}\
  \bibnamefont {Vermersch}}, \bibinfo {author} {\bibfnamefont {Andreas}\
  \bibnamefont {Elben}}, \bibinfo {author} {\bibfnamefont {Lukas~M}\
  \bibnamefont {Sieberer}}, \bibinfo {author} {\bibfnamefont {Norman~Y}\
  \bibnamefont {Yao}}, \ and\ \bibinfo {author} {\bibfnamefont {Peter}\
  \bibnamefont {Zoller}},\ }\bibfield  {title} {\enquote {\bibinfo {title}
  {Probing scrambling using statistical correlations between randomized
  measurements},}\ }\href@noop {} {\bibfield  {journal} {\bibinfo  {journal}
  {Physical Review X}\ }\textbf {\bibinfo {volume} {9}},\ \bibinfo {pages}
  {021061} (\bibinfo {year} {2019})}\BibitemShut {NoStop}%
\bibitem [{\citenamefont {Zhang}\ \emph {et~al.}(2017)\citenamefont {Zhang},
  \citenamefont {Pagano}, \citenamefont {Hess}, \citenamefont {Kyprianidis},
  \citenamefont {Becker}, \citenamefont {Kaplan}, \citenamefont {Gorshkov},
  \citenamefont {Gong},\ and\ \citenamefont {Monroe}}]{zhang2017observation}%
  \BibitemOpen
  \bibfield  {author} {\bibinfo {author} {\bibfnamefont {Jiehang}\ \bibnamefont
  {Zhang}}, \bibinfo {author} {\bibfnamefont {Guido}\ \bibnamefont {Pagano}},
  \bibinfo {author} {\bibfnamefont {Paul~W}\ \bibnamefont {Hess}}, \bibinfo
  {author} {\bibfnamefont {Antonis}\ \bibnamefont {Kyprianidis}}, \bibinfo
  {author} {\bibfnamefont {Patrick}\ \bibnamefont {Becker}}, \bibinfo {author}
  {\bibfnamefont {Harvey}\ \bibnamefont {Kaplan}}, \bibinfo {author}
  {\bibfnamefont {Alexey~V}\ \bibnamefont {Gorshkov}}, \bibinfo {author}
  {\bibfnamefont {Z-X}\ \bibnamefont {Gong}}, \ and\ \bibinfo {author}
  {\bibfnamefont {Christopher}\ \bibnamefont {Monroe}},\ }\bibfield  {title}
  {\enquote {\bibinfo {title} {Observation of a many-body dynamical phase
  transition with a 53-qubit quantum simulator},}\ }\href@noop {} {\bibfield
  {journal} {\bibinfo  {journal} {Nature}\ }\textbf {\bibinfo {volume} {551}},\
  \bibinfo {pages} {601--604} (\bibinfo {year} {2017})}\BibitemShut {NoStop}%
\bibitem [{\citenamefont {Barrett}\ \emph {et~al.}(2003)\citenamefont
  {Barrett}, \citenamefont {DeMarco}, \citenamefont {Schaetz}, \citenamefont
  {Meyer}, \citenamefont {Leibfried}, \citenamefont {Britton}, \citenamefont
  {Chiaverini}, \citenamefont {Itano}, \citenamefont
  {Jelenkovi\ifmmode~\acute{c}\else \'{c}\fi{}}, \citenamefont {Jost},
  \citenamefont {Langer}, \citenamefont {Rosenband},\ and\ \citenamefont
  {Wineland}}]{barrett2003sypathetic}%
  \BibitemOpen
  \bibfield  {author} {\bibinfo {author} {\bibfnamefont {M.~D.}\ \bibnamefont
  {Barrett}}, \bibinfo {author} {\bibfnamefont {B.}~\bibnamefont {DeMarco}},
  \bibinfo {author} {\bibfnamefont {T.}~\bibnamefont {Schaetz}}, \bibinfo
  {author} {\bibfnamefont {V.}~\bibnamefont {Meyer}}, \bibinfo {author}
  {\bibfnamefont {D.}~\bibnamefont {Leibfried}}, \bibinfo {author}
  {\bibfnamefont {J.}~\bibnamefont {Britton}}, \bibinfo {author} {\bibfnamefont
  {J.}~\bibnamefont {Chiaverini}}, \bibinfo {author} {\bibfnamefont {W.~M.}\
  \bibnamefont {Itano}}, \bibinfo {author} {\bibfnamefont {B.}~\bibnamefont
  {Jelenkovi\ifmmode~\acute{c}\else \'{c}\fi{}}}, \bibinfo {author}
  {\bibfnamefont {J.~D.}\ \bibnamefont {Jost}}, \bibinfo {author}
  {\bibfnamefont {C.}~\bibnamefont {Langer}}, \bibinfo {author} {\bibfnamefont
  {T.}~\bibnamefont {Rosenband}}, \ and\ \bibinfo {author} {\bibfnamefont
  {D.~J.}\ \bibnamefont {Wineland}},\ }\bibfield  {title} {\enquote {\bibinfo
  {title} {Sympathetic cooling of ${}^{9}{\mathrm{be}}^{+}$ and
  ${}^{24}{\mathrm{mg}}^{+}$ for quantum logic},}\ }\href {\doibase
  10.1103/PhysRevA.68.042302} {\bibfield  {journal} {\bibinfo  {journal} {Phys.
  Rev. A}\ }\textbf {\bibinfo {volume} {68}},\ \bibinfo {pages} {042302}
  (\bibinfo {year} {2003})}\BibitemShut {NoStop}%
\bibitem [{\citenamefont {Korenblit}\ \emph {et~al.}(2012)\citenamefont
  {Korenblit}, \citenamefont {Kafri}, \citenamefont {Campbell}, \citenamefont
  {Islam}, \citenamefont {Edwards}, \citenamefont {Gong}, \citenamefont {Lin},
  \citenamefont {Duan}, \citenamefont {Kim}, \citenamefont {Kim},\ and\
  \citenamefont {Monroe}}]{korenblit2012quantum}%
  \BibitemOpen
  \bibfield  {author} {\bibinfo {author} {\bibfnamefont {Simcha}\ \bibnamefont
  {Korenblit}}, \bibinfo {author} {\bibfnamefont {Dvir}\ \bibnamefont {Kafri}},
  \bibinfo {author} {\bibfnamefont {Wess~C}\ \bibnamefont {Campbell}}, \bibinfo
  {author} {\bibfnamefont {Rajibul}\ \bibnamefont {Islam}}, \bibinfo {author}
  {\bibfnamefont {Emily~E}\ \bibnamefont {Edwards}}, \bibinfo {author}
  {\bibfnamefont {Zhe-Xuan}\ \bibnamefont {Gong}}, \bibinfo {author}
  {\bibfnamefont {Guin-Dar}\ \bibnamefont {Lin}}, \bibinfo {author}
  {\bibfnamefont {Lu-Ming}\ \bibnamefont {Duan}}, \bibinfo {author}
  {\bibfnamefont {Jungsang}\ \bibnamefont {Kim}}, \bibinfo {author}
  {\bibfnamefont {Kihwan}\ \bibnamefont {Kim}}, \ and\ \bibinfo {author}
  {\bibfnamefont {Chris}\ \bibnamefont {Monroe}},\ }\bibfield  {title}
  {\enquote {\bibinfo {title} {Quantum simulation of spin models on an
  arbitrary lattice with trapped ions},}\ }\href@noop {} {\bibfield  {journal}
  {\bibinfo  {journal} {New Journal of Physics}\ }\textbf {\bibinfo {volume}
  {14}},\ \bibinfo {pages} {095024} (\bibinfo {year} {2012})}\BibitemShut
  {NoStop}%
\bibitem [{\citenamefont {Teoh}\ \emph {et~al.}(2020)\citenamefont {Teoh},
  \citenamefont {Drygala}, \citenamefont {Melko},\ and\ \citenamefont
  {Islam}}]{teoh2020machine}%
  \BibitemOpen
  \bibfield  {author} {\bibinfo {author} {\bibfnamefont {Yi~Hong}\ \bibnamefont
  {Teoh}}, \bibinfo {author} {\bibfnamefont {Marina}\ \bibnamefont {Drygala}},
  \bibinfo {author} {\bibfnamefont {Roger~G}\ \bibnamefont {Melko}}, \ and\
  \bibinfo {author} {\bibfnamefont {Rajibul}\ \bibnamefont {Islam}},\
  }\bibfield  {title} {\enquote {\bibinfo {title} {Machine learning design of a
  trapped-ion quantum spin simulator},}\ }\href@noop {} {\bibfield  {journal}
  {\bibinfo  {journal} {Quantum Science and Technology}\ }\textbf {\bibinfo
  {volume} {5}},\ \bibinfo {pages} {024001} (\bibinfo {year}
  {2020})}\BibitemShut {NoStop}%
\bibitem [{\citenamefont {Else}\ \emph {et~al.}(2020)\citenamefont {Else},
  \citenamefont {Machado}, \citenamefont {Nayak},\ and\ \citenamefont
  {Yao}}]{else2020improved}%
  \BibitemOpen
  \bibfield  {author} {\bibinfo {author} {\bibfnamefont {Dominic~V.}\
  \bibnamefont {Else}}, \bibinfo {author} {\bibfnamefont {Francisco}\
  \bibnamefont {Machado}}, \bibinfo {author} {\bibfnamefont {Chetan}\
  \bibnamefont {Nayak}}, \ and\ \bibinfo {author} {\bibfnamefont {Norman~Y.}\
  \bibnamefont {Yao}},\ }\bibfield  {title} {\enquote {\bibinfo {title}
  {Improved lieb-robinson bound for many-body hamiltonians with power-law
  interactions},}\ }\href {\doibase 10.1103/PhysRevA.101.022333} {\bibfield
  {journal} {\bibinfo  {journal} {Phys. Rev. A}\ }\textbf {\bibinfo {volume}
  {101}},\ \bibinfo {pages} {022333} (\bibinfo {year} {2020})}\BibitemShut
  {NoStop}%
\bibitem [{\citenamefont {Zhou}\ \emph {et~al.}(2020)\citenamefont {Zhou},
  \citenamefont {Xu}, \citenamefont {Chen}, \citenamefont {Guo},\ and\
  \citenamefont {Swingle}}]{zhou2020operator}%
  \BibitemOpen
  \bibfield  {author} {\bibinfo {author} {\bibfnamefont {Tianci}\ \bibnamefont
  {Zhou}}, \bibinfo {author} {\bibfnamefont {Shenglong}\ \bibnamefont {Xu}},
  \bibinfo {author} {\bibfnamefont {Xiao}\ \bibnamefont {Chen}}, \bibinfo
  {author} {\bibfnamefont {Andrew}\ \bibnamefont {Guo}}, \ and\ \bibinfo
  {author} {\bibfnamefont {Brian}\ \bibnamefont {Swingle}},\ }\bibfield
  {title} {\enquote {\bibinfo {title} {Operator l\'evy flight: Light cones in
  chaotic long-range interacting systems},}\ }\href {\doibase
  10.1103/PhysRevLett.124.180601} {\bibfield  {journal} {\bibinfo  {journal}
  {Phys. Rev. Lett.}\ }\textbf {\bibinfo {volume} {124}},\ \bibinfo {pages}
  {180601} (\bibinfo {year} {2020})}\BibitemShut {NoStop}%
\bibitem [{\citenamefont {Martyn}\ and\ \citenamefont
  {Swingle}(2019)}]{martyn2019product}%
  \BibitemOpen
  \bibfield  {author} {\bibinfo {author} {\bibfnamefont {John}\ \bibnamefont
  {Martyn}}\ and\ \bibinfo {author} {\bibfnamefont {Brian}\ \bibnamefont
  {Swingle}},\ }\bibfield  {title} {\enquote {\bibinfo {title} {Product
  spectrum ansatz and the simplicity of thermal states},}\ }\href@noop {}
  {\bibfield  {journal} {\bibinfo  {journal} {Physical Review A}\ }\textbf
  {\bibinfo {volume} {100}},\ \bibinfo {pages} {032107} (\bibinfo {year}
  {2019})}\BibitemShut {NoStop}%
\bibitem [{\citenamefont {Cirac}\ and\ \citenamefont
  {Zoller}(1995)}]{cirac1995quantum}%
  \BibitemOpen
  \bibfield  {author} {\bibinfo {author} {\bibfnamefont {J.~I.}\ \bibnamefont
  {Cirac}}\ and\ \bibinfo {author} {\bibfnamefont {P.}~\bibnamefont {Zoller}},\
  }\bibfield  {title} {\enquote {\bibinfo {title} {Quantum computation with
  cold trapped ions},}\ }\href@noop {} {\bibfield  {journal} {\bibinfo
  {journal} {Phys. Rev. Lett.}\ }\textbf {\bibinfo {volume} {74}},\ \bibinfo
  {pages} {4091--4094} (\bibinfo {year} {1995})}\BibitemShut {NoStop}%
\bibitem [{\citenamefont {Mølmer}\ and\ \citenamefont
  {Sørensen}(1999)}]{molmer_multiparticle_1999}%
  \BibitemOpen
  \bibfield  {author} {\bibinfo {author} {\bibfnamefont {Klaus}\ \bibnamefont
  {Mølmer}}\ and\ \bibinfo {author} {\bibfnamefont {Anders}\ \bibnamefont
  {Sørensen}},\ }\bibfield  {title} {\enquote {\bibinfo {title} {Multiparticle
  {Entanglement} of {Hot} {Trapped} {Ions}},}\ }\href {\doibase
  10.1103/PhysRevLett.82.1835} {\bibfield  {journal} {\bibinfo  {journal}
  {Phys. Rev. Lett.}\ }\textbf {\bibinfo {volume} {82}},\ \bibinfo {pages}
  {1835--1838} (\bibinfo {year} {1999})}\BibitemShut {NoStop}%
\bibitem [{\citenamefont {Monroe}\ \emph {et~al.}(2019)\citenamefont {Monroe},
  \citenamefont {Campbell}, \citenamefont {Duan}, \citenamefont {Gong},
  \citenamefont {Gorshkov}, \citenamefont {Hess}, \citenamefont {Islam},
  \citenamefont {Kim}, \citenamefont {Pagano}, \citenamefont {Richerme} \emph
  {et~al.}}]{monroe2019programmable}%
  \BibitemOpen
  \bibfield  {author} {\bibinfo {author} {\bibfnamefont {C}~\bibnamefont
  {Monroe}}, \bibinfo {author} {\bibfnamefont {WC}~\bibnamefont {Campbell}},
  \bibinfo {author} {\bibfnamefont {L-M}\ \bibnamefont {Duan}}, \bibinfo
  {author} {\bibfnamefont {Z-X}\ \bibnamefont {Gong}}, \bibinfo {author}
  {\bibfnamefont {AV}~\bibnamefont {Gorshkov}}, \bibinfo {author}
  {\bibfnamefont {P}~\bibnamefont {Hess}}, \bibinfo {author} {\bibfnamefont
  {R}~\bibnamefont {Islam}}, \bibinfo {author} {\bibfnamefont {K}~\bibnamefont
  {Kim}}, \bibinfo {author} {\bibfnamefont {G}~\bibnamefont {Pagano}}, \bibinfo
  {author} {\bibfnamefont {P}~\bibnamefont {Richerme}},  \emph {et~al.},\
  }\bibfield  {title} {\enquote {\bibinfo {title} {Programmable quantum
  simulations of spin systems with trapped ions},}\ }\href@noop {} {\bibfield
  {journal} {\bibinfo  {journal} {arXiv preprint arXiv:1912.07845}\ } (\bibinfo
  {year} {2019})}\BibitemShut {NoStop}%
\bibitem [{\citenamefont {Zhu}\ \emph {et~al.}(2006)\citenamefont {Zhu},
  \citenamefont {Monroe},\ and\ \citenamefont {Duan}}]{zhu2006trapped}%
  \BibitemOpen
  \bibfield  {author} {\bibinfo {author} {\bibfnamefont {Shi-Liang}\
  \bibnamefont {Zhu}}, \bibinfo {author} {\bibfnamefont {C.}~\bibnamefont
  {Monroe}}, \ and\ \bibinfo {author} {\bibfnamefont {L.-M.}\ \bibnamefont
  {Duan}},\ }\bibfield  {title} {\enquote {\bibinfo {title} {Trapped ion
  quantum computation with transverse phonon modes},}\ }\href
  {https://journals.aps.org/prl/abstract/10.1103/PhysRevLett.97.050505}
  {\bibfield  {journal} {\bibinfo  {journal} {Phys. Rev. Lett.}\ }\textbf
  {\bibinfo {volume} {97}},\ \bibinfo {pages} {050505} (\bibinfo {year}
  {2006})}\BibitemShut {NoStop}%
\bibitem [{\citenamefont {Debnath}\ \emph {et~al.}(2016)\citenamefont
  {Debnath}, \citenamefont {Linke}, \citenamefont {Figgatt}, \citenamefont
  {Landsman}, \citenamefont {Wright},\ and\ \citenamefont
  {Monroe}}]{DebnathQC:2016}%
  \BibitemOpen
  \bibfield  {author} {\bibinfo {author} {\bibfnamefont {S.}~\bibnamefont
  {Debnath}}, \bibinfo {author} {\bibfnamefont {N.~M.}\ \bibnamefont {Linke}},
  \bibinfo {author} {\bibfnamefont {C.}~\bibnamefont {Figgatt}}, \bibinfo
  {author} {\bibfnamefont {K.~A.}\ \bibnamefont {Landsman}}, \bibinfo {author}
  {\bibfnamefont {K.}~\bibnamefont {Wright}}, \ and\ \bibinfo {author}
  {\bibfnamefont {C.}~\bibnamefont {Monroe}},\ }\bibfield  {title} {\enquote
  {\bibinfo {title} {Demonstration of a small programmable quantum computer
  with atomic qubits},}\ }\href {https://www.nature.com/articles/nature18648}
  {\bibfield  {journal} {\bibinfo  {journal} {Nature}\ }\textbf {\bibinfo
  {volume} {563}},\ \bibinfo {pages} {63} (\bibinfo {year} {2016})}\BibitemShut
  {NoStop}%
\bibitem [{\citenamefont {Landsman}\ \emph
  {et~al.}(2019{\natexlab{b}})\citenamefont {Landsman}, \citenamefont {Wu},
  \citenamefont {Leung}, \citenamefont {Zhu}, \citenamefont {Linke},
  \citenamefont {Brown}, \citenamefont {Duan},\ and\ \citenamefont
  {Monroe}}]{Landsman2019arb}%
  \BibitemOpen
  \bibfield  {author} {\bibinfo {author} {\bibfnamefont {K.~A.}\ \bibnamefont
  {Landsman}}, \bibinfo {author} {\bibfnamefont {Y.}~\bibnamefont {Wu}},
  \bibinfo {author} {\bibfnamefont {P.~H.}\ \bibnamefont {Leung}}, \bibinfo
  {author} {\bibfnamefont {D.}~\bibnamefont {Zhu}}, \bibinfo {author}
  {\bibfnamefont {N.~M.}\ \bibnamefont {Linke}}, \bibinfo {author}
  {\bibfnamefont {K.~R.}\ \bibnamefont {Brown}}, \bibinfo {author}
  {\bibfnamefont {L.}~\bibnamefont {Duan}}, \ and\ \bibinfo {author}
  {\bibfnamefont {C.}~\bibnamefont {Monroe}},\ }\bibfield  {title} {\enquote
  {\bibinfo {title} {Two-qubit entangling gates within arbitrarily long chains
  of trapped ions},}\ }\href {\doibase 10.1103/PhysRevA.100.022332} {\bibfield
  {journal} {\bibinfo  {journal} {Phys. Rev. A}\ }\textbf {\bibinfo {volume}
  {100}},\ \bibinfo {pages} {022332} (\bibinfo {year}
  {2019}{\natexlab{b}})}\BibitemShut {NoStop}%
\bibitem [{\citenamefont {Wright}\ \emph {et~al.}(2019)\citenamefont {Wright}
  \emph {et~al.}}]{Wright:2019}%
  \BibitemOpen
  \bibfield  {author} {\bibinfo {author} {\bibfnamefont {K.}~\bibnamefont
  {Wright}} \emph {et~al.},\ }\bibfield  {title} {\enquote {\bibinfo {title}
  {Benchmarking an 11-qubit quantum computer},}\ }\href
  {https://www.nature.com/articles/s41467-019-13534-2} {\bibfield  {journal}
  {\bibinfo  {journal} {Nat. Commun.}\ }\textbf {\bibinfo {volume} {10}},\
  \bibinfo {pages} {5464} (\bibinfo {year} {2019})}\BibitemShut {NoStop}%
\bibitem [{\citenamefont {Debnath}(2016)}]{debnath2016programmable}%
  \BibitemOpen
  \bibfield  {author} {\bibinfo {author} {\bibfnamefont {Shantanu}\
  \bibnamefont {Debnath}},\ }\emph {\bibinfo {title} {A Programmable Five Qubit
  Quantum Computer Using Trapped Atomic Ions}},\ \href@noop {} {Ph.D. thesis},\
  \bibinfo  {school} {University of Maryland, College Park} (\bibinfo {year}
  {2016})\BibitemShut {NoStop}%
\bibitem [{\citenamefont {Smith}\ \emph {et~al.}(2016)\citenamefont {Smith},
  \citenamefont {Lee}, \citenamefont {Richerme}, \citenamefont {Neyenhuis},
  \citenamefont {Hess}, \citenamefont {Hauke}, \citenamefont {Heyl},
  \citenamefont {Huse},\ and\ \citenamefont {Monroe}}]{smith2016many}%
  \BibitemOpen
  \bibfield  {author} {\bibinfo {author} {\bibfnamefont {J}~\bibnamefont
  {Smith}}, \bibinfo {author} {\bibfnamefont {A}~\bibnamefont {Lee}}, \bibinfo
  {author} {\bibfnamefont {P}~\bibnamefont {Richerme}}, \bibinfo {author}
  {\bibfnamefont {B}~\bibnamefont {Neyenhuis}}, \bibinfo {author}
  {\bibfnamefont {PW}~\bibnamefont {Hess}}, \bibinfo {author} {\bibfnamefont
  {P}~\bibnamefont {Hauke}}, \bibinfo {author} {\bibfnamefont {M}~\bibnamefont
  {Heyl}}, \bibinfo {author} {\bibfnamefont {DA}~\bibnamefont {Huse}}, \ and\
  \bibinfo {author} {\bibfnamefont {C}~\bibnamefont {Monroe}},\ }\bibfield
  {title} {\enquote {\bibinfo {title} {Many-body localization in a quantum
  simulator with programmable random disorder},}\ }\href@noop {} {\bibfield
  {journal} {\bibinfo  {journal} {Nature Physics}\ }\textbf {\bibinfo {volume}
  {12}},\ \bibinfo {pages} {907--911} (\bibinfo {year} {2016})}\BibitemShut
  {NoStop}%
\bibitem [{\citenamefont {Hensinger}\ \emph {et~al.}(2006)\citenamefont
  {Hensinger}, \citenamefont {Olmschenk}, \citenamefont {Stick}, \citenamefont
  {Hucul}, \citenamefont {Yeo}, \citenamefont {Acton}, \citenamefont
  {Deslauriers}, \citenamefont {Monroe},\ and\ \citenamefont
  {Rabchuk}}]{hensinger2006t}%
  \BibitemOpen
  \bibfield  {author} {\bibinfo {author} {\bibfnamefont {WK}~\bibnamefont
  {Hensinger}}, \bibinfo {author} {\bibfnamefont {S}~\bibnamefont {Olmschenk}},
  \bibinfo {author} {\bibfnamefont {D}~\bibnamefont {Stick}}, \bibinfo {author}
  {\bibfnamefont {D}~\bibnamefont {Hucul}}, \bibinfo {author} {\bibfnamefont
  {M}~\bibnamefont {Yeo}}, \bibinfo {author} {\bibfnamefont {M}~\bibnamefont
  {Acton}}, \bibinfo {author} {\bibfnamefont {L}~\bibnamefont {Deslauriers}},
  \bibinfo {author} {\bibfnamefont {C}~\bibnamefont {Monroe}}, \ and\ \bibinfo
  {author} {\bibfnamefont {J}~\bibnamefont {Rabchuk}},\ }\bibfield  {title}
  {\enquote {\bibinfo {title} {T-junction ion trap array for two-dimensional
  ion shuttling, storage, and manipulation},}\ }\href@noop {} {\bibfield
  {journal} {\bibinfo  {journal} {Applied Physics Letters}\ }\textbf {\bibinfo
  {volume} {88}},\ \bibinfo {pages} {034101} (\bibinfo {year}
  {2006})}\BibitemShut {NoStop}%
\bibitem [{\citenamefont {Kaufmann}\ \emph {et~al.}(2017)\citenamefont
  {Kaufmann}, \citenamefont {Ruster}, \citenamefont {Schmiegelow},
  \citenamefont {Luda}, \citenamefont {Kaushal}, \citenamefont {Schulz},
  \citenamefont {von Lindenfels}, \citenamefont {Schmidt-Kaler},\ and\
  \citenamefont {Poschinger}}]{kaufmann2017fast}%
  \BibitemOpen
  \bibfield  {author} {\bibinfo {author} {\bibfnamefont {H.}~\bibnamefont
  {Kaufmann}}, \bibinfo {author} {\bibfnamefont {T.}~\bibnamefont {Ruster}},
  \bibinfo {author} {\bibfnamefont {C.~T.}\ \bibnamefont {Schmiegelow}},
  \bibinfo {author} {\bibfnamefont {M.~A.}\ \bibnamefont {Luda}}, \bibinfo
  {author} {\bibfnamefont {V.}~\bibnamefont {Kaushal}}, \bibinfo {author}
  {\bibfnamefont {J.}~\bibnamefont {Schulz}}, \bibinfo {author} {\bibfnamefont
  {D.}~\bibnamefont {von Lindenfels}}, \bibinfo {author} {\bibfnamefont
  {F.}~\bibnamefont {Schmidt-Kaler}}, \ and\ \bibinfo {author} {\bibfnamefont
  {U.~G.}\ \bibnamefont {Poschinger}},\ }\bibfield  {title} {\enquote {\bibinfo
  {title} {Fast ion swapping for quantum-information processing},}\ }\href
  {\doibase 10.1103/PhysRevA.95.052319} {\bibfield  {journal} {\bibinfo
  {journal} {Phys. Rev. A}\ }\textbf {\bibinfo {volume} {95}},\ \bibinfo
  {pages} {052319} (\bibinfo {year} {2017})}\BibitemShut {NoStop}%
\bibitem [{\citenamefont {Schuster}\ and\ \citenamefont
  {Yao}(2022)}]{schuster2021operator}%
  \BibitemOpen
  \bibfield  {author} {\bibinfo {author} {\bibfnamefont {Thomas}\ \bibnamefont
  {Schuster}}\ and\ \bibinfo {author} {\bibfnamefont {Norman~Y.}\ \bibnamefont
  {Yao}},\ }\bibfield  {title} {\enquote {\bibinfo {title} {Operator growth in
  open quantum systems},}\ }\href@noop {} {\bibfield  {journal} {\bibinfo
  {journal} {(forthcoming)}\ } (\bibinfo {year} {2022})}\BibitemShut {NoStop}%
\bibitem [{\citenamefont {Brun}\ \emph {et~al.}(2006)\citenamefont {Brun},
  \citenamefont {Devetak},\ and\ \citenamefont {Hsieh}}]{brun2006correcting}%
  \BibitemOpen
  \bibfield  {author} {\bibinfo {author} {\bibfnamefont {Todd}\ \bibnamefont
  {Brun}}, \bibinfo {author} {\bibfnamefont {Igor}\ \bibnamefont {Devetak}}, \
  and\ \bibinfo {author} {\bibfnamefont {Min-Hsiu}\ \bibnamefont {Hsieh}},\
  }\bibfield  {title} {\enquote {\bibinfo {title} {Correcting quantum errors
  with entanglement},}\ }\href@noop {} {\bibfield  {journal} {\bibinfo
  {journal} {science}\ }\textbf {\bibinfo {volume} {314}},\ \bibinfo {pages}
  {436--439} (\bibinfo {year} {2006})}\BibitemShut {NoStop}%
\bibitem [{\citenamefont {Swingle}\ \emph {et~al.}(2016)\citenamefont
  {Swingle}, \citenamefont {Bentsen}, \citenamefont {Schleier-Smith},\ and\
  \citenamefont {Hayden}}]{swingle2016measuring}%
  \BibitemOpen
  \bibfield  {author} {\bibinfo {author} {\bibfnamefont {Brian}\ \bibnamefont
  {Swingle}}, \bibinfo {author} {\bibfnamefont {Gregory}\ \bibnamefont
  {Bentsen}}, \bibinfo {author} {\bibfnamefont {Monika}\ \bibnamefont
  {Schleier-Smith}}, \ and\ \bibinfo {author} {\bibfnamefont {Patrick}\
  \bibnamefont {Hayden}},\ }\bibfield  {title} {\enquote {\bibinfo {title}
  {Measuring the scrambling of quantum information},}\ }\href@noop {}
  {\bibfield  {journal} {\bibinfo  {journal} {Physical Review A}\ }\textbf
  {\bibinfo {volume} {94}},\ \bibinfo {pages} {040302} (\bibinfo {year}
  {2016})}\BibitemShut {NoStop}%
\bibitem [{\citenamefont {Yao}\ \emph {et~al.}(2016)\citenamefont {Yao},
  \citenamefont {Grusdt}, \citenamefont {Swingle}, \citenamefont {Lukin},
  \citenamefont {Stamper-Kurn}, \citenamefont {Moore},\ and\ \citenamefont
  {Demler}}]{yao2016interferometric}%
  \BibitemOpen
  \bibfield  {author} {\bibinfo {author} {\bibfnamefont {Norman~Y}\
  \bibnamefont {Yao}}, \bibinfo {author} {\bibfnamefont {Fabian}\ \bibnamefont
  {Grusdt}}, \bibinfo {author} {\bibfnamefont {Brian}\ \bibnamefont {Swingle}},
  \bibinfo {author} {\bibfnamefont {Mikhail~D}\ \bibnamefont {Lukin}}, \bibinfo
  {author} {\bibfnamefont {Dan~M}\ \bibnamefont {Stamper-Kurn}}, \bibinfo
  {author} {\bibfnamefont {Joel~E}\ \bibnamefont {Moore}}, \ and\ \bibinfo
  {author} {\bibfnamefont {Eugene~A}\ \bibnamefont {Demler}},\ }\bibfield
  {title} {\enquote {\bibinfo {title} {Interferometric approach to probing fast
  scrambling},}\ }\href@noop {} {\bibfield  {journal} {\bibinfo  {journal}
  {arXiv preprint arXiv:1607.01801}\ } (\bibinfo {year} {2016})}\BibitemShut
  {NoStop}%
\bibitem [{\citenamefont {John}\ \emph {et~al.}(2007)\citenamefont {John},
  \citenamefont {Angelov}, \citenamefont {{\"O}nc{\"u}l},\ and\ \citenamefont
  {Th{\'e}venin}}]{john2007techniques}%
  \BibitemOpen
  \bibfield  {author} {\bibinfo {author} {\bibfnamefont {V}~\bibnamefont
  {John}}, \bibinfo {author} {\bibfnamefont {I}~\bibnamefont {Angelov}},
  \bibinfo {author} {\bibfnamefont {AA}~\bibnamefont {{\"O}nc{\"u}l}}, \ and\
  \bibinfo {author} {\bibfnamefont {D}~\bibnamefont {Th{\'e}venin}},\
  }\bibfield  {title} {\enquote {\bibinfo {title} {Techniques for the
  reconstruction of a distribution from a finite number of its moments},}\
  }\href@noop {} {\bibfield  {journal} {\bibinfo  {journal} {Chemical
  Engineering Science}\ }\textbf {\bibinfo {volume} {62}},\ \bibinfo {pages}
  {2890--2904} (\bibinfo {year} {2007})}\BibitemShut {NoStop}%
\bibitem [{\citenamefont {Else}\ \emph {et~al.}(2018)\citenamefont {Else},
  \citenamefont {Machado}, \citenamefont {Nayak},\ and\ \citenamefont
  {Yao}}]{else2018improved}%
  \BibitemOpen
  \bibfield  {author} {\bibinfo {author} {\bibfnamefont {Dominic~V}\
  \bibnamefont {Else}}, \bibinfo {author} {\bibfnamefont {Francisco}\
  \bibnamefont {Machado}}, \bibinfo {author} {\bibfnamefont {Chetan}\
  \bibnamefont {Nayak}}, \ and\ \bibinfo {author} {\bibfnamefont {Norman~Y}\
  \bibnamefont {Yao}},\ }\bibfield  {title} {\enquote {\bibinfo {title} {An
  improved lieb-robinson bound for many-body hamiltonians with power-law
  interactions},}\ }\href@noop {} {\bibfield  {journal} {\bibinfo  {journal}
  {arXiv preprint arXiv:1809.06369}\ } (\bibinfo {year} {2018})}\BibitemShut
  {NoStop}%
\bibitem [{\citenamefont {Tran}\ \emph {et~al.}(2020)\citenamefont {Tran},
  \citenamefont {Chen}, \citenamefont {Ehrenberg}, \citenamefont {Guo},
  \citenamefont {Deshpande}, \citenamefont {Hong}, \citenamefont {Gong},
  \citenamefont {Gorshkov},\ and\ \citenamefont {Lucas}}]{tran2020hierarchy}%
  \BibitemOpen
  \bibfield  {author} {\bibinfo {author} {\bibfnamefont {Minh~C}\ \bibnamefont
  {Tran}}, \bibinfo {author} {\bibfnamefont {Chi-Fang}\ \bibnamefont {Chen}},
  \bibinfo {author} {\bibfnamefont {Adam}\ \bibnamefont {Ehrenberg}}, \bibinfo
  {author} {\bibfnamefont {Andrew~Y}\ \bibnamefont {Guo}}, \bibinfo {author}
  {\bibfnamefont {Abhinav}\ \bibnamefont {Deshpande}}, \bibinfo {author}
  {\bibfnamefont {Yifan}\ \bibnamefont {Hong}}, \bibinfo {author}
  {\bibfnamefont {Zhe-Xuan}\ \bibnamefont {Gong}}, \bibinfo {author}
  {\bibfnamefont {Alexey~V}\ \bibnamefont {Gorshkov}}, \ and\ \bibinfo {author}
  {\bibfnamefont {Andrew}\ \bibnamefont {Lucas}},\ }\bibfield  {title}
  {\enquote {\bibinfo {title} {Hierarchy of linear light cones with long-range
  interactions},}\ }\href@noop {} {\bibfield  {journal} {\bibinfo  {journal}
  {arXiv preprint arXiv:2001.11509}\ } (\bibinfo {year} {2020})}\BibitemShut
  {NoStop}%
\bibitem [{\citenamefont {Gu}\ \emph {et~al.}(2017)\citenamefont {Gu},
  \citenamefont {Qi},\ and\ \citenamefont {Stanford}}]{gu2017local}%
  \BibitemOpen
  \bibfield  {author} {\bibinfo {author} {\bibfnamefont {Yingfei}\ \bibnamefont
  {Gu}}, \bibinfo {author} {\bibfnamefont {Xiao-Liang}\ \bibnamefont {Qi}}, \
  and\ \bibinfo {author} {\bibfnamefont {Douglas}\ \bibnamefont {Stanford}},\
  }\bibfield  {title} {\enquote {\bibinfo {title} {Local criticality, diffusion
  and chaos in generalized sachdev-ye-kitaev models},}\ }\href@noop {} {\
  \textbf {\bibinfo {volume} {2017}},\ \bibinfo {pages} {125} (\bibinfo {year}
  {2017})}\BibitemShut {NoStop}%
\bibitem [{\citenamefont {Bentsen}\ \emph {et~al.}(2019)\citenamefont
  {Bentsen}, \citenamefont {Gu},\ and\ \citenamefont
  {Lucas}}]{bentsen2019fast}%
  \BibitemOpen
  \bibfield  {author} {\bibinfo {author} {\bibfnamefont {Gregory}\ \bibnamefont
  {Bentsen}}, \bibinfo {author} {\bibfnamefont {Yingfei}\ \bibnamefont {Gu}}, \
  and\ \bibinfo {author} {\bibfnamefont {Andrew}\ \bibnamefont {Lucas}},\
  }\bibfield  {title} {\enquote {\bibinfo {title} {Fast scrambling on sparse
  graphs},}\ }\href@noop {} {\bibfield  {journal} {\bibinfo  {journal}
  {Proceedings of the National Academy of Sciences}\ }\textbf {\bibinfo
  {volume} {116}},\ \bibinfo {pages} {6689--6694} (\bibinfo {year}
  {2019})}\BibitemShut {NoStop}%
\bibitem [{\citenamefont {Salberger}\ \emph {et~al.}(2017)\citenamefont
  {Salberger}, \citenamefont {Udagawa}, \citenamefont {Zhang}, \citenamefont
  {Katsura}, \citenamefont {Klich},\ and\ \citenamefont
  {Korepin}}]{salberger2017deformed}%
  \BibitemOpen
  \bibfield  {author} {\bibinfo {author} {\bibfnamefont {Olof}\ \bibnamefont
  {Salberger}}, \bibinfo {author} {\bibfnamefont {Takuma}\ \bibnamefont
  {Udagawa}}, \bibinfo {author} {\bibfnamefont {Zhao}\ \bibnamefont {Zhang}},
  \bibinfo {author} {\bibfnamefont {Hosho}\ \bibnamefont {Katsura}}, \bibinfo
  {author} {\bibfnamefont {Israel}\ \bibnamefont {Klich}}, \ and\ \bibinfo
  {author} {\bibfnamefont {Vladimir}\ \bibnamefont {Korepin}},\ }\bibfield
  {title} {\enquote {\bibinfo {title} {Deformed fredkin spin chain with
  extensive entanglement},}\ }\href@noop {} {\bibfield  {journal} {\bibinfo
  {journal} {Journal of Statistical Mechanics: Theory and Experiment}\ }\textbf
  {\bibinfo {volume} {2017}},\ \bibinfo {pages} {063103} (\bibinfo {year}
  {2017})}\BibitemShut {NoStop}%
\bibitem [{\citenamefont {Alexander}\ \emph {et~al.}(2018)\citenamefont
  {Alexander}, \citenamefont {Evenbly},\ and\ \citenamefont
  {Klich}}]{alexander2018exact}%
  \BibitemOpen
  \bibfield  {author} {\bibinfo {author} {\bibfnamefont {Rafael~N}\
  \bibnamefont {Alexander}}, \bibinfo {author} {\bibfnamefont {Glen}\
  \bibnamefont {Evenbly}}, \ and\ \bibinfo {author} {\bibfnamefont {Israel}\
  \bibnamefont {Klich}},\ }\bibfield  {title} {\enquote {\bibinfo {title}
  {Exact holographic tensor networks for the motzkin spin chain},}\ }\href@noop
  {} {\bibfield  {journal} {\bibinfo  {journal} {arXiv preprint
  arXiv:1806.09626}\ } (\bibinfo {year} {2018})}\BibitemShut {NoStop}%
\bibitem [{\citenamefont {Plugge}\ \emph {et~al.}(2020)\citenamefont {Plugge},
  \citenamefont {Lantagne-Hurtubise},\ and\ \citenamefont
  {Franz}}]{plugge2020revival}%
  \BibitemOpen
  \bibfield  {author} {\bibinfo {author} {\bibfnamefont {Stephan}\ \bibnamefont
  {Plugge}}, \bibinfo {author} {\bibfnamefont {{\'E}tienne}\ \bibnamefont
  {Lantagne-Hurtubise}}, \ and\ \bibinfo {author} {\bibfnamefont {Marcel}\
  \bibnamefont {Franz}},\ }\bibfield  {title} {\enquote {\bibinfo {title}
  {Revival dynamics in a traversable wormhole},}\ }\href@noop {} {\bibfield
  {journal} {\bibinfo  {journal} {Physical review letters}\ }\textbf {\bibinfo
  {volume} {124}},\ \bibinfo {pages} {221601} (\bibinfo {year}
  {2020})}\BibitemShut {NoStop}%
\bibitem [{\citenamefont {Schuster}\ \emph {et~al.}(2021)\citenamefont
  {Schuster}, \citenamefont {Kobrin}, \citenamefont {Gao}, \citenamefont
  {Cong}, \citenamefont {Khabiboulline}, \citenamefont {Linke}, \citenamefont
  {Lukin}, \citenamefont {Monroe}, \citenamefont {Yoshida},\ and\ \citenamefont
  {Yao}}]{zenodo}%
  \BibitemOpen
  \bibfield  {author} {\bibinfo {author} {\bibfnamefont {Thomas}\ \bibnamefont
  {Schuster}}, \bibinfo {author} {\bibfnamefont {Bryce}\ \bibnamefont
  {Kobrin}}, \bibinfo {author} {\bibfnamefont {Ping}\ \bibnamefont {Gao}},
  \bibinfo {author} {\bibfnamefont {Iris}\ \bibnamefont {Cong}}, \bibinfo
  {author} {\bibfnamefont {Emil~T}\ \bibnamefont {Khabiboulline}}, \bibinfo
  {author} {\bibfnamefont {Norbert~M}\ \bibnamefont {Linke}}, \bibinfo {author}
  {\bibfnamefont {Mikhail~D}\ \bibnamefont {Lukin}}, \bibinfo {author}
  {\bibfnamefont {Christopher}\ \bibnamefont {Monroe}}, \bibinfo {author}
  {\bibfnamefont {Beni}\ \bibnamefont {Yoshida}}, \ and\ \bibinfo {author}
  {\bibfnamefont {Norman~Y}\ \bibnamefont {Yao}},\ }\bibfield  {title}
  {\enquote {\bibinfo {title} {Many-body quantum teleportation via operator
  spreading in the traversable wormhole protocol},}\ }\href
  {https://doi.org/10.5281/zenodo.5532758} {\bibfield  {journal} {\bibinfo
  {journal} {Zenodo, 10.5281/zenodo.5532757}\ } (\bibinfo {year}
  {2021})}\BibitemShut {NoStop}%
\bibitem [{\citenamefont {Zhuang}\ \emph {et~al.}(2019)\citenamefont {Zhuang},
  \citenamefont {Schuster}, \citenamefont {Yoshida},\ and\ \citenamefont
  {Yao}}]{zhuang2019scrambling}%
  \BibitemOpen
  \bibfield  {author} {\bibinfo {author} {\bibfnamefont {Quntao}\ \bibnamefont
  {Zhuang}}, \bibinfo {author} {\bibfnamefont {Thomas}\ \bibnamefont
  {Schuster}}, \bibinfo {author} {\bibfnamefont {Beni}\ \bibnamefont
  {Yoshida}}, \ and\ \bibinfo {author} {\bibfnamefont {Norman~Y}\ \bibnamefont
  {Yao}},\ }\bibfield  {title} {\enquote {\bibinfo {title} {Scrambling and
  complexity in phase space},}\ }\href@noop {} {\bibfield  {journal} {\bibinfo
  {journal} {Physical Review A}\ }\textbf {\bibinfo {volume} {99}},\ \bibinfo
  {pages} {062334} (\bibinfo {year} {2019})}\BibitemShut {NoStop}%
\bibitem [{\citenamefont {Kobrin}\ \emph {et~al.}(2021)\citenamefont {Kobrin},
  \citenamefont {Yang}, \citenamefont {Kahanamoku-Meyer}, \citenamefont
  {Olund}, \citenamefont {Moore}, \citenamefont {Stanford},\ and\ \citenamefont
  {Yao}}]{kobrin2021many}%
  \BibitemOpen
  \bibfield  {author} {\bibinfo {author} {\bibfnamefont {Bryce}\ \bibnamefont
  {Kobrin}}, \bibinfo {author} {\bibfnamefont {Zhenbin}\ \bibnamefont {Yang}},
  \bibinfo {author} {\bibfnamefont {Gregory~D}\ \bibnamefont
  {Kahanamoku-Meyer}}, \bibinfo {author} {\bibfnamefont {Christopher~T}\
  \bibnamefont {Olund}}, \bibinfo {author} {\bibfnamefont {Joel~E}\
  \bibnamefont {Moore}}, \bibinfo {author} {\bibfnamefont {Douglas}\
  \bibnamefont {Stanford}}, \ and\ \bibinfo {author} {\bibfnamefont {Norman~Y}\
  \bibnamefont {Yao}},\ }\bibfield  {title} {\enquote {\bibinfo {title}
  {Many-body chaos in the sachdev-ye-kitaev model},}\ }\href@noop {} {\bibfield
   {journal} {\bibinfo  {journal} {Physical Review Letters}\ }\textbf {\bibinfo
  {volume} {126}},\ \bibinfo {pages} {030602} (\bibinfo {year}
  {2021})}\BibitemShut {NoStop}%
\bibitem [{\citenamefont {Aaronson}\ and\ \citenamefont
  {Gottesman}(2004)}]{aaronson2004improved}%
  \BibitemOpen
  \bibfield  {author} {\bibinfo {author} {\bibfnamefont {Scott}\ \bibnamefont
  {Aaronson}}\ and\ \bibinfo {author} {\bibfnamefont {Daniel}\ \bibnamefont
  {Gottesman}},\ }\bibfield  {title} {\enquote {\bibinfo {title} {Improved
  simulation of stabilizer circuits},}\ }\href@noop {} {\bibfield  {journal}
  {\bibinfo  {journal} {Physical Review A}\ }\textbf {\bibinfo {volume} {70}},\
  \bibinfo {pages} {052328} (\bibinfo {year} {2004})}\BibitemShut {NoStop}%
\bibitem [{\citenamefont {Li}\ \emph {et~al.}(2019)\citenamefont {Li},
  \citenamefont {Chen},\ and\ \citenamefont {Fisher}}]{li2019measurement}%
  \BibitemOpen
  \bibfield  {author} {\bibinfo {author} {\bibfnamefont {Yaodong}\ \bibnamefont
  {Li}}, \bibinfo {author} {\bibfnamefont {Xiao}\ \bibnamefont {Chen}}, \ and\
  \bibinfo {author} {\bibfnamefont {Matthew~PA}\ \bibnamefont {Fisher}},\
  }\bibfield  {title} {\enquote {\bibinfo {title} {Measurement-driven
  entanglement transition in hybrid quantum circuits},}\ }\href@noop {}
  {\bibfield  {journal} {\bibinfo  {journal} {Physical Review B}\ }\textbf
  {\bibinfo {volume} {100}},\ \bibinfo {pages} {134306} (\bibinfo {year}
  {2019})}\BibitemShut {NoStop}%
\end{thebibliography}%

\newpage
\widetext
\appendix

\section{Precise bound for the peaked size regime}\label{app: bounds}

In this Appendix, we provide a precise mathematical bound guaranteeing that the teleportation correlator obeys the peaked-size prediction [Eq.~(\ref{phase assumption}), Section \ref{peaked sizes}] when the size distribution is sufficiently tightly peaked.
We apply this bound to two examples where the size distribution is known exactly: late times in all scrambling systems (Section \ref{late times}), and the large-$q$ SYK model (Sections \ref{sec: SYK infinite temperature} and \ref{SYK finite}).
Notably, in the latter we find that our bound applies \emph{only} at infinite temperature, despite the profile of the size distribution (e.g. its ratio of size width to average size) behaving similarly at all temperatures.
The discrepancy arises instead because the correlator magnitude, $(G_\beta)^p$, decreases exponentially in the encoding size $p$ at all finite temperatures.

\subsection{Precise bound}

As in the main text, we decompose a time-evolved finite temperature operator into a sum of Pauli strings:
\be
Q_A(t) \rho^{1/2}  = \sum_{\String} c_{\String}(t) S \\
\ee
In this basis, for qubit systems the correlator takes the form
\be
C_{Q} = \bra{\tfd} \tilde{Q}_{A,r}^{\dagger}(-t) e^{igV} Q_{A,l}^{}(t) \ket{\tfd} = e^{i g + i\pi\Size[Q_A(t=0)]} \sum_{\String} e^{-i \constd g \Size[\String]/N}  c_{\String}^2(t) = e^{i g + i\pi\Size[Q_A(t=0)]} \sum_n e^{i \constd g n / N} f(n)
\ee
where again $\tilde{Q}_{A,r}^{\dagger} = D Q_{A,r}^{\dagger} D^{\dagger}$ for the decoding operation $D  = Y \otimes \ldots \otimes Y$, and we use $\bra{\tfd} \tilde{Q}_{A,r}^{\dagger}(-t) = e^{i\pi \Size[Q_A]} \bra{\epr} Q_{A,l}(t) \rho^{1/2}$ for qubit Pauli operators $Q_A$.
Here we define the winding size distribution~\cite{brown2019quantum,brown2020quantum}
\be
f(n) \equiv \sum_{S : \Size[\String] = n} c_{\String}^2(t).
\ee
At finite temperature, this size wavefunction is distinct from the size distribution:
\be
P(n) \equiv \sum_{S : \Size[\String] = n} |c_{\String}(t) |^2,
\ee
which is a real, normalized probability distribution probed by the \emph{one}-sided correlator~\cite{qi2019quantum}
\be
\bra{\tfd} Q_{A,l}^{\dagger}(t) e^{igV} Q_{A,l}^{}(t) \ket{\tfd} = e^{ig} \sum_{\String} e^{-i \constd g \Size[\String]/N}  |c_{\String}|^2(t) = \sum_n e^{i \constd g n / N} P(n).
\ee
Nevertheless, the size distribution bounds the size wavefunction magnitude via the triangle inequality:
\be
| f(n) | \leq P(n),
\ee
with equality achieved when all Pauli operators of size $n$ contribute the same phase to $f(n)$. 

The average size and size variance are easily found from the size distribution as
\be
\Size = \int_0^{\infty} \, dn \, n \, P(n), \,\,\,\,\,\, \delta \Size^2 + \Size^2 = \int_0^{\infty} \, dn \, n^2 \, P(n)
\ee
where we work in the continuum limit replacing sums over the size by integrals for simplicity.
We now define the \emph{asymptotic size width with error} $\varepsilon$ as the minimal width $W_\varepsilon$ about the average size such that
\be \label{asymptotic width}
1 - \int_{\Size - W_\varepsilon}^{\Size + W_\varepsilon} \, dn \, P(n) \leq \varepsilon,
\ee
i.e. a fraction $1-\varepsilon$ of the size distribution's support is contained in the interval $I = [ \Size - W_\varepsilon, \Size + W_\varepsilon]$ (the lower limit of the integral should be bounded by zero; for simpler notation we'll deal with this by instead defining $P(n) = f(n) = 0$ for $n < 0$).
We can now separate the correlator into two pieces, one arising from sizes in the interval $I$ and the other from the interval's complement $\bar{I} = [ -\infty, \Size - W_\varepsilon] \cup  [ \Size + W_\varepsilon, \infty] $:
\be
C_Q = \int_I \, dn \, f(n) e^{i \constd g n / N} + R
\ee
where the remainder $R = \int_{\bar{I}} \, dn \, f(n) e^{i \constd g n / N}$ is strictly smaller than $\varepsilon$:
\be
\begin{split}
| R | & =   \left| \int_{\bar{I}} \, dn \, f(n) e^{i \constd g n / N} \right| \\
& \leq  \int_{\bar{I}} \, dn \, \left| f(n) e^{i \constd g n / N} \right| \\
& \leq  \int_{\bar{I}} \, dn \, \left| P(n) \right| \\
& \leq \varepsilon \\
\end{split}
\ee
Peaked size teleportation occurs in the regime where $g W_\varepsilon / N \ll 1$. In this limit, we can expand
\be
e^{i \constd g n / N} = e^{i \constd g \Size / N} \left[ 1 + E(n) \right]
\ee
where the deviation for $n\in I$ is bounded by
\be
| E(n) | \leq \text{max}_{n \in I} \bigg| 1 - e^{i \constd g (n - \Size) / N} \bigg| = \bigg| \sin( \constd g W_\varepsilon / N ) \bigg|,
\ee
which holds as long as $g W_\varepsilon / N \leq \pi/2$.
We then have
\be
\begin{split}
C_Q & = \int_I \, dn \, f(n) e^{i \constd g \Size / N} \left[ 1 + E(n) \right] + R \\
& = e^{i \constd g \Size / N} G_\beta(Q_A) + R + R' + R'' \\
\end{split}
\ee
where $G_\beta(Q_A) = \int_0^\infty dn \, f(n) = \tr(Q_A^\dagger \rho^{1/2} Q_A \rho^{1/2})$ is the imaginary time two-point function, and the error $R' = e^{i g \Size / N} \int_I \, dn \, f(n) E(n)$ is bounded by
\be
\begin{split}
| R' | & = \left| \int_I \, dn \, f(n) E(n) \right| \\
& \leq \int_I \, dn \, | f(n) | | E(n) | \\
& \leq \bigg| \sin( \constd g W_\varepsilon / N ) \bigg| \int_I \, dn \, | f(n) | \\
& \leq \bigg| \sin( \constd g W_\varepsilon / N ) \bigg|
\end{split}
\ee
and the second error $R'' = G_\beta(Q_A) - \int_I \, dn \, f(n)$ is bounded by
\be
|R''| = \left| G_\beta(Q_A) - \int_I \, dn \, f(n) \right| = \left| \int_{\bar{I}} \, dn \, f(n) \right| \leq \varepsilon.
\ee
We therefore conclude that whenever $\constd g W_\varepsilon / N \leq \pi/2$, the deviation of $C_Q$ from the peaked size value is controlled by the upper bound
\be\label{bound}
\left| C_Q - e^{i \constd g \Size / N} G_\beta(Q_A) \right| \leq 2 \varepsilon + \bigg| \sin( \constd g W_\varepsilon / N ) \bigg|\equiv \mB.
\ee
Practically speaking, the lowest value of $g$ for successful peaked-size teleportation is $\constd g \Size / N = \pi$.
Therefore, for a given size distribution, we can guarantee that peaked-size teleportation is possible if we find $\varepsilon$ such that $\mB \ll G_\beta(Q_A)$, i.e. the error in the correlator is small compared to the correlator magnitude.

\subsection{Application to late times}

We illustrate this with some examples, in the few cases where we can exactly solve for operators' full size distribution.
First, consider a thermalized system at late times, which we will approximate by setting the size distribution of $Q_A(t)$ to be that of a random Pauli string. For large $n, N$ is a Gaussian distribution with mean $\Size = 3N/4$ and variance $\delta \Size^2 = 3N/16$:
\be
P(n) = (3/4)^n (1/4)^{N-n} \approx \frac{1}{\sqrt{2\pi} \delta \Size} \exp \left( - (n - \Size)^2 / 2 \delta \Size^2 \right).
\ee
We therefore have
\be
1 - \int_{\Size - W_\varepsilon}^{\Size + W_\varepsilon} \, dn \, P(n) = 2 \,\text{erfc} \left( \frac{ W_\varepsilon }{ \sqrt{2}\delta \Size}  \right) = \varepsilon.
\ee
The error function decays exponentially in its argument, so even for exponentially small $\varepsilon$ we require only $W_\varepsilon = A \delta \Size$ for some constant $A \sim \mathcal{O}(1)$.
Setting $g$ equal to its minimal value, $\constd g \Size / N = \pi$, we have both $\varepsilon \ll 1$ and $\big| \sin( \constd g W_\varepsilon / N ) \big| \approx A \delta \Size / \Size \sim 1/\sqrt{N} \ll 1$, and so peaked size teleportation is guaranteed.

\subsection{Application to the large-$q$ SYK model}

We can also use this method to guarantee peaked-size teleportation in the large-$q$ SYK model at infinite temperature. 

We begin by writing down the size distribution for the large-$q$ SYK model in detail, quoting the results of Ref.~\cite{qi2019quantum}. The generating function for the size distribution is:
\begin{equation}
\sum_{n}P(n)e^{-\mu n}=\f{e^{-\mu p}}{(1+(1-e^{-\mu q})\sinh^{2}Jt)^{2p/q}}=\sum_{n}\f{\D_{n}}{n!}x^{n}(1-x)^{\D}e^{-\mu(qn+p)}
\end{equation}
where we define
\begin{equation}
\D_{n}\equiv\f{\Gamma(\D+n)}{\Gamma(\D)},\quad x\equiv\f{\sinh^{2}Jt}{1+\sinh^{2}Jt},\quad\D\equiv2p/q.
\end{equation}
From this, we can identify the size distribution:
\begin{equation}
P(qn+p)=\f{\D_{n}}{n!}x^{n}(1-x)^{\D}.
\end{equation}
The size and size width are
\begin{equation}
\Size = \overline{n}=\sum_{n}n\f{\D_{n}}{n!}x^{n}(1-x)^{\D}=\f{\D x}{1-x},\qquad\delta \Size = \sqrt{\overline{n^2} - \overline{n}^2} = \f{\sqrt{\D x}}{1-x}.
\end{equation}
Therefore, the ratio of size width to average size is
\begin{equation}
\delta \Size/\Size=\sqrt{\f x{\D}}\f 1{1+x},
\end{equation}
which approaches zero when $p\rightarrow\infty$ ($\Delta \rightarrow \infty$).

To apply the upper bound Eq.~\eqref{bound}, we need to integrate (i.e. sum) the tail of the size distribution in order to compute its asymptotic width [Eq.~(\ref{asymptotic width})]. In this example, the discrete tail can be summed explicitly and we define
\begin{equation}
I(k)\equiv \sum_{n=k}^{\infty}P(qk+p) = \sum_{n=k}^{\infty}\f{\D_{n}}{n!}x^{n}(1-x)^{\D}=\f{B_{x}(k,\D)}{B(k,\D)}
\end{equation}
where $B_x(a,b)$ and $B(a,b)$ are incomplete and ordinary beta function respectively. Let us take $k=\bar{n}(1\pm\zeta)$
for some small $\zeta$ representing the asymptotic width
\begin{equation}
W_{\varepsilon}=\bar{n}\zeta q.
\end{equation}
This width corresponds to an error
\begin{equation}
\varepsilon=1-I(\bar{n}(1-\zeta))+I(\bar{n}(1+\zeta)).
\end{equation}
Taking $g\Size/N=\pi$, the upper bound is
\begin{align}
\mB &=2[1-I(\bar{n}(1-\zeta))+I(\bar{n}(1+\zeta))]+\sin\f{2\pi\zeta x}{1+x}\nonumber \\
 & =2\left(1-\f{B_{x}(\f{\D x(1-\zeta)}{1-x},\D)}{B(\f{\D x(1-\zeta)}{1-x},\D)}+\f{B_{x}(\f{\D x(1+\zeta)}{1-x},\D)}{B(\f{\D x(1+\zeta)}{1-x},\D)}\right)+\sin\f{2\pi\zeta x}{1+x}.
\end{align}
At infinite temperature $G_\beta(Q_A)=1$, we need to show that the minimum of $\mB$ tends to zero when $\Delta \rightarrow \infty$. 

For early time $\sinh Jt\sim \mathcal{O}(1)$, $1-x$ is an order 1 number, and we take $\D\rightarrow\infty$ limit to get
\begin{equation}
\f{B_{x}(\f{\D x(1-\zeta)}{1-x},\D)}{B(\f{\D x(1-\zeta)}{1-x},\D)}\rightarrow1,\qquad\f{B_{x}(\f{\D x(1+\zeta)}{1-x},\D)}{B(\f{\D x(1+\zeta)}{1-x},\D)}\rightarrow0
\end{equation}
The bound becomes
\begin{equation}
\mB\rightarrow\sin\f{2\pi\zeta x}{1+x}
\end{equation}
This basically means that the integrated probability between $\bar{n}(1-\zeta)$ and $\bar{n}(1+\zeta)$
for any finite $\zeta$ is 1. One can thus take $\zeta\rightarrow0$ with speed slower
than $1/\D \rightarrow 0$ in order to have the bound vanish. This computation applies for
$x\in(0,1)$, which means that the peaked size always holds for early
time. This is physically reasonable as the operator has not yet been scrambled
extensively. However, since the size is small at such early times, in order for teleportation
to work we must choose $g\sim N$. 

For intermediate times, such that $\sinh^{2}Jt\sim N$ and $\D\ll N\sim 1/(1-x)$, we must take the $x\rightarrow 1$ limit first. Using the fact that
\be
\f{B_{x}(\f{\D x(1-\zeta)}{1-x},\D)}{B(\f{\D x(1-\zeta)}{1-x},\D)}=1-\f{(1-x)^{\D}x^{\f{\D x(1-\zeta)}{1-x}}\Gamma(\f{\D(1-x\zeta)}{1-x})}{\Gamma(\f{\D x(1-\zeta)}{1-x})\Gamma(1+\D)}F(1,\f{\D(1-x\zeta)}{1-x};\D+1;1-x)\label{eq:18}
\ee
where $F$ is Gauss hypergeometric function, in $x\rightarrow1$ limit the right portion of Eq.~(\ref{eq:18}) tends to
\begin{equation}
F(1,\f{\D(1-x\zeta)}{1-x};\D+1;1-x)\rightarrow{}_{1}F_{1}(1;\D+1;\D(1-\zeta))=\D^{1-\D}e^{\D(1-\zeta)}(1-\zeta)^{-\D}(\Gamma(\D)-\Gamma(\D,\D(1-\zeta)))
\end{equation}
where $\Gamma(x,a)$ is incomplete gamma function. Meanwhile, the left portion of the second
term of Eq.~(\ref{eq:18}) gives
\begin{equation}
\f{(1-x)^{\D}x^{\f{\D x(1-\zeta)}{1-x}}\Gamma(\f{\D(1-x\zeta)}{1-x})}{\Gamma(\f{\D x(1-\zeta)}{1-x})\Gamma(1+\D)}\rightarrow\f{\D^{\D}(1-\zeta)^{\D}e^{-\D(1-\zeta)}}{\Gamma(1+\D)}
\end{equation}
under $x\rightarrow1$. Combining the two, we have
\begin{equation}
\lim_{x\rightarrow1}\f{B_{x}(\f{\D x(1-\zeta)}{1-x},\D)}{B(\f{\D x(1-\zeta)}{1-x},\D)}=\f{\Gamma(\D,\D(1-\zeta))}{\Gamma(\D)}.
\end{equation}
It follows that the upper bound is
\begin{equation}
\mB=2\left(1-\f{\Gamma(\D,\D(1-\zeta))}{\Gamma(\D)}+\f{\Gamma(\D,\D(1+\zeta))}{\Gamma(\D)}\right)+\sin\pi\zeta
\end{equation}
This function has a unique minimum for $\zeta\in[0,1/2]$ and this minimum
decreases as $\D$ increases. Taking derivative with respect to $\zeta$, we get
\begin{align}
\partial_{\zeta}\mB&=\pi\cos\pi\zeta-\f{2\D^{\D}}{\Gamma(\D)}\left[(1+\zeta)^{\D-1}e^{-\D(1+\zeta)}+(1-\zeta)^{\D-1}e^{-\D(1-\zeta)}\right] \nonumber\\
&\rightarrow\pi\cos\pi\zeta-\sqrt{\f{2\D}{\pi}}\left[(1+\zeta)^{\D-1}e^{-\D\zeta}+(1-\zeta)^{\D-1}e^{\D\zeta}\right]
\end{align}
where in the second step we have taken large $\D$ limit. Solving $\partial_{\zeta} \mB =0$ in this limit, we find the minimum at
\begin{equation}
\zeta\approx\sqrt{\f 1{\D}\log\f{8\D}{\pi^{3}}}\rightarrow0
\end{equation}
which in turn gives the limit value of $\mB$ to be zero. This proves that at
infinite temperature, teleportation exactly matches the peaked-size prediction for both early and intermediate times. For late times $t\gg \f{1}{2J}\log N$ the size distribution above breaks down, as can be seen since $P(n)$ is dominated by some $n>N$, which is unphysical since $N$ is the total number of fermions. 

In contrast, we can also show that the above bound does \emph{not} apply at low temperatures for large-$q$ SYK, as expected from the main text.
At low temperature, the upper bound $\mB$ needs to be much smaller than the two-sided correlation function $G_\beta(Q_A)\sim (\beta J)^{-2\Delta}$ in order to guarantee peaked-size teleportation. The low temperature size distribution is essentially the same as at infinite temperature, requiring only the replacement~\cite{qi2019quantum}:
\begin{equation}
x\rightarrow\f{\sinh^{2}\pi t/\beta}{(\pi/\beta J)^{2}+\sinh^{2}\pi t/\beta}\in[0,1]
\end{equation}
and adding $e^{-\mu N \delta_\beta}$ to the distribution, which shifts the initial size by a constant amount $N\delta_\beta$ (accounting for the size of the thermal density matrix~\cite{qi2019quantum}). Following a similar computation to above, one can show that $\mB$ still asymptotes to zero, but now with a \emph{slower} speed than $G_\beta(Q_A)$. For example, in the early time and large $\Delta$ limits, $\mB\sim \exp(-\Delta C(x,\zeta))/\sqrt{\Delta}$ where $C(x,\zeta)$ is order 1, while $G_\beta (Q_A)\sim \exp(-2\Delta\log(\beta J))$ is exponentially smaller for large $\beta J$. Therefore, the upper bound $\mB$ fails to guarantee peaked-size teleportation. This is consistent with the fact that the correlation function $C_Q(t)$ in Eq. \eqref{ping correlator full} in low temperature is far from being a pure phase.

\section{The Hayden-Preskill recovery protocol}\label{app: YK}

In this appendix we review the HPR protocol following Refs.~\cite{yoshida2017efficient,yoshida2019disentangling} and derive its equivalence to the TW protocol in the case of infinite temperature teleportation of a single qubit (introduced in Section \ref{sec: YK}). 
This single-qubit variant of the HPR protocol was experimentally implemented in Ref.~\cite{landsman2019verified}, although an explicit derivation of its quantum circuit was not provided.

There are two variants of the HPR protocol: a probabilistic variant, which teleports successfully only with some finite probability, and a deterministic variant, which uses an analog of Grover's search algorithm and succeeds with unit probability, but involves a more complex decoding operation. 
Both protocols take the general form,
\begin{align}
\figbox{1.0}{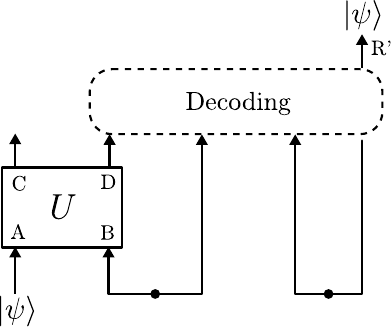} \centering \label{YK-decoding}
\end{align}
shown for teleportation of a quantum state $\ket{\psi}$ (the generalization to EPR teleportation is straightforward).
We now outline the interpretation of each aspect of the above protocol in the context of the Hayden-Preskill thought experiment.
For consistency with past literature, we have used \emph{different} subsystem labels than introduced in the main text---most notably, subsystem D now denotes the coupled qubits, and subsystem C denotes its complement.
Subsystem B represents an eternal black hole that is maximally entangled with its past Hawking radiation subsystem B', as represented by a dimension $d_B = d_B'$ EPR pair between the two subsystems.
Subsystem A contains the initial state $\ket{\psi}$ of an observer Alice's diary.
Upon falling into the black hole, the diary's information is scrambled by the unitary time-evolution $U$ acting on the left subsystem $l \equiv$ AB $=$ CD.
Far from destroying the information of Alice's diary, scrambling by $U$ in fact allows an outside observer Bob to decode the diary if he has access to \emph{any} few qubits of new Hawking radiation D, along with the past Hawking radiation B' and an ancillary EPR pair between A' and R', where $d_A' = d_A$.
This decoding relies on OTOCs between subsystem A and D being minimal, a general feature of thermalizing time-evolution after the scrambling time.
We describe each of the decoding protocols of Ref.~\cite{yoshida2017efficient} in detail below.

%

\subsubsection{Probabilistic decoding: intuition}

Although our main focus will be on the deterministic teleportation protocol, we review the probabilistic protocol here for completeness, and as a convenient platform to introduce the intuition connecting operator spreading to the success of teleportation.
The decoding operation of the probabilistic HPR protocol consists of projection onto EPR pairs on a subsystems D, D':
\begin{align}
\figbox{1.0}{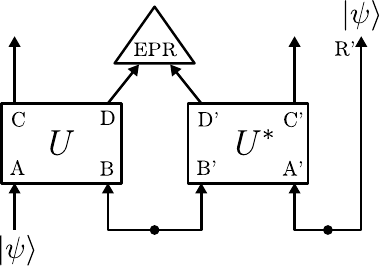} \centering \label{eq: YK probabilistic state}
\end{align}
%
Perfect teleportation requires $d_D \geq d_A$, and succeeds with probability $1/d_A^2$ when $U$ is maximal scrambling. The non-unity success probability signifies that the decoding protocol becomes exponentially more complex with the number of qubits to be teleported.

To provide intuition for the protocol's success, we analyze the action of EPR projection on the initial states $Q_{A,l}(t)\ket{\epr}$.
We restrict to infinite temperature, i.e. EPR pairs in place of the TFD state, in keeping with the original introduction of the HPR protocol in Ref.~\cite{yoshida2017efficient}.
We write $Q_{A}(t)$ as a sum of Pauli strings $S$ on the entire system:
\be
Q_A(t) = \sum_{\String} c_{\String}(t) S.
\ee
Denoting the EPR projector on subsystems D, D' as $P_{\epr,D}$ and writing each Pauli string as a tensor product $\String = \String_C \otimes \String_D$ of Paulis on subsystems D and C, we have
\be
P_{\epr,D} \String_{l} \ket{\epr}  = \delta_{\String_D,\mathbbm{1}} \String_l  \ket{\epr},
\ee
since $\bra{\epr_{D,D'}} S_{D,l} \ket{\epr_{D,D'}} = \tr_D (\String_D)/d_D = \delta_{\String_D,\mathbbm{1}}$. 
Perfect teleportation is achieved when all input Pauli operators on subsystem A have spread to subsystem D, such that every Pauli string $S$ composing $Q_A(t)$ has non-identity support on subsystem D, for all non-identity $Q_A$. 
%
%
In this situation, the EPR projector has eigenvalue 1 on the thermofield double state and eigenvalue 0 in \emph{all} perturbed states:
 \be \label{YK P action}
P_{\epr,D}  \ket{\epr}  =  \ket{\epr}  \,\,\,\, , \,\,\,\, P_{\epr,D} \, Q_{A,l}(t) \ket{\epr}  =  0.
\ee
However, this is no different than projecting onto EPR pairs between subsystems A and A' before time-evolution by $U_l U^*_r$! %
This projection would, of course, have an action 
 \be
P_{\epr}  \ket{\epr}  =  \ket{\epr} \,\,\,\, , \,\,\,\, P_{\epr} \, Q_{A,l} \ket{\epr}  = \tr( Q_A ) =  0.
\ee
Expressed diagrammatically, this equivalence is:
\begin{align}
\figbox{1.0}{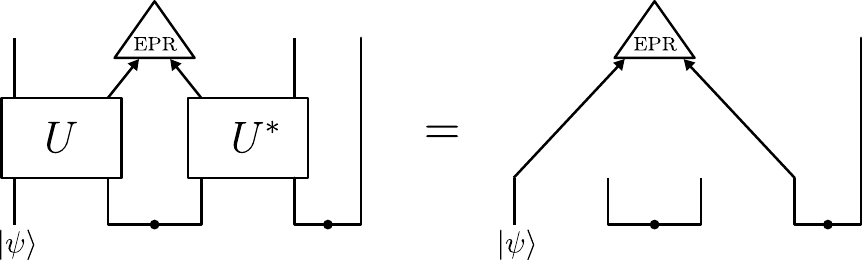} \centering \label{eq: YK probabilistic state equality}
\end{align}
for all initial states $\psi$.
However, performing EPR projection between subsystems A, A' before time-evolution is precisely the standard quantum teleportation protocol, applied to subsystems A, A', and R'.
The scrambling dynamics of $U$ allow one to perform this teleportation via coupling \emph{any} subsystem D of the system's qubits.

\subsubsection{Deterministic decoding}

After scrambling, the probability of successful EPR projection on subsystem D, $\mathcal{O}(1/d_A^2)$, is exponentially small in the size of subsystem A, the state to be teleported.
In contrast to standard teleportation, non-successful EPR projection (i.e. projection onto a different maximally entangled state, not $\ket{\epr_{D,D'}}$) \emph{cannot} be corrected via an additional decoding operation.
This exponential decrease in success probability is overcome in the deterministic HPR protocol, which uses an analog of Grover's search algorithm to search for an EPR pair between subsystems D, D'.
The protocol requires $\mathcal{O}(d_A)$ steps for completion, again exponential in the number of qubits to be teleported (albeit with half the exponent of the probabilistic decoding).

Grover's search algorithm involves two operations: the first applies a minus sign to the state one is searching for, and the second applies a minus sign to the system's initial state. We will search for an EPR pair on subsystem $D$, so for the first step we apply $W_D \equiv 1 - 2 P_{\epr,D} = e^{i \pi P_{\epr,D}}$:
\begin{align}
\figbox{1.0}{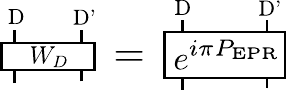} \centering \label{eq: WD}
\end{align}
In the second step, we flip the sign of the initial state (the time-evolved EPR pair between A' and the reference qubit R') by applying $\widetilde{W}_A \equiv U^* W_A U^T$:
\begin{align}
\figbox{1.0}{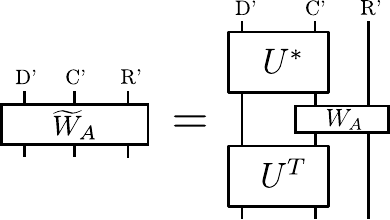} \centering
\end{align}
where $W_A = 1- 2P_{\epr,A}$ acts on A', R' to apply a minus sign if the two are in an EPR pair. 

The entire Grover protocol is identical to the probabilistic protocol, but with EPR measurement replaced by repeated applications of the two above steps until the EPR pair is found. Displaying, for instance, only the first two iterations:
\begin{align}
\figbox{1.0}{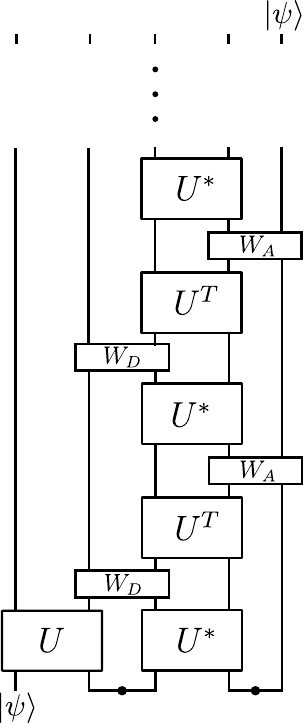} \centering
\end{align}
After $\mathcal{O}(d_A)$ iterations, the state $\ket{\psi}$ is found on subsystem R'.

\subsubsection{Single qubit deterministic decoding}

Two important simplifications occur to the deterministic HPR protocol in the case of single qubit teleportation, $d_A = 2$.
The first is that the Grover operator $W_A$ is equal to a SWAP operator composed with single-qubit $Y$ operations.
To see this, we expand $W_A$ in terms of Pauli operators:
\be
\bs
W_A  & = 1 - 2 P_{\epr,A} \\
& = 1 - \frac{2}{d_A^2} \sum_{P_A} P_{A,l} \, P^{*}_{A,r} \\
& = \frac{1}{2} - \frac{1}{2} X_l X_r + \frac{1}{2} Y_l Y_r - \frac{1}{2} Z_l Z_r \\
& = \frac{1}{2} Y_l \left[ 1 +  X_l X_r + Y_l Y_r + Z_l Z_r \right] Y_l \\
& = Y_l \, (\SWAP) \, Y_l, \\
& = Y_l Y_r \, (\SWAP) \\
\end{split}
\ee
where in the final equality we used $Y_r \SWAP = \SWAP Y_l$, and in the second equality we used the Pauli decomposition for the swap operator between two $d_A$-dimensional boson systems:
\be \label{SWAP decompose}
\SWAP = \frac{1}{d_A} \sum_{P_A} P_{A,l} P^{\dagger}_{A,r}.
\ee
Expressed graphically, we have
\begin{align}
\figbox{1.0}{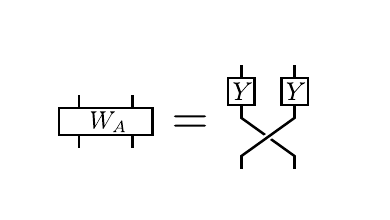} \centering
\end{align}

The second simplification is that Grover's search for an EPR pair D, D' succeeds after only one step; this is a general result for Grover's search in a $d_D^2=4$-dimensional database~\cite{nielsen2002quantum}. It implies that the Grover protocol can teleport one qubit through the circuit:
\begin{align}
\figbox{1.0}{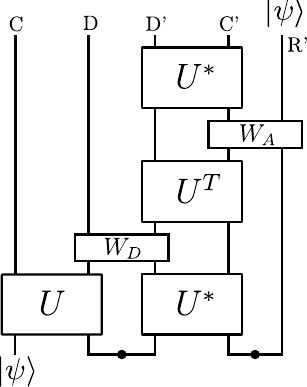} \centering \label{eq: Grover one step}
\end{align}
If we only care about the fidelity of the teleported state, we can neglect the final application of $U^*$. Performing the SWAP gate explicitly, and neglecting the action of the final $Y$ operator on R', we have:
\begin{align}
\figbox{1.0}{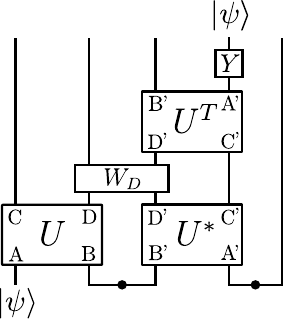} \centering \label{eq: Grover one step simplified WD}
\end{align}
This exact circuit has been performed in trapped ion experiment~\cite{landsman2019verified}. 
We now make a small cosmetic adjustment, and move the reference qubit R' from the far right to the far left, 
\begin{align}
\figbox{1.0}{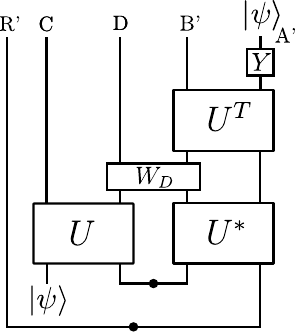} \centering \label{eq: Grover one step simplified Rtoleft}
\end{align}
Sliding $U^*$ to the left side using Eq.~(\ref{eq: O slide}), we have:
\begin{align}
\figbox{1.0}{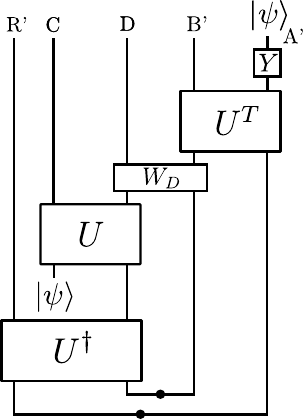} \centering \label{eq: Grover one step slide}
\end{align}
%
This is the same circuit appearing the teleportation protocol of Ref.~\cite{brown2019quantum,brown2020quantum}, modulo the precise form of the coupling.
In the case of EPR teleportation, we would instead have
\begin{align}
\figbox{1.0}{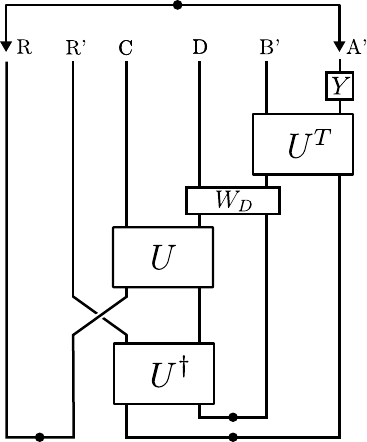} \centering \label{eq: Grover one step slide EPR}
\end{align}
where subsystems R' and A' are in an EPR pair when teleportation is successful. This is the circuit appearing in Ref.~\cite{gao2019traversable}, modulo the form of the coupling as well as the $Y$ decoding operation. The lack of a $Y$ decoding operation for fermionic teleportation is discussed in Appendix~\ref{app: fermions}.

\section{State teleportation fidelity} \label{app: state fidelity}

In Section \ref{sec: rigorous}, we provided a detailed derivation of the teleportation fidelity's relation to the teleportation correlators in the case where one teleports one half of an EPR pair. This allowed us to lower bound the fidelity in Eq.~(\ref{F bound}) and calculate the peaked-size fidelity in Eq.~(\ref{F phases}).
In this appendix we do the same for teleportation of a quantum state, as shown in Fig.~\ref{fig: 1}(a) and outlined in Section \ref{intro diagrams}.
Our results provide a rigorous foundation for the arguments of Section \ref{sec: requirements}, in particular the insertion of the state $\bra{\phi}$ and the subsequent replacement of $\dyad{\psi}{\phi}$ with a Pauli operator $Q_A$.

We begin with the insertion of $\bra{\phi}$ into the protocol Eq.~(\ref{TWH protocol state swap}).
We do so by inserting the resolution of the identity $\frac{1}{d_A} \sum_{\ket{\phi}} \dyad{\phi} = \mathbbm{1}$ into the  ancillary qubit leg of the diagram for the state teleportation fidelity. We find:
\begin{align}
\figbox{.7}{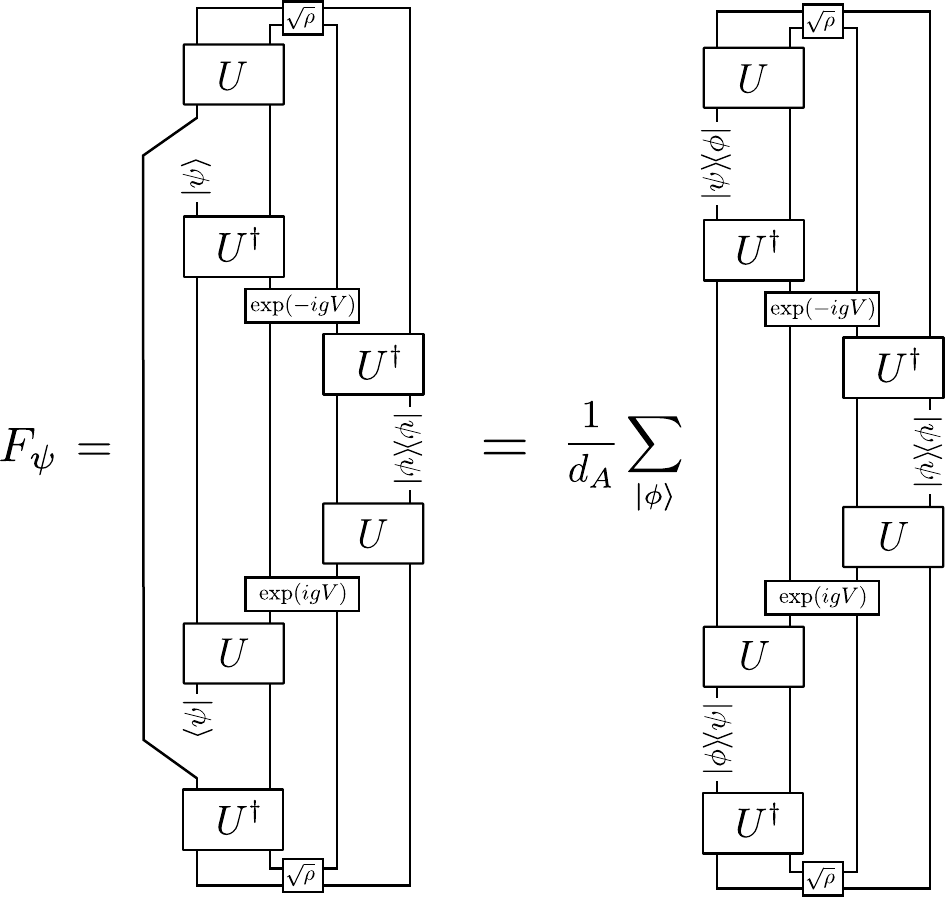} \centering \label{eq: state-fidelity}
\end{align}
%
Plugging Eq.~(\ref{eq-correlator-inner-product}) into this diagram provides unit teleportation fidelity, as described in the main text.
When teleportation is successful each of the $d_A$ terms of the sum must succeed individually, so the right input state $\ket{\phi}$ will not affect the success of the teleportation.

As with EPR distillation [Eq.~(\ref{eq: EPR-fidelity})], we can relate the state teleportation fidelity to correlators of Pauli operators by decomposing the SWAP operator. Diagramatically, 
\begin{align}
\figbox{.7}{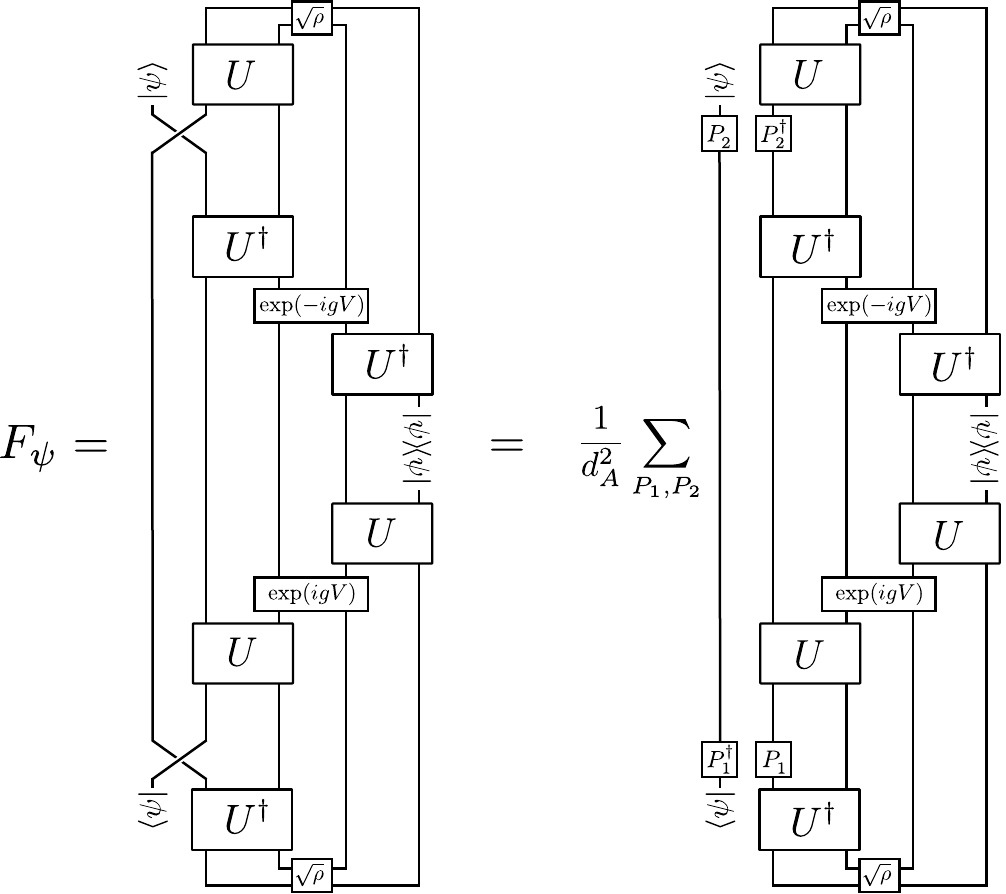} \centering \label{eq: state-fidelity},
\end{align}
and in equation form,
\be
F_\psi  = \frac{1}{d_A^2} \sum_{P_1,P_2} \bra{\psi} P _2 P_1^{\dagger} \ket{\psi} \cdot \bra{\tfd} P_{2,l}^{\dagger}(t)e^{-igV} \dyad{\psi}_r\!(-t) \,\, e^{igV}P_{1,l}(t) \ket{\tfd}.
\ee
When the correlators are maximal with phases $e^{i\theta_{P}}$, i.e. when $e^{igV} P_{1,l}(t) \ket{\tfd} = e^{i\theta_{P}} P_{1,r}(-t) \ket{\tfd}$, we can simplify this expression as
\be
\begin{split}
F_\psi  & \approx \frac{1}{d_A^2} \sum_{P_1,P_2} \bra{\psi} P _2 P_1^{\dagger} \ket{\psi} \cdot \bra{\tfd} P_{2,r}^\dagger(-t) \dyad{\psi}_r\!(-t) \,\, P_{1,r}(-t) \ket{\tfd} \\
& = \frac{1}{d_A^2} \sum_{P_1,P_2} e^{i(\theta_{P_1}-\theta_{P_2})} \cdot \bra{\psi} P_2 P_1^{\dagger} \ket{\psi} \cdot \tr \bigg( \rho  \, P_{2}^{\dagger} \dyad{\psi} P_{1} \bigg) \\
& = \frac{1}{d_A^2} \sum_{P_1,P_2} e^{i(\theta_{P_1}-\theta_{P_2})} \cdot \tr \bigg( P_1^{\dagger} \dyad{\psi} P _2 \bigg) \cdot \tr \bigg( \rho  \, P_{2}^{\dagger} \dyad{\psi} P_{1}^{} \bigg). \\
\end{split}
\ee
As expected, when the phases $e^{i\theta_{P}}$ are the same for all operators, this gives unit fidelity:
\be
\begin{split}
F_\psi  & = \frac{1}{d_A^2} \sum_{P_1,P_2}  \tr( P_1^{\dagger} \dyad{\psi} P_2 ) \cdot \tr( \rho  \, P_{2}^{\dagger} \dyad{\psi} P_{1}^{} ) \\
& = \frac{1}{d_A} \sum_{P_1}  \tr( P_1^{\dagger} \dyad{\psi} \dyad{\psi} P_{1}^{} \rho ) \\
& =  \tr( \dyad{\psi} ) \tr(\rho ) \\
& = 1, \\
\end{split} 
\ee
using properties of Pauli operators as a 1-design~\cite{zhuang2019scrambling}. Differing phases $e^{i\theta_{P}}$ cause the terms in the sum to interfere with each other, giving lower fidelity.
At finite temperature, the fidelity of peaked-size teleportation is again limited. For instance, if $\ket{\psi}$ is a single-qubit eigenstate of the Pauli $Z$ operator, we have:
\be
\begin{split}
F_{\epr}  & = \frac{1}{2^2} \sum_{P_1,P_2} \bra{\psi} P_2 P_1^{\dagger} \ket{\psi} \cdot \bra{\tfd} P_{2,l}^{\dagger}(t)e^{-igV} [Y\dyad{\psi}\!Y]_r(-t) \,\, e^{igV}P_{1,l}(t) \ket{\tfd} \\
& = \frac{1}{2^2} \sum_{P_1,P_2} \bra{\psi} P _2 P_1^{\dagger} \ket{\psi} \cdot \tr( \dyad{\psi} \, \rho^{1/2}P_{2}^{\dagger} P_{1} \rho^{1/2} ) \\
& = \sum_{P} \bra{\psi} P^{\dagger} \ket{\psi} \cdot \tr( \dyad{\psi} \, \rho^{1/2}P \rho^{1/2} ) \\
& = 2 \tr( \dyad{\psi} \, \rho^{1/2} \dyad{\psi} \rho^{1/2} ) \\
& \approx \frac{1}{2} \tr( (\mathbbm{1} + Z) \, \rho^{1/2} (\mathbbm{1} + Z) \rho^{1/2} ) \\
& \approx \frac{1}{2} + \frac{1}{2} G(t' - t + i \beta/2) + \expval{Z}_\beta, \\
\end{split}
\ee
where $\expval{Z}_\beta = \tr ( Z \rho)$, which averages to zero for different initial states $\ket{\psi}$.

\section{Rydberg numerical simulations}
For the numerical results shown in Fig.~\ref{fig:rydberg-implementation} and \ref{fig:rydberg_scaling}, we simulate the full TW protocol with time evolution generated by the analog Rydberg Hamiltonian [Eq.~\eqref{eq:ryd-h}].
In particular, we implement the \emph{one-sided} version of state teleportation, which is obtained by replacing the EPR measurement in Fig.~\ref{fig:one-sided}(b) with a measurement of a two-qubit state $\ket{\psi} \otimes \ket{\psi^*}$.
The many-body unitary corresponds to $U=e^{-iHt}$, where $H$ is given in Eq.~\eqref{eq:ryd-h} with $\Omega_i = .9$, $\Delta_i = -1.5$, $J_0 = 1$ and open boundary conditions.
The teleported state $\ket \psi$ is inserted in the middle qubit, and the remaining $K=N-1$ qubits serve as `measured' qubits, with $\hat O_i = \hat Z_i$ (see Section \ref{sec:one-sided}).

%

More explicitly, the numerical procedure is given as follows: 
($i$) begin in a random initial state, $\ket {o_1 \cdots o_N}$;
($ii$) evolve forward for time $t$ under the Rydberg Hamiltonian;
($iii$) apply the operator $\dyad{\phi}{\psi}$ onto the middle qubit;
($iv$) evolve backward in time, apply $\hat{V}_i = e^{i g o_i \hat{Z}_i / K}$ to each of the $K = N-1$ `measured' qubits (where $o_i$ is determined by the initial state), and evolve forward again; ($v$) measure the projector $\dyad{\psi}$ on the middle qubit.
We repeat this procedure for $\ket{\phi} \in \{\ket{0},\ket{1}\}$ [see Eq.~\eqref{eq: state-fidelity}] and starting from $\sim 100$ random initial states. 
Moreover, to compute the \emph{average} state fidelity, we average $\ket{\psi}$ over all single-qubit states in a 2-design~\cite{landsman2019verified}, i.e.~$\ket{\psi} \in \{\ket{0},\ket{1},\frac {1}{\sqrt{2}}(\ket{0}\pm \ket{1}),\frac {1}{\sqrt{2}}(\ket{0}\pm i\ket{1}\})$.
Lastly, we note that the time evolution is implemented with Krylov subspace methods, an iterative technique that is amenable to parallelization and is more computationally efficient than exact diagonalization~\cite{dynamite,kobrin2021many}.
%



\section{Random unitary circuit numerics} \label{app: RUC-numerics}
In this section, we provide additional details and numerical data from our random unitary circuit simulations (Section \ref{sec: intermediate time} B, C).

\subsection{Algorithm}
Our goal for the RUC simulations is to compute the Haar-averaged EPR fidelity and operator size distribution for the circuit layouts shown in Fig.~\ref{fig: RUC}.
%
%
Crucially, the relevant diagrams for computing these quantities---Eq.~\eqref{eq: EPR-fidelity} for the EPR fidelity, and Eq.~\eqref{V OTOC} for the operator size distribution---contain at most three copies of $U$ and $U^\dagger$.
Together with the fact that Clifford unitaries form a 3-designs for qubits, this implies that  can compute the averaged quantities by replacing each Haar-random gate with a random \emph{Clifford} gate \cite{webb2015clifford,kueng2015qubit,zhu2015multiqubit}. 
This dramatic simplification has been exploited in prior studies of operator growth in random unitary circuits \cite{nahum2018operator,von2018operator}; here, we adapt these same techniques for computing the full size distribution and the teleportation fidelity. 

In more detail, our algorithm consists of the following three ingredients.
First, following a standard approach \cite{aaronson2004improved,nahum2018operator}, we represent an initial $n$-qubit Pauli operator, $Q$, as a binary string $v=x_1 x_2 \cdots x_n z_1 z_2 \cdots z_n$ of length $2n$:
\begin{equation}
Q = \prod_{i=1}^{n} Q_i(x_{i},z_{i})
\end{equation}
where $Q_i(0,0) = I_i$, $Q_i(1,0) = X_i$, $Q_i(0,1) = Z_i$, and $Q_i(1,1) = Y_i$ denote individual Pauli operators within the Pauli string.
For example, the operator $\mathbbm{1} \otimes \mathbbm{1} \otimes \mathbbm{Z} \otimes \mathbbm{1} \otimes \mathbbm{1}$ is represented as $x = 00000$ and $z = 00100$.
Normally, one would also keep track of the overall phase of $Q$, but for our purposes the phase will be irrelevant and is dropped in the above notation.

Second, we evolve $Q$ under a random Clifford unitary $U$ to obtain $Q(t) = U Q U^\dagger$.
We consider the circuits shown in Fig.~\ref{fig: RUC}, which are composed of random 2-qubit Clifford unitaries laid out in a ``brick-layer'' fashion. 
Each of the 2-qubit unitaries is sampled uniformly from the set of 2-qubit Clifford unitaries.
While an algorithm exists to perform this sampling directly \cite{li2019measurement}, in practice we find it more convenient to pre-compute and enumerate the entire 2-qubit Clifford set (which consists of 11520 distinct unitaries)\footnote{We are grateful to Maxwell Block for sharing code to generate the full set of 2-qubit Clifford unitaries.}.
In the binary notation, each 2-qubit Clifford unitary corresponds to a map which acts on the relevant components $v$, i.e.~a unitary with support on the $j$th and $k$th qubits updates the values of $(x_j,z_j,x_k,z_k)$.
The time complexity of applying the full circuit thus scales linearly with the number of 2-qubit gates and does not otherwise depend on the number of qubits $n$.
As a reference point, simulating a 0D circuit until the scrambling time with $10^8$ qubits for a single realization takes approximately one day on a standard single-core processor.

Third, we compute the average operator size distribution and EPR fidelity of the time-evolved operators.
For the former, we simply count the size, i.e.~number of non-identity terms, of a time-evolved operator $Q(t)$ for a single circuit realization and determine the distribution of sizes with respect to $\sim 10^3$ circuit realizations. 
Depending on the simulation, we either initialize $Q$ with support on a single site (i.e.~$p$=1) or as a $p$-body operator.
In either case, the specific terms in $Q$ (e.g.~whether each site is initialized as $X$, $Y$, or $Z$) is arbitrary since the averaged quantities are basis independent.

Computing the averaged EPR fidelity requires an additional average over the initial operators.
In particular, for a single circuit realization $U$, we compute the EPR fidelity using [Eq.~\eqref{F phases}]:
\begin{equation}\label{F_EPR}
F_{\epr}  
 = \bigg| \frac{1}{d_A^2} \sum_{Q_A} e^{i\theta_{Q_A}} \bigg|^2 \\
\end{equation}
where
\begin{equation}
\theta_{Q_A} = \constd g \Size_K[U Q_A U^\dagger] / K + \pi \Size[Q_A].
\end{equation}
and $\constd \equiv 1/(1-1/d^2)$, as defined in Section \ref{peaked sizes}.
Note that the first term in $\theta_{Q_A}$ corresponds to the phase applied by the coupling, while the second term accounts for minus signs associated with transposition and decoding (see Section \ref{decomposition}).
The sum in Eq.~\eqref{F_EPR} is over the complete basis of Pauli operators in subsystem A. 
For single-qubit teleportation, this consists of three non-trivial Pauli operators and the identity (for which $\theta = 0$), and the sum can be performed explicitly. 
However, for teleporting many qubits, the number of terms quickly becomes intractable, and we instead approximate the sum by sampling $Q_A$ (e.g.~$\sim 100$ randomly selected operators).
To compute the average EPR fidelity, we repeat this procedure for $\sim 100$ realizations of $U$.
Finally, we note that the coupling strength $g$ enters the fidelity calculation in a computationally efficient manner; in particular, upon determining the distribution of sizes for a particular circuit realization, we can compute the teleportation fidelity for arbitrary values of $g$ ``offline'' with no additional simulation cost.

\begin{figure*}[h]
\centering
\includegraphics[width=.8\textwidth]{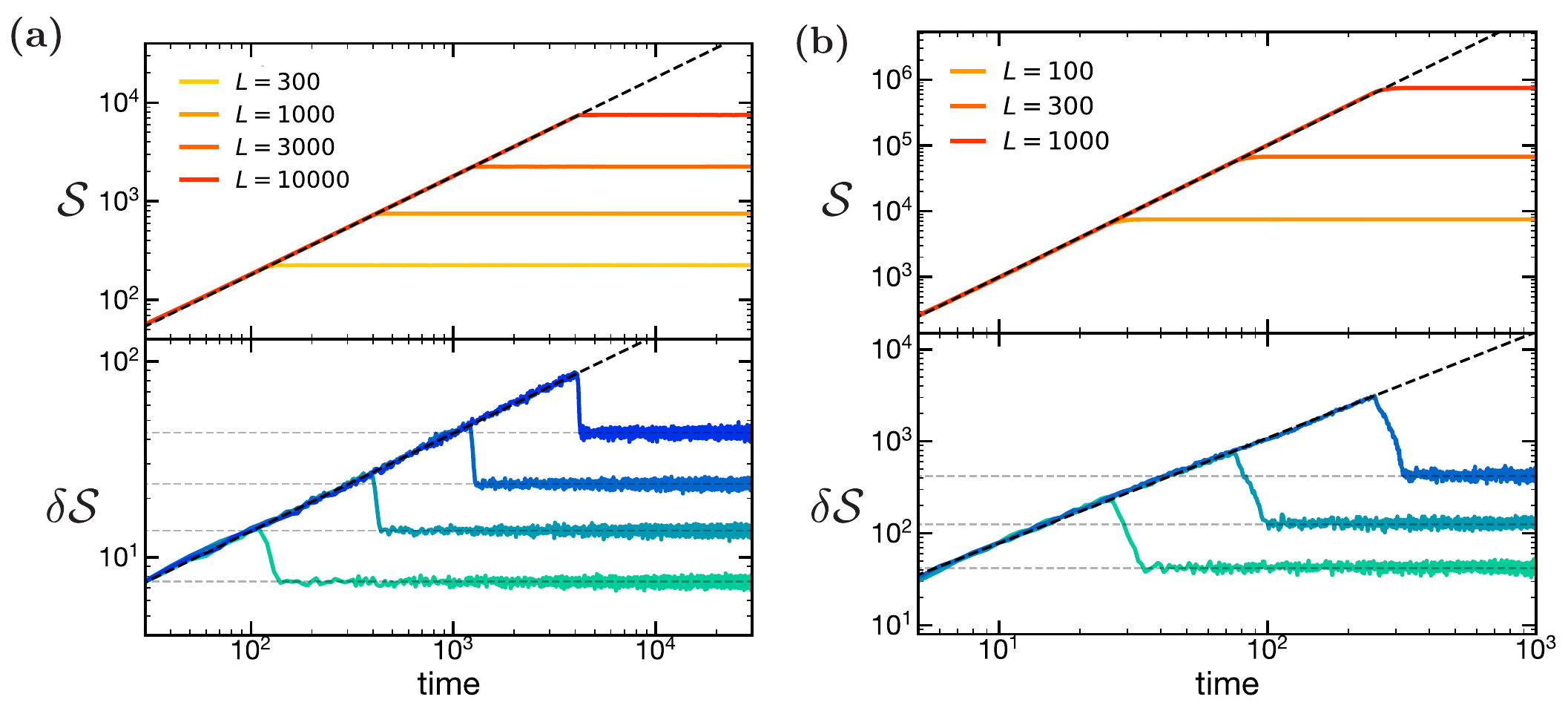}
\caption{
Extended data for average operator size and size width in 1D \textbf{(a)} and 2D \textbf{(b)} RUCs.
The average size grows ballistically $\sim t^{d}$ (dashed line) and saturates at $t_\textrm{scr} \sim L \sim N^{1/d}$. The size width matches the predictions from the KPZ universality class (dashed lines) and allows us to extract the prefactors in Eq.~\eqref{eq:kpz_1d} and \eqref{eq:kpz_2d}.
In particular, we determine $\alpha_\textrm{bulk}$ and $\beta_\textrm{bulk}$ from the saturation values (light gray), and  $\alpha_\textrm{boundary}$ and $\beta_\textrm{boundary}$ from the initial growth rate (dark gray). 
} 
\label{fig:RUC_supp}
\end{figure*}

\subsection{Extended data for 1D and 2D RUCs}

\emph{Size distribution---} The average size and size width for time-evolved operators in 1D and 2D for various system sizes are shown in Fig.~\ref{fig:RUC_supp}.
In each case, we apply periodic boundary conditions and begin with a single-qubit operator. 
These results match the functional forms predicted by the KPZ universality class [Eq.~\eqref{eq:kpz_1d} and \eqref{eq:kpz_2d}] and allow us to extract the growth parameters $\{\alpha_\textrm{bulk},\alpha_\textrm{boundary},\beta_\textrm{bulk},\beta_\textrm{boundary}\} = \{0.66,0.70, 1.2,4.5\}$.

\vspace{4mm}

\begin{figure*}
\centering
\includegraphics[width=.9\textwidth]{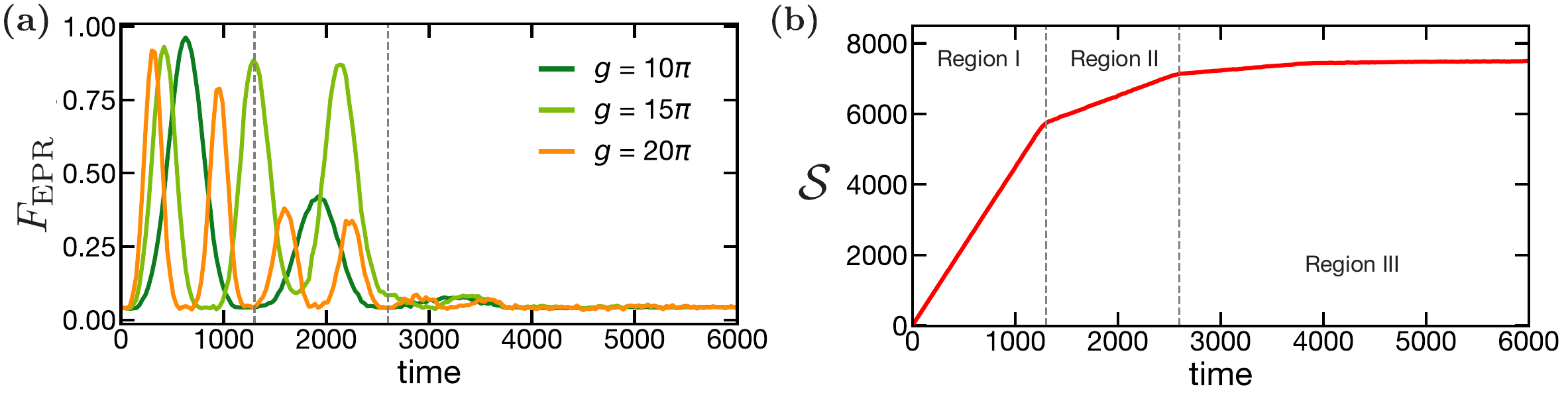}
\caption{
Teleporting multiple qubits ($n=5$) in 1D, where the input qubits are evenly spaced in the system ($N = 10000$).  \textbf{(a)} Teleportation is achieved with high fidelity for $t \le 1300$ (Region I). This corresponds to the regime in which the light cones of the operators are non-overlapping. Interestingly, order-one fidelity can also occur for $1300 < t < 2600$ (Region II), when adjacent light cones have overlapped, but only for certain values of $g$. No multi qubit teleportation is possible for $t \ge 2600$ (Region III), as expected from the lack of size addition. \textbf{(b)} The three Regions can be detected by changes in the slope of the operator size as a function of time. In particular, the growth rate decreases when nearest neighbor light cones, then next nearest neighbor light cones, etc.~begin to overlap. 
} 
\label{fig:1d_multiple_qubits}
\end{figure*}

\emph{Multiple qubits---}In Fig.~\ref{fig:1d_multiple_qubits}, we present numerical evidence to support our claim that  multiple qubits can be teleported in $\ge 1$D short-range models if their operator light cones are non-overlapping (Section \ref{geq1D}). 
In particular, we simulate the teleportation of $n=5$ qubits that are initially evenly spaced in a 1D RUC with periodic boundary conditions.
At early times ($t < 1300$, Region I), we confirm that high-fidelity teleportation is possible for a wide range of coupling strengths, and by measuring the average operator size we infer that during this time the operator light cones have not overlapped. 
In contrast, after the light cones have overlapped, we generally observe a large suppression in the teleportation fidelity. 

Interestingly, there is one noticeable exception to this reasoning: When only adjacent light cones have overlapped (i.e.~$1300 < t < 2600$, Region II), high-fidelity teleportation can still occur for specific values of $g$.
This situation corresponds to when the multi-qubit size is a multiple of $2\pi K / \constd g$ off from the size addition value, e.g. $\Size[Q_1(t) Q_2(t)] = \Size[Q_1(t)] + \Size[Q_1(t)] - 2\pi m (K/\constd g)$, where $m$ is an integer value.
Therefore, strictly speaking, it is possible to satisfy the conditions for many-body teleportation \emph{without} size addition; nevertheless, it is a non-generic effect that requires finely tuned values of $g$ and evenly spaced input qubits.

\subsection{Channel capacity for 0D RUCs}

\begin{figure*}
\centering
\includegraphics[width=1\textwidth]{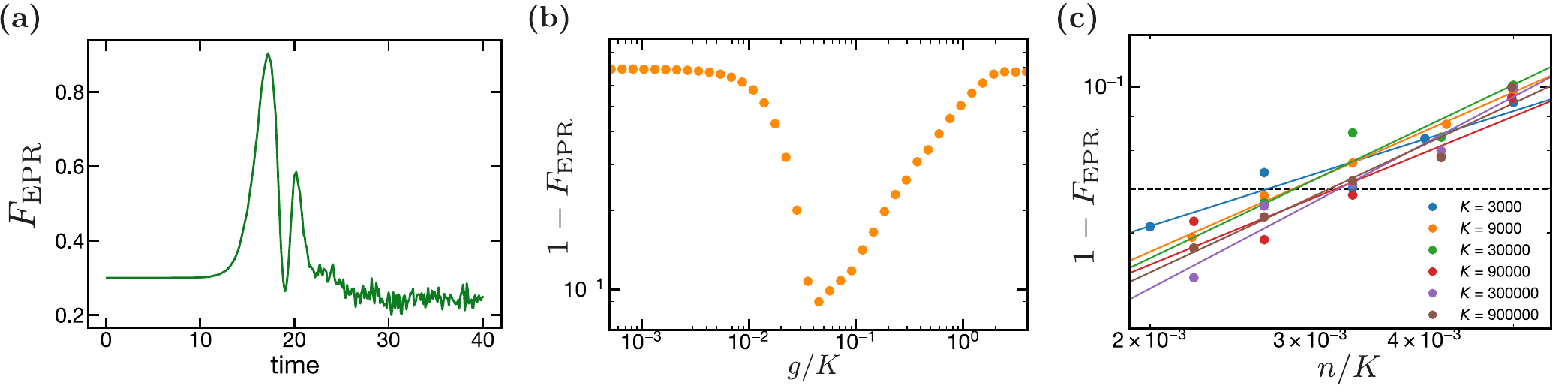}
\caption{
Procedure for determining the channel capacity in 0D RUCs. \textbf{(a-b)} For fixed $n$ and $K$, we compute the per qubit fidelity while sweeping both the evolution time and coupling strength $g$. (a) The fidelity as a function of evolution time with coupling strength fixed is optimized at the first local maximum, which corresponds to $\constd g \mathcal{S}/N = \pi$. (b) After optimizing the evolution time, the fidelity as a function of the coupling strength $g$ is maximal when $g$ (and correspondingly the average operator size $\mathcal{S}$) is tuned to balance errors due to size addition and the finite number of couplings (see Section \ref{0D RUCs} for details). The data shown correspond to $n = 38$ and $K = 9000$. \textbf{(c)} The channel capacity is defined as the maximum number of qubits that can be teleported while maintaining the fidelity per qubit above a fixed threshold, i.e.~$1-F^{(1)}_\textrm{EPR} \le 0.07$ (dashed line). To determine this number, we fit the optimal fidelity as a function of $n$ (for each $K$) with a linear fit in log space and compute the intercept of the fit with the threshold  fidelity. The fits approximately collapse with respect to $n/K$, indicating that the channel capacity is linear in $K$. } 
\label{fig:0d_channel_procedure}
\end{figure*}

An important result of our numerical simulations is substantiating the claim that 0D RUCs exhibit a channel capacity that scales linearly with the number of coupled qubits $K$.
To this end, we first recall that our working definition for the channel capacity is based on setting a threshold for the \emph{per qubit fidelity}.
The most direct way to compute this fidelity would be to take the $n$-th root of the many-body EPR fidelity; in practice, however, this approach is numerically unstable for large $n$.
Thus, we instead consider a modified protocol for estimating the per qubit fidelity where one attempts to send $n$ qubits but only measures the fidelity of one of the $n$ qubits.  
At infinite temperature and generalizing from one to $m$ qubits, this fidelity is given by:
\begin{align}
F^{(m)}_{\textrm{EPR}} &= \frac{1}{d_A^4} \sum_{Q_1,Q_2} \bra{\tfd} Q_{2,l}^\dagger(t)\,e^{-igV} \,  \tilde{Q}_{2,r}^{m}(-t)
\tilde{Q}_{1,r}^{m\dagger}(-t) \, e^{igV}\,Q_{1,l}(t) \ket{\tfd} \, \cdot \tr( Q_1^{u\dagger} Q_2^u ) \nonumber \\
&= 
\frac{1}{d_m^4 d_u^2} \sum_{Q_1, Q_2} e^{i(\theta_{Q_1}-\theta_{Q_2})} \delta_{Q_1^{u},Q_2^{u}}\\ \nonumber
\end{align}
where $Q = Q^m \otimes Q^u$ and $d_A = d_m d_u$, such that $Q^{m}$ acts on the measured qubit(s), and $Q^{u}$ acts on the unmeasured qubits.
This can be derived diagrammatically via
\begin{align}
\centering
\figbox{.51}{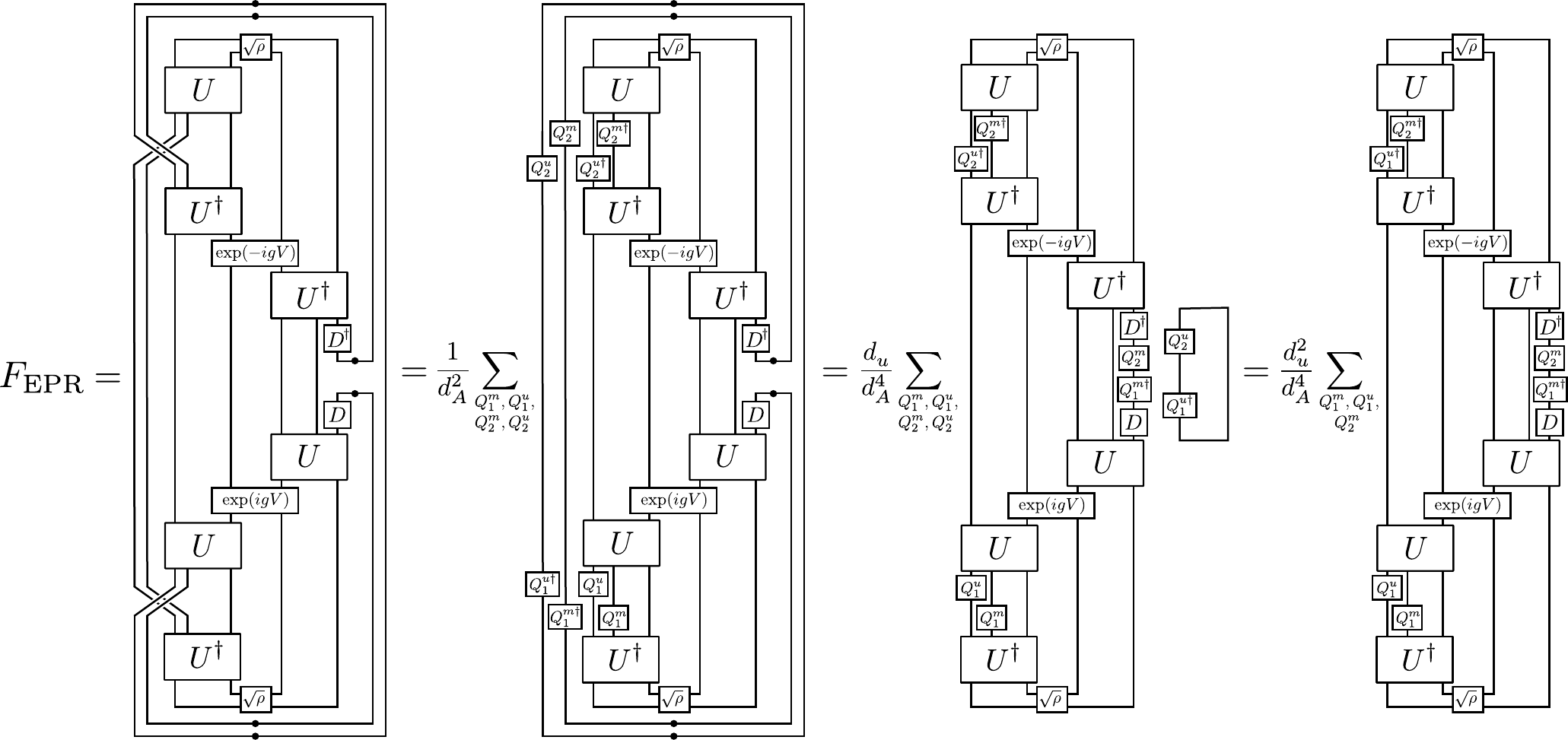} \centering \label{eq: state-fidelity-swap-EPR-fermions}
\end{align}
Hence, computing the per qubit fidelity, $F^{(1)}_\textrm{EPR}$, is nearly identical to computing the full many-body fidelity, except we sample only over pairs of Pauli operators $(Q_1, Q_2)$ which are identical on every qubit except for one.

We next discuss how to determine the channel capacity from the teleportation fidelity. 
Specifically, we compute the maximum number of qubits $n_\textrm{max}$ that can be teleported above a certain teleportation fidelity, where we fix the number of coupled qubits $K$ and optimize over the evolution time $t$ and the coupling strength $g$.
We consider each of these parameters in turn.
First, when sweeping the evolution time and holding all other parameters fixed, the maximum fidelity occurs during the first peak in the time profile; this corresponds to a size $\constd g \mathcal{S} = \pi/N$.
After optimizing the evolution time (but holding $n$ and $K$ fixed), the fidelity is non-monotonic with respect to $g$, owing to the competition among errors due the size addition and finite $K$. 
Finally, after optimizing evolution time and $g$, we determine the maximum number of qubits that can be teleported while maintaining a per qubit fidelity above a fixed threshold value, i.e.~$1-F^1_\textrm{EPR} \geq 0.07$.
Our results from this procedure are shown in Fig.~\ref{fig:0d_channel_procedure} and demonstrate that the channel capacity follows a linear trend in $K$ across two orders of magnitude, in agreement with our analytical predictions.

\section{Random circuit calculations} \label{app: RUC}

Here we provide more detailed technical calculations of the size overlap and $K$-size distribution of random Pauli operators of a fixed size.
The former is relevant to 0D RUCs (Section \ref{0D RUCs}), as the vanishingly small overlap of random Pauli strings with size much less than the system size underlies the circuit's ability to teleport multiple qubits at intermediate times.
The latter is applicable to all systems when the $K$ coupled qubits are chosen randomly, and quantifies the width introduced to the $K$-size by this random sampling (Section \ref{peaked sizes}).
In the appropriate limits, these calculations reproduce the intuitive binomial scalings we argued for in the main text.

\subsection{Distribution of the overlap of two random Pauli strings}

We are interested in the probability distribution of the size of the overlap, $p$ (not to be confused with the large-$p$ encoding, which we do not reference in this Appendix) of two randomly samples Pauli strings of fixed size $R_1, R_2$, in a system of $N$ qubits.
We expect this to quantify errors to the size addition formula, Eq.~(\ref{eq: size decompose}) in Section \ref{sec: intermediate time}, for 0D RUCs with large-$p$ encoding (Section \ref{0D RUCs}), where the assumption of random Pauli strings of a fixed size is appropriate.
Our precise derivation is necessarily quite technical, however our final result matches that obtained by intuitive arguments in Section \ref{sec: intermediate time} [see beneath Eq.~(\ref{size addition approx})].

This probability distribution is computed exactly as a product of various factorials:
\begin{equation}
P[p] = \frac{ C^N_p C^{N-p}_{R_1-p} C^{N-R_1}_{R_2-p} }{ C^N_{R_1} C^N_{R_2} } = \frac{1}{p!} \frac{R_1!}{(R_1-p)!} \frac{R_2!}{(R_2-p)!} \frac{(N-R_1)!(N-R_2)!}{N! (N-R_1-R_2+p)!}
\end{equation}
The numerator computes the number of distinct configurations with Pauli strings of size $R_1$, $R_2$ and overlap $p$, while the denominator computes the number of distinct Pauli strings of size $R_1$, $R_2$ regardless of the overlap.
We are interested in the case where all variables are extensive (scale with $N$), but $N \gg R_1,R_2 \gg p$.
We will proceed by applying Stirling's approximation to each term above, which holds as long as all quantities are large compared to 1.
For instance, for dummy variables $n, k$, we have:
\begin{equation}
 \frac{n!}{(n-k)!} \approx \sqrt{\frac{n}{n-k}} \frac{n^n}{(n-k)^{n-k}} e^{-k} = n^k \left(1 - \frac{k}{n} \right)^{-n+k-1/2} e^{-k} 
\end{equation}
 or, taking the logarithm,
\begin{equation}
\log  \frac{n!}{(n-k)!} \approx k \log(n) -  \left(n-k+\frac{1}{2}\right)\log(1 - \frac{k}{n}) - k.
\end{equation}
We will apply this to a few pairs of factorials in our original expression for $P[p]$.
For convenience, we only keep track of the $p$-dependence of the probability, and neglect overall constants which serve to normalize the distribution.
Anticipating that the average $p$ will be $R_1 R_2 / N$, we expand $p = R_1 R_2 / N + \delta$ and work to second order in $\delta$.
At the end we will show that this is justified.
We have:
\begin{equation}
\begin{aligned}
\log  \frac{R_1!}{(R_1-p)!} & \approx p \log(R_1) -  \left(R_1 - \frac{R_1 R_2}{N} + \frac{1}{2} \right) \log(1 - \frac{R_2}{N} - \frac{\delta}{R_1}) - \frac{R_1 R_2}{N} - \delta \\
\end{aligned}
\end{equation}
Expanding the logarithm using
\begin{equation}
\log( 1- y - x) \approx \log(1-y) - \frac{x}{1-y} - \frac{1}{2} \frac{x^2}{(1-y)^2} + \mathcal{O}(x^3) \\
\end{equation}
we have
\begin{equation}
\begin{aligned}
\log  \frac{R_1!}{(R_1-p)!} & \approx p \log(R_1) -  \left(R_1 - \frac{R_1 R_2}{N} - \delta + \frac{1}{2} \right) \left[ \log(1 - \frac{R_2}{N}) - \frac{\delta/R_1}{1 - R_2/N} - \frac{(\delta/R_1)^2}{(1 - R_2/N)^2} \right]  - \delta + \ldots \\
& \approx p \log(R_1) + \delta \log(1 - \frac{R_2}{N}) - \frac{1}{2} \delta^2 \left[ \frac{1}{R_1} \frac{1}{1 - R_2/N}  \right] + \mathcal{O}(\delta/R) + \mathcal{O}(\delta^3/R^2) + \ldots.
\end{aligned}
\end{equation}
This gives
\begin{equation}
\begin{aligned}
\log  \frac{R_1!}{(R_1-p)!} \frac{R_2!}{(R_2-p)!} & \approx p \log(R_1 R_2) + \delta  \log\left( (1 - \frac{R_2}{N})(1-\frac{R_1}{N}) \right)  - \frac{1}{2} \delta^2 \left[ \frac{1}{R_1} \frac{1}{1 - R_2/N}+ \frac{1}{R_2} \frac{1}{1 - R_1/N}  \right] \\
& \,\,\,\,\,\,\, + \mathcal{O}(\delta/R) + \mathcal{O}(\delta^3/R^2) + \ldots.
\end{aligned}
\end{equation}
The last piece is
\begin{equation}
\begin{aligned}
\log  \frac{N!}{(N - R_1 - R_2 + p)!} & \approx - p \log(N) -  \left(N - R_1 - R_2 + \frac{R_1 R_2}{N} + \delta + \frac{1}{2} \right) \log(1 - \frac{R_1}{N} - \frac{R_2}{N} + \frac{R_1 R_2}{N^2} + \frac{\delta}{N}) + \delta + \ldots \\
& \approx - p \log(N) -  \left(N - R_1 - R_2 + \frac{R_1 R_2}{N} + \delta + \frac{1}{2} \right) \times \\
& \,\,\,\,\,\, \left[ \log\left( (1 - \frac{R_1}{N})(1 - \frac{R_2}{N}) \right)  + \frac{\delta/N}{(1 - \frac{R_1}{N})(1 - \frac{R_2}{N})} - \frac{\delta^2/N^2}{(1 - \frac{R_1}{N})^2(1 - \frac{R_2}{N})^2} \right]+ \delta + \ldots \\
& \approx - p \log(N) - \delta \left[ \log\left( (1 - \frac{R_1}{N})(1 - \frac{R_2}{N}) \right)  \right]  - \frac{1}{2} \delta^2 \left[ \frac{1}{N} \frac{1}{(1 - R_1/N)(1-R_2/N)} \right] +  \\
& \,\,\,\,\,\,\,\,\, \mathcal{O}(\delta/N) + \mathcal{O}(\delta^3/N^2)
\end{aligned}
\end{equation}
Combining these together, we have
\begin{equation}
\begin{split}
\log P[p] \approx & -\log(p!) + p \log( \frac{R_1 R_2}{N} ) -  \\
& \frac{1}{2} \delta^2 \left[ \frac{1}{R_1} \frac{1}{1 - R_2/N}+ \frac{1}{R_2} \frac{1}{1 - R_1/N} + \frac{1}{N} \frac{1}{(1 - R_1/N)(1-R_2/N)} \right] + \mathcal{O}(\delta/R) + \mathcal{O}(\delta^3/R^2). \\
\end{split}
\end{equation}
Exponentiating, 
\begin{equation}
P[p] \approx \frac{1}{p!} \left(\frac{R_1 R_2}{N} \right)^p \exp\left( - \frac{1}{2} \left( p - \frac{R_1 R_2 }{N} \right)^2 \left[ \frac{R_1 R_2}{R_1+R_2} + \mathcal{O}(1/N) \right]^{-1} + \mathcal{O}(\delta/R) + \mathcal{O}(\delta^3/R^2) \right).
\end{equation}
The first two terms are precisely a Poisson distribution, which has mean $R_1 R_2 / N$ and width $\sqrt{R_1 R_2 / N}$.
The exponential is a Gaussian with the same mean $R_1 R_2 / N$, and a larger width $\sqrt{R_1 R_2 / (R_1 + R_2)}$.
The smaller width determines the width of the product of the two functions, so we conclude:
\begin{equation}
\expval{p} = \frac{R_1 R_2}{N}, \,\,\,\,\,\,\,\,\, \expval{p^2} - \expval{p}^2 \approx \frac{R_1 R_2}{N}.
\end{equation}
This is what we would expect for drawing $p$ random sites out of $N$, where each site has independent probability $R_i/N$ of being in either Pauli string (Section \ref{0D RUCs}).
The width is subextensive, $\delta \sim \varepsilon \sqrt{N}$, justifying the higher order terms we neglected along the way.

\subsection{Distribution of the $K$-size}

Here we are interested in the probability distribution of the $K$-size of a Pauli string of fixed total size $\Size$, with $K$ randomly chosen couplings.
Our results substantiate those obtained by intuitive arguments beneath Eq.~(\ref{eigV expansion}) in Section \ref{peaked sizes} in the main text.

This objective is in fact an identical problem to calculating the overlap: the $K$-size is the overlap of the $K$ coupled qubits with the $\Size$ qubits acted on by the operator of interest. We should just replace $R_1 \rightarrow K$, $R_2 \rightarrow \Size$, $p \rightarrow n$ above, where $n$ is the $K$-size.
This is confirmed by comparing the factorial expressions:
\begin{equation}
\begin{aligned}
P[n] & = \frac{ C^\Size_n C^{N-\Size}_{K-n}  }{ C^N_{K} } = \frac{1}{n!} \frac{\Size!}{(\Size-n)!} \frac{K!}{(K-n)!} \frac{(N-\Size)!(N-K)!}{N! (N-\Size-K+n)!}
\end{aligned}
\end{equation}
where the numerator computes the number of distinct configurations with $n$ qubits overlapping the Pauli operator support of size $\Size$ and $K-n$ qubits not overlapping, and the denominator computes the number of distinct configurations of the $K$ coupled qubits.
There are two regimes of interest: when $K$ and $\Size$ are both extensive, and when $\Size$ is extensive but $K$ is not.
The former provides a more accurate measure of the full operator size ($K \rightarrow N$), while the latter is relevant for probing the channel capacity.
Both regimes share the same mean $K$-size $\Size_K$ and $K$-size width $\delta \Size_K$:
\begin{equation}
\Size_K \equiv \expval{n} = \frac{\Size K}{N}, \,\,\,\,\,\,\,\,\, \delta \Size_K^2 \equiv \expval{n^2} - \expval{n}^2 \approx \frac{\Size K}{N} = \Size_K.
\end{equation}
This matches our prediction in Section \ref{peaked sizes}, which was based on a simple scenario of picking $K$ sites, each with a $\Size / N$ chance of being in the support of the Pauli operator.

\section{Teleportation of fermions}\label{app: fermions}

Here we generalize the teleportation protocol to Majorana fermion systems, as discussed in the main text for the SYK model. 
This involves a few small modifications, stemming from ($i$) a different definition of fermionic EPR (FEPR) pairs, and ($ii$) a different relation between FEPR pair projector and the SWAP gate.
These modifications explain the results of Ref.~\cite{gao2019traversable}, which find that late time teleportation in the SYK model occurs with less than unit fidelity even at infinite temperature (where we would generally expect perfect fidelity, from late time peaked-size teleportation [Section \ref{finite temperature}, VI]).
In particular, we find that the encoding procedure for the late time fermionic protocol must be modified for teleportation to succeed with perfect fidelity, due to modification ($ii$) above.

Consider two complex fermions $\chi_l$ and $\chi_r$, decomposed into pairs of Majorana fermions via $\chi_l = \psi_l^1 + i \psi_l^2$, $\chi_r = \psi_r^1 + i \psi_r^2$.
The number operators of the original fermions are Majorana bilinears, e.g.
$
i \psi_l^1 \psi_l^2 = 2 \hat{N}_l - 1 = (-1)^{\hat{N}_l}.
$
We define a single FEPR pair as the positive eigenstate of $i \psi_l^1 \psi_r^1$ and $i \psi_l^2 \psi_r^2$.
In the number operator basis of the original complex fermions, this is the maximally entangled state $(\ket{10} - i \ket{01})/\sqrt{2}$.
%
%
Multiple fermion EPR pairs are formed as a tensor product of single FEPR pairs.

This definition leads to some simple relations when `sliding' fermion operators around FEPR bras and kets in diagrammatic calculations.
We have:
\be \label{slide}
\bs
\psi_l^j \ket{\fepr} & = i \psi_r^j \ket{\fepr} \\
\bra{\fepr} \psi_l^j & = - i \bra{\fepr} \psi_r^j, \\
\end{split}
\ee
diagrammatically,
\begin{align}
\centering
\figbox{.45}{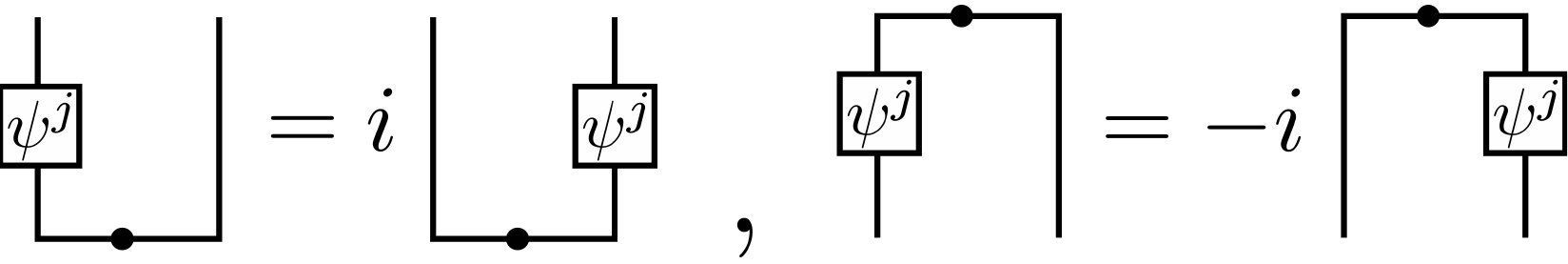} \centering \label{eq: fermion slide}
\end{align}
As in bosonic systems, the thermofield double state is obtained by applying $\rho^{1/2}$ to one side, $\ket{\tfd} = \rho_l^{1/2} \ket{\fepr}$.
Since the SYK Hamiltonian is composed of 4-fermion terms, we have
\be
\bs
H_l \ket{\tfd} = (i)^4 H_r \ket{\tfd} = H_r \ket{\tfd}.  \\
\end{split}
\ee
As in bosonic systems, the coupling for Majorana systems [Eq.~(\ref{V fermions})] measures the size of Majorana strings.

There are two options teleportation in fermionic system.
First, we could teleport an ordinary bosonic qubit by encoding it into Majorana fermion operators, for instance:
\be
\begin{split}
X &  \equiv i \psi_1 \psi_2 \\
Y & \equiv i \psi_2 \psi_3 \\
Z & \equiv i \psi_1 \psi_3. \\
\end{split}
\ee
At infinite temperature before coupling, each of the above operators has a correlator equal to $-1$,
which is exactly the result for bosonic systems, but without a need for the decoding operation $Y$.
At late times, the coupling $e^{igV}$ applies a relative phase between the identity and non-identity Paulis, giving correlator phases:
\begin{center}
\begin{tabular}{ |c|c| } 
\hline
$\mathbbm{1}$ & $e^{ig\expval{V}}$  \\ 
$i \psi_1 \psi_2$ & $-1$  \\ 
$i \psi_2 \psi_3$ & $-1$  \\ 
$i \psi_1 \psi_3$ & $-1$  \\ 
 \hline
\end{tabular}
\end{center}
When $g \expval{V} = \pi$ all correlators have the same phase, and peaked-size teleportation succeeds with perfect fidelity at infinite temperature.
At intermediate times, peaked-size teleportation of multiple bosonic qubits will succeed just as in bosonic systems.

The second option is to send a fermionic qubit, for instance by inserting half of an ancillary FEPR pair.
Here we begin with intermediate times, and discuss a modification necessary for late time teleportation afterwards.
We represent a single complex fermion with two Majorana operators $\psi_1,\psi_2$, and suppose that the operators' size distributions are tightly peaked, and the size of $i \psi_1 \psi_2$ is twice that of the individual Majorana sizes, denoted $\Size$ (this assumption of size addition is appropriate in all-to-all coupled systems, e.g. SYK, but would not necessarily hold for e.g. a 1D Majorana chain).
The relevant operator correlators after coupling with a bilinear fermionic interaction, as in Eq.~\eqref{V fermions}, are:
\begin{center}
\begin{tabular}{ |c|c| } 
\hline
$\mathbbm{1}$ & $1$  \\ 
$\psi_1$ & $-i \cdot e^{ i g \Size/qN} $  \\ 
$\psi_2$ & $-i \cdot  e^{ i g \Size/qN}$  \\ 
$i \psi_1 \psi_2$ & $-1 \cdot  e^{ i g 2 \Size/qN}$  \\ 
 \hline
\end{tabular}
\end{center}
At $g \Size/qN = \pi/2$ we have perfect teleportation.
This generalizes straightforwardly to multiple fermionic qubits: a $p$-fermion operator will gain a phase $i^p$ from sliding across the FEPR pair, and a phase $e^{ i g p \Size/qN}$ from coupling.

At late times, the sizes of initial single-body and two-body Majorana operators are equal, since they have saturated the size of the system, and the above operator correlators do not have the same phase.
We now show that an alteration of the encoding procedure can rectify this and lead to perfect late time teleportation.
This alteration is explained by the HPR protocol, and we derive it by reexamining the equivalence between the HPR and TW protocols in the case of fermionic qubits.
Here, the relevant difference between bosons and fermions is that the fermionic SWAP gate is \emph{not} related to the Grover search operation $1- 2P_{\fepr}$ by single-qubit rotations.
Since fermions gain a minus sign upon exchange, the fermionic SWAP gate takes the form
\be
\SWAP_F = 
\begin{pmatrix}
1 & 0 & 0 & 0 \\
0 & 0 & 1 & 0 \\
0 & 1 & 0 & 0 \\
0 & 0 & 0 & -1 \\
\end{pmatrix} =
\frac{ i \psi_{1,l} \psi_{2,l} + i \psi_{1,r} \psi_{2,r} +  i\psi_{1,l} \psi_{2,r}  - i\psi_{2,l} \psi_{1,r} }{2}.
\ee
This is a \emph{two-qubit} controlled-phase (CZ) gate away from $1- 2P_{\fepr}$:
\be \label{eq: CZ SWAP}
1 - 2 P_{\fepr} = 
\begin{pmatrix}
1 & 0 & 0 & 0 \\
0 & 0 & i & 0 \\
0 & -i & 0 & 0 \\
0 & 0 & 0 & 1 \\
\end{pmatrix}
= 
\frac{ 1 -  i \psi_{1,l} \psi_{1,r}  - i \psi_{2,l} \psi_{2,r} - (i \psi_{1,l} \psi_{1,r})(i \psi_{2,l} \psi_{2,r})}{2} =
 \SWAP_F \cdot \CZ,
\ee
where the CZ gate is defined as
\be
\begin{split}
\CZ & = 
\begin{pmatrix}
1 & 0 & 0 & 0 \\
0 & i & 0 & 0 \\
0 & 0 & -i & 0 \\
0 & 0 & 0 & -1 \\
\end{pmatrix}
=  (1 + i) \frac{ \psi_{1,l} \psi_{2,l} +   i \psi_{1,r} \psi_{2,r}}{2} \\ 
& = \exp( i \frac{\pi}{4}  ) \cdot \exp( -i \frac{\pi}{2} [ i \psi_{1,l} \psi_{2,l}] ) \cdot \exp( i \frac{\pi}{4} [ i \psi_{1,l} \psi_{2,l}] [ i \psi_{1,r} \psi_{2,r}]  ). \\
\end{split}
\ee
The single-fermion $\exp( -i \frac{\pi}{2} [ i \psi_{1,l} \psi_{2,l}] )$ gate occurs on the swapped-out fermion and may be neglected.
Inserting this in place of the second Grover search operation gives the appropriate teleportation protocol:
\begin{align}
\centering
\figbox{.6}{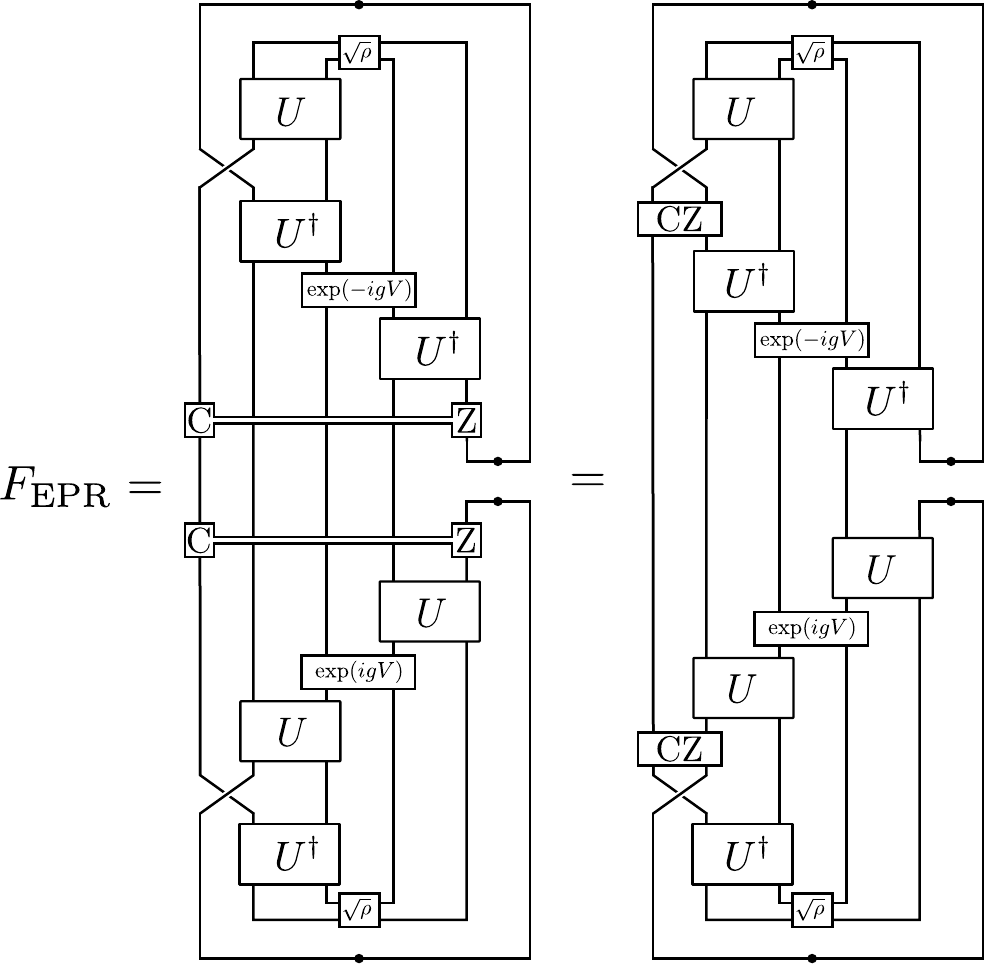} \centering \label{eq: state-fidelity-swap-EPR-fermions}
\end{align}
In the second diagram we have slid the action of each side of the CZ gate such that the gate acts at the same time and on the same qubits as the initial SWAP gate.
%
%

We can relate the fidelity of teleportation to operator correlators by decomposing the encoding gate as
\be
\CZ \cdot \SWAP_F = \frac{1}{2}\sum_{j=1}^4 S^L_{j,l} S^R_{j,r}
\ee
where we define the operators:
\begin{center}
\begin{tabular}{ |c|c|c|c| } 
\hline
$j$ & $S^L_{j}$ & $S^R_{j}$ & $S^R_{j} S^L_{j}$    \\ 
\hline
1 & $\mathbbm{1}$ & $\mathbbm{1}$ & $\mathbbm{1}$ \\ 
2 & $i \psi_1 \psi_2$ & $i \psi_1 \psi_2$ & $\mathbbm{1}$ \\ 
3 & $i \psi_{1}$ &  $\psi_{1}$ & $i \mathbbm{1}$ \\ 
4 & $i \psi_{2}$ &  $\psi_{2}$ & $i \mathbbm{1}$  \\ 
 \hline
\end{tabular}
\end{center}
according to Eq.~(\ref{eq: CZ SWAP}).
The final column displays the product $S^L_{j} S^R_{j}$, where both act on the same qubit, which will be useful shortly.
We find a fidelity:
\begin{align}
\centering
\figbox{.6}{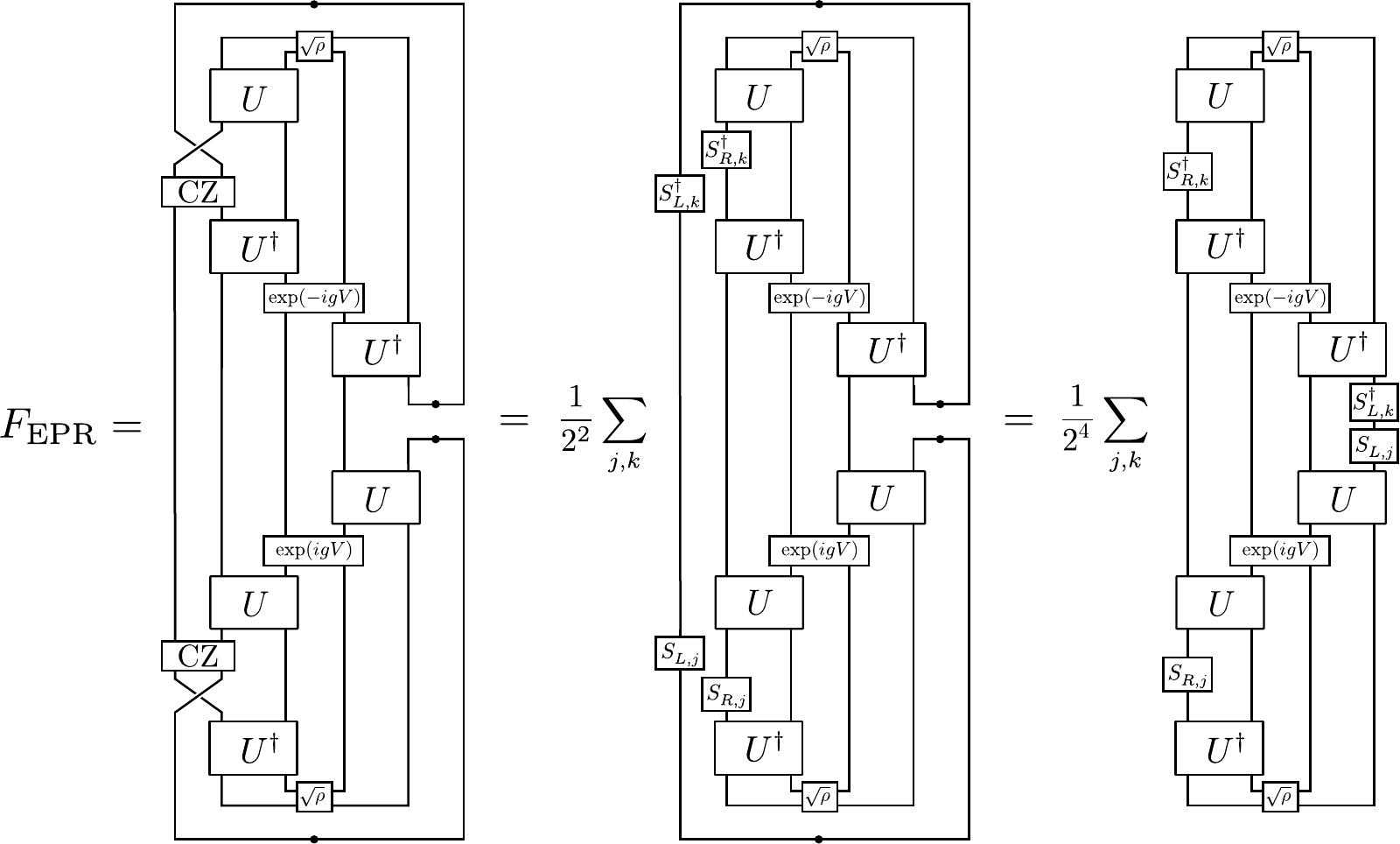} \centering \label{eq: state-fidelity-swap-EPR-fermions}
\end{align}
In the peaked-size regime with correlator phases $\theta_{R,j}$, we have
\be
\begin{split}
F_{\epr}  & = \frac{1}{2^4} \sum_{j,k} \bra{\tfd} S_{R,j,l}(t)\,e^{-igV} \, [S_{L,j,r} S_{L,k,r}^{\dagger} ](-t') \, e^{igV}\,S_{R,k,l}^{\dagger}(t) \ket{\tfd} \\
& = \frac{1}{2^4} \sum_{j,k} \exp( -i [ \theta_{R,j} - \theta_{R,k} ] ) \tr ( S_{R,j}(t-t') \, \rho^{1/2} \, S_{L,j}(0)  \, S_{L,k}^{\dagger}(0) \, \rho^{1/2} \, S_{R,k}^{\dagger}(t-t')  ) \\
\end{split}
\ee
At infinite temperature, late times, and $g \expval{V} = \pi$, we have correlator phases $\theta_{R,j} = 0$ for the identity and two-bosonic operator and $\theta_{R,j} = \pi/2$ for single-body fermionic operators, and find perfect teleportation fidelity:
\be
\begin{split}
F_{\epr} & = \frac{1}{2^4} \sum_{j,k} \exp( -i [ \theta_{R,j} - \theta_{R,k}  ] ) \tr ( S_{R,j} \, S_{L,j}  \, S_{L,k}^{\dagger} \, S_{R,k}^{\dagger}  ) \\
& = \frac{1}{2^4} \sum_{j,k} \exp( -i [ \theta_{R,j} - \theta_{R,k}  ] ) \cdot i^{F_j} \cdot  (-i)^{F_k} \cdot  \tr ( i \psi_1 \psi_2  \, i \psi_1 \psi_2  ) \\
& = \frac{1}{2^4} \sum_{j,k} \exp( -i [ \theta_{R,j} - \theta_{R,k}  ] ) \cdot i^{F_j} \cdot  (-i)^{F_k} \\
& = \frac{1}{2^4} \sum_{j,k} (-i)^{F_j} \cdot i^{F_k} \cdot i^{F_j} \cdot  (-i)^{F_k} \\
& = 1, \\
\end{split}
\ee
where we define $F_j = 1$ if $S_{L/R,j}$ is fermionic, and $0$ if bosonic.

We note that for state, as opposed to EPR, teleportation, the above CZ gate turns out not to be necessary. 
Since coherent superpositions of different fermion parity cannot be created by physical Hamiltonians, which contain only even combinations of fermionic operators, we should only consider teleporting states of definite fermion parity. 
The CZ gate applies only an overall phase on these states, and so does not affect the success of teleportation.

We can also briefly analyze the low temperature results of Ref.~\cite{gao2019traversable} through the lens of operator correlator phases.
Here, state teleportation is found to succeed perfectly at low temperatures only when the initial operators are encoded in $p$-body Majoranas, with $p = q/2 + 1$, despite the operator correlators having maximal magnitude for any value of $p$.
At the semiclassical gravity teleportation time, the correlators have phases:
\begin{center}
\begin{tabular}{ |c|c| } 
\hline
$\mathbbm{1}$ & $1$  \\ 
$\psi_1$ & $i^p (i)^{2p/q}$  \\ 
$\psi_2$ & $i^p (i)^{2p/q}$  \\ 
$i \psi_1 \psi_2$ & $(-1)^p (i)^{4p/q}$  \\ 
 \hline
\end{tabular}
\end{center}
For single-body Majoranas, $p=1$, the correlators clearly do not have the same phase---in fact, their phases are nearly identical to their phases at infinite temperature with no coupling---so state teleportation is not possible.
When $p = q/2 + 1$, in the large-$q$ limit, these phases are $1, \pm 1, \pm 1, 1$, respectively, where the sign is determined by whether $p = 1,3 \text{ mod } 4$. 
When the sign is odd, it can be corrected via the decoding operation $i \psi_1 \psi_2 = (-1)^N$, which applies a minus sign when conjugating fermionic operators.
Either case can therefore achieve perfect teleportation.

\section{Teleportation and inelastic scattering at infinite temperature}\label{app: inelastic}

In Section~\ref{sec:stringy}, we found that strong stringy corrections to a bulk theory of gravity led to peaked-size teleportation as well as a deeply inelastic scattering amplitude.
We will now demonstrate that these two phenomena---peaked-size teleportation and inelastic scattering---also coincide at infinite temperature, for arbitrary functional forms of the wavefunctions and scattering amplitudes.
As we argued before, for a UV complete theory of quantum gravity, strong stringy (and in general deep inelastic) effects are expected to dominate only at high temperatures, $\beta \rightarrow 0$. 

At infinite temperature, the form of the correlator is constrained by the equality 
\begin{equation}\label{correlator conjugate infinite temperature}
    C_\psi(t;g)^* = -C_\psi(t;-g).
\end{equation}
This implies that $C_\psi(t)$ can be written as a real function of $ig$ multiplied by the two-point function:
\be
C_\psi(t)=\langle \psi_l \psi_r \rangle e^{-F(ig,t)}.
\ee
When $g=0$, $C_\psi(t)$ is equal to $\langle \psi_l \psi_r \rangle$, implying
\be
F(ig)=ig f_1(t) + \mathcal{O}(g^2),
\ee
where $f_1(t)$ is a real function.
Therefore, at this order in $g$, the infinite temperature correlator is simply the two-point function multiplied by a pure phase, matching peaked-size teleportation [Eq. \eqref{C2}].

To justify that higher order terms in $g$ are subleading, we need an additional assumption: that the wavefunction of $\psi(t)$ has a saddle point at some momentum $k$. 
This is analogous to the boundary assumption that operator sizes are tightly peaked.
At early times, this saddle will not be significantly changed by the coupling, since the derivative of the scattering matrix with respect to $k$ will be suppressed by $G_N$, and at early times the time-dependence of the wavefunction will not be strong enough to compensate for this suppression (for example, in semiclassical AdS$_2$, we observed competition between $e^{2\pi t/\beta}$ and $1/G_N$). 
In such cases, it is easy to see that Eq.~\eqref{eq:77} becomes $\langle \psi_l \psi_r\rangle$ times a pure phase linear in $g$, with higher powers of $g$ suppressed by $G_N$.

Infinite temperature also implies purely inelastic scattering, i.e. the scattering amplitude, $e^{i\delta}=1-S(k,s)$, is automatically real.
To see this, we first rewrite the correlator in terms of the in-falling momentum operators, $\hat{P}$ and $\hat{K}$, for $\psi_1$ and $\psi(t)$ respectively.
For instance, for the former we have:
\be
\begin{split}
\Psi_{1,r}(s,0)\Psi_{1,l}^*(s,0)&= \langle \psi_{1,l}(0)|s\rangle \langle s|\psi_{1,r}(0)\rangle\\
&=\int \frac{da}{2\pi} \langle \psi_{1,l}(0) e^{-ia \hat{P}}\psi_{1,r}(0)\rangle e^{ias}.
\end{split}
\ee
As $\psi(t)$ and $\psi_1$ are in principle independent operators, we have $[\hat{K},\hat{P}]=0$. 
Using this, we can rewrite Eq.~\eqref{eq:77} as
\be
C_\psi(t) = \langle \psi_r(-t) \exp \left( -ig S(\hat{K},\hat{P}) i\psi_{1,l}\psi_{1,r} \right) \psi_l(t) \rangle.
\ee
Taking the complex conjugate gives
\be
\begin{split}
C_\psi(t)^* & = \langle \psi_l(t) \exp \left( ig S(\hat{K},\hat{P})^* (-i)\psi_{1,r} \psi_{1,l} \right) \psi_r(-t) \rangle \\
& =-\langle \psi_r(-t) \exp \left( ig S(\hat{K},\hat{P})^* i\psi_{1,l}\psi_{1,r} \right) \psi_l(t) \rangle
\end{split}
\ee
where we used the fact that $\hat{K},~\hat{P}$ are Hermitian and that at infinite temperature $\psi_l(t) \ket{\tfd} = \psi_r(-t) \ket{\tfd}$.
Combining this with Eq.~(\ref{correlator conjugate infinite temperature}) then enforces $S(\hat{K},\hat{P})^* = S(\hat{K},\hat{P})$, i.e. purely inelastic scattering.

\end{document}